%% file: mccauley_thesis.tex
\documentclass[12pt,a4paper]{book}   

\usepackage{fancyhdr}			
\usepackage{float}                                	
\usepackage{anyfontsize}
\usepackage[]{amsmath}
\usepackage{amssymb}
\usepackage{graphicx}
\usepackage{nccmath}
\usepackage{bm}
\usepackage{mathtools}
\usepackage{appendix}

\usepackage[authoryear]{natbib}	
\usepackage[colorlinks,citecolor=blue,linkcolor=blue,breaklinks,pagebackref]{hyperref}
\usepackage[all]{hypcap}
\usepackage{etoolbox}

\usepackage{breakurl}					
\usepackage{fixltx2e}
\usepackage[newlinetospace,small]{titlesec}
\usepackage{siunitx}
\usepackage{url}
\usepackage{ulem}
\usepackage{expl3}
\usepackage[version=4]{mhchem}
\usepackage{soul}
\usepackage{subfig}
\usepackage{makecell}
\usepackage{caption}
\usepackage{enumitem} 
\usepackage{booktabs}

\usepackage{gensymb}

\usepackage{mbenotes}

\usepackage{setspace}

\providecommand{\e}[1]{\ensuremath{\times 10^{#1}}}
\providecommand{\lsr}[1]{\textit{V}\tsb{LSR}}

\providecommand{\rsolar}[1]{#1 R\tsb{$\odot$}}
\providecommand{\kms}{km s\tsp{-1}}

\DeclareMathOperator*{\argminA}{arg\,min}

\makeatletter
\newcommand\tsb[1]{\@textsubscript{\selectfont#1}}
\def\@textsubscript#1{{\m@th\ensuremath{_{\mbox{\fontsize\sf@size\z@#1}}}}}
\newcommand\tsp[1]{\@textsuperscript{\selectfont#1}}
\def\@textsuperscript#1{{\m@th\ensuremath{^{\mbox{\fontsize\sf@size\z@#1}}}}}

\patchcmd{\NAT@citex}
  {\@citea\NAT@hyper@{%
     \NAT@nmfmt{\NAT@nm}%
     \hyper@natlinkbreak{\NAT@aysep\NAT@spacechar}{\@citeb\@extra@b@citeb}%
     \NAT@date}}
  {\@citea\NAT@nmfmt{\NAT@nm}%
   \NAT@aysep\NAT@spacechar\NAT@hyper@{\NAT@date}}{}{}

\patchcmd{\NAT@citex}
  {\@citea\NAT@hyper@{%
     \NAT@nmfmt{\NAT@nm}%
     \hyper@natlinkbreak{\NAT@spacechar\NAT@@open\if*#1*\else#1\NAT@spacechar\fi}%
       {\@citeb\@extra@b@citeb}%
     \NAT@date}}
  {\@citea\NAT@nmfmt{\NAT@nm}%
   \NAT@spacechar\NAT@@open\if*#1*\else#1\NAT@spacechar\fi\NAT@hyper@{\NAT@date}}
  {}{}

\providecommand\eg{\textit{e.g.}}

\titlespacing*{\chapter}{0pt}{0pt}{*4}

\begin{document} 

 \pagenumbering{alph}
 \pagestyle{plain}

 \belowpdfbookmark{Title}{title}			
 \input{front_matter/title}

 \frontmatter
\belowpdfbookmark{Dedication}{dedication}
  \input{front_matter/dedication.tex} \clearpage \thispagestyle{empty} \cleardoublepage
 \belowpdfbookmark{Declaration of originality}{original}
  \input{front_matter/original.tex} \clearpage \clearpage \thispagestyle{empty} \cleardoublepage
 \belowpdfbookmark{Included papers}{includes}
  \input{front_matter/includes.tex} \clearpage \thispagestyle{empty} \cleardoublepage
   \belowpdfbookmark{Additional publications}{other_papers}
  \input{front_matter/other_papers.tex} \clearpage \thispagestyle{empty} \cleardoublepage
 \belowpdfbookmark{Acknowledgements}{ack}
  \input{front_matter/acknowledgment.tex}  \clearpage \thispagestyle{empty} \cleardoublepage
 \belowpdfbookmark{Abstract}{abstract} 
  \input{front_matter/abstract.tex}  \clearpage \thispagestyle{empty} \cleardoublepage 

 \belowpdfbookmark{Contents}{contents}
  \tableofcontents \color{black} \clearpage \thispagestyle{empty} \cleardoublepage 
      
 \mainmatter
\fancyhead{}
  \pagestyle{fancy} 
\fancyhead[OR]{}\fancyhead[EL]{}   

   \input{chapter1/chapter1.tex}

   \input{chapter2/chapter2.tex}

  \input{chapter3/chapter3.tex}

   \input{chapter4/chapter4.tex}
   \input{chapter5/chapter5.tex}

   \input{bibliography/bibliography.tex}

\end{document}

%% file: front_matter/title.tex
\thispagestyle{empty}
\begin{center}
\vspace*{\stretch{1}}

\parbox{120mm}{\fontsize{30}{30}\bf\centering 

                Radio Burst and Circular Polarization Studies of the Solar Corona at Low Frequencies}


\fontsize{15}{20}
\vspace{\stretch{2}}
{\sl A  thesis submitted to fulfill requirements for the degree of}\\
{\sl Doctor of Philosophy}\\

\vspace{\stretch{1}}
{\sl by}\\

\vspace{\stretch{1}}
{\Large Patrick I. McCauley}\\

\vspace{\stretch{2}}
{\sl School of Physics}\\
{\sl Faculty of Science}\\
{\sl University of Sydney}\\
{\sl Australia}\\[3ex]
{\large October, $2019$}

\vspace*{\stretch{1}}
\end{center}

\clearpage \thispagestyle{empty} \cleardoublepage

%% file: front_matter/dedication.tex
\parbox{150mm}{\begin{center}
\textit{Dedicated to Julie. Thanks for everything.}
\end{center}}

%% file: front_matter/original.tex
\vspace*{20mm}
\noindent {\LARGE \bf Declaration of originality}
\vspace{10mm}

\noindent To the best of my knowledge, this thesis contains no copy or 
paraphrase of work published by another person, except where 
duly acknowledged in the text. This thesis contains no material 
which has previously been presented for a degree at the University of 
Sydney or any other university. I certify that the 
authorship attribution statements on the next page are correct.

\vspace{10mm}
\vspace{10mm}
\hspace{\fill}{\large Patrick I. McCauley}

\vspace{10mm}

\noindent As supervisor for the candidature upon which this thesis is based, I confirm that the originality and authorship attribution statements above and on the next page are correct.

\vspace{10mm}
  \vspace{5mm}
\hspace{\fill}{\large Iver H. Cairns}

%% file: front_matter/includes.tex
\vspace*{20mm}
\noindent {\LARGE \bf Included Papers and Attribution}
\vspace{10mm}

\noindent Chapters \ref{ch2}\,--\,\ref{ch4} are peer-reviewed journal articles for which I was primarily responsible. 
They are reproduced as published aside from minor typographical changes for consistency. 
In addition to general manuscript comments, the main contributions of my coauthors were as follows. 
Cairns: general supervision and guidance, advice on experimental design and interpretation. 
Morgan: assistance with data reduction and advice on instrumental effects. 
Gibson: assistance with observation\,--\,model comparisons. 
Harding: assistance with flux calibration. 
Lonsdale \& Oberoi: facilitation of data collection and advice on data reduction. 
White: advice on results interpretation. 
Mondal: independent data validation. 
Lenc: advice on polarimetry and instrumental effects. 


\vspace{4ex}\noindent
\parbox[t]{0.17\textwidth}{\large Chapter~\ref{ch2}}
\parbox[t]{0.81\textwidth}{
\textbf{Type III Solar Radio Burst Source Region Splitting due to a Quasi-separatrix Layer}\\
Patrick I. McCauley, Iver H. Cairns, John Morgan, Sarah E. Gibson, James C. Harding, Colin Lonsdale, and Divya Oberoi \\
Published in \textit{The Astrophysical Journal}, \href{http://ui.adsabs.harvard.edu/abs/2017ApJ...851..151M}{851:151 (2017)}}




\vspace{4ex}\noindent
\parbox[t]{0.17\textwidth}{\large Chapter~\ref{ch3}}
\parbox[t]{0.81\textwidth}{
\textbf{Densities Probed by Coronal Type III Radio Burst Imaging}\\
Patrick I. McCauley, Iver H. Cairns, and John Morgan \\
Published in \textit{Solar Physics},  \href{http://ui.adsabs.harvard.edu/abs/2018SoPh..293..132M}{293:132 (2018)}}\\




\vspace{4ex}\noindent
\parbox[t]{0.17\textwidth}{\large Chapter~\ref{ch4}}
\parbox[t]{0.81\textwidth}{
\textbf{The Low-Frequency Solar Corona in Circular Polarization}\\
Patrick I. McCauley, Iver H. Cairns, Stephen M. White, Surajit Mondal, Emil Lenc, John Morgan, and Divya Oberoi \\
Published in \textit{Solar Physics},  \href{https://ui.adsabs.harvard.edu/abs/2019SoPh..294..106M}{294:106 (2019)}}\\

\vspace{3ex}

\noindent(Blue and red text denote internal and external hyperlinks, respectively.)



%% file: front_matter/other_papers.tex
\newcommand{\pmc}{\textbf{McCauley, P.}}
\newcommand{\pimc}{\textbf{McCauley, P. I.}}
\newcommand{\mcite}[6]{#1, \textit{#3}. \textbf{#4}, \href{#2}{#5}, \textbf{#6}}

\vspace*{20mm}
\noindent {\LARGE \bf Other Works}
\vspace{10mm}

\noindent The following is a reverse-chronological list of other peer-reviewed articles, 
excluding conference proceedings, that I have authored or coauthored. 
Items 1\,--\,8 were published or submitted during my PhD candidature. 

\vspace{5mm}

{\footnotesize{}

\begin{enumerate}

\item{
\mcite
{Beardsley, A. P., et al. (57 coauthors, including \pimc)}
{https://ui.adsabs.harvard.edu/abs/2019arXiv191002895B}
{Science with the Murchison Widefield Array: Phase I Results and Phase II Opportunities}
{Publications of the Astronomical Society of Australia}
{\textit{in press}}
{2019}
}

\item{
\mcite
{Cairns, I. H., et al. (15 coauthors, including \pmc)}
{https://ui.adsabs.harvard.edu/abs/2019arXiv191003319C}
{Comprehensive Characterization of Solar Eruptions with Remote and In Situ Observations, and Modeling: The Major Solar Events on 4 November 2015}
{Solar Physics}
{\textit{in press}}
{2019}
}

\item{
\mcite
{Mohan, A., \pimc, Oberoi, D., \& Mastrano, A.}
{https://ui.adsabs.harvard.edu/abs/2019ApJ...883...45M}
{A Weak Coronal Heating Event Associated with Periodic Particle Acceleration Episodes}
{The Astrophysical Journal}
{883:45}
{2019}
}


\item{
\mcite
{Rahman, M., \pimc, \& Cairns, I. H.}
{http://ui.adsabs.harvard.edu/abs/2019SoPh..294....7R}
{On the Relative Brightness of Coronal Holes at Low Frequencies}
{Solar Physics}
{294:7}
{2018}
}


\item{
\mcite
{Kleint, L., Wheatland, M. S., Mastrano, A., \& \pimc}
{http://ui.adsabs.harvard.edu/abs/2018ApJ...865..146K}
{Nonlinear Force-Free Modeling of Magnetic Field Changes at the Photosphere and Chromosphere Due to a Flare}
{The Astrophysical Journal}
{865:146}
{2018}
}

\item{
\mcite
{Jones, G. H., et al. (19 coauthors, including \pmc)}
{http://ui.adsabs.harvard.edu/abs/2018SSRv..214...20J}
{The Science of Sungrazers, Sunskirters, and Other Near-Sun Comets}
{Space Science Reviews}
{214:20}
{2018}
}

\item{
\mcite
{Cairns, I. H., Lobzin, V. V., Donea, A., Tingay, S. J., \pimc, et al. (46 coauthors)}
{http://ui.adsabs.harvard.edu/abs/2018NatSR...8.1676C}
{Low Altitude Solar Magnetic Reconnection, Type III Solar Radio Bursts, and X-ray Emissions}
{Scientific Reports}
{8:1676}
{2018}
}


\item{
\mcite
{Lenc, E., et al. (28 coauthors, including \pimc)}
{http://ui.adsabs.harvard.edu/abs/2017PASA...34...40L}
{The challenges of low-frequency polarimetry: lessons from the Murchison Widefield Array}
{Publications of the Astronomical Society of Australia}
{34:e040}
{2017}
}

\item{
\mcite
{Janvier, M., Savcheva, A., Pariat, E., Tassev, S., Millholland, S., Bommier, V., \pmc, McKillop, S., \& Dougan, F.}
{http://ui.adsabs.harvard.edu/abs/2016arXiv160407241J}
{Evolution of Flare Ribbons, Electric Currents and Quasi-separatrix Layers During an X-class Flare}
{Astronomy and Astrophysics}
{591:A141}
{2016}
}

\item{
\mcite
{Savcheva, A., Pariat, E., McKillop, S., \pmc, Hanson, E., Su, Y., \& DeLuca, E. E.}
{http://ui.adsabs.harvard.edu/abs/2016ApJ...817...43S}
{The Relation between CME Topologies and Observed Flare Features II:
Dynamical Evolution}
{The Astrophysical Journal}
{817:43}
{2016}
}

\item{
\mcite
{Savcheva, A., Pariat, E., McKillop, S., \pmc, Hanson, E., Su, Y., Werner, E., \& DeLuca, E. E.}
{http://ui.adsabs.harvard.edu/abs/2015ApJ...810...96S}
{The Relation between CME Topologies and Observed Flare Features I:
Flare Ribbons}
{The Astrophysical Journal}
{810:96}
{2015}
}

\item{
\mcite
{Su, Y., van Ballegooijen, A. A., \pimc, Ji, H., Reeves, K. K., \& DeLuca, E. E.}
{http://ui.adsabs.harvard.edu/abs/2015ApJ...807..144S}
{Magnetic Structure and Dynamics of the Erupting Solar Polar Crown Prominence on 2012 March 12}
{The Astrophysical Journal}
{807:144}
{2015}
}

\item{
\mcite
{Reeves, K. K., \pimc, \& Tian, H.}
{http://ui.adsabs.harvard.edu/abs/2015ApJ...807....7R}
{Direct Observations of Magnetic Reconnection Outflow and CME Triggering in a Small Erupting Solar Prominence}
{The Astrophysical Journal}
{807:7}
{2015}
}

\item{
\mcite
{\pimc, Su, Y., Schanche, N., Evans, K. E., Su, C., McKillop, S., \& Reeves, K. K.}
{http://ui.adsabs.harvard.edu/abs/2015SoPh..290.1703M}
{Prominence and Filament Eruptions Observed by the Solar Dynamics Observatory: Statistical Properties, Kinematics, and Online Catalog}
{Solar Physics}
{290:1703}
{2015}
}

\item{
\mcite
{Tian, H., et al. (27 coauthors, including \pmc)}
{http://ui.adsabs.harvard.edu/abs/2014Sci...346A.315T}
{Prevalence of Micro-Jets from the Network Structures of the Solar Transition Region and Chromosphere}
{Science}
{346:6207}
{2014}
}

\item{
\mcite
{Raymond, J. C., \pimc, Cranmer, S., \& Downs, C.}
{http://ui.adsabs.harvard.edu/abs/2014ApJ...788..152R}
{The Solar Corona as Probed by Comet Lovejoy (C/2011 W3)}
{The Astrophysical Journal}
{788:152}
{2014}
}

\item{
\mcite
{Masson, S., \pmc, Golub, L., Reeves, K. K., \& DeLuca, E.}
{http://ui.adsabs.harvard.edu/abs/2014ApJ...787..145M}
{Dynamics of the Transition Corona}
{The Astrophysical Journal}
{787:145}
{2014}
}

\item{
\mcite
{Savcheva, A. S., McKillop, S. C., \pimc, Hanson, E. M., \& DeLuca, E. E.}
{http://ui.adsabs.harvard.edu/doi/10.1007/s11207-013-0469-3}
{New Sigmoid Catalog from Hinode and the Solar Dynamics Observatory: Statistical Properties and Evolutionary Histories}
{Solar Physics}
{289:3297}
{2014}
}

\item{
\mcite
{\pimc, Saar, S. H., Raymond, J. C., Ko, Y. -K., \& Saint-Hilaire, P.}
{http://ui.adsabs.harvard.edu/abs/2013ApJ...768..161M}
{Extreme-Ultraviolet and X-Ray Observations of Comet Lovejoy (C/2011) in the Lower Corona}
{The Astrophysical Journal}
{768:161}
{2013}
}

\item{
\mcite
{Cirtain, J. W., et al. (13 coauthors, including \pmc)}
{http://ui.adsabs.harvard.edu/abs/2013Natur.493..501C}
{Energy Release in the Solar Corona from Spatially Resolved Magnetic Braids}
{Nature}
{493:501}
{2013}
}

\item{
\mcite
{\pmc, Mangum, J. G., \& Wootten, A.}
{http://ui.adsabs.harvard.edu/abs/2011ApJ...742...58M}
{Formaldehyde Densitometry Of Galactic Star-Forming Regions Using The H2CO 3(12)-3(13) And 4(13)-4(14) Transitions}
{The Astrophysical Journal}
{742:58}
{2011}
}

\end{enumerate}

}

%% file: front_matter/acknowledgment.tex
\vspace*{20mm}
\noindent {\LARGE \bf Acknowledgements}
\vspace{10mm}

\noindent This thesis contains work done between 2016 and 2019, but 
I view it as the culmination of my scientific career to date. 
I have been very fortunate to interact with many wonderful people along the way, 
all of whom have helped me in large and small ways to achieve this goal. I will take this opportunity to acknowledge all of them, to the best of my memory, back to my earliest research experiences. 

First, I am most grateful for the consistent support and encouragement from my family and friends. Particularly from my partner, Julie, my parents, Ken and Carolyn, and my brothers, Nick and Petey. I love you all. 

Next, I am extremely grateful to all of my immediate supervisors. I first thank my thesis advisor at the University of Sydney, Iver Cairns. Iver consistently demonstrates great care and respect for all of his students. 
He was always supportive and kindly afforded me great freedom in my research direction. 
I next thank Kathy Reeves and the solar group at the Harvard-Smithsonian Center for Astrophysics. Kathy was a wonderful boss, and the whole group could not have been more supportive. They are responsible for my love of solar physics. I am also grateful to Jeff Mangum at the National Radio Astronomy Observatory. Jeff kindly brought me back to NRAO to finish a summer research project that I had failed to complete as an undergraduate. I would not have continued in research without that opportunity. Finally, I thank Harold Butner and William Alexander for supporting me through my first research and outreach experiences at James Madison University, which ultimately led to everything else. 

In addition to my supervisors, I have been fortunate to collaborate with and receive assistance from many excellent scientists over the years. In alphabetical order, I am grateful to Alisdair Davey, Ed DeLuca, Kaitlin Evans, Sarah Gibson, Leon Golub, Natasha Hurley-Walker, Kelly Korreck, Kamen Kozarev, Emil Lenc, Colin Lonsdale, Alpha Mastrano, Don Melrose, Atul Mohan, Surajit Mondal, John Morgan, Nick Murphy, Aimee Norton, Divya Oberoi, John Raymond, Steve Saar, Pascal Saint-Hilaire, Jon Sattelberger, Antonia Savcheva, Mike Stevens, Yingna Su, Bill Thompson, Hui Tian, Mark Weber, Mike Wheatland, Stephen White, and Trae Winter. 


I have also made many great friends in my scientific journey, including several people from the previous list. My favorite part of the PhD experience was unquestionably getting to know my wonderful officemates in Sydney: Samira Tasnim, James Harding, Ron Maj, Mozibur Rahman, Fiona Schleyer, Tom Hanly, and Will Trevett.
I will miss all of you. In Boston, my fellow support scientists Sean McKillop, Nicole Schanche, and Trish Jibben likewise made my time there infinitely more enjoyable. The same goes for my fellow summer students in 
Charlottesville: Katy Wyman, Missie Louie, Michael Lam, Jen Shitanishi, Francillia Samuel, and Brian Roper. And from the Astro Lab in Harrisonburg: Dan Simonson, Colin Wilson, Chris Wolfe, and Dillon Trewlany. 

Finally, I am very grateful to the Australian government for supporting this work through an Endeavour Postgraduate Scholarship.

%% file: front_matter/abstract.tex
\vspace*{20mm}
\noindent {\LARGE \bf Abstract}
\vspace{10mm}

\noindent 
This thesis presents low-frequency (80\,--\,240 MHz) radio observations of the solar corona 
using the \textit{Murchison Widefield Array} (MWA). 
It represents the first attempt to process large amounts of solar MWA data with 
supercomputing facilities.
This wavelength regime has been under-explored in recent years, and a number of 
discoveries are reported. 
A brief review of the solar corona and associated observations is followed by three 
research chapters that focus on Type III solar radio bursts and circularly-polarized emission 
from the quiescent corona. 
Finally, conclusions are presented with an eye toward future work, including a discussion of preliminary 
results that compare the observed polarization structure to model predictions and that 
report novel coronal mass ejection (CME) observations. 

The first research chapter details new dynamics in a particular set of Type III bursts. 
The source region for each burst splits from one dominant component at higher frequencies 
into two increasingly-separated components at lower frequencies. 
For channels below $\sim$132 MHz, the two components repetitively diverge at high speeds (0.1\,--\,0.4 c) 
along directions tangent to the limb, with each episode lasting just  $\sim$2 s. 
Both effects are argued to result from the strong magnetic field connectivity gradient that 
the burst-driving electron beams move into, which is supported by extreme ultraviolet (EUV) jet
observations that outline characteristic magnetic field structures associated with coronal null points. 
Electrons are accelerated along neighboring field lines that are immediately adjacent in the 
flare site but diverge with height, causing the beams to reach the requisite height to produce 
radio emission at slightly different times. 
This produces an apparent motion that is nearly perpendicular 
to that of the electron beams themselves.
A method for flux calibration is also developed, the structure of 
the quiescent corona is compared to model predictions, and a 
coronal hole is reported to transition from being relatively dark at 
higher frequencies to relatively bright at lower frequencies. 

The second chapter uses Type III bursts observed at the limb to probe the coronal density structure. 
Assuming harmonic plasma emission, they imply 
2.4\,--\,5.4$\times$ enhancements over canonical background levels.
High densities inferred from Type III source heights can be explained by assuming that the exciting 
electron beams travel along overdense fibers or by radio propagation effects that may cause a source 
to appear at a larger height than the true emission site. 
The arguments for both scenarios are reviewed in light of recent results. 
A comparison of the extent of the quiescent corona versus model predictions 
is then used to conclude that propagation effects can largely but not entirely explain the 
apparent density enhancements for these events. 

The third chapter presents the first spectropolarimetric imaging of the quiescent corona at these frequencies, 
including a survey of circular polarization features detected in over 100 observing runs near solar maximum. 
Around 700 compact polarized sources are detected with polarization fractions ranging 
from less than 0.5\% to nearly 100\%.
They are interpreted as a continuum of plasma emission noise storm sources 
down to intensities and polarization fractions that were not previously observable. 
A characteristic ``bullseye" structure is observed for many low-latitude coronal holes in
which a central polarized component is surrounded by a ring of the opposite sense. 
The central component does not match the sign expected from thermal bremsstrahlung emission, 
which may be due to propagation effects or an alternative emission mechanism. 
The large-scale polarimetric structure at the lowest frequencies is shown to be reasonably well-correlated 
with the line-of-sight (LOS) magnetic field component inferred from a global potential field model, with 
the boundaries between opposite circular polarization signs being generally aligned with polarity inversion 
lines in the model. 
This is not true at the highest frequencies, however, where the LOS magnetic field direction and 
polarization sign are often not straightforwardly correlated.

The last chapter summarizes conclusions from the previous chapters and 
outlines future work on a number of open questions. 
These include steps toward a general understanding of Type III burst source motions, 
explaining the peculiar low-frequency signatures of coronal holes, 
using low-frequency spectropolarimetry to constrain global magnetic field models, 
and exploring the behavior of CMEs in low-frequency observations. 
Preliminary results are shared that compare observations to 
model predictions of circularly-polarized bremsstrahlung emission, yielding
good agreement in both the qualitative structure and quantitative polarization fractions. 
Radio CME observations are also presented, revealing an intense arc of emission that 
is morphologically similar and aligned to the CME front seen in white light, which 
has not been observed before.

%% file: chapter1/chapter1.tex
\pagestyle{fancy}

\chapter{Introduction and Literature Review} \label{ch1}

\providecommand{\alert}[1]{\textbf{\color{red}{#1}}}

\fancyhead[OR]{~}\fancyhead[EL]{\bf Ch. \thechapter~Introduction} 

\section{The Solar Corona}
\label{sec:intro}

\begin{figure}\graphicspath{{chapter1/}}
    \centering
    \includegraphics[width = 1.0\textwidth]{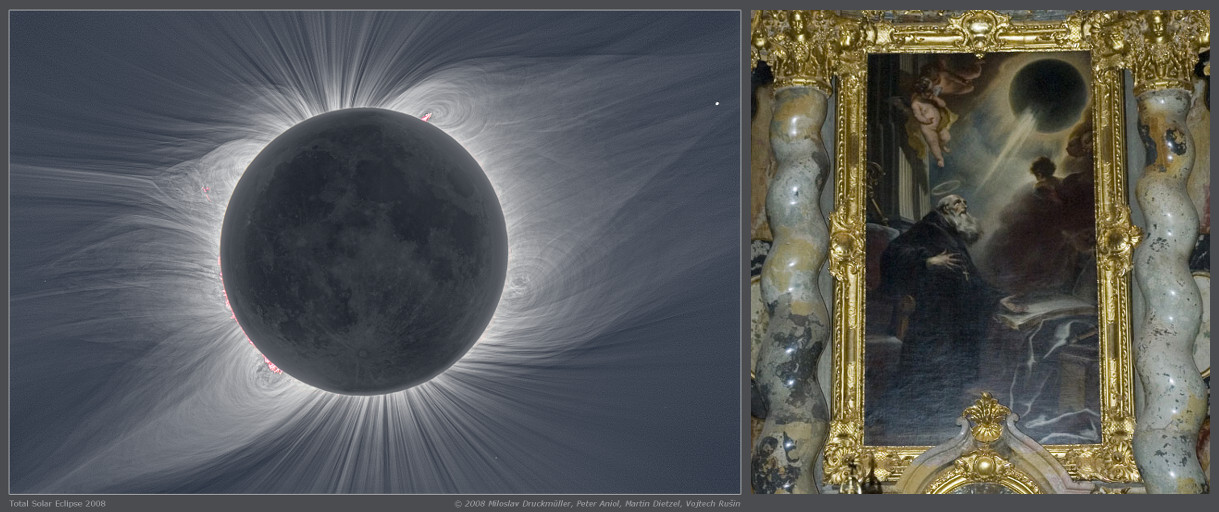}
    \caption{\footnotesize{}A modern image from the 2008 total solar eclipse (\textit{left}) alongside 
    a 1735 eclipse painting by Cosmas Damian Asam (\textit{right}). 
    The modern image is highly processed to accentuate the fine ray and 
    loop structures, still best-observed during eclipses, and the 
    painting is arguably the earliest surviving ``accurate" representation of the corona. 
    Image credits: 
    Miloslav Druckm{\"u}ller, Peter Aniol, Martin Dietzel, and Vojtech Ru\v{s}in 
    (\href{http://www.zam.fme.vutbr.cz/~druck/eclipse/Ecl2008m/Tse2008_1250_mo1/0-info.htm}{\textit{left}}); 
    Jay Pasachoff 
    (\href{https://apod.nasa.gov/apod/ap080128.html}{\textit{right}}).}
    \label{fig:eclipse}
\end{figure}

The \textit{solar corona} is the outer atmosphere of the Sun, which begins thousands of kilometers above the surface 
and extends to the outer solar system where particles from the Sun meet the interstellar medium. 
It can be observed by the naked eye only for brief periods during total solar eclipses because it is over a million times 
fainter than the Sun's apparent visible surface, and the technology required to reveal 
the corona's highly dynamic nature has been developed only fairly recently. 
Figure~\ref{fig:eclipse} shows a modern eclipse image alongside what is arguably  
the corona's earliest surviving accurate depiction. 
The first definitive written reference to the corona was by the famed astronomer Johannes Kepler in his 1604 book titled 
\textit{Astronomiae Pars Optica} (``The Optical Part of Astronomy"), although plausible references exist thousands 
of years earlier in ancient Babylonian, Greek, and Chinese texts \citep{Golub10}. 
Kepler believed the corona to be a feature of the Moon, and others would later attribute it to both the Sun 
and effects related to Earth's atmosphere. 
Two technologies of the 1800s, photography and spectroscopy, would ultimately prove the 
corona to be part of the Sun. 

Photographs of the same eclipse from multiple sites demonstrated that the structure of the corona 
does not vary between terrestrial viewing locations, implying that it cannot be an atmospheric effect \citep{Rue64}, 
and images of consecutive eclipses suggested that variation in 
the corona's appearance from year to year is tied to the sunspot cycle \citep{Darwin89}.
In the late 1800s, the first spectroscopic measurements of the Sun were made, resulting in the discovery 
of helium \citep{Lockyer20} and the identification of the characteristic solar Fraunhofer lines in the coronal spectrum \citep{Janssen73}. 
This latter discovery demonstrated the presence of reflected sunlight and further suggested that the corona is part of the Sun. 
Emission from helium comes from the Sun's lower atmosphere, but early 
spectroscopic observations of the corona also revealed puzzling new spectral lines \citep{Young95}. 
Like helium, these were not consistent with any known element, leading astronomers to again posit the 
existence of a new element, this time called ``coronium." 
That hypothesis was short-lived, as laboratory experiments in the 1930s would soon demonstrate that the 
coronium lines actually come from highly-ionized forms of known elements such as iron and calcium \citep{Edlen45}.
This presented a new mystery in that the temperatures 
required to produce these high ionization states are in excess of one million Kelvin, 
much hotter than the Sun's surface. 
``The coronal heating problem" remains one of the longest-standing mysteries in astrophysics and space physics \citep{Klimchuk06,DeMoortel15}. 

\begin{figure}\graphicspath{{chapter1/}}
    \centering
    \includegraphics[width = 0.7\textwidth]{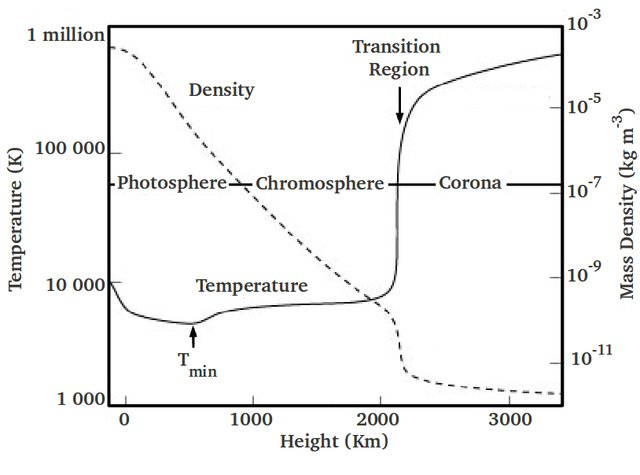}
    \caption{\footnotesize{}The mean variation in temperature and density as a function of height above the 
    photosphere based on the popular VAL model \citep{Vernazza73,Vernazza76,Vernazza81}. Image credit: \citet{Lang01}.}
    \label{fig:val}
\end{figure}

Temperature is now generally seen as the primary feature that delineates the Sun's outer layers \citep{Aschwanden05,Golub10,Priest14}. 
Light generated in the Sun's interior escapes through a thin layer called the \textit{photosphere}, commonly referred to as the visible surface, 
which is composed primarily of neutral gas with a temperature of around 6,000 K.
Above the photosphere is the \textit{chromosphere}, which is partially ionized with temperatures upwards of 20,000\,--\,50,000 K. 
Then, over a few hundred kilometers known as the \textit{transition region}, the temperature 
increases dramatically to beyond 1,000,000 K in the fully-ionized corona. 
The height of the transition region varies with location but typically 
occurrs between 2,000 and 3,000 km above the photosphere.  
A schematic of the temperature and density profile from the photosphere into the corona is 
shown in Figure~\ref{fig:val}. 

From its base, the corona expands outward to form the \textit{solar wind}, a variable but persistent flow 
of plasma that creates a cavity in the interstellar medium known as the \textit{heliosphere}. 
As the boundary between the corona and the solar wind is not well-defined, a natural definition 
for the outermost edge of the corona is simply the \textit{heliopause}, where the pressure from the solar wind is 
balanced by that of interstellar gas \citep{Holzer89,Golub10}. 
However, most researchers use the term \textit{corona} to refer to structures within less than roughly 5\,--\,10 
solar radii (\rsolar{}) and the terms \textit{solar wind} or \textit{interplanetary medium} 
to refer to larger heliocentric distances. 
The observations presented in this thesis will primarily be confined to $\lesssim$ \rsolar{2} from Sun-center. 

This chapter reviews some basic concepts and provides context for the research chapters to follow. 
Section~\ref{observationsi} discusses the different types of observations used to study the corona, and 
Section~\ref{mechanisms} reviews the known solar radio emission mechanisms.
Section~\ref{activity} outlines the basic forms of solar activity, and 
Section~\ref{bursts} describes the different types of solar radio bursts. 
Section~\ref{interferometry} introduces the concepts of interferometry and aperture synthesis, 
along with describing the radio interferometer used in this thesis. 
Finally, Section~\ref{aims} describes the main research aims and outlines the subsequent chapters. 
Note that the references given throughout this chapter are not meant to be exhaustive; early and 
highly-cited papers, along with review articles and textbooks, are emphasized. 

\section{Observing the Corona}
\label{observationsi}

Prior to the 1930s, the corona could be observed only when the photosphere was completely occulted by the Moon during a 
total solar eclipse. 
These events occur just once every 18 months on average and are visible only from particular locations on Earth that change for 
each eclipse, posing obvious barriers to observation. 
Eclipses continue to be important to coronal research to this day (\eg~\citealp{Phillips00,Habbal13}), but a number of advances over the last century, 
beginning with the invention of the coronagraph, have made routine observations of the corona possible. 
This section discusses the main observation types 
and associated emission mechanisms considered in this thesis, 
along with a few additional notes on 
other relevant observations. 

\subsection{White Light Observations}

Visible light from the corona is divided into three main components, each of which arise 
from different mechanisms \citep{Rusin00,Golub10}.
The K- (kontinuierlich) corona has the continuous emission spectrum 
of thermal blackbody radiation from the photosphere that has been 
Thomson scattered by coronal electrons, which produces a large degree of polarization 
that can be used to estimate the coronal electron density \citep{vandeHulst50,Hayes01}. 
The inner corona seen in Figure~\ref{fig:eclipse} is dominated by this source. 
Beyond around 2 solar radii, the F- (Fraunhofer) corona begins to exceed the brightness of the K-corona. 
Like the K-corona, the F-corona is also produced by scattering of photospheric light, but instead by 
dust particles in the ecliptic plane. 
The F-corona is therefore not directly-related to the corona as we understand it today and may 
instead be considered inner zodiacal light.  
The E- (emission) corona is comprised of isolated spectral lines emitted by ions in the high-temperature 
coronal plasma, making it the only component that is produced directly by the corona itself. 
Although much fainter in integrated light than both the K- and F-coronas, 
the E-corona can be observed using narrowband filters around the spectral line of interest. 

Observing coronal white light outside of an eclipse requires the use of a coronagraph. 
A coronagraph is a telescope that incorporates an occulting disk to block light coming from the photosphere,  
creating an artificial eclipse. 
While this is conceptually straightforward, it is very difficult in practice to block enough photospheric light 
from reaching the detector because of scattering within the telescope after light strikes the occulting disk. 
Bernard Lyot was the first to succeed in sufficiently limiting internal scattering, which 
revolutionized the study of the corona \citep{Lyot39}. 
Coronagraph observations on Earth are limited by the natural atmospheric scattering of 
sunlight and can be made only from high-altitude observatories. 
This limitation can be overcome by placing the telescope in space, which was first done successfully in 1963 
using a sounding rocket and later made routine with satellite observations in 1971 \citep{Koutchmy88}.
One of the most prolific modern coronagraphs is 
the \textit{Large Angle and Spectrometric Coronagraph} (LASCO; \citealp{Brueckner95}) 
onboard the space-based \textit{Solar and Heliospheric Observatory} (SOHO; \citealp{Domingo95}). 
LASCO data is shown in Section~\ref{cmes} and will be used in Chapter~\ref{ch3}. 

\subsection{Radio Observations}

Coronagraph observations are limited in that they cannot observe the corona directly above the photosphere. 
Fortunately, as a consequence of its high temperature, the corona produces 
significant emission at radio, extreme ultraviolet (EUV), and X-ray wavelengths. 
Unlike visible light, this radiation far exceeds that produced by the photosphere, 
allowing the corona to be observed directly without requiring an eclipse or occulting disk. 

Solar radio emission was first reported in the scientific literature by \citet{Reber44}, which 
corresponded to high-frequency microwave emission from the chromosphere. 
Intense low-frequency (metric) radio burst emission from the corona was actually detected 
two years earlier by British radar operators but was not publicly reported 
until after World War II \citep{Hey46}. 
One of the early achievements of solar radio astronomy was to independently verify 
the high temperature of the corona, which was identified previously through optical 
spectroscopy but remained controversial. 
Beginning with wavelengths below $\approx$ 1 cm, 
the quiescent solar radio emission starts to exceed that of the Sun's blackbody spectrum \citep{Pawsey46}. 
\citet{Ginzburg46} showed that this excess radiation is due to 
thermal bremsstrahlung emission from the hot coronal plasma. 
This mechanism, along with the others known to generate solar radio emission, 
will be reviewed in Section~\ref{mechanisms}. 

\begin{figure}\graphicspath{{chapter1/}}
    \centering
    \includegraphics[width = 1.0\textwidth]{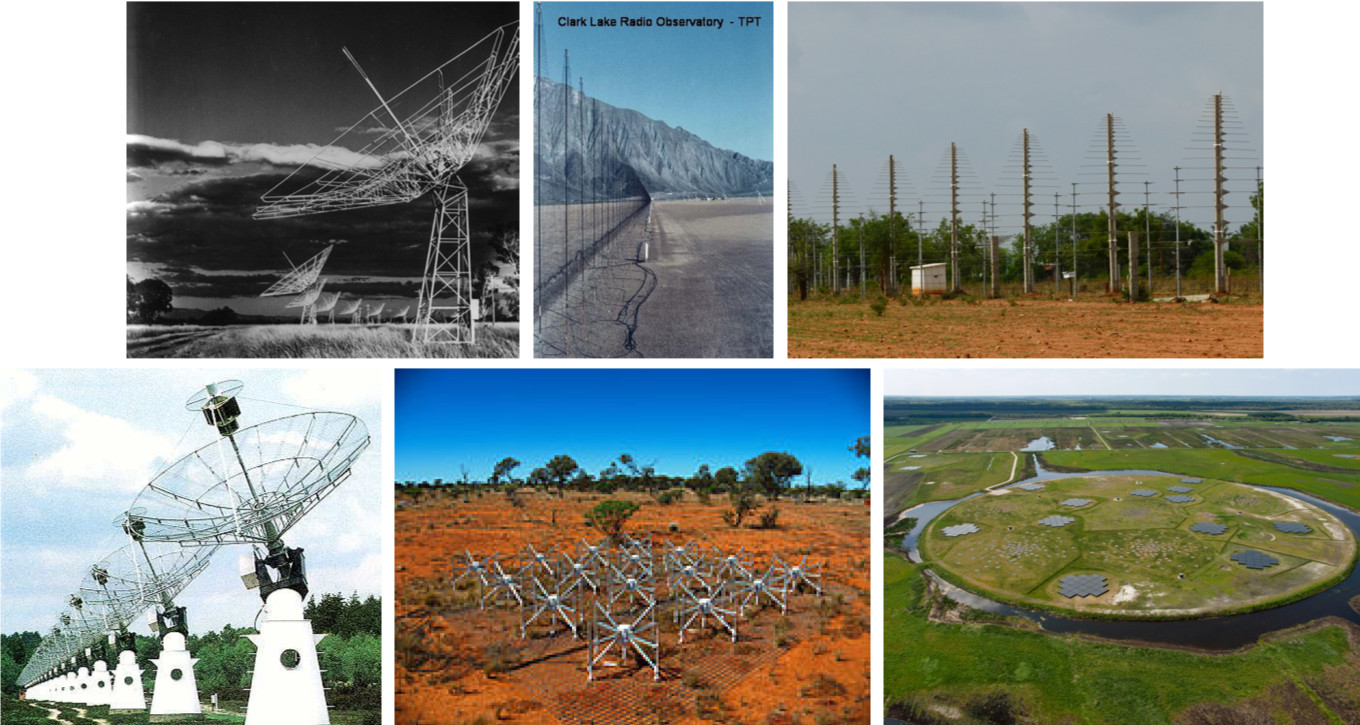}
    \caption{\footnotesize{}Antennas from different low-frequency ($\lesssim$ 300 MHz) radio imaging arrays, past and present. 
    From left to right, top to bottom: the \textit{Culgoora Radioheliograph}, \textit{Clark Lake Radioheliograph}, 
    \textit{Guaribidanur Radioheliograph}, \textit{Nan\c{c}ay Radioheliograph}, 
    \textit{Murchison Widefield Array}, and \textit{Low Frequency Array}. }
    \label{fig:arrays}
\end{figure}

Prior to 1950, most instruments observed at a single frequency \citep{Pick08}.  
Simultaneous observations of radio bursts at several frequencies showed 
that the onset times varied, suggesting that the bursts were related 
to disturbances that propagate outward through the corona to excite emission 
at different frequencies as they move through plasma of different densities \citep{Payne47}.
This motivated the development of solar \textit{radiospectrographs} that could 
continuously observe the Sun over a range of frequencies.
These instruments produce \textit{dynamic spectra} that show the solar emission 
as a function of both time and frequency.
Much of the nomenclature the characterizes the Sun's behavior at radio wavelengths 
is based on the appearance of features in dynamic spectra, such as the 
classification of solar radio bursts by \citet{Wild50}. 
Examples of dynamic spectra will be shown with the discussion of radio 
bursts in Section~\ref{bursts}.  

Radiospectrograph data are limited in that they cannot spatially localize 
or track features. 
Generating a radio image of the Sun with adequate time resolution requires 
an interferometer.
\textit{Interferometry} combines the signals received from multiple 
antennas and will be described in Section~\ref{interferometry}. 
In the 1950s, a number of simple interferometers with small numbers of elements  
provided limited tracking of radio bursts, but 
routine imaging of the corona began with the 
\textit{Culgoora Radioheliograph} (43\,--\,327 MHz; \citealp{Wild70,Sheridan72,Sheridan83}).
\textit{Radioheliographs} are interferometers that are dedicated to solar observing, 
although instruments primarily used for other astrophysical sources 
may also target the Sun.  
In addition to improved tracking of radio bursts, these instruments allow the 
quiescent structure to be observed and compared to that at other 
wavelengths. 

In recent decades, two of the most notable radio imagers dedicated to solar physics have been 
the \textit{Nan\c{c}ay Radioheliograph} (150\,--\,450 MHz; \citealp{Kerdraon97}) 
and the \textit{Nobeyama Radioheliograph} (17\,--\,34 GHz; \citealp{Nakajima94}). 
General astrophysical instruments like the \textit{Very Large Array} (VLA; currently 1\,--\,50 GHz; \citealp{Perley11})
and the \textit{Atacama Large Millimeter Array} (ALMA; 84\,--\,950 GHz; \citealp{Wedemeyer16}) 
have also made important solar physics contributions. 
Since the decommissioning in the 1980s of the \textit{Culgoora Radioheliograph}
and later the \textit{Clark Lake Radioheliograph} (20\,--\,125 MHz; \citealp{Kundu83}), 
there has been little investment in new low-frequency ($\lesssim$ 300 MHz) imaging instrumentation. 
A notable exception to this is the \textit{Guaribidanur Radioheliograph}, which has been 
operating since 1997 at 80 MHz \citep{Ramesh98,Ramesh05}.

In 2012, two new low-frequency interferometers were commissioned, 
the \textit{Murchison Widefield Array} (MWA; 80\,--\,300 MHz; \citealp{Tingay13})
and the \textit{Low Frequency Array} (LOFAR; 10\,--\,240 MHz; \citealp{van13}). 
While they are not dedicated solar telescopes, these instruments can  
produce the most sensitive images of the corona to date and represent a 
significant advance over previous observational capabilities, with improvements in 
sensitivity, simultaneous frequency coverage, and both 
spatial and time resolution. 
This thesis presents results from the MWA, which will be discussed further 
in Section~\ref{mwai}, and
Figure~\ref{fig:arrays} shows images of antennas from the low-frequency 
arrays listed above.
Two additional instruments for which solar studies are planned are 
the \textit{Long Wavelength Array} (LWA; 10\,--\,88 MHz; \citealp{Ellingson09}) and the 
\textit{Owens Valley Long Wavelength Array} (OVRO-LWA; 24\,--\,82 MHz).

\subsection{Extreme Ultraviolet and Soft X-Ray Observations}

Due to its high temperature, the corona's radiation energy comes primarily at 
extreme ultraviolet (EUV) and soft X-ray (SXR) wavelengths \citep{Golub10}. 
Although the Sun and its corona are composed almost entirely of hydrogen 
and helium, these elements do not contribute significantly to coronal emission  
because they are completely ionized at coronal temperatures. 
Instead, the corona radiates energy predominantly through isolated 
spectral lines of trace heavy elements such as carbon, oxygen, silicon, and iron. 
These elements are stripped of their outer shell electrons but still 
retain some or many of their inner shell electrons. 
Several effects, primarily collisions with free electrons, may then either 
excite one of the electrons to a higher energy level or remove it from the atom, 
producing a higher ionization state. 
When an electron recombines with an ion or a bound electron decays to a 
lower energy state, a photon is released to balance the energy lost by the atom in 
that transition.  
This process predominantly generates EUV and SXR photons because 
of the ionization states typical of the corona.  

\begin{figure}\graphicspath{{chapter1/}}
    \centering
    \hspace{-0.8cm}
    \includegraphics[width = 1.0\textwidth]{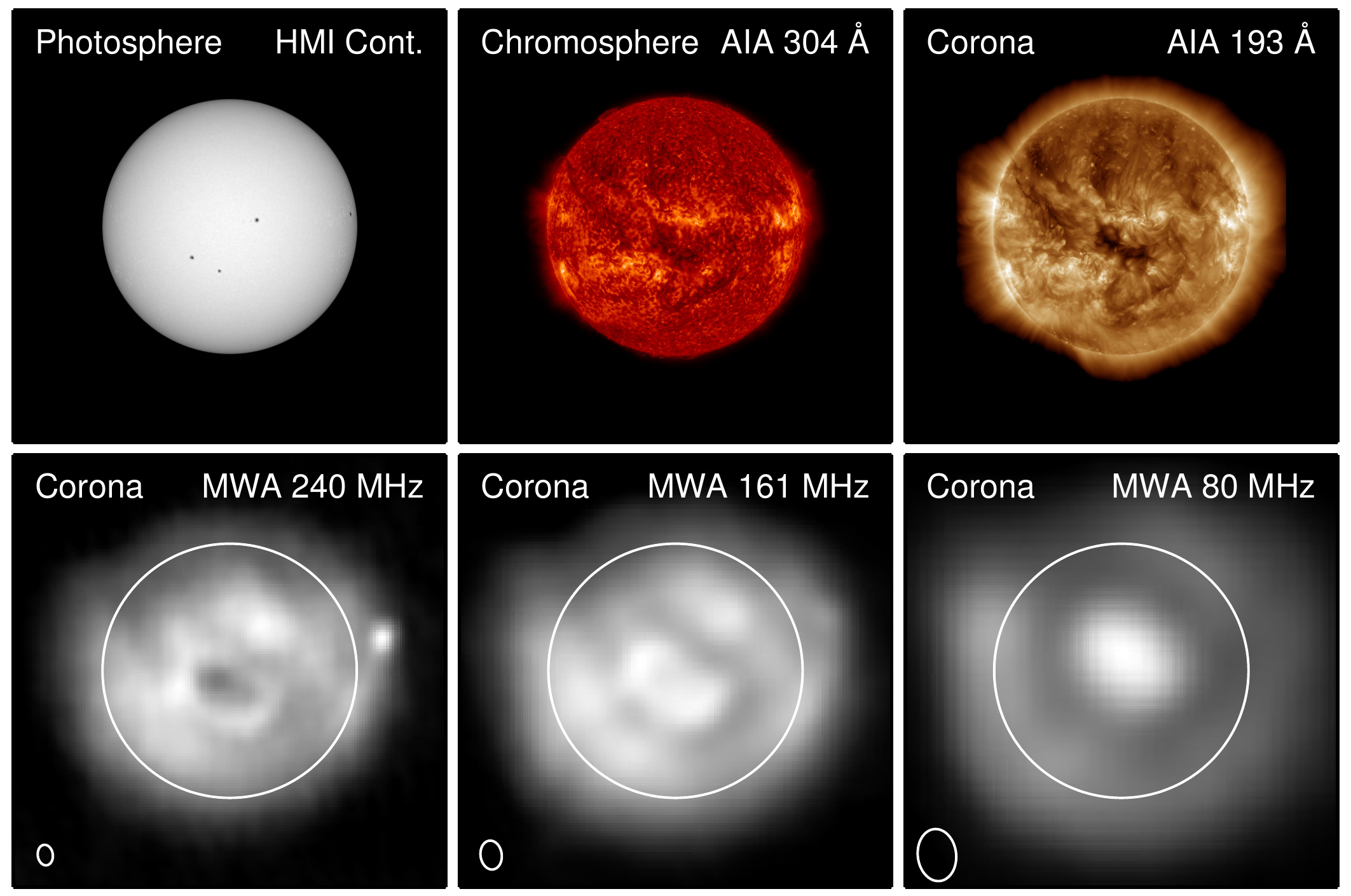}
    \caption{\footnotesize{}\textit{Left-to-right, upper row}: A white light observation of the 
    photosphere, an EUV observation of the chromosphere, and an EUV 
    observation of the corona. 
    \textit{Lower row}: Radio observations of the corona at three  
    frequencies; lower frequencies correspond to larger heights above 
    the photosphere. 
    The upper and lower rows come from the SDO
    and MWA, respectively.}
    \label{fig:obs_compare}
\end{figure}

Observations at these wavelengths may either be narrowband, targeting 
just one or more of these spectral line transitions, or broadband, incorporating a 
range of spectral lines. 
EUV and X-ray emission is blocked by Earth's atmosphere, meaning that the observations 
must be conducted from space and thus began after ground-based white-light and radio observations. 
Attempts to detect X-rays from space began with the U.S. rocketry program after WWII, and 
techniques for producing high-resolution images developed substantially in the 1960s \citep{Vaiana73}. 
A number of satellite missions culminated in the EUV and X-ray instrumentation flown on the \textit{Skylab} space 
station \citep{Vaiana73b,Tousey73b}, 
which is arguably the most productive space-based solar physics mission to date \citep{Golub10}. 

Many of the terms used to describe the general features of the solar corona, which 
will be described in Section~\ref{features}, were developed based on \textit{Skylab} 
observations and those that immediately preceded it. 
Since \textit{Skylab}, there have been several EUV and SXR telescopes, such as those on  
the \textit{Yohkoh} \citep{Ogawara91}, \textit{Hinode} \citep{Kosugi07}, 
SOHO \citep{Domingo95}, STEREO \citep{Kaiser08}, and PROBA-2 \citep{Berghmans06} satellites. 
The most widely-used instrument of this type today is the \textit{Atmospheric Imaging Assembly} (AIA; \citealp{Lemen12}) 
onboard the \textit{Solar Dynamics Observatory} (SDO; \citealp{Scherrer12}), which will feature in each of the 
research chapters. 
EUV and radio images of the corona are compared in Figure~\ref{fig:obs_compare}, along with 
corresponding images of the photosphere and chromosphere. 

\subsection{High-Energy, \textit{In situ}, and Related Observations}

The observations from the previous subsections are the most relevant to this thesis, but 
a number of other types are also important to coronal studies. 
While the temperatures in the corona are not high enough to produce 
appreciable hard X-ray (HXR) radiation during quiescent periods, 
intense solar flares can produce significant high energy emission up to 
and including gamma rays \citep{Ramaty75}. 
Observations at these wavelengths have been critical to understanding 
the physical processes of solar flares and their impacts. 
The most significant recent mission in this regime was the 
\textit{Reuven Ramaty High Energy Solar Spectroscopic Imager} (RHESSI; \citealp{Lin02}), 
which was launched in 2002 and decommissioned in 2018. 

In addition to remote sensing, coronal studies benefit from \textit{in situ} 
observations, which refer to those from instruments that are 
immersed in the interplanetary medium and can directly probe the 
plasma of the solar wind. 
Several instruments have been placed near Earth at the L1 Lagrangian point, 
where satellites may orbit the Sun at a fixed location with respect to Earth. 
These instruments provide routine measurements of Earth's ``space weather" environment. 
Examples include SOHO, 
the \textit{Advanced Composition Explorer} (ACE; \citealp{Stone98}), and the \textit{Wind} spacecraft \citep{Ogilvie97}. 
Similar measurements have been made elsewhere in the heliosphere, 
notably by the \textit{Helios} \citep{Schwenn75} and \textit{Ulysses} \citep{Bame92} missions near the Sun 
and the \textit{Voyager} spacecraft in the outer solar system \citep{Bridge77}. 
\textit{In situ} observations have contributed many important results, but one of 
particular interest here was the direct detection of the electrons and plasma oscillations that 
were theorized to produce Type III solar radio bursts \citep{Frank72,Lin73,Gurnett76}, which will be 
described in Section~\ref{typeiii}.

\begin{figure}\graphicspath{{chapter1/}}
    \centering
    \includegraphics[width = 1.0\textwidth]{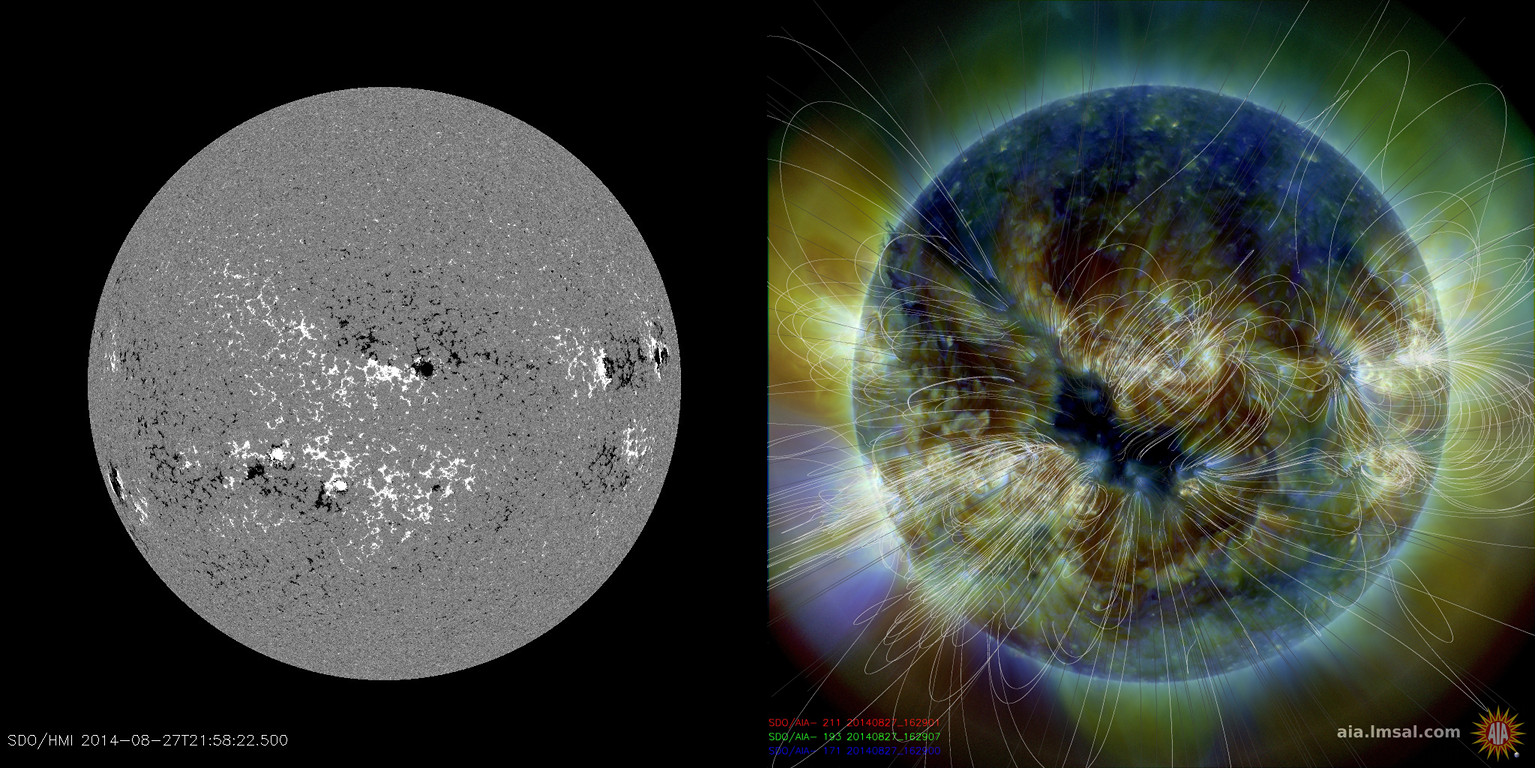}
    \caption{\footnotesize{}Magnetogram observation (\textit{left}) showing the photospheric magnetic field 
    strength next to an EUV observation (\textit{right}) of the corona overlaid with 
    magnetic field lines extrapolated from magnetograms.  
    The EUV image is a red-green-blue composite of three spectral channels, the time period 
    shown is the same as in Figure~\ref{fig:obs_compare}, and the observations come from the SDO. 
    Image credit: Lockheed Martin Solar and Astrophysics Laboratory (\href{aia.lmsal.com}{LMSAL}).}
    \label{fig:magnetogram}
\end{figure}

Finally, while they will not be reviewed here, observations of other parts of the 
Sun are of course also important to understanding the corona. 
Of particular importance to this thesis are observations of the photospheric magnetic field. 
The photospheric field can be measured using the Zeeman effect \citep{Hale08,Babcock53}, which is the 
splitting of spectral lines in the presence of a magnetic field, and such 
observations are commonly-referred to as \textit{magnetograms}. 
Today, the most wildely-used magnetograms are from the SOHO's \textit{Michelson Doppler Imager} (MDI; \citealp{Scherrer95})
and the SDO's \textit{Helioseismic and Magnetic Imager} (HMI; \citealp{Scherrer12}).
Magnetograms are extremely important for coronal studies because they can 
be used to estimate the coronal magnetic field by extrapolation from the photosphere, 
and the coronal field is what largely determines both the structure of 
and activity within the corona. 
Magnetogram data is presented directly in Chapter~\ref{ch4} and indirectly through 
models in Chapters~\ref{ch2} and \ref{ch3}.

Extrapolations from the photospheric field are needed because the coronal magnetic field 
is very difficult to measure directly. 
Coronal magnetic field measurements are possible from near-infrared and optical 
coronagraph observations of the Zeeman and Hanle effects, 
the latter of which refers to polarization state changes 
due to the presence of a magnetic field, and a number of advances 
have been made in recent years on this topic (\eg~\citealp{Tomczyk08,Kramar16}). 
Polarimetric radio observations may also be used to measure the coronal magnetic field at 
specific locations (\eg~\citealp{Dulk78,White97}), 
but it has not yet been possible to probe the global coronal field structure 
with radio observations because of limited instrument sensitivity. 
Chapter~\ref{ch4} \citep{McCauley19} will present the first circular polarization images of the quiescent corona at 
low frequencies, which will be used to probe the magnetic field in future studies. 

\section{Solar Radio Emission}
\label{mechanisms}

The Sun produces radio emission largely by converting the energy of moving electrons into radiation 
via several mechanisms that operate in different contexts with varying levels of complexity. 
\textit{Incoherent emission} refers to the summation of radiation
generated independently by many individual particles, which may be 
accelerated by Coulomb collisions, as in bremsstrahlung emission, or by gyration in a magnetic field, 
as in gyromagnetic emission. 
\textit{Coherent emission} refers to mechanisms that convert electron energy into radiation more efficiently 
due to electrons emitting in phase or 
through the development of instabilities that amplify particular wave modes. 
The coherent mechanisms relevant for solar physics are plasma emission and electron-cyclotron maser emission, 
both of which require particles to be accelerated by energetic events like solar flares. 

\begin{figure}\graphicspath{{chapter1/}}
    \centering
    \includegraphics[width = 0.8\textwidth]{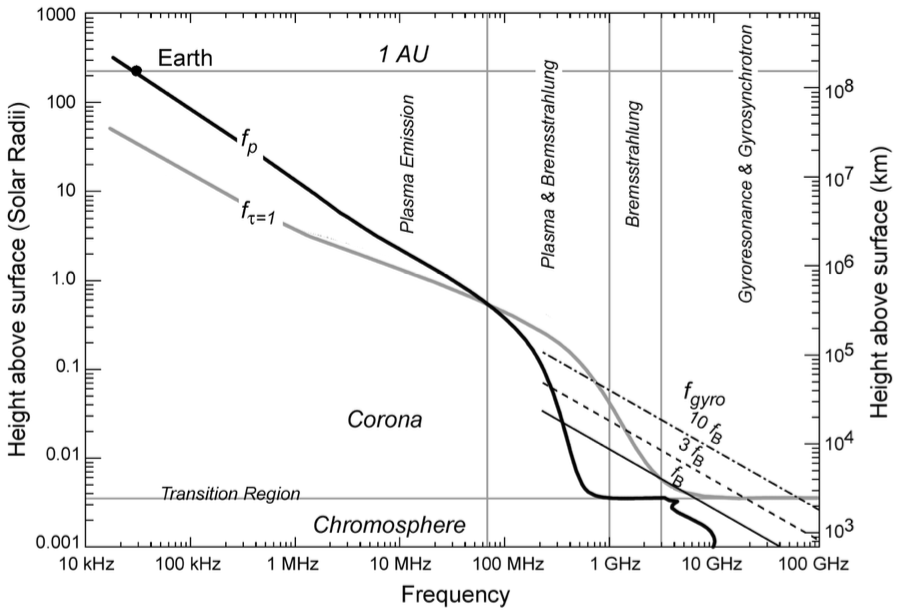}
    \caption{\footnotesize{}The solar atmosphere's characteristic frequencies based 
    on standard models; low heights 
    correspond to active regions. 
    The uppermost curve at any location indicates the dominant emission mechanism based on 
    the relative sizes of $f_{\rm p}$, $f_B$, 
    and the frequency at which bremsstrahlung emission reaches an optical depth of one [$f_{\tau=1}$].
    Image credit: \citet{Gary05}}
    \label{fig:emission_mechanisms}
\end{figure}

Incoherent and coherent mechanisms are often distinguishable by their 
brightness temperatures [$T_{\rm B}$], a common measure of intensity in radio astronomy.
A \textit{brightness temperature} is the temperature that a blackbody would need to have 
to reproduce an observed intensity from a source of a particular size and at a particular frequency. 
A \textit{blackbody} is a theoretical construct that exists in thermal equilibrium at a single temperature, 
absorbs all incident radiation perfectly, and emits radiation as a function only of its temperature 
in accordance with Planck's Law. 
For incoherent emission, the brightness temperature is equal to or less than the actual source 
temperature (e.g. the electron temperature), 
depending primarily on the density and temperature of emitting material along the line of sight. 
Coherent emission can have brightness temperatures that far exceed the 
source temperature, implying that the emission mechanism must be nonthermal. 
However, coherent mechanisms may also produce weak signals that are not 
distinguishable from incoherent emission using brightness temperature alone. 

The emission frequency for each mechanism is determined by the plasma's 
characteristic frequencies, which depend on parameters such as density and magnetic field strength. 
Two of the most important quantities are the \textit{electron plasma frequency},

\begin{equation}
\label{eq:fpi}
f_{\rm p} = \sqrt{\frac{e^2n_{\mathrm e}}{\pi{}m_{\mathrm e}}} \approx 0.009\sqrt{n_{\mathrm e}}~\rm{MHz}, 
\end{equation}

\noindent and the \textit{electron gyrofrequency}, also called the \textit{electron-cyclotron frequency},

\begin{equation}
\label{eq:fbi}
f_{B} = \frac{eB}{2\pi{}m_{\rm e}c} \approx 2.8B~\rm{MHz}, 
\end{equation}

\noindent where $n_{\rm e}$ is the electron density in cm\tsp{-3}, 
$B$ is the magnetic field strength in G, 
$e$ is the electron charge, 
$m_{\rm e}$ is the electron mass, 
and $c$ is the speed of light. 
The relative sizes of these two frequencies largely determines the dominant 
emission mechanism in a particular environment. 
Figure~\ref{fig:emission_mechanisms} shows the 
characteristic frequencies of the solar atmosphere and their associated 
emission mechanisms, which will be described in the following 
subsections. 

\subsection{The Magnetoionic Modes}
\label{magnetoionic}

Before describing the emission mechanisms, it is important to first introduce  
the magnetoionic theory that is commonly used to describe 
the propagation of electromagnetic waves in an ionized 
medium under the presence of an external magnetic field. 
More detailed reviews of the following discussion can be found in several 
textbooks (\textit{e.g.} \citealp{Ginzburg70,Melrose86,Aschwanden05,Koskinen11}).
The solar corona is most often treated with the cold plasma approach, which assumes 
that the characteristic velocities of the waves are much faster than the thermal 
velocity of the plasma particles, allowing thermal effects to be neglected. 
This approach also generally ignores the motions of ions and assumes that the 
particles do not interact through collisions. 

\begin{figure}\graphicspath{{chapter1/}}
    \centering
    \includegraphics[width = 1.0\textwidth]{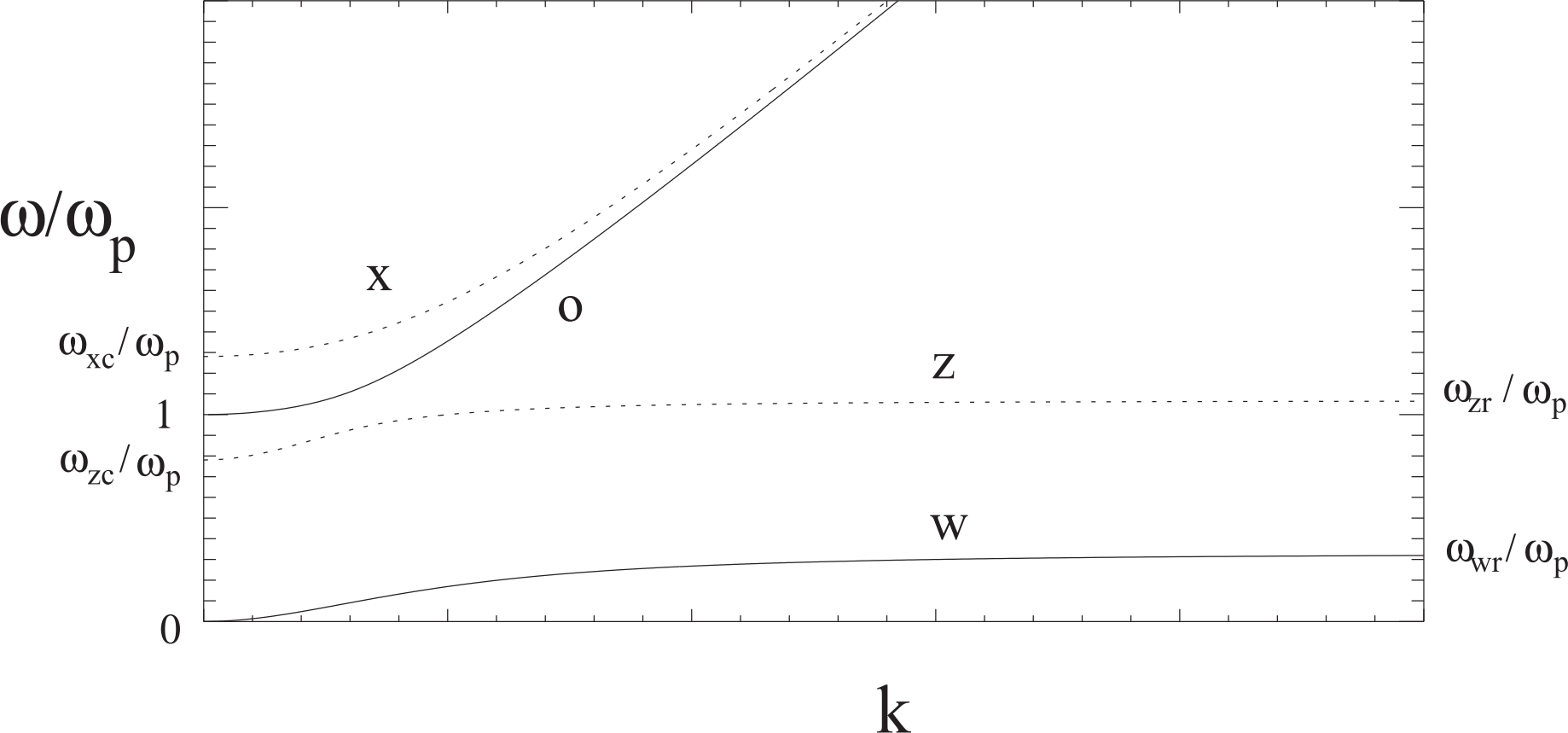}
    \caption{\footnotesize{}Dispersion relations for the magnetoionic modes in a magnetized plasma for which 
    the electron cyclotron frequency [$\Omega{}_{\rm e}$] is smaller than the 
    plasma frequency [$\omega{}_{\rm p}$]. 
    The two free-space (electromagnetic radiation) modes are at the top, and 
    $\omega{}_{x\rm{c}}$ refers to the $x$-mode cutoff, the frequency above which $x$-mode 
    radiation begins. 
    This plot uses angular unit notation as opposed to the 
    $f_{B}$ and $f_{\rm p}$ in the text 
    for $\Omega{}_{\rm e}$ and $\omega{}_{\rm p}$, respectively.      
    Image credit: Iver Cairns}
    \label{fig:magnetoionic}
\end{figure}

Under these approximations, the dispersion equation for electromagnetic waves has 
the four solutions shown in Figure~\ref{fig:magnetoionic} for the case where $f_{\rm p} > f_B$. 
The two uppermost curves correspond to the two modes that can escape the plasma 
as radiation (radio waves).  
These are referred to as the \textit{ordinary} [$o$] and \textit{extraordinary} [$x$] modes. 
The ordinary mode is so-named because the plasma response is the same as 
if there were no magnetic field, while the $x$-mode has a somewhat different refractive index. 
Each mode is polarized in opposite senses that depend on the angle with respect 
to the magnetic field. 
In most cases, a so-called quasi-circular approximation applies \citep{Melrose86}, and 
the two modes are 100\% circularly-polarized with opposite senses. 
For the $x$-mode, the electric field vector of the wave rotates in the same direction 
as the gyromotion of electrons around the magnetic field, whereas an $o$-mode wave's
electric field vector rotates in the opposite direction. 

A net circular polarization arises when the two modes are received with unequal intensities, which is 
characterized by the degree of circular polarization relative to the total intensity. 
The polarization degree depends on the emission mechanism, plasma parameters, 
and several effects that may modulate the polarization state during propagation to the observer. 
These dependencies make the degree of polarization a powerful diagnostic, and 
Chapter~\ref{ch4} \citep{McCauley19} will present the first circular polarization observations 
of the low-frequency corona that are sensitive enough to detect the weak polarization signals of 
thermal bremsstrahlung emission and very weak plasma emission sources. 

Circularly-polarized radio emission is most prevalent in the corona, but there are 
circumstances and processes that may produce linear polarizations.
However, linearly-polarized radiation propagating through a magnetized plasma experiences 
Faraday rotation that, during propagation, rotates the polarization plane as a function 
of frequency and magnetic field strength. 
The magnetic field strength in the corona is large enough that over a typical observing bandwidth, 
Faraday rotation will produce many turns of the electric field vector, thereby washing out 
any linear polarization signal (\textit{e.g.} \citealp{Bastian10,Gibson16}).
Linear polarizations have not been reported from the radio Sun except for a few cases at  
GHz frequencies \citep{Alissandrakis94,Segre01}.
It is possible, however, to observe linearly-polarized background astrophysical sources 
that are occulted and Faraday-rotated by the corona, which can be used as a magnetic 
field diagnostic at large heliocentric distances \citep{Spangler07,Ingleby07,Ord07}.

\subsection{Bremsstrahlung (Free-Free) Emission}

Bremsstrahlung emission, from the German ``braking radiation," refers to electromagnetic waves produced 
by the acceleration of charged particles, which converts some of the particles' kinetic energy into radiation. 
The term \textit{bremsstrahlung} was introduced in 1909 in reference to emission generated by electrons in 
cathode ray tube experiments, and a historical review of the concept is given by \citet{Wheaton09}.
Of primary importance are free electrons that are deflected by the Coulomb fields of ions. 
In a fully-ionized medium like the corona, this is often referred to as \textit{free-free} emission because 
it does not involve particles transitioning between bound states in an atom. 
\textit{Thermal bremsstrahlung} refers to radiation produced by a plasma in thermal equilibrium, and 
this is the dominant source of quiescent emission from the corona at low frequencies \citep{Ginzburg46,Dulk85,Aschwanden05}. 

The emission frequency is tied to the plasma's electron density through the electron plasma
frequency [$f_{\rm p}$] from Equation~\ref{eq:fpi}. 
Only emission at frequencies at or below $f_{\rm p}$ can be produced by a plasma with the corresponding density.  
This limit corresponds to the region below $\omega{}_{\rm p}$ in Figure~\ref{fig:magnetoionic} 
and may be understood in terms of the refractive index, which is imaginary for frequencies 
smaller than the plasma frequency, indicating that those waves cannot propagate in the medium \citep{Melrose86,Aschwanden05}.
The density of the corona generally decreases with height above the photosphere, meaning that 
lower frequency emission corresponds to larger heights and that the corona appears larger with 
decreasing frequency. 
This effect is illustrated by the increasing height of the $f_{\rm p}$ curve with decreasing frequency in 
Figure~\ref{fig:emission_mechanisms}. 
Very dense coronal structures may generate bremsstrahlung emission with frequencies into the GHz range, 
but canonical background coronal density models correspond to frequencies bellow $\approx$ 300 MHz (\textit{e.g.} \citealp{Newkirk61,Saito77}).
 
Early radio astronomers quickly recognized that the solar brightness temperature at wavelengths longer 
than $\approx$ 1 cm is significantly greater than a blackbody with the Sun's surface temperature of $\approx$ 5,800 K \citep{Appleton45,Martyn46,Pawsey46}. 
\citet{Ginzburg46} showed that this excess could be explained by thermal bremsstrahlung emission 
from a much hotter corona, the existence of which remained controversial after such high temperatures 
were first identified using optical spectroscopy. 
The physics of bremsstrahlung emission in the solar context has since been reviewed by many authors 
(\textit{e.g.} \citealp{Dulk85,McLean85,Aschwanden05}).

\subsection{Gyromagnetic Emission}
\label{gyro}

Like bremsstrahlung, gyromagnetic emission converts the kinetic energy of charged particles, 
mainly electrons, into radiation. 
In this case, the presence of a magnetic field produces a spiral 
gyromotion in a particle's trajectory along a particular magnetic field line, resulting in a centripetal acceleration \citep{Dulk85}.
Different terminology is used for the same basic phenomenon depending on the 
particle's rotation speed about the magnetic field. 
\textit{Gyroresonance} emission refers to non-relativistic speeds and is sometimes also called \textit{cyclotron} 
or \textit{magneto-bremsstrahlung} emission. 
\textit{Gyrosynchrotron} refers to mildly relativistic speeds, where the particles rotate at a small but significant 
fraction of the speed of light. 
\textit{Synchrotron} emission refers to the relativistic case where the speeds approach that of light. 

For solar radio emission, we are mainly concerned with gyroresonance and gyrosynchrotron emission, 
though synchrotron emission may be important in certain contexts (\textit{e.g.} \citealp{Winske83,Bastian07}). 
In each case, emission occurs near the electron gyrofrequency [$f_B$] from Equation~\ref{eq:fbi} 
or one of its harmonics, which 
depend primarily on the magnetic field strength, divided by the Lorentz factor [$\gamma$].
In the low-frequency observations discussed in this thesis, gyromagnetic emission does not 
contribute significantly. 
However, in dense regions of the corona where the magnetic field strength is strong, higher-frequency 
observations of gyroresonance emission can be used to measure the magnetic field strength \citep{Akhmedov82,White97,White05}. 
Gyroresonance emission also dominates over bremsstrahlung throughout the chromosphere, 
where the densities and magnetic field strengths are higher than in the 
corona such that $f_B > f_{\rm p}$ (see Figure~\ref{fig:emission_mechanisms}).
Gyrosynchrotron emission is also the accepted mechanism for certain microwave radio bursts from the chromosphere 
and is thought to contribute significantly to specific energetic events in the corona,  
namely Type IV bursts and coronal mass ejections (\textit{e.g.} \citealp{Melrose80,Alissandrakis86,Nindos08}). 

\subsection{Plasma Emission}
\label{plasma}

\begin{figure}\graphicspath{{chapter1/}}
    \centering
    \includegraphics[width = 0.8\textwidth]{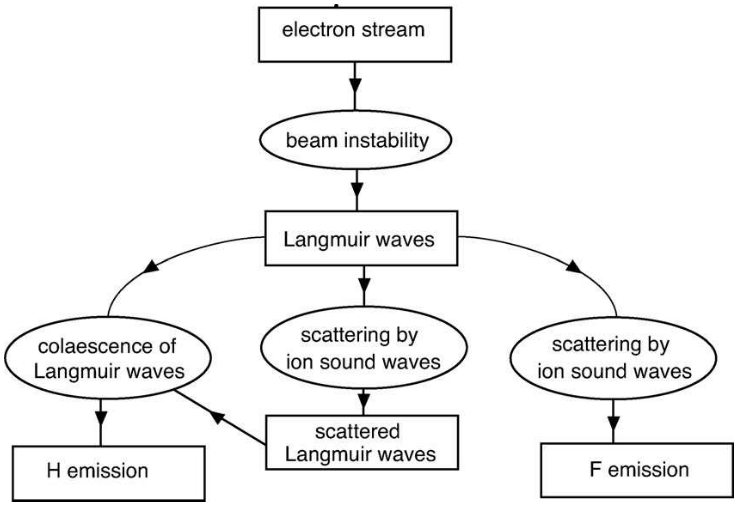}
    \caption{\footnotesize{}A flow chart outlining the stages of plasma emission, which is 
    responsible for most types of solar radio bursts. 
    Image credit: \citet{Melrose09}}
    \label{fig:plasma_emission}
\end{figure}

The most common form of coherent radio emission from the Sun is 
\textit{plasma emission}, which refers to a set of related processes 
that partially convert the energy of Langmuir waves into radiation \citep{Melrose09}. 
A flowchart of the basic plasma emission stages is 
shown in Figure~\ref{fig:plasma_emission}. 
\textit{Langmuir waves}, also referred to as \textit{electron plasma waves} 
or simply \textit{plasma oscillations}, 
are oscillations of a plasma's electron density \citep{Tonks29}. 
They occur when a plasma is perturbed such that an electron population 
is displaced with respect to the ions, and the Coulomb force then pulls  
the electrons back, leading them to oscillate back and forth. 

The process that produces Langmuir waves in the solar corona 
is generally assumed to be 
an instability driven by a beam of nonthermal electrons that move through 
the background plasma after being accelerated by magnetic reconnection 
or a shock wave \citep{Robinson00}.
Langmuir waves are produced by an electron beam through the \textit{two-stream instability}, 
which is often referred to as the \textit{bump-on-tail instability} 
in cases where an electron stream is injected into a plasma, creating a 
``bump" on the high-energy tail of the plasma's particle velocity distribution. 
This bump produces a region of wavevector space in which the number of electrons transferring 
energy to Langmuir waves exceeds the reverse case, leading to exponential 
Langmuir wave growth. 
A small fraction of the energy in the Langmuir waves can then be converted 
into electromagnetic radiation through interactions with other wave 
modes, namely ion sound waves \citep{Cairns87b,Cairns87c,Robinson00,Melrose09}. 
Depending on the wave interactions  
outlined by the flowchart in Figure~\ref{fig:plasma_emission}, 
radio emission may be produced either at $f_{\rm p}$ or its harmonic [2$f_{\rm p}$]. 

The theory of plasma emission was first proposed by \citet{Ginzburg58} to address 
observations of radio bursts for which the emission frequency drifts to lower values, 
which was interpreted in terms of disturbances that propagate outward through the corona 
to excite radio emission at the local plasma frequency (\textit{e.g.} \citealp{Payne47}). 
Plasma emission theory has since been developed by many authors (see reviews by \citealp{Robinson00,Melrose09}) 
and is thought to operate in different contexts to produce most types of coronal radio bursts (\textit{e.g.} \citealp{Dulk85}), 
which will be reviewed in Section~\ref{bursts}. 

\subsection{Electron-Cyclotron Maser Emission}
\label{ecm}

The word \textit{maser} is an acronym for ``microwave amplification by stimulated emission of radiation," 
which originally referred to a device that produces intense radiation of a 
specific frequency.
\textit{Stimulated emission} is a process by which a population of atoms or molecules 
are moved into energy levels above those of thermal equilibrium, which is referred to as 
a population inversion. The inverted population can then be stimulated to 
emit photons of a specific wavelength corresponding to a particular energy level transition, 
producing radiation that has a brightness temperature greater than the source temperature. 

After the first laboratory maser was built in 1953, astronomers in the 1960s and 70s 
identified intense molecular spectral line sources in interstellar space 
that would also be dubbed \textit{masers} because they are attributed to 
population inversions that occur naturally \citep{Reid81}.
When emission with very high brightness temperatures from the Sun and 
planetary magnetospheres was identified near $f_B$ and its harmonics, 
the term \textit{maser} was also adopted. 
This is somewhat of a misnomer, however, because electron-cyclotron maser emission (ECME)  
does not involve population inversions of atomic energy levels, but rather involves a plasma instability \citep{Treumann06}. 

The injection of nonthermal, semi-relativistic (fast) electrons into a plasma produces a  
population inversion analogous to that of a maser 
in the sense that a high-energy population was added to an 
equilibrium distribution. 
If the plasma density is low and/or the magnetic field strength is high such that 
$f_B > f_{\rm p}$, then the excess energy of the nonthermal electrons is not most  
efficiently converted into Langmuir waves, as in the previous section, and instead direct 
emission of radiation at $f_B$ via a plasma instability becomes favorable \citep{Treumann06}. 
This is expressed analytically as a negative absorption coefficient (or positive growth rate) for a particular 
particle distribution, the most famous of which is the loss-cone distribution (\textit{e.g.} \citealp{Wu79,Dulk85,Melrose09}).

As ECME requires special conditions, namely semi-relativistic nonthermal electrons and $f_B > f_p$, 
it is less broadly applicable than the other solar radio emission mechanisms. 
Microwave spike bursts emanating from the chromosphere are the most commonly-accepted 
example of ECME from the Sun \citep{Dulk85}, but the applicability of this mechanism 
to the corona is less certain. 
Nevertheless, high-density coronal regions with large magnetic field strengths at 
relatively low heights can support ECME \citep{Morosan16}, which is sometimes invoked 
to explain radio burst features that cannot be easily explained by 
plasma emission or gyrosynchrotron emission (\textit{e.g.} \citealp{Winglee86,Aschwanden88,Tang13,Wang15,Liu18}). 

\subsection{Propagation Effects}
\label{propagation}

The observed properties of solar radio emission, particularly at low frequencies, 
are greatly influenced by propagation effects that occur after the radiation is emitted. 
These effects depend on the medium that the radiation moved through en route 
to the observer. 
The dominant effects occur in the corona and Earth's ionosphere, though 
very long wavelengths may encounter similar effects in the interplanetary medium. 
The corona is highly structured, often with large density contrasts between 
adjacent regions (\eg~\citealt{Woo07,Raymond14,Hahn16}). 
Once emitted, a radio wave may be reflected many times by neighboring high-density regions, 
whether steady-state or turbulent, before 
the ambient density becomes low enough for the wave to propagate freely. 
The process of undergoing this successive combination of reflection and propagation 
is often called \textit{scattering}, and it has many important implications. 

Scattering increases the apparent size of a source, which is referred to as 
\textit{angular broadening} (\eg~\citealp{Steinberg85,Bastian94,Ingale15}). 
This broadening affects both compact burst sources and the entire Sun, and 
it has a side effect of decreasing the apparent brightness temperature (\eg~\citealp{Melrose88,Alissandrakis94,Thejappa08}).
In addition to angular broadening, scattering dramatically increases the cone-angle over 
which directed emission may be observed, which can even allow low-frequency detections of events 
originating from the far side of the Sun \citep{Dulk85b}.  

Random scattering may also systematically shift the observed location of a radio burst to 
larger heights because the structures responsible for scattering are not randomly 
arranged, instead consisting of high-density fibers that are aligned with the magnetic field 
and are generally radial \citep{Robinson83,Poquerusse88}. 
This is analogous to \textit{ducting}, which refers to the guiding of radio waves through a 
coherent low-density structure through successive reflections against the high-density 
``walls" of the duct \citep{Duncan79}. 
We will discuss these effects further in the context of Type III bursts in Chapter~\ref{ch3} \citep{McCauley18}. 
Finally, scattering tends to depolarize emission and is thought to be responsible 
for the fact that many radio bursts have much lower circular polarization fractions 
than are expected by standard theories (\eg~\citealp{Wentzel86,Melrose06,Kaneda17}).

In addition to scattering, simple refraction is also important. 
The density of the corona generally decreases radially, which means 
that radio waves will tend to refract toward the radial direction \citep{Stewart76,Mann18}.
The $o$- and $x$-modes described in the previous section also have slightly different 
refractive indices. 
This difference means that refraction may separate the two modes, influencing the sense 
and/or degree of polarization.
For ground-based observations, refraction becomes important again when the 
radiation passes through Earth's ionosphere, which may 
significantly shift the apparent source location depending on how the 
instrument was calibrated. 

Propagation effects related to the magnetic field may also modulate the polarization 
state through \textit{mode coupling}, which refers to how the polarization of the $o$- and $x$-modes 
are changed by different plasma conditions. 
Regions for which the magnetic field orientation is nearly perpendicular to the 
ray propagation direction are referred to as \textit{quasi-transverse} (QT) regions \citep{Zheleznyakov70,Ryabov04}. 
Passing through a QT region may cause the circular polarization sign to flip 
if the emission frequency is below a certain threshold \citep{Cohen60,Zheleznyakov64,Melrose94}.
This concept is vital to the interpretation of polarization reversals in 
microwave observations (\eg~\citealp{Ryabov99,Sharykin18}) and may also be relevant in certain 
low-frequency radio burst contexts (\eg~\citealp{Suzuki80,White92,Kaneda15,Kong16}). 

\section{Solar Activity} 
\label{activity}

\subsection{The Solar Cycle}

\begin{figure}\graphicspath{{chapter1/}}
    \centering
    \includegraphics[width = 1.0\textwidth]{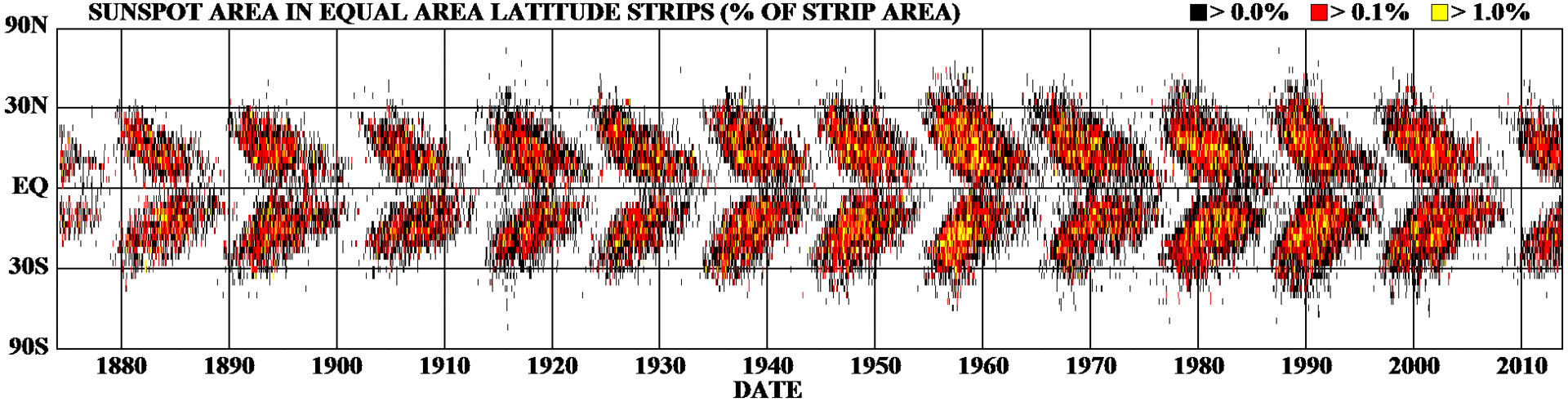}
    \caption{\footnotesize{}``Butterfly diagram" showing the total sunspot area 
    as a function of time and latitude from 1875 to the present day. 
    Each set of ``wings" corresponds to one solar cycle. 
    Image credit: 
    \href{https://en.wikipedia.org/wiki/Solar_cycle}{NASA Marshall Spaceflight Center}.}
    \label{fig:butterfly}
\end{figure}

\begin{figure}\graphicspath{{chapter1/}}
    \centering
    \includegraphics[width = 1.0\textwidth]{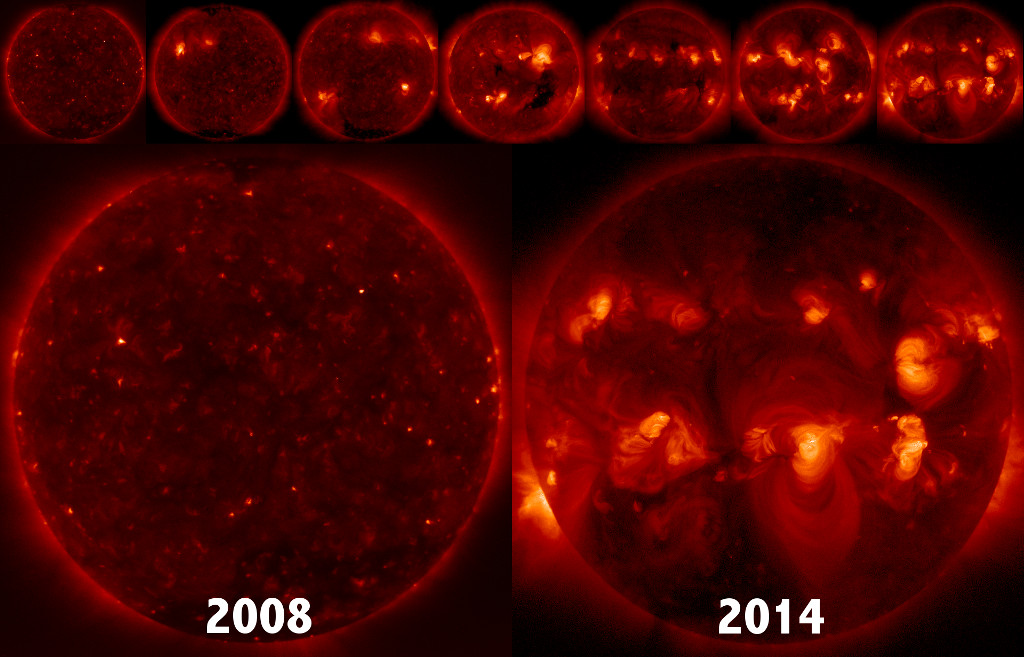}
    \caption{\footnotesize{}The corona in soft X-rays across the solar cycle. 
    There is one image in the upper row 
    for each year from last sunspot minimum (2008) to the 
    last maximum (2014). 
    Data are from the \textit{Hinode X-Ray Telescope} (XRT). 
    Image credit: Patrick McCauley, 
    \href{https://xrt.cfa.harvard.edu/xpow/20140725.html}{Smithsonian Astrophysical Observatory (SAO)}}
    \label{fig:xrt_cycle}
\end{figure}

The earliest known form of solar activity is sunspots, which 
are dark regions on the photosphere. 
Sunspots are dark because they are cooler than the surrounding 
material due to the presence of strong magnetic fields that 
inhibit the convective flow of energy from the interior. 
Very large sunspots can be seen with the naked eye, and their 
presence was routinely documented by early Chinese astronomers, 
with the first written record appearing in the \textit{Book of Changes}
in the late 9th century BC \citep{Xu90}.
In the mid 1800s, it was discovered that the number of 
sunspots varies on an approximately 11-year cycle \citep{Schwabe44} and that their 
latitudes drift from high to low over the course of the cycle \citep{Carrington58,Maunder03}.
This is famously represented by the ``butterfly diagram," which 
shows the number and distribution of sunspots over time \citep{Maunder04}. 
An example covering the period from 1875 to the present is shown 
in Figure~\ref{fig:butterfly}. 

In 1908, sunspots were discovered to be regions with strong 
magnetic fields through the first astrophysical observations 
of the Zeeman effect \citep{Hale08}. 
This implied that the sunspot cycle was really a cycle in the Sun's magnetic field \citep{Hale19}, 
and it was later discovered that the Sun's magnetic poles reverse with each cycle \citep{Babcock59}. 
Because the structure of and activity within the corona is largely determined 
by the magnetic field, solar cycle variations in the corona are dramatic.
Figure~\ref{fig:xrt_cycle} shows soft X-ray images of the corona between 2008 and 2014, 
which were the years of the most recent sunspot minimum and maximum, respectively. 
The observations presented in this thesis were taken in 2014 and 2015, near the maximum phase of the solar cycle. 
There are many more important and interesting details on solar cycle variations 
and their implications that can be found in 
several textbooks (\eg~\citealp{Foukal04,Aschwanden05,Golub10,Priest14}) and review articles \citep{Hathaway15}. 
The next section will introduce the process that is thought to drive the cycle and ultimately 
give rise to all of the Sun's activity. 

\subsection{The Solar Dynamo}

\begin{figure}\graphicspath{{chapter1/}}
    \centering
    \includegraphics[width = 1.0\textwidth]{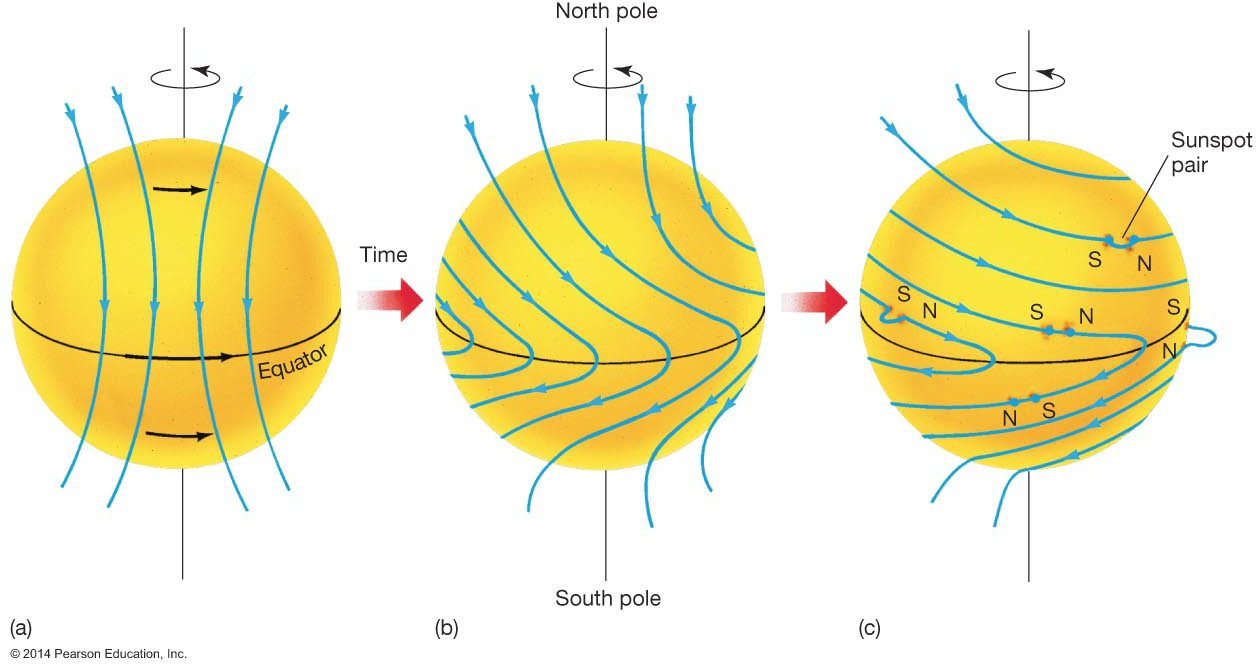}
    \caption{\footnotesize{}Differential rotation's effect on the Sun's magnetic field. 
    Poloidal field on the left, representing the start of the solar cycle, 
    is warped into toroidal field as low latitudes rotate faster than high 
    latitudes. This is the \textit{omega effect}, which has a less-easily-illustrated corollary 
    in the \textit{alpha effect} that regenerates poloidal field to renew the cycle. 
    Image credit: Pearson Eduction, Inc.}
    \label{fig:omega}
\end{figure}

The solar cycle and solar activity in general are ultimately byproducts of the  
dynamo process that generates the Sun's magnetic field, which is 
both the scaffolding for structures in the corona 
and the energy reservoir that powers eruptions. 
A dynamo is a process that generates a magnetic field through induction 
by the rotation of a convecting, turbulent, and electrically-conductive fluid. 
Different versions of this process can exist in a variety of different contexts at vastly different spatial scales, 
from the dynamos that generate Earth's magnetic field to the large-scale magnetic fields of galaxies. 
To introduce the solar dynamo, it is helpful to first outline the Sun's interior structure. 

There are three basic layers to the solar interior: the \textit{core}, \textit{radiative zone}, and 
\textit{convection zone}. 
The core is the innermost region where mass is converted into energy by nuclear 
fusion of hydrogen into helium. 
Energy from the core is first transported toward the exterior by radiative transfer 
in the radiative zone, where photons are absorbed and reemitted many times before 
reaching the base of the convection zone, after which energy is transported outward primarily by a 
cycle of hot plasma bubbles that rise to the surface, cool, and sink back down.
The modern picture of the Sun's interior came initially from models based on the transport 
of fusion energy from the core, the existence of which was proposed in 1920 \citep{Eddington20} 
and confirmed experimentally through the detection of solar neutrinos \citep{Davis64,Bahcall76}. 
Additional advances came after the discovery of global oscillations visible at the surface \citep{Leighton62}, 
which led to the development of \textit{helioseismology} (see reviews by \citealp{Turck93,Basu16}).

Perhaps the most critical insight of helioseismology has been the discovery of a thin 
layer at the base of the convection zone called the \textit{tachocline}, 
where the Sun's large-scale magnetic field is now thought to be primarily generated \citep{Dziembowski89,Spiegel92,Tomczyk95}. 
Beneath the tachocline, the Sun rotates nearly as a rigid body, and above the tachocline, different latitudes 
exhibit significantly different rotation rates. 
This is referred to as \textit{differential rotation}. 
At the surface, material at the equator has a rotation rate of 
around 25 days compared to around 35 days near the poles, and this pattern continues smoothly in 
the interior down to the base of the convection zone. 
The onset of differential rotation in the tachocline introduces a strong shear 
that is thought to be a key ingredient in the cyclic nature of the solar dynamo. 

At the beginning of the solar cycle, the Sun's magnetic field is largely \textit{poloidal}, 
where the field is perpendicular to the rotational flow direction and resembles that of a bar magnet. 
As the cycle progresses, differential rotation converts poloidal field into \textit{toroidal} field, 
where the field is instead aligned with the rotational flow. 
This is referred to as the \textit{omega effect}, 
and it occurs because the plasma pressure inside the 
Sun is larger than the magnetic pressure, so the magnetic field may be dragged along 
by the differential plasma motion. 
The omega effect, illustrated by Figure~\ref{fig:omega}, is countered by the \textit{alpha effect}, which regenerates the poloidal field 
from the toroidal field to renew the cycle. 
This basic picture is commonly-accepted, but many challenges remain in developing 
a dynamo model that can reproduce all of the observed solar cycle features. 
A recent review of solar dynamo models is given by \citet{Charbonneau10}. 

\begin{figure}\graphicspath{{chapter1/}}
    \centering
    \includegraphics[width = 0.6\textwidth]{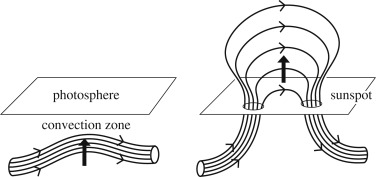}
    \caption{\footnotesize{}Magnetic buoyancy and flux emergence illustration. 
    Concentrations of strong toroidal field  
    rise to the surface, forming a bipolar sunspot with 
    a magnetic field extending into the corona. 
    Image credit: \citet{Priest19}}
    \label{fig:flux_emergence}
\end{figure}

Once the toroidal field generated in the tachocline becomes strong enough in a particular region, 
the magnetized plasma becomes buoyant and rises to the surface \citep{Babcock61,Parker75,Fan93}. 
\textit{Magnetic buoyancy} thereby transports toroidal flux from the interior to the surface, where it emerges to 
form a bipolar sunspot with a magnetic field extending into the corona. 
This process, illustrated by Figure~\ref{fig:flux_emergence}, 
is referred to as \textit{flux emergence} \citep{Golub81,Archontis08,Schmieder14}.
The coronal magnetic field then develops as new flux continues to emerge 
and interact with preexisting field structures, 
along with other processes such as photospheric flows that may alter the coronal magnetic field topology.
This constant evolution of the magnetic field gives rise to both the observed structure of the corona 
and energetic events such as solar flares and coronal mass ejections, which allow 
the abrupt release of energy stored in the magnetic field. 

\subsection{Features of the Corona}
\label{features}

The ever-evolving features of the corona are themselves indicators of solar activity. 
This section briefly outlines the basic features that are relevant for this thesis, though a number 
of important features like prominences are excluded because 
they are not important for the radio observations that will be presented. 
Complete reviews of the corona's features can be found in several textbooks (\eg~\citealp{Aschwanden05,Golub10,Priest14}). 
The low corona is frequently divided into three main region types: \textit{active regions}, \textit{coronal holes}, 
and the \textit{quiet Sun}.
This nomenclature developed largely around the interpretation of early soft X-ray observations.
 
Active regions are, as their name implies, sites of enhanced activity, where energetic events like 
flares and coronal mass ejections are most likely to originate. 
They have strong magnetic field concentrations and are easily distinguished as bright 
features in EUV and soft X-ray observations \citep{Golub10}. 
In radio observations, active regions are also bright at higher frequencies ($\gtrsim$ 300 MHz) where 
gyroresonance emission from the strong magnetic fields 
either contributes significantly or dominates \citep{White97,Lee07,Shibasaki11}. 
With decreasing frequency, corresponding to increasing heights, they become less distinguishable from 
the background but are often sites of persistent 
nonthermal emission referred to as noise storms \citep{LeSqueren63,Gergely75,Elgar77,Alissandrakis85}.

Active regions are the atmospheric counterparts of photospheric sunspots, although it is possible to have 
an active region without a visible sunspot if the magnetic flux is relatively weak.
As described in the previous section and illustrated by Figure~\ref{fig:flux_emergence}, 
active regions are produced by the generation of toroidal 
field in the interior that then rises to the surface in a process called flux emergence. 
New regions typically emerge with a bipolar field configuration \citep{Harvey93}, including a pair of sunspots 
with opposite polarities. 
However, more complicated configurations are also possible either intrinsically or through 
additional flux emergence and interactions with preexisting structures \citep{McIntosh90,Jaeggli16}, 
and more magnetically complex regions tend to be more active \citep{McAteer05}. 
Regions of opposite polarity at the surface are often visibly linked by closed coronal loops, 
which are distinct magnetic field structures that radiate when populated by sufficient plasma \citep{Reale14}. 
Once emerged, the magnetic flux dissipates
until the region is no longer distinguishable as an active region, which typically takes a few weeks, 
though very large active regions may persist for 1-2 months \citep{van98,Hathaway08}.

Coronal holes are regions where the magnetic field is open to interplanetary space, allowing material to 
flow freely away from the Sun to form the fast solar wind \citep{Cranmer09}. 
They were first discovered in X-ray images from sounding rocket experiments in 1970 and 
characterized in greater detail shortly thereafter by \textit{Skylab} data \citep{Timothy75}, which definitively associated 
coronal holes with high speed solar wind streams \citep{Krieger73,Zirker77}. 
The term \textit{hole} was chosen because they appear dark in EUV and X-ray images due to 
the plasma not being confined by closed fields, resulting in much lower densities and therefore emissivities. 
Coronal holes also appear dark in higher-frequency radio observations of the corona, but for reasons 
that are not yet clear, they often 
become increasingly bright structures with decreasing frequency below around 120 MHz \citep{Lantos99,McCauley17,Rahman19}. 
Coronal holes form when a large swath of the Sun becomes dominated by a single magnetic polarity, 
leading to a unipolar open-field region. 
Near solar maximum, this configuration can arise at lower latitudes 
from the right combination of decayed active regions, 
and the resulting coronal hole may last for weeks to months \citep{Wang10,Petrie13}.
Around solar minimum and for much of the solar cycle, coronal holes are also continuously present 
at both poles \citep{Waldmeier81,Harvey02}. 

The \textit{quiet Sun} is a somewhat ill-defined term that is generally 
used to refer to regions that are absent of features such 
as active regions, coronal holes, and others not discussed here such as X-ray bright points and filaments, 
although it may also include quiescent features like coronal holes. 
At the surface, quiet Sun regions have weaker, granular magnetic fields that are not dominated by a single 
polarity as in coronal holes \citep{Lin99,Bellot19}.
In X-ray images, quiet Sun regions are often populated with large-scale diffuse structures that 
are the lower portions of large loops connected to the remnants of decayed active regions \citep{Golub10}. 
At radio wavelengths, quiet Sun regions are typically taken to be those that are absent of obvious nonthermal emission \citep{Smerd50,Kundu77}. 
Although they are quiet in comparison to active regions, 
there is considerable magnetic energy in the quiet Sun \citep{Trujillo04} and small-scale 
energetic events are ubiquitous \citep{Habbal92,Harrison77}.

Finally, coronal \textit{streamers} are the most prominent features seen in white light images of 
the corona. 
They are long-lived, high-density structures that extend radially outward to several 
solar radii from the surface \citep{Koutchmy92}. 
Like all coronal structures, streamers are related to the magnetic field configuration 
and generally represent regions of large-scale closed fields adjacent to open-field regions. 
Helmet streamers refer to the often particularly large and symmetric streamers that connect 
regions of opposite polarity in active regions. 
The closed-field structure of streamers generally constrains the outflow of material 
rather than freely permitting it, 
as in the open fields of coronal holes, although they also generate intermittent 
but persistent outflows due to dynamical interactions with neighboring field structures \citep{Wang00}. 
Streamers are associated with moderately enhanced quiescent radio emission \citep{Thejappa94,Ramesh00}, 
but they are not easily distinguished from the quiet Sun in radio images of the corona. 
There are, however, many reports of spatial associations between streamers 
and radio bursts \citep{Trottet82,Kundu84,Gopalswamy87,Mugundhan18}, 
which will also be found in Chapter~\ref{ch3} \citep{McCauley18}. 

\subsection{Solar Flares}

Solar flares are sudden brightenings of the solar atmosphere caused by an explosive 
release of energy stored in the local magnetic field.
The first flare observations were reported from white light observations of a 
sunspot region in 1859 \citep{Carrington59,Hodgson59}. 
Radiation levels are enhanced across the electromagnetic spectrum, but only 
the largest flares produce a significant white light signature at the photosphere \citep{Neidig89}. 
The primary response comes from the corona at X-ray wavelengths, and flares 
are now generally classified by their peak X-ray flux observed by the 
\textit{Geostationary Operational Environmental Satellites} (GOES), 
although an earlier classification scheme also exists based on optical H$\alpha$ observations \citep{Golub10}.
In ascending order, the GOES flare classes are A, B, C, M, and X. 
Each class is an order of magnitude more intense than the previous, with A-class 
flares beginning at 10\tsp{-8} W m\tsp{-2} and X-class flares beginning at 10\tsp{-4} 
W m\tsp{-2}. 

\begin{figure}\graphicspath{{chapter1/}}
    \centering
    \includegraphics[width = 0.7\textwidth]{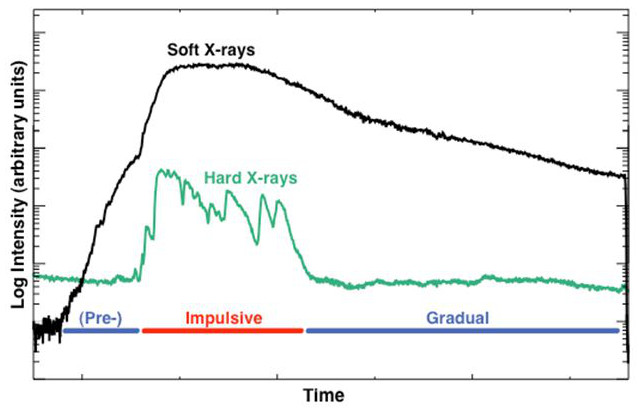}
    \caption{\footnotesize{}The three basic phases of a solar flare: the pre-flare, 
    impulsive rise, and gradual decay phases. 
    The impulsive phase also commonly includes nonthermal 
    radio burst emission. 
    Image credit: Amir Caspi}
    \label{fig:flare_phases}
\end{figure}

Flares generally exhibit three distinct phases: the pre-flare, impulsive rise, 
and gradual decay phases \citep{Aschwanden05,Golub10,Priest14}. 
These are illustrated by Figure~\ref{fig:flare_phases}. 
Many, but not all, flares are preceded by minor activity at EUV and X-ray 
wavelengths from the flare site during the pre-flare phase, after which 
the soft X-ray flux abruptly spikes during the impulsive phase. 
The impulsive phase represents an abrupt release of stored magnetic energy 
into plasma heating, particle acceleration, and bulk outflow. 
The soft X-ray peak is largely due to thermal emission from the heated plasma.  
Particle acceleration during the impulsive phase leads to intense and 
rapidly-fluctuating bursts of radio, hard X-ray, and sometimes gamma radiation. 
The impulsive phase lasts only a few minutes, after which the bursty emissions 
subside and the soft X-ray 
radiation gradually decreases during the decay phase as 
the plasma cools back to the pre-flare temperatures.

The physical processes that drive solar flares are actively debated and are not 
very well understood. 
There is general consensus, however, that the magnetic field configuration 
in the corona develops into an unstable configuration that leads to \textit{magnetic reconnection}, 
which produces an explosive release of energy. 
In the corona, the ratio of the plasma pressure to the magnetic pressure 
(the plasma beta) is very small. 
This means that the motion of coronal plasma is tightly constrained 
by the magnetic field [\textbf{B}], primarily being only 
along particular magnetic field structures except for certain plasma 
drifts perpendicular to \textbf{B} (i.e., the \textbf{E}$\times$\textbf{B}, $\nabla$\textbf{B}, and curvature drifts).
The opposite is true in the photosphere, where the density is much larger 
and the magnetic field can be dragged by plasma motions. 
As the magnetic field of the corona is rooted in the photosphere, 
the footpoints of coronal loops may be shifted around by photospheric flows. 
Along with continued flux emergence and interactions between neighboring field structures, 
photospheric flows can lead to the 
development of sheared and/or twisted magnetic field structures in the corona \citep{Hagyard84,Wang93,van99}.
These are referred to as \textit{non-potential} field configurations, whereas a \textit{potential} 
field corresponds to the lowest-energy configuration in which magnetic field lines straightforwardly 
connect opposite polarities. 
Energy is gradually stored in the magnetic field as it becomes increasingly non-potential, 
and that energy can then be catastrophically released through magnetic reconnection. 

\begin{figure}\graphicspath{{chapter1/}}
    \centering
    \includegraphics[width = 0.9\textwidth]{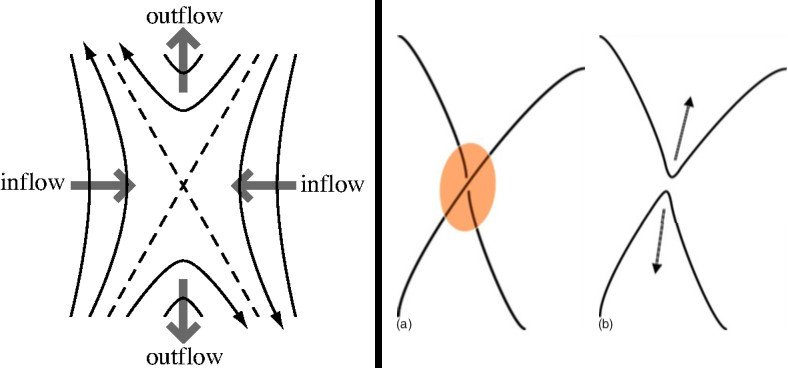}
    \caption{\footnotesize{}Simple magnetic reconnection cartoons. 
    The left panel illustrates how inflows force magnetic field 
    lines of opposite polarity toward an \textit{x-point}, where they 
    \textit{reconnect} to form new field lines that relax to produce outflows. 
    The right panel illustrates how the field connectivity changes 
    at the x-point. 
    Image credits: \citealp{Brown02} (\textit{left});
    \citealp{Yamada10} (\textit{right})}
    \label{fig:reconnection}
\end{figure}

Magnetic reconnection is process that converts stored magnetic energy into kinetic 
and thermal energy through the reconfiguration of the magnetic field topology \citep{Biskamp00,Priest00,Yamada10}.
In the simplest form, two opposite-polarity field lines are forced together 
by oppositely-directed inflows. 
The field lines collide at a central diffusion region, where they ``reconnect" to form two 
new field lines that are accelerated out of the diffusion region by the magnetic tension 
force, resulting in outflows that are perpendicular to the initial inflow. 
Cartoons of this process are shown in Figure~\ref{fig:reconnection}.
In addition to bulk outflows, collimated beams of particles are accelerated out of the 
diffusion region due primarily to the presence of very strong electric fields \citep{Reames99,Petrosian99,Schlickeiser03}. 
The theory of magnetic reconnection still faces a number of challenges to account 
for all of the observed properties of solar flares. 
For instance, the predicted reconnection rates are too slow and the observed widths 
of current sheets are too large, 
requiring the introduction of effects such as turbulence and/or various 
plasma instabilities (\eg~\citealp{Shibata01,Lazarian15,Cairns18}). 
Nevertheless, reconnection is widely-believed to be the driver of nearly all impulsive activity 
on the Sun. 

\begin{figure}\graphicspath{{chapter1/}}
    \centering
    \includegraphics[width = 1.0\textwidth]{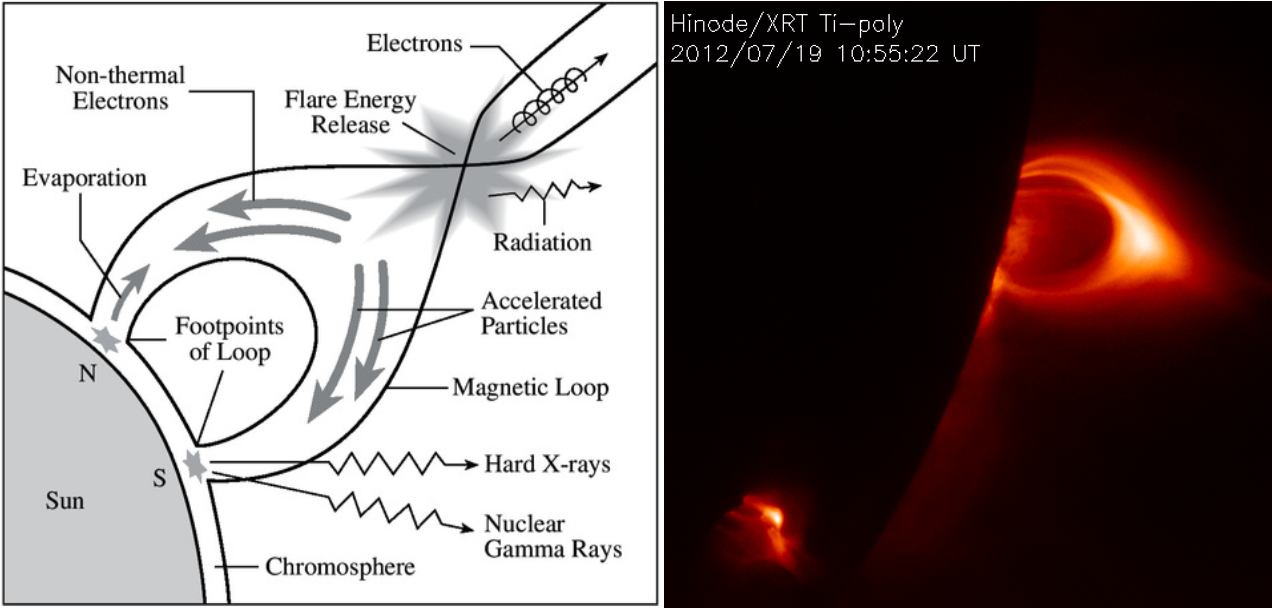}
    \caption{\footnotesize{}Simplified schematic of the ``standard" (CSHKP) flare 
    model (\textit{left}) next to a \textit{Hinode} XRT image exhibiting  
    the classic cusp loop structure in soft X-rays (\textit{right}). 
    Image credits: \citealp{Lang06} (\textit{left}); 
    Patrick McCauley, \href{http://xrt.cfa.harvard.edu/xpow/20120831.html}{SAO} (\textit{right})}
    \label{fig:flare_model}
\end{figure}

Figure~\ref{fig:flare_model} shows a simplified schematic of the CSHKP flare model \citep{Carmichael64,Sturrock66,Hirayama74,Kopp76}, 
which places reconnection in the flare context and can explain many commonly-observed features. 
Reconnection occurs relatively high in the corona above the cusp-shaped loop, accelerating 
particles both toward and away from the Sun. 
Particles accelerated toward the Sun impact the chromosphere, which radiates hard X-rays 
and sometimes gamma rays. 
The chromospheric material is rapidly heated and expands into the corona, a process called 
\textit{chromospheric evaporation}, which populates the cusp-shaped loop with hot plasma that 
produces intense EUV, soft X-ray, and microwave radiation. 
An observation of such a cusp loop is also shown in Figure~\ref{fig:flare_model}. 
Particles, mainly electrons, accelerated away from the Sun stimulate waves 
in the background plasma as they stream through it, resulting in radio bursts that can 
cover a broad range of frequencies depending on the height at which the electron beams start. 
Finally, plasma that is suspended above the reconnection point can be accelerated away 
from the Sun to produce coronal mass ejections, which will be discussed in the 
next section. 

The CSHKP model has proven to be so successful that it is often referred to simply 
as the ``standard model," though other flare models exist and other reconnection 
scenarios are important in different contexts, 
such as breakout reconnection \citep{Antiochos99}, 
interchange reconnection \citep{Edmondson12}, or
reconnection driven by emerging flux \citep{Shibata92}.
More complete descriptions of solar reconnection models 
can be found in solar physics textbooks \citep{Aschwanden05,Golub10,Priest14}.
 
\subsection{Coronal Mass Ejections}
\label{cmes}

\textit{Coronal mass ejections} (CMEs) are large expulsions of plasma and magnetic fields 
that were first discovered in the 1970s using white light coronagraph observations \citep{Tousey73,Gosling74}. 
CMEs are the most impactful form of space weather and can have a number of potentially 
severe consequences, from threatening the health of astronauts to causing widespread 
power blackouts \citep{2008sswe}. 
Prior to the discovery of CMEs, a connection between geomagnetic storms and solar flares 
had long been recognized (\eg~\citealp{Sabine52,Hale31,McLean59}), implying that 
flares must sometimes be associated with the expulsion of solar material that later interacts 
with Earth's magnetosphere. 

CMEs are related to solar flares in that they are both driven by the restructuring of the 
coronal magnetic field, likely as a result of magnetic reconnection.
Whereas a flare represents the conversion of magnetic energy largely into thermal energy, a 
CME moves energy stored in the magnetic field into the kinetic energy of a macroscopic outflow.
While most large flares are accompanied by a CME and their production is incorporated 
into the ``standard" flare model discussed in the previous subsection, is important to note 
that CMEs may occur without a corresponding flare and vice versa \citep{Kahler92,Yashiro09}. 
A CME may also begin before the associated flare, which is contrary  
to the standard model \citep{Harrison91}. 
There is therefore no perfectly clear causal relationship between the two phenomenon, which may 
instead be understood as different manifestations of the same or similar processes that lead to abrupt 
changes in the coronal magnetic field \citep{Hudson95,Golub10}.

\begin{figure}\graphicspath{{chapter1/}}
    \centering
    \includegraphics[width = 1.0\textwidth]{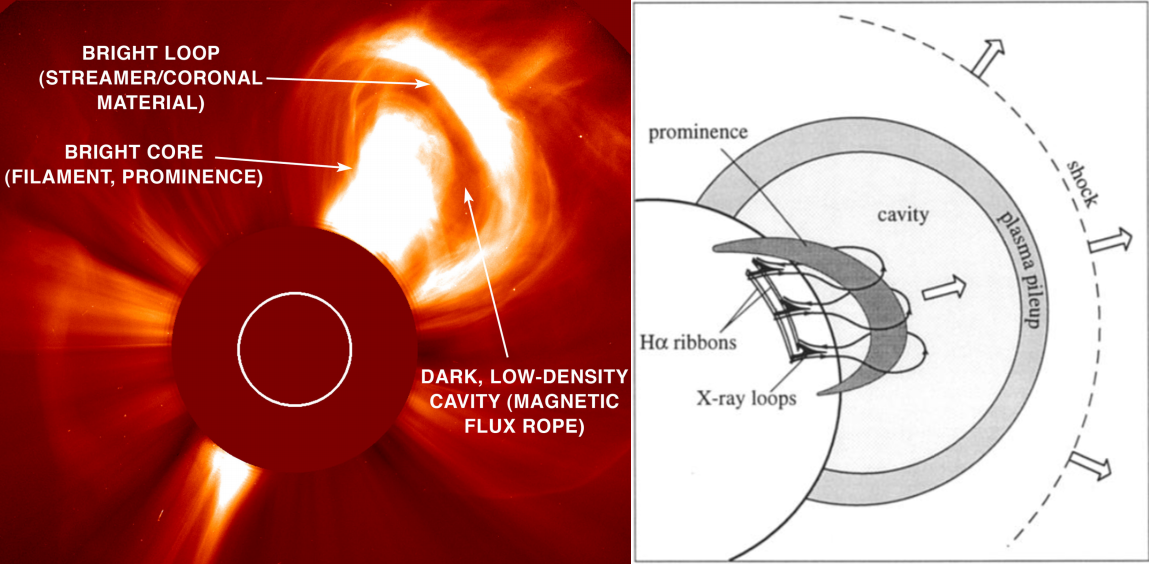}
    \caption{\footnotesize{}LASCO coronagraph image (\textit{left}) and  
    schematic (\textit{right}) of a CME 
    exhibiting the classic three-part structure of a bright plasma pileup 
    followed by a dark cavity and bright core. 
    The schematic also shows the leading shock wave along 
    with X-ray and H$\alpha$ features back at the Sun. 
    The white circle in the LASCO image represents the photosphere at 
    the center of the coronagraph's occulting disk.  
    Image credits: \citealp{Muller13} (\textit{left}); 
    \citealp{Forbes00} (\textit{right})}
    \label{fig:cme_structure}
\end{figure}

CMEs often exhibit a three-part structure in coronagraph observations that consists of 
a bright plasma pileup at the leading edge followed by a dark cavity and dense core \citep{Hundhausen99,Schwenn06}. 
The Sun produces around 1 CME per day on average, though the CME rate exhibits a strong 
solar cycle dependence, with less than one per day near solar minimum and as a high as 5 per day near solar 
maximum \citep{Yashiro04}. 
They range in speed from less then 100 \kms{} to greater than 1500 \kms{} \citep{Yurchyshyn05}, averaging around 
300 \kms{} near solar minimum and 600 \kms{} near solar maximum \citep{Gopalswamy09}. 
As these speeds generally exceed the sound speed and often exceed the Alfv\'{e}n speed, shock waves 
may develop just ahead of the CME in the low corona and/or interplanetary medium.  
These shocks accelerate electrons, which may produce Type II radio bursts through the 
plasma emission process (\eg~\citealp{Wild50,Nelson85,Gopalswamy06,Cairns11}), along with 
heavier particles referred to as \textit{solar energetic particles} (SEPs)
that may be observed directly with \textit{in situ} observations \citep{Reames13}. 
Plasma emission was discussed in Section~\ref{plasma}.  
Type II bursts are described in Section~\ref{typeii} along with Type IV bursts, 
which are also associated with CMEs. 

CME cores may be directly imaged in high-frequency microwave observations \citep{Gopalswamy03,Alissandrakis13}, 
but the primary low-frequency signatures of CMEs are the radio bursts associated 
with the shock-accelerated electrons \citep{Nindos08}. 
Gyrosychrotron emission from the expanding magnetic fields of a CME may also be observed in 
radio observations, though this has so far been rare \citep{Bastian01,Maia07}. 
Chapter~\ref{ch5} includes a preliminary account of novel radio CME observations 
that were reduced in the course of this thesis but have yet to be analyzed in detail. 

\section{Solar Radio Bursts} 
\label{bursts}

\begin{figure}\graphicspath{{chapter1/}}
    \centering
    \includegraphics[width = 1.0\textwidth]{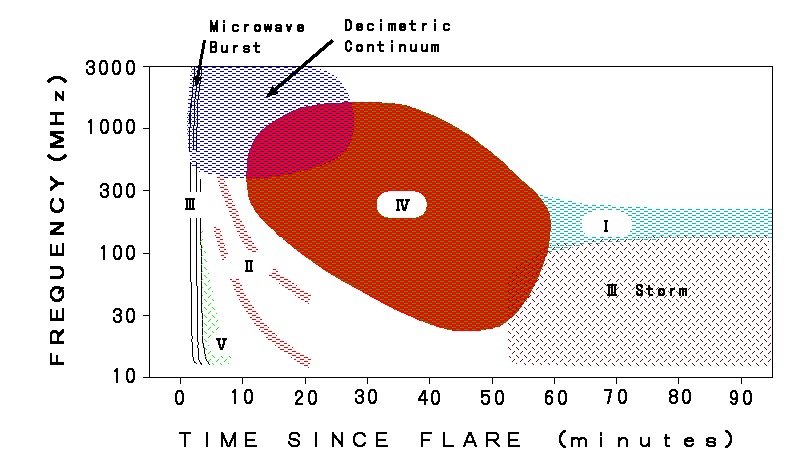}
    \caption{\footnotesize{}Dynamic spectrum schematic of radio burst types for 
    an idealized event. 
    Note that all types are seldom neatly observed for a single event. 
    Real dynamic spectra are shown in Figure~\ref{fig:types_i_ii_iii}. 
    Image credit: \href{http://Sunbase.nict.go.jp/solar/denpa/hiras/types.html}{nict.go.jp}}
    \label{fig:burst_types}
\end{figure}

Solar radio bursts are brief periods during which the Sun's radio emission is elevated above 
the thermal background level. 
Bursts may exceed the background level only slightly or by many orders of magnitude depending 
on the amount of energy released, along with a variety of other effects related to the structure of the source region, 
the viewing geometry, and the medium through which the radiation propagates en route to the observer. 
There are several types of radio bursts, most of which are produced by plasma emission operating in different contexts, 
although some types are attributed to (gyro)synchrotron and/or electron-cyclotron maser emission. 
These emission processes were reviewed in Section~\ref{mechanisms}. 

The classification of radio bursts is largely based on how they appear in dynamic spectra produced by radiospectrographs, 
which record the Sun's intensity as a function of both time and frequency. 
\citet{Wild50} defined the first three types using the earliest radiospectrograph observations of metric bursts, 
primarily based on variations in their apparent drifts in the emission frequency over time, 
and Types IV and V were added later. 
Since they were initially defined, numerous subtypes have been added and many of the associated physical processes have 
been identified. 
What follows are brief descriptions of the five main classes of solar radio bursts, along with references to a few additional types. 
An idealized schematic of how radio bursts appear in dynamic spectra is shown in Figure~\ref{fig:burst_types}. 

\subsection{Type I}
\label{typei}

	\textit{Type I bursts} are spikes of enhanced radio emission that last around one second and 
	occur over a relatively narrow frequency range ($\Delta f/f \approx 0.025$), with 
	little-to-no discernible drift in the frequency. 
	They generally occur in groups, referred to as \textit{noise storms}, that are superimposed 
	on enhanced continuum emission of the same frequency range (see reviews by \citealp{Elgar77,Klein98}). 
	While each individual spike does not drift in frequency, a chain of Type I bursts may 
	slowly drift from higher to lower 
	frequencies over a few minutes. 
	An example of a Type I burst noise storm is shown in the upper panel of 
	Figure~\ref{fig:types_i_ii_iii}. 
	
	Noise storms are associated with active regions, they may last anywhere from hours to weeks, 
	and they are most commonly observed 
	at relatively low frequencies ($\approx$ 50\,--\,500 MHz). 
	Although the association with active regions has been known for 
	decades (\eg~\citealp{LeSqueren63,Gergely75,Alissandrakis85}), 
	it is still not entirely clear what conditions within 
	active regions lead to noise storms. 
	Not all active regions that exhibit activity at other wavelengths generate noise storms, 
	and unlike other radio burst types, non-radio 
	signatures are often scant \citep{Willson05,Iwai12,Li17}. 
	
	Type I bursts are generally attributed to fundamental plasma emission, 
	largely due to their often high circular polarizations, but there is not yet a  
	consensus on what accelerates the electrons. 
	Minor reconnection events \citep{Benz81} or shocks associated with different types 
	of upward-propagating waves \citep{Spicer82} are the two leading ideas.  
	Recent work has favored reconnection in different contexts, either between open and closed fields at the 
	boundaries of active regions \citep{Del11,Mandrini15} or driven by 
	moving magnetic features at the photosphere \citep{Bentley00,Li17}. 
	We will present polarization measurements of noise storm continua in Chapter~\ref{ch4} \citep{McCauley19}, including 
	much weaker sources than have been previously reported. 	
	
\subsection{Type II}	
\label{typeii}		
	
\begin{figure}\graphicspath{{chapter1/}}
    \centering
    \includegraphics[width = 0.85\textwidth]{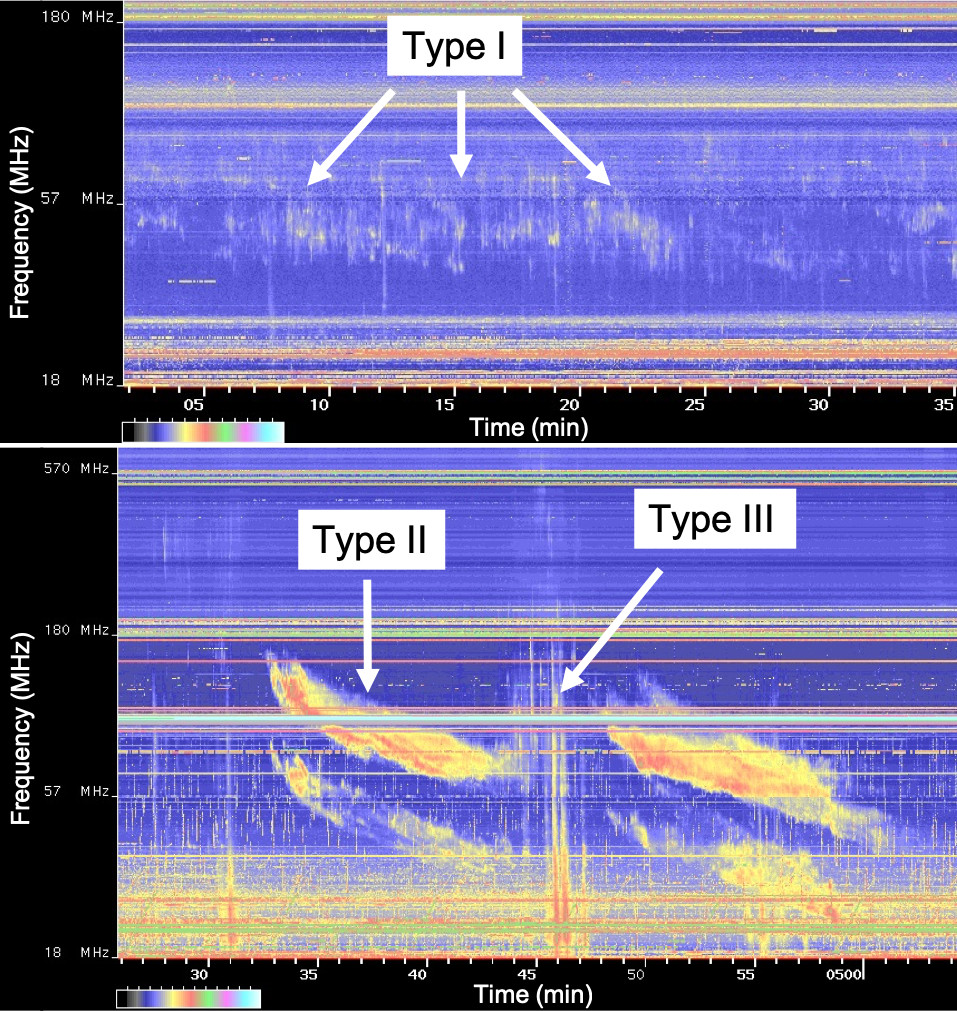}
    \caption{\footnotesize{}Learmonth solar radiospectrograph dynamic spectra 
    showing Type I bursts (\textit{top}), along with Types II and III bursts (\textit{bottom}).
    Adapted from spectra published by the Bureau of Meteorology's
    \href{www.sws.bom.gov.au}{Space Weather Service}.}
    \label{fig:types_i_ii_iii}
\end{figure}

	\textit{Type II bursts} drift slowly ($\approx$ 1 MHz s$^{-1}$) from high to low frequencies, typically lasting a few minutes. 
	They often exhibit two distinct bands of emission that are 
	interpreted as being fundamental--harmonic pairs of plasma emission from a single source region \citep{Roberts59,Sturrock61}.
	Examples of Type II bursts exhibiting such a structure are shown in the lower panel of Figure~\ref{fig:types_i_ii_iii}. 
	Type II bursts are associated with coronal mass ejections and are believed to be caused by electrons accelerated by a shock 
	wave at the leading edge of the CME (\eg~\citealp{Cane84,Gopalswamy01,Cairns11}). 
	The emission frequency drifts down to lower frequencies because it depends on the local density, which generally decreases 
	as a function of radial distance from the Sun. 
	By assuming a density model, the frequency drift rate may be converted to a physical speed that then refers to the 
	speed of the disturbance moving outward through the corona. 
	For Type II bursts, this procedure typically yields speeds of around 1000 km s$^{-1}$, which adequately matches that of CME shocks. 
	
	Although Type II bursts are also attributed to plasma emission, they do not exhibit the high circular polarizations seen in 
	Type I bursts, instead exhibiting little-to-no polarized intensities (\eg~\citealp{Komesaroff58,Akabane61}). 
	The reason for the low polarization is not entirely understood, but dispersion effects related to the inhomogeneous magnetic 
	field near a magnetohydrodynamic shock wave is a leading hypothesis \citep{McLean85}. 
	Type II bursts also sometimes exhibit short-lived fine structures called herringbone bursts that emanate from the 
	main burst and extent to lower frequencies, suggesting that beams of shock-accelerated electrons where able to 
	escape far beyond the shock region \citep{Cairns87}.
	A review of the theory and space weather implications of Type II bursts is given by \citet{Cairns03}. 

\subsection{Type III}	
\label{typeiii}
	
	\textit{Type III bursts} are similar to Type IIs in that they also drift from high to low frequencies and are attributed 
	to plasma emission. 
	However, Type IIIs drift at a much faster rate ($\approx$ 100 MHz s$^{-1}$) and must therefore be excited by disturbances 
	moving outward through the corona more quickly than the shocks responsible for Type II bursts.
	Examples of Type III bursts are shown in the lower panel of Figure~\ref{fig:types_i_ii_iii}.  
	Type IIIs are attributed to semi-relativistic electron beams accelerated by magnetic reconnection, the process that underpins solar 
	flares. 
	Correspondingly, Type III bursts are strongly associated with X-ray flares. 
	Nearly all large flares have associated non-thermal radio emission, typically Type III bursts \citep{Benz05,Benz07}. 
	However, not all small-to-moderate X-ray flares have associated Type III bursts and vice versa due to the different 
	conditions required to produce the high- and low-energy emission (\eg~\citealp{Alissandrakis15,Reid17,Cairns18}). 
	
	Type III bursts may occur in isolation, small groups, or chains lasting many minutes that are referred to as Type III storms. 
	They exhibit moderate circular polarization fractions, typically less than 50\%.
	Like Type IIs, the observed polarization fractions of Type III bursts are much lower than theoretical predictions (\eg~\citealp{Wentzel84}). 
	Scattering by density inhomogeneities and other propagation effects are likely responsible for this discrepancy, though 
	this has yet to be fully resolved (\eg~\citealp{Wentzel86,Melrose06,Kaneda17}).	
	A review of recent Type III burst literature is provided by \citet{Reid14}, and a review focused on the theoretical 
	aspects is given by \citet{Robinson00}. 
	Type III bursts will be discussed in more detail in Chapters~\ref{ch2}~and~\ref{ch3} \citep{McCauley17,McCauley18}, 
	which focus on their relationship with the coronal magnetic field and density structures, respectively. 
	
\subsection{Type IV}		
\label{typeiv}	
	
	\textit{Type IV bursts} are broad-band continuum emissions that come in a few distinct varieties that are 
	associated with different phenomena and different emission mechanisms.  
	\textit{Moving} Type IV bursts were the first to be defined and require an interferometer to be 
	detected (i.e. imaging observations), as they are characterized by an outward-moving continuum source \citep{Boischot58,McLean85}. 
	The emission mechanism is unclear and is generally attributed to gyrosynchrotron emission, plasma emission, or some 
	combination thereof associated with fast electrons trapped within the magnetic fields of an erupting CME \citep{Dulk85,Morosan19}. 
	As they are also related to CMEs, Type IV bursts are often preceded by a Type II burst driven by the shock at 
	the leading edge of the eruption.  
	
	\textit{Stationary} Type IV bursts are more common and are broad-band continuum emissions that are 
	either associated with solar flares or Type I bursts \citep{McLean85,Dulk85}. 
	Flare-associated Type IVs are referred to as \textit{flare continuum} 
	bursts, typically beginning at or shortly after the impulsive phase of the flare. 
	These bursts are thought to be caused by plasma emission generated 
	by electrons trapped in large, closed magnetic loops \citep{Dulk85}. 
	The flare continuum may be followed by a second phase referred to as 
	a \textit{storm continuum} burst that commonly occurs with larger flares  \citep{Pick61}. 
	The storm continuum lasts for hours to days, progressively becoming 
	an ordinary Type I noise storm if the duration is long enough \citep{Pick08}. 
	Both phases are thought to be caused by plasma emission, although the contexts 
	must be somewhat different because the storm continuum tends to exhibit a 
	much larger degree of circular polarization \citep{Dulk85}.

\subsection{Type V}		
\label{typev}
	
	\textit{Type V bursts} are continuum emissions that last for one to a few minutes, immediately following 
	a group of Type III bursts and generally occurring at frequencies below $\approx$ 120 MHz \citep{Suzuki85,Dulk85,Reid14}.
	This radiation is much less common than Type III bursts and is 
	generally thought to be caused by harmonic plasma emission associated with the same electron streams 
	responsible for the associated Type III bursts \citep{Zheleznyakov68,Robinson78,Dulk85}, perhaps moving along 
	different field structures to explain the sometimes large positional offsets from the associated Type III bursts \citep{Weiss65,Robinson77}. 
	Type V emission may last longer because it is generated by a slower electron population that is less collimated 
	than the Type III bursts, which contributes to the broader-band emission and also leads to a reversal in the 
	sense of the circular polarization from that of the Type IIIs due to the different angular distribution of Langmuir waves \citep{Dulk80,Dulk85}. 	
	Alternatively, some authors have suggested that Type V bursts may be an example of electron cyclotron 
	maser emission \citep{Winglee86,Tang13}.

\subsection{Other}
\label{typeo}		

In addition to Types I\,--\,V, there are a number of additional classes of radio bursts. 
These include variants of the standard types, fine structure within a particular type, and wholly distinct phenomena. 
Examples of variants include Types J and U bursts, which are Type III bursts for which 
the frequency drift reverses, indicating that the associated electron beams reverse direction and 
travel back toward the Sun along closed magnetic field lines \citep{Reid14}. 
Examples of fine structure bursts include the zebra patterns \citep{Slottje72} and fibre bursts \citep{Rausche08} that may 
be observed in association with Type IV bursts, along with the herringbone bursts that sometimes 
accompany Type II bursts \citep{Cairns87}. 
Additional distinct classes include Type S bursts \citep{Ellis69,Morosan15}, which last just milliseconds, along with a 
variety of microwave bursts such as microwave Type IV bursts,
impulsive bursts, postbursts, and spike bursts \citep{Kundu82}.

\section{Interferometry and Aperture Synthesis}
\label{interferometry}

\textit{Interferometry} is the superposition of waves, usually electromagnetic, to 
generate interference patterns that can be used to extract information about the 
source that generated the waves and/or the medium through which the waves propagated. 
Since the development of the first interferometers in the late 1800s \citep{Michelson87}, 
variations of this concept have been implemented in many fields of science. 
Two types of interferometry are relevant to the study of 
the solar corona, Fabry-P\'{e}rot interferometry \citep{Fabry99} and 
aperture synthesis, sometimes called synthesis imaging. 
Fabry-P\'{e}rot interferometers, which will not be discussed further,  
are commonly-used to develop 
narrowband filters that precisely isolate 
particular spectral lines for imaging telescopes or tuneable spectrographs (\eg~\citealp{Brueckner95,Puschmann12,Prabhakar19}).

In the astrophysical context, \textit{interferometry} usually refers 
to \textit{aperture synthesis}, which is a type of interferometry that combines 
signals from multiple telescopes. 
An array of telescopes can be used to synthesize the aperture of a much 
larger telescope, permitting an angular resolution up to 
that of a telescope with a diameter equal to the largest separation 
between interferometer elements. 
This is particularly important for radio astronomy because the long
wavelengths mean that impractically large telescopes would be 
needed to achieve high angular resolutions without interferometry. 
Interferometers are also much easier to construct at radio wavelengths 
because the signals can be combined electronically, whereas infrared and 
optical interferometers require precise optics that have been developed 
only fairly recently and may still support only a small number of elements.

This section will introduce a few basic concepts and terms, including a 
description of the main instrument used in this thesis and some notes 
on data reduction. 
More details on interferometry and the associated mathematics can be 
found in several textbooks (\eg~\citealp{Taylor99,Thompson17}). 

\subsection{Basic Concepts}

\textit{Synthesis imaging} is the reconstruction of an image from measurements 
of the Fourier transforms of its brightness distribution \citep{Thompson17}. 
The fundamental measurements of interferometers are called 
\textit{visibilities}, which are cross correlations (interference patterns) 
of the signals received from each pair of antennas in the array. 
The relative position between any pair of antennas is referred to as a \textit{baseline}, and 
each baseline represents a single point in the \textit{$u$-$v$ plane}, where 
$u$ is the east-west separation and $v$ is the north-south separation. 
The length of a baseline determines its sensitivity to different spatial 
scales on the sky, with longer baselines being sensitive to smaller 
spatial scales and shorter baselines being sensitive to larger scales. 
Arrays with longer baselines therefore have finer spatial resolutions, while more compact 
arrays are more sensitive to large-scale diffuse structures. 

The combination of all the baselines in an array is referred to its 
$u$-$v$ coverage. 
This represents the sampling of a target's brightness distribution, with 
the visibilities [$V(u,v)$] being the two-dimensional Fourier 
transform of the brightness on the sky [$T(x,y)$]. 
Because the $u$-$v$ coverage can never be complete given the finite 
number of array elements, the reconstruction of a source's brightness 
distribution (image) is an approximation that improves 
with the density of $u$-$v$ coverage (i.e. the array density). 

The initial inversion of $V(u,v)$ is usually done by interpolating the data onto 
a regular grid and applying the fast Fourier transform algorithm, which results in  
a so-called \textit{dirty image}. 
The image is ``dirty" in the sense that it is the brightness distribution 
convolved with an instrumental response function referred to as 
the \textit{synthesized beam} or \textit{point spread function} (PSF), which is 
analogous to the PSF of an optical telescope. 
The size and orientation of the synthesized beam will be shown in 
one corner of every MWA image in this thesis, and it is effectively 
the image's resolution element; features smaller than the 
synthesized beam cannot be spatially resolved. 

The PSF cannot be straightforwardly deconvolved out because it 
contains null points due to the incomplete sampling of the $u$-$v$ plane. 
Instead, a number of deconvolution algorithms have been developed. 
The most widely-used and the one implemented here is \textsf{CLEAN} \citep{Hogbom74,Schwarz78,Clark80,Schwab84}, 
the basic steps of which are as follows: find the location of peak intensity, 
subtract a multiple of the PSF centered at the peak, and repeat until the next peak 
location is below some threshold, often specified by the user. 
The resultant image is then convolved with a \textsf{CLEAN} beam, which is generally 
a two-dimensional Gaussian fit to the central component of the PSF, and the 
residuals of the dirty image are added back in. 

The deconvolved \textsf{CLEAN} image now represents the brightness distribution as 
viewed by that particular array, which has a particular response to the sky that 
determines the instrument's field of view and is 
referred to as its \textit{primary beam}.
The primary beam is effectively how the interferometer ``sees" 
the sky, and its pattern varies significantly with antenna type and 
array configuration. 
A model of the primary beam can then be divided out to 
obtain the ``true" instrument-independent brightness distribution. 
The beam model may be derived either from an analytic description 
of the combined antenna response or empirical measurements of known sources. 
Imperfections in the beam model have important consequences
for polarimetry that will be discussed in Chapter~\ref{ch4} \citep{McCauley19}, which 
includes the development of an algorithm to mitigate beam-related errors. 

The description above is a very simplified overview of aperture synthesis and its 
fundamental terminology. 
An important caveat is that the consideration of $V(u,v)$ as the 
Fourier transform of $T(x,y)$ is valid only for fairly small angles on the 
sky. 
Widefield instruments such as the MWA require specialized algorithms 
to invert $V(u,v)$, such as $w$-stacking, which is implemented in the 
\textsf{WSCLEAN} software used to reduce the data presented in this thesis \citep{Offringa14}. 


\begin{figure}\graphicspath{{chapter1/}}
    \centering
    \includegraphics[width = 0.8\textwidth]{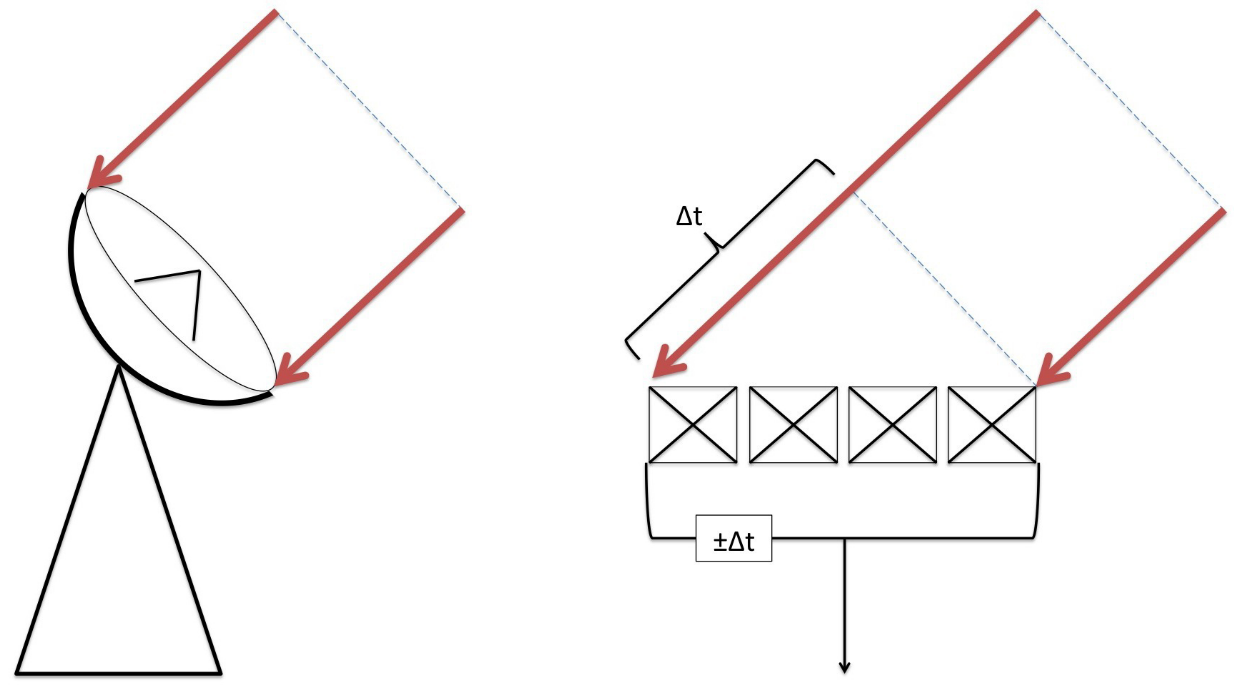}
    \caption{\footnotesize{}Illustration of the electronic ``steering" of an 
    aperture array (\textit{right}) versus the mechanical steering 
    of a parabolic dish (\textit{left}). 
    The time lag [$\Delta{}t$] between 
    radiation arriving at different antennas is specific 
    to a specific location on the sky, and 
    shifting the signals by $\pm\Delta{}t$ before they are correlated 
    focuses the telescope on that location. 
    Image credit: \citep{Carroll16}}
    \label{fig:aperture_array}
\end{figure}

\subsection{The \textit{Murchison Widefield Array} (MWA)}
\label{mwai}

The MWA is a low-frequency (80\,--\,300 MHz) interferometer 
located in the Murchison Shire of Western Australia, which is an 
exceptional site for radio astronomy because of limited 
radio frequency interference (RFI) from the small 
human population. 
A technology demonstrator and 
precursor telescope for the \textit{Square Kilometre Array} (SKA; \citealp{Dewdney09}), 
which is planned to be the world's largest 
telescope, the MWA 
has four main science themes \citep{Bowman13}. 

These are 1) attempting to detect redshifted 21-cm 
emission from the early Universe's Epoch of Reionization (EoR), 
2) conducting galactic and extragalactic surveys, 
3) searching for and localizing various radio transients (i.e. time-domain 
astrophysics), and 4) solar, heliospheric, and ionospheric (SHI) studies. 
The latter category includes direct observations of the Sun, which is 
the subject of this thesis. 
The MWA can also probe the solar wind through observations of interplanetary 
scintillation, which refers to brightness fluctuations (twinkling) exhibited  
by astronomical sources as their radiation passes through the solar wind plasma \citep{Morgan18}.  
These observations are also sensitive to the passage of CMEs \citep{Kaplan15}, and 
a major goal of the SHI collaboration is to make Faraday rotation measurements 
of linearly-polarized background sources occulted by a CME, which can be used 
to deduce the strength and orientation of the CME's magnetic field. 

The MWA is comprised of many individual dual-polarization dipole antennas arranged in 4$\times$4 
grids called ``tiles," one of which is pictured in the lower-middle panel of Figure~\ref{fig:arrays}. 
A prototype array with 32 tiles operated between 2009 and 2011 \citep{Lonsdale09}, followed by 
the commencement of Phase I operations with 128 tiles in 2013 \citep{Tingay13}. 
The data presented in this thesis are from Phase I. 
Phase II operations began in 2018 with 256 tiles, half of which can be used simultaneously 
in different configurations \citep{Wayth18}. 
Each MWA tile is an individual \textit{aperture array}, which is a collection of antennas that 
receive signals from the entire sky but are ``steered" electronically to target specific regions. 
This is in contrast to parabolic dish antennas, which resemble satellite dishes 
and are mechanically pointed to reflect radiation from a specific region toward a 
receiver placed in front of the dish. 

Figure~\ref{fig:aperture_array} illustrates how an aperture array is pointed. 
Radiation is received by different elements in the array at slightly different 
times, with a specific time delay corresponding to a specific location on the sky. 
The array can be focused to a particular location by adding the 
characteristic time delay for that location to the signals received by each 
antenna. 
In addition to providing a wide and flexible field of view (FOV), steering
the telescope in this way requires no moving parts, which greatly reduces 
hardware costs. 
However, electronic steering and wide FOVs require complicated  
and computationally-expensive signal processing and data reduction methods. 
Parabolic dishes have dominated radio astronomy since the late 1960s 
partly for this reason, but aperture arrays are currently experiencing a revival 
due to advances in signal processing, digital electronics, and high-performance 
computing that have made possible large arrays like the MWA and LOFAR \citep{Garrett12}. 

The MWA's novel hardware, signal processing backend, and observational capabilities 
are described by \citet{Tingay13}. 
It has an instantaneous bandwidth of 30.72 MHz that can be distributed in different 
configurations between 80 and 300 MHz. 
For solar observations, data are typically recorded with a 0.5 s time resolution and 
40 kHz spectral resolution. 
More details on the solar observing configuration and data reduction 
methods used in this thesis, along with solar science results from other studies, 
are given in Chapters~\ref{ch2}\,--\,\ref{ch4}. 


\subsection{Data Reduction}  
\label{data_reduce}

As mentioned in the previous section, reducing 
aperture array data is computationally expensive, 
and high-performance computers are required to 
turn the MWA's enormous volume of raw visibilities  
into science-ready images. 
The observations that are presented here were processed 
using the Pawsey Supercompting Centre. 
Incidentally, this facility is named for Joseph Pawsey, who made 
several important early contributions to solar radio astronomy, such as the 
identification of thermal emission from a million-degree corona and the localization of 
radio bursts to sunspot groups using sea interferometry \citep{Pawsey46b,Pawsey46,Pawsey53}. 

The data reduction methods used in this thesis were adapted from 
those used by the MWA's astronomical surveys, which required some modifications to suit 
solar observations. 
Chapter~\ref{ch2} \citep{McCauley17} will describe this procedure, which is further developed for 
polarimetry in Chapter~\ref{ch4} \citep{McCauley19}. 
The development and implementation of a semi-automated data processing pipeline, 
along with associated visualization and analysis software, occupied a 
significant fraction of my candidature, and this subsection 
briefly describes some aspects that are not documented elsewhere. 

Because of the computational requirements, observation periods with activity 
identified by other instruments were initially targeted for data reduction. 
A metadata catalog\footnote{\url{http://www.physics.usyd.edu.au/~pmcc8541/mwa/catalog/}} of solar observations was developed that associates 
each contiguous MWA observing window with radio bursts, flares, CMEs, 
and other events that are published in various event catalogs such as 
the NOAA event reports\footnote{\url{https://www.swpc.noaa.gov/products/solar-and-geophysical-event-reports}}, 
Heliophysics Event Knowledgebase (HEK; \citealp{Hurlburt12}), 
and CACTus CME catalog \citep{Robbrecht09}. 

120 five-minute observation periods were reduced from 82 different days in 2014 and 2015. 
Around half of these were chosen to include isolated Type III bursts and were 
imaged at the full 0.5-sec time resolution, while the other half 
targeted different observing days with the same ``picket fence" observing mode 
and were generally sampled at a 4-sec cadence.  
Including all of the frequency channels and polarization states, 
an archive of over 4.5 million images was compiled, 
corresponding to around 50,000 individual time steps. 
This was a significant computational expenditure, requiring around 1.5 million 
core hours on the Pawsey system. 
A further 750,000 core hours were awarded 
through the National Computational Merit Allocation Scheme (NCMAS) in 2019. 
This allocation is currently being used to target CMEs using the same pipeline, 
including further processing of serendipitous detections made with the initial archive, 
and some preliminary CME results are shown in Chapter~\ref{ch5}. 

\section{Research Aims and Outline}
\label{aims}

This thesis represents the first attempt to reduce and analyze large 
amounts of solar MWA data. 
As described in the previous section, the development of a 
data processing pipeline led to an archive of millions of images. 
Broadly speaking, my thesis aims to exploit this dataset, 
and the enclosed projects developed somewhat organically 
from what presented itself in the observations. 
Much more science can be supported 
by the data already reduced, 
along with the much larger volume of unprocessed data, 
and the archive developed for this thesis has already 
facilitated published and ongoing research beyond what is presented here. 

Observation periods that included isolated Type III bursts were 
initially targeted because they are common and of general interest 
to researchers at the University of Sydney, who have developed 
state-of-the-art theoretical simulations over many years (\eg~\citealp{Cairns00,Li06,Li08}). 
New Type III burst behavior was discovered soon after 
compiling several observations.
The characterization and interpretation of this behavior 
is the subject of Chapter~\ref{ch2} \citep{McCauley17}, which among other things, 
demonstrates the usefulness of these observations for probing 
magnetic field connectivities in the corona. 
In the course of developing a rough flux calibration method for 
Chapter~\ref{ch2}, previously-known but rarely-observed coronal hole 
behavior was also detected. 
Additional serendipitous coronal hole observations were 
then identified in the archive and analyzed by \citet{Rahman19}, whose 
results are summarized in Chapter~\ref{ch4}. 

Having collected a sample of Type III burst imaging observations, 
a small number of events at the limb with uncomplicated dynamics 
were employed to probe the coronal density structure. 
Chapter~\ref{ch3} \citep{McCauley18} is an updated repetition of classic experiments conducted  
using some of the earliest two-dimensional burst measurements. 
A novel addition is to relate Type III 
source heights to the increased extent of the quiescent corona 
over that of modern model predictions. 

Examination of the image archive also revealed the  
initial MWA detections of circularly-polarized sources on the Sun.
However, it was also immediately obvious that the polarimetric 
images frequently suffered from contamination that would need 
to be mitigated before the images could be used. 
While the calibration artefacts responsible for this effect were 
known from other MWA studies, the existing mitigation techniques  
could not be directly applied to solar observations. 
Different methods were explored and one ultimately proved successful, 
leading to the first low-frequency spectropolarimetric imaging sensitive enough to  
detect the weak polarization signals from thermal bremsstrahlung emission. 
Chapter~\ref{ch4} \citep{McCauley19} introduces the mitigation strategy and surveys the 
range of circular polarization features found in over 100 observing runs. 

Conclusions from each chapter are summarized in Chapter~\ref{ch5}, as well as 
discussions of open questions and future work ideas 
that can be addressed with the existing dataset produced for this thesis.
These include natural extensions of the work on Type III bursts, 
coronal holes, and polarimetry, along with a preview of a 
CME discovery that has yet to be reported in the literature.

%% file: chapter2/chapter2.tex
\pagestyle{fancy}

\providecommand{\edit}[1]{{\color{black}{#1}}}

\chapter[Type III Solar Radio Burst Source Region Splitting Due to a Quasi-Separatrix Layer]{Type III Solar Radio Burst \\ Source Region Splitting Due to a \\ Quasi-Separatrix Layer}\label{ch2}

\fancyhead[OR]{~}\fancyhead[EL]{\bf Ch. \thechapter~Type III Burst Source Splitting} 

{\large{}Published as \citet{McCauley17}, \href{http://ui.adsabs.harvard.edu/abs/2017ApJ...851..151M}{\textit{Astrophys. J.}, 851:151}}

\section{Abstract}

We present low-frequency (80--240 MHz) radio imaging of Type III solar radio bursts observed  
by the \textit{Murchison Widefield Array} (MWA) on 2015/09/21.
The source region for each burst splits from one dominant component at higher frequencies into two increasingly-separated 
components at lower frequencies. 
For channels below $\sim$132 MHz, the two components repetitively diverge 
at high speeds (0.1--0.4 c) along directions tangent to the limb, with each episode 
lasting just $\sim$2 s.
We argue that both effects result from the strong magnetic field connectivity gradient that the 
burst-driving electron beams move into. 
Persistence mapping of extreme ultraviolet (EUV) jets observed by 
the \textit{Solar Dynamics Observatory} reveals quasi-separatrix layers (QSLs) associated with coronal null points, 
including separatrix dome, spine, and curtain structures. 
Electrons are accelerated at the flare site toward an open QSL, where the beams follow diverging field lines to produce the source splitting, with larger separations at larger heights (lower frequencies). 
The splitting motion within individual frequency bands is interpreted as a projected time-of-flight effect, whereby electrons 
traveling along the outer field lines take slightly longer to excite emission at adjacent positions. 
Given this interpretation, we estimate an average beam speed of 0.2 c. 
We also qualitatively describe the quiescent corona, noting in particular that 
a disk-center coronal hole transitions from being dark at higher 
frequencies to bright at lower frequencies, turning over around 120 MHz. 
These observations are compared to synthetic images based on the 
Magnetohydrodynamic Algorithm outside a Sphere (MAS) model, which we use to flux-calibrate the burst data. 


\section{Introduction}
\label{intro}

Type III solar radio bursts are among the principal signatures of magnetic reconnection, the process thought to  
underlie solar flares. 
Their high brightness temperatures demand a coherent, nonthermal emission 
mechanism that is generally attributed to plasma emission stimulated by semi-relativistic electron beams.
Electrons accelerated at the reconnection site generate Langmuir waves (plasma oscillations) in the ambient plasma through the 
bump-on-tail beam instability. 
Those Langmuir waves then shed a small fraction of their energy in radio emission near the 
fundamental plasma frequency ($f_{\rm p}$) or its second harmonic. 
This theory was proposed by \citet{Ginzburg58} and has since been developed by many authors (see reviews by \citealt{Robinson00,Melrose09}). 

Radio bursts are classified by their frequency drift rates,  
and Type IIIs are so named because they drift faster than Types I and II \citep{Wild50}. 
A recent review of Type III literature is provided by \citet{Reid14}. 
Starting frequencies are typically in the 100s of MHz, and because 
the emission frequency is proportional to the 
square of the ambient electron density ($f_{\rm p}~\propto{}~\sqrt{n_{e}}$), 
standard Type III radiation drifts to lower frequencies as the accelerated electrons stream outward. 
\textit{Coronal} Type III bursts refer to those that drift down to tens of MHz or higher. 
Beams that 
escape along open field lines may continue to stimulate Langmuir waves in the solar wind plasma, 
producing \textit{interplanetary} Type III bursts that may reach 20 kHz and below around 1 AU and beyond. 
We will focus on coronal bursts for which some fraction of the electrons do escape to produce 
an interplanetary Type III.  

X-ray flares and Type III bursts have been linked by many studies. 
Various correlation rates have been found, with a general trend toward increased association with better instrumentation.  
Powerful flares ($\ge{}$C5 on the GOES scale) almost always generate coherent radio emission, generally meaning a  
Type III burst or groups thereof \citep{Benz05, Benz07}. 
Weaker flares may or may not have associated Type IIIs depending on the magnetic field configuration \citep{Reid17}, 
and Type IIIs may be observed with no GOES-class event if, for instance,  
the local X-ray production does not sufficiently enhance the 
global background \citep{Alissandrakis15}. 
Flares that produce X-ray or extreme ultraviolet (EUV) jets are frequently    
associated with Type III emission \citep{Aurass94,Kundu95,Raulin96,Trottet03,Chen13b,Innes16,Mulay16,Hong17,Cairns18}. 
Such jets are collimated thermal plasma ejections that immediately follow,  
are aligned with, and are possibly heated by 
the particle acceleration responsible for radio bursts \citep{Saint-Hilaire09,Chen13}. 
We will exploit the alignment between EUV jets and Type III 
electron beams to develop an understanding of radio source region behavior  
that, to our knowledge, has not been previously reported. 

This is the first Type III imaging study to use the full 128-tile 
\textit{Murchison Widefield Array} (MWA; \citealt{Lonsdale09,Tingay13}), which 
follows from Type III imaging presented by \citet{Cairns18} using the 32-tile prototype array. 
The MWA's primary 
science themes are outlined by \citet{Bowman13}, 
and potential solar science is further highlighted by \citet{Tingay13b}. 
The first solar images using the prototype array and later the full array are detailed by \citet{Oberoi11} 
and \citet{Oberoi14}, respectively.
\citet{Suresh17} present a statistical study of single-baseline dynamic spectra, 
which exhibit the lowest-intensity solar radio bursts ever reported. 
We present the first time series imaging. 

Along with the \textit{Low Frequency Array} (LOFAR; \citealt{van13,Morosan14}), the MWA 
represents a new generation of low frequency interferometers capable of solar imaging.
Previous imaging observations at the low end of our frequency range were made by the 
decommissioned \textit{Culgoora} \citep{Sheridan72,Sheridan83} 
and \textit{Clark Lake} \citep{Kundu83} radioheliographs, along with the 
still-operational \textit{Gauribidanur Radioheliograph} \citep{Ramesh98, Ramesh05}. 
The high end of the MWA's frequency range overlaps with 
the \textit{Nan\c{c}ay Radioheliograph} (NRH; \citealt{Kerdraon97}), which has 
facilitated a number of Type III studies referenced here.
 
This paper is structured as follows. 
Section~\ref{obs} describes our observations and data reduction procedures. 
Our analyses and results are detailed in Section~\ref{analysis1}. 
Section~\ref{forward} considers the quiescent corona outside burst periods, which we compare 
to synthetic images used to flux calibrate the burst data in Section~\ref{flux}.
Section~\ref{kinematics} characterizes the Type III source region structure and motion, 
and the local magnetic field configuration is inferred using EUV observations in Section~\ref{persistence}. 
In Section~\ref{discussion1}, our results are combined to produce an interpretation of the  
radio source region behavior. 
Section~\ref{conclusion1} provides concluding remarks. 


\section{Observations}
\label{obs}

We focus on a brief series of Type III bursts associated with a C8.8 flare that peaked at 05:18 UT on 
2015/09/21. 
The flare occurred in Active Region 12420\footnote{AR 12420 summary: \url{https://www.solarmonitor.org/index.php?date=20150921&region=12420}} 
on the east limb. 
This investigation began by associating MWA observing periods that utilize the 
mode described in Section~\ref{mwa1} with isolated Type III bursts logged   
in the National Oceanic and Atmospheric Administration (NOAA) solar event 
reports\footnote{NOAA event reports: \url{http://www.swpc.noaa.gov/products/solar-and-geophysical-event-reports}}. A small sample of bursts detected from 80 to 240 MHz were selected, 
and we chose this event for a case study because of the unusual source structure and motion. 
A survey of other Type III bursts is ongoing. 


  \begin{figure}\graphicspath{{chapter2/}}
\centering
\includegraphics[width = 0.6\textwidth]{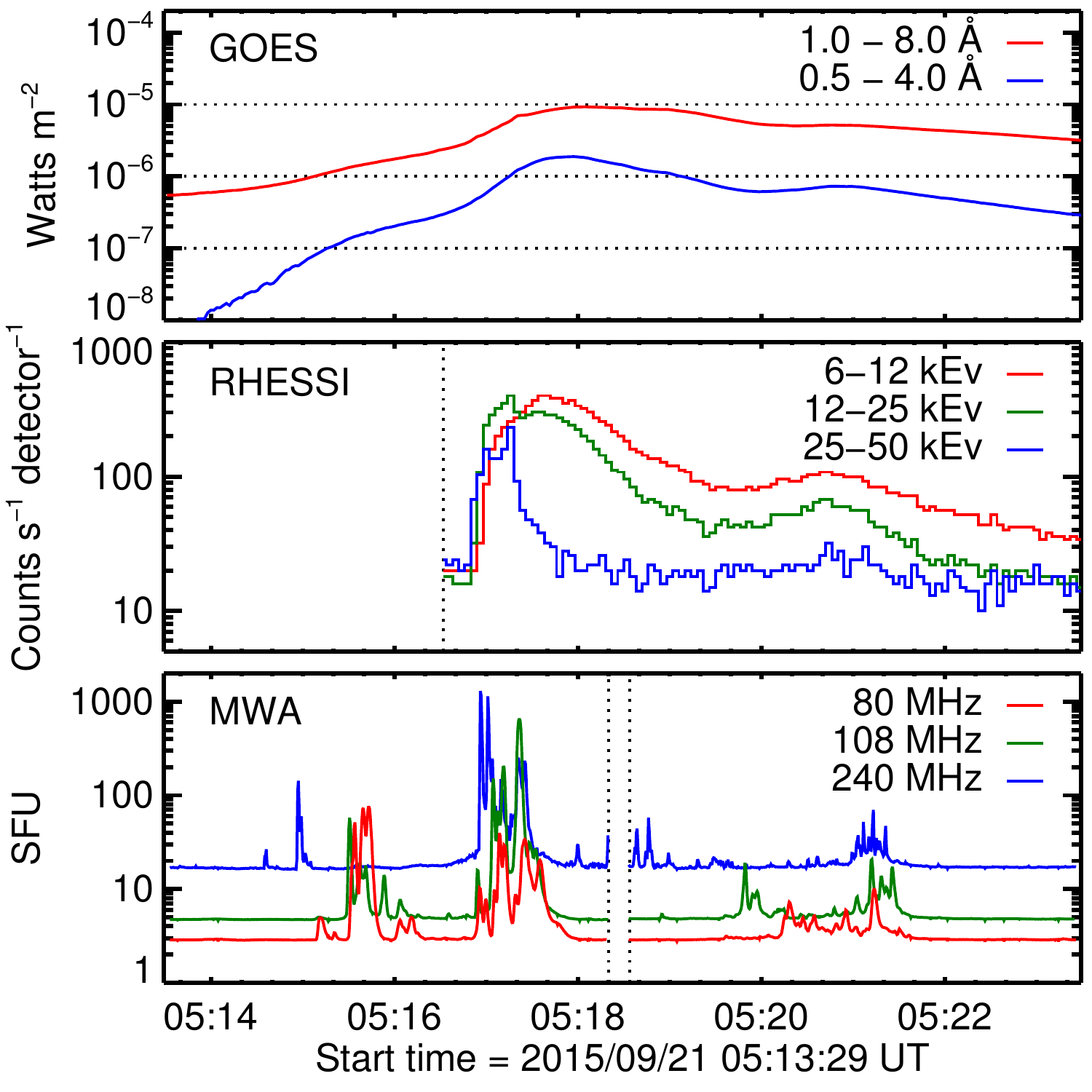}
 \caption{\footnotesize{}
\textit{Top}: GOES soft X-ray light curves, showing the C8.8 flare that peaked at 05:18 UT. Dotted lines 
from bottom to top indicate the B, C, and M-class thresholds. 
\textit{Middle:} RHESSI count rates from 6--50 kEv. The dotted line indicates the end of RHESSI's night (Earth-eclipse) period. 
\textit{Bottom:} MWA light curves at 80, 108, and 240 MHz. Dotted lines indicate the transition between continuous 
observing periods. 
}
 \label{fig:lightcurves}
 \end{figure}

Figure~\ref{fig:lightcurves} shows the soft X-ray (SXR) light curves from the 
\textit{Geostationary Operational Environmental Satellite} (GOES\footnote{GOES X-ray flux: \url{http://www.swpc.noaa.gov/products/goes-x-ray-flux}})
for our MWA observation period, along 
with those from the \textit{Reuven Ramaty High-Energy Solar Spectroscopic Imager} (RHESSI; \citealt{Lin02}). 
The corresponding MWA light curves, as derived in Section~\ref{mwa1} and Section~\ref{forward}, show that the radio bursts occur primarily around the hard X-ray (HXR, 25--50 keV) peak and just before the SXR peak, with some minor radio bursts scattered throughout the SXR rise and decay phases. 
HXR and Type III emissions are known to be approximately coincident in time \citep{Arzner05} and  
are generally attributed to oppositely-directed particle acceleration, with HXR 
production resulting from heating by the sunward component. 
The same process may underlie both, however small differences in the timing, along with  
large differences in the requisite electron populations, suggest there may be multiple related acceleration processes (e.g. \citealt{Brown77,Krucker07,White11,Cairns18}). 
In contrast, SXR emission is associated with thermal plasma below the reconnection site, generally peaking somewhat later with a 
more gradual profile as in Figure~\ref{fig:lightcurves}.

\begin{figure}\graphicspath{{chapter2/}}
\centering
\includegraphics[width = 0.65\textwidth]{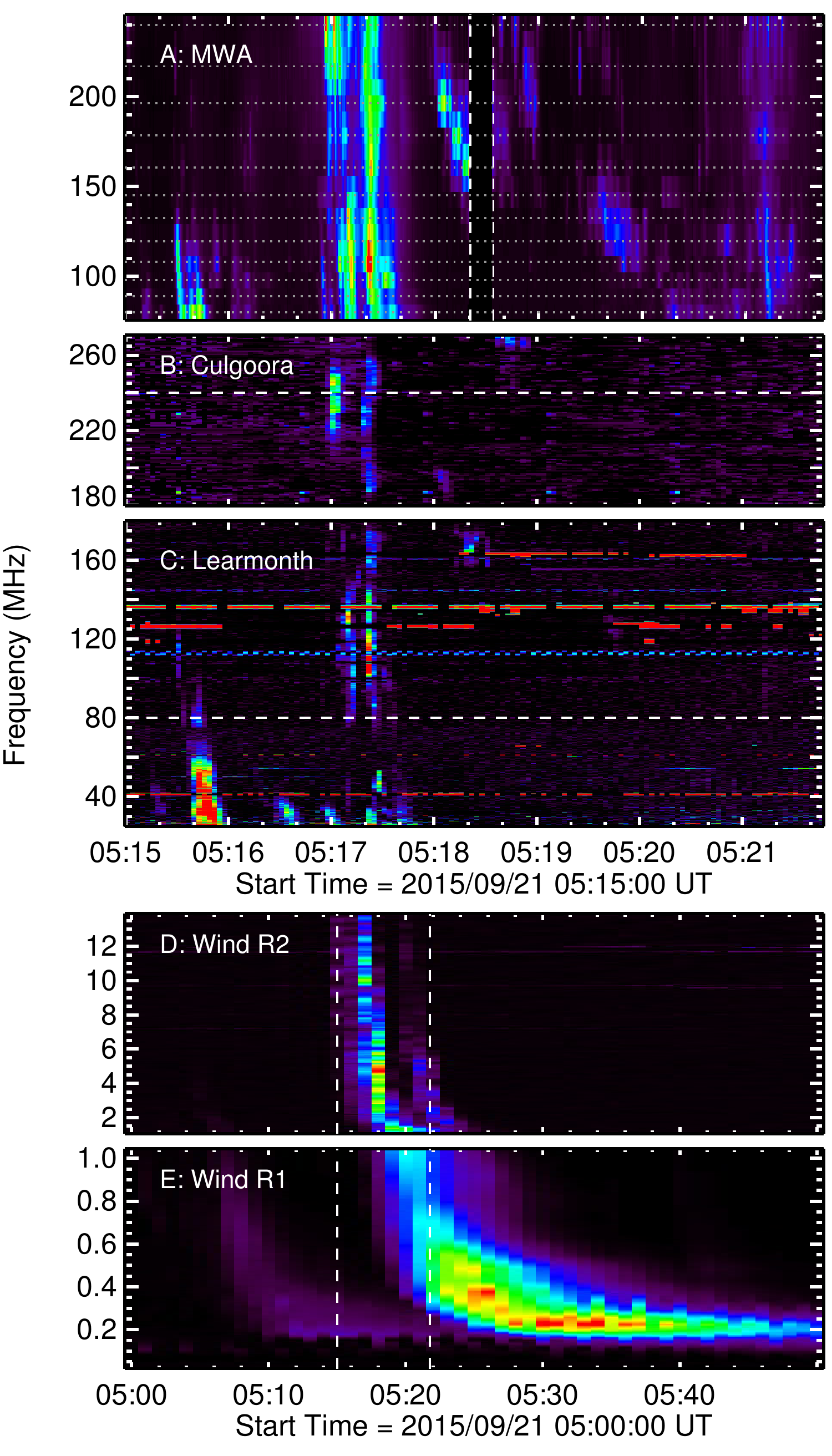}
 \caption{\footnotesize{}
\edit{\textit{A}}: MWA dynamic spectrum (DS) produced from total image intensities 
and interpolated to a spectral resolution equal to the minimum separation between observing bandwidths (see Section~\ref{mwa1}.) 
Dashed vertical lines indicate the transition between continuous observing periods, and dotted horizontal lines mark 
the 12, 2.56 MHz-wide frequency channels.
\edit{\textit{B--C}}: Culgoora and Learmonth DS. 
Dashed lines indicate the MWA frequency coverage bounds (80--240 MHz).
\edit{\textit{D--E}}: \textit{Wind}/WAVES RAD\edit{2 and RAD1} DS. Note that the time axis is expanded to show the 
low-frequency tail. The dashed \edit{lines indicate the period covered by panels A--C}.
All DS are log-scaled and then background-subtracted.
A corresponding movie is available in the 
\href{https://iopscience.iop.org/article/10.3847/1538-4357/aa9cee}{online material}.
}
 \label{fig:spectra1}
 \end{figure}

Our initial radio burst detections relied on observations from the \textit{Learmonth} and \textit{Culgoora} solar radio spectrographs. 
Part of the global \textit{Radio Solar Telescope Network}\footnote{RSTN data: \url{ftp://ftp.ngdc.noaa.gov/STP/space-weather/solar-data/solar-features/solar-radio/rstn-spectral}} (RSTN; \citealt{Guidice81}), the \textit{Learmonth} spectrograph covers 25 to 180 MHz 
in two 401-channel bands that run from 25--75 and 75--180 MHz. Additional technical details are 
provided by \citet{Kennewell03}. 
The \textit{Culgoora} spectrograph\footnote{Culgoora data: \url{ftp://ftp-out.sws.bom.gov.au/wdc/wdc_spec/data/culgoora/}} \citep{Prestage94} 
has broader frequency coverage (18--1800 MHz) over four 501-channel bands.
Only the 180--570 MHz band is relevant here, and we show just a portion of it because the \textit{Learmonth} spectrograph 
is more sensitive where they overlap. 
Both instruments perform frequency sweeps every 3 s. 
Dynamic spectra are plotted in Figure~\ref{fig:spectra1}, each being log-scaled and   
background-subtracted by 5-min boxcar averages. 


Figure~\ref{fig:spectra1} also includes \edit{dynamic spectra} from the 
\textit{Radio and Plasma Wave Investigation} (WAVES; \citealt{Bougeret95}) 
on the \textit{Wind} spacecraft. These data demonstrate an 
interplanetary component to the coronal Type III bursts, which requires there be connectivity to open field lines 
along which electrons escaped the corona. This will be important to our interpretation of the 
magnetic field configuration in Section~\ref{discussion1}. 


\subsection{\textit{Murchison Widefield Array} (MWA)}
\label{mwa1}

\begin{figure*}\graphicspath{{chapter2/}}
\centering
\includegraphics[width = 1.0\textwidth]{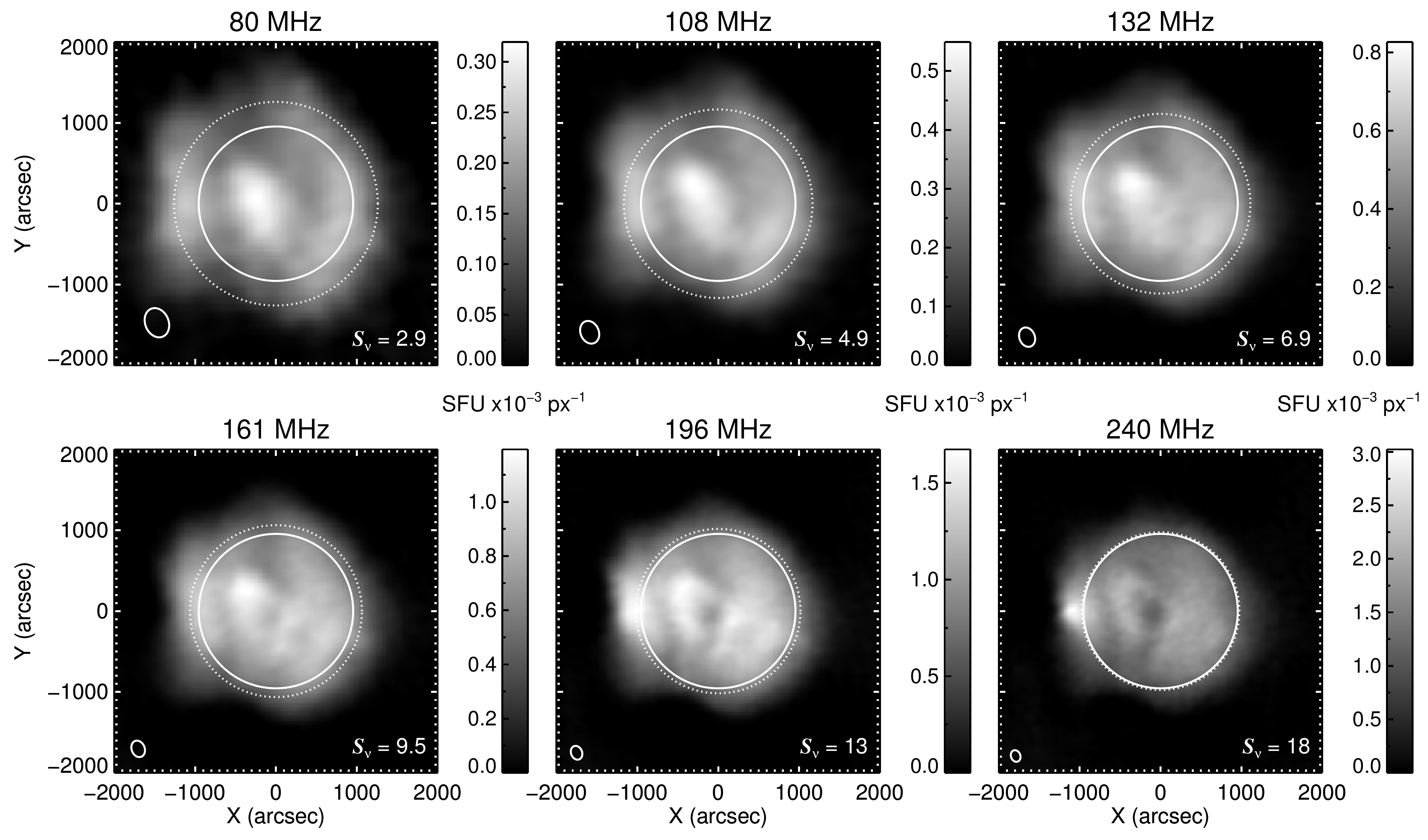}
 \caption{\footnotesize{}
MWA Stokes I images for 6 of the 12 frequency bands during a quiescent period at 2015/09/21 05:13:33.20 UT.
The solid inner circles denote the optical disk, and the dotted outer circles denote the 
Newkirk-model \citep{Newkirk61} limb for a given frequency. Ellipses in the bottom-left corners 
represent the synthesized beams. Values in the bottom-right corners are full-Sun integrated flux densities ($S_{\nu}$) in SFU, 
and the color bars represent the flux density enclosed by each \edit{20''} pixel in SFU$\times{}10^{-3}$ (see Section~\ref{flux} for details). 
A movie showing the full time series for all 12 bands 
is available in the 
\href{https://iopscience.iop.org/article/10.3847/1538-4357/aa9cee}{online material}. 
 }
 \label{fig:preburst}
 \end{figure*}

\begin{figure*}\graphicspath{{chapter2/}}
\centering
\includegraphics[width = 1.0\textwidth]{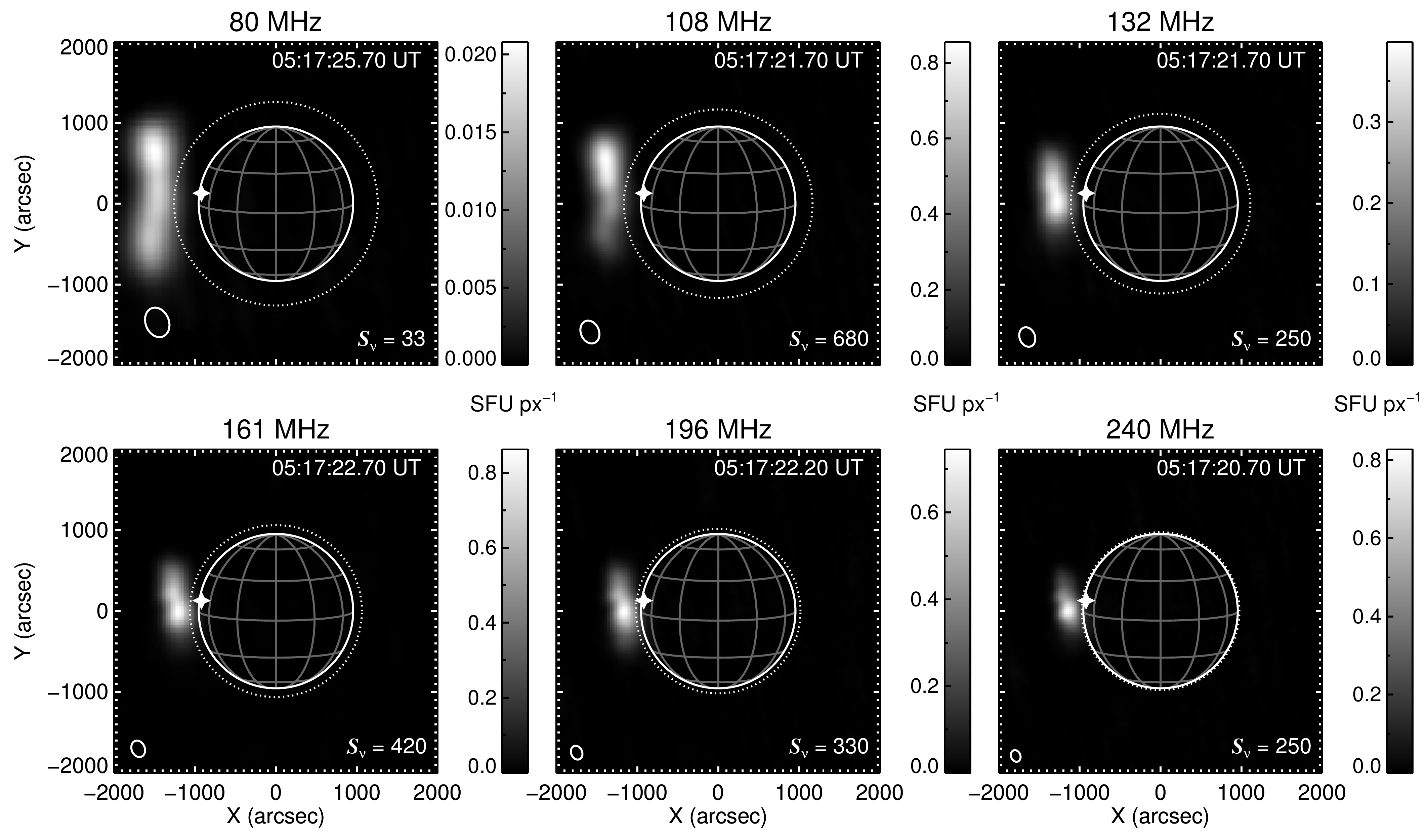}
 \caption{\footnotesize{}
Same as Figure~\ref{fig:preburst} but for the \edit{frequency-specific peak intensity times associated with 
the event from 05:17:20 to 05:17:25 UT, which may comprise multiple overlapping bursts (see Section~\ref{kinematics} \& Section~\ref{discussion1}}). 
Color bar units \edit{are} in SFU\edit{ px\tsp{-1}, and} stars mark the X-ray flare site.
 }
 \label{fig:burst}
 \end{figure*}

The MWA is a low-frequency radio interferometer in Western Australia that consists of 
128 aperture arrays (``tiles"), each comprised of 16 dual-polarization dipole antennas \citep{Tingay13}.
It has an instantaneous bandwidth of 30.72 MHz that can be spread flexibly from 80 to 300 MHz. 
Our data employ a ``picket fence" observing mode, whereby twelve 2.56 MHz bands are 
distributed between 80 and 240 MHz with gaps of 9--23 MHz between them. 
This configuration is chosen to maximize spectral coverage while 
avoiding radio frequency interference (RFI).
Data are recorded with a time resolution of 0.5 s and a spectral resolution of 40 kHz, which we 
average across the 2.56 MHz bandwidths to produce images centered at 
80, 89, 98, 108, 120, 132, 145, 161, 179, 196, 217, and 240 MHz.
Figures \ref{fig:preburst} and \ref{fig:burst} 
show images at six frequencies during quiescent and burst phases, respectively, and 
a movie showing all twelve bands over the full time series is available in the 
\href{https://iopscience.iop.org/article/10.3847/1538-4357/aa9cee}{online material}\footnote{Movies available at \url{https://iopscience.iop.org/article/10.3847/1538-4357/aa9cee}}. 

Visibilities were produced using the standard MWA correlator \citep{Ord15} and \textsf{cotter} \citep{Offringa15}. 
For our calibrator observations, this included 8-s time averaging and RFI flagging 
using the \textsf{aoflagger} algorithm \citep{Offringa12}. 
RFI flagging was disabled for the solar observations, as it tends to flag out burst data. 
Calibration solutions for the complex antenna gains were obtained 
with standard techniques \citep{Hurley14} using observations 
of a bright and well-modelled calibrator source (Centaurus A) 
made $\sim$2 hours after the solar observations. 
To improve the calibration solutions, the calibrator was imaged and ten loops  
of self-calibration were performed in the manner described by \citet{Hurley17}. 

This last step is typically performed on science target images, but we  
apply it instead to the calibrator for two reasons. 
First, we find that day-time observations generally produce inferior calibration solutions 
compared to analogous night-time data. 
We attribute this to contamination of the calibrator field 
by sidelobe emission from the Sun, but ionospheric and temperature effects 
may also be important. 
Second, the \textsf{clean} algorithm 
essential to the self-calibration process works best when the field is dominated by compact, 
point-like sources, which is not the case for the Sun.
The same steps performed on our solar images tended to degrade the overall quality of the calibration 
solutions and bias the flux distribution of the final images. 
However, we find that it is best to self-calibrate on the field source to obtain 
quality polarimetry because transferring calibration solutions from a lower-elevation 
pointing typically produces overwhelming Stokes $I$ leakage into the other Stokes portraits. 
For this reason, we do not include polarimetry here. 
Progress has been made on producing reliable polarimetric images of the Sun with the MWA, as well as 
improving the dynamic range, but that is beyond the scope of this paper.


Once calibrated, imaging for each 0.5 s integration is accomplished using \textsf{WSClean} \citep{Offringa14} with the 
default settings except where noted below. 
Frequencies are averaged over each 2.56 MHz bandwidth, excluding certain fine channels 
impacted by instrumental artefacts. 
To emphasize spatial resolution, we use the Briggs -2 weighting scheme \citep{Briggs95}. 
Cleaning is performed with $\sim$10 pixels across the synthesized beam, 
yielding 16--36'' px\tsp{-1} from 240--80 MHz. 
We use a stopping threshold of 0.01, which is roughly the average 
RMS noise level in arbitrary units obtained for quiescent images cleaned with no threshold. 
Major clean cycles are used with a gain of 0.85 (\textsf{-mgain 0.85}), and peak finding uses the 
quadrature sum of the instrumental polarizations (\textsf{-joinpolarizations}). 
Finally, Stokes I images are produced using the primary 
beam model described by \citet{Sutinjo15}. 

To compare MWA data with other solar imaging observations, we introduce the \textsf{mwa\_prep} 
routine, now available in the 
SolarSoftWare libraries for IDL 
(SSW\footnote{SSW: \url{https://www.lmsal.com/solarsoft/}}, \citealt{Freeland98}). 
\textsf{WSClean} and the alternative MWA imaging tools produce FITS images using the   
\textsc{sin}-projected celestial coordinates standard in radio astronomy. 
Solar imaging data typically use ``helioprojective-cartesian" coordinates, which is a \textsc{tan} projection 
aligned to the solar rotation axis with its origin at Sun-center \citep{Thompson06}. 
To convert between the two coordinate systems, \textsf{mwa\_prep} rotates the image about 
Sun-center by the solar P angle, interpolates onto a slightly different grid to account for the 
difference between the \textsc{sin} and \textsc{tan} projections, and scales 
the images to a uniform spatial scale (20'' px\tsp{-1}). 
By default, the final images are cropped to 6$\times$6 $\rm{R}_{\odot}$, yielding 289$\times$289 pixels. 
FITS headers are updated accordingly, after which the various SSW mapping tools can 
be used to easily overplot data from different instruments.

We will consider quiescent radio structures in Section~\ref{forward} against corresponding 
model images that are used for flux calibration in Section~\ref{flux}. 
Burst structure and dynamics are discussed 
in Section~\ref{kinematics}.  


  \begin{figure*}\graphicspath{{chapter2/}}
\centering
\includegraphics[width = 1.0\textwidth]{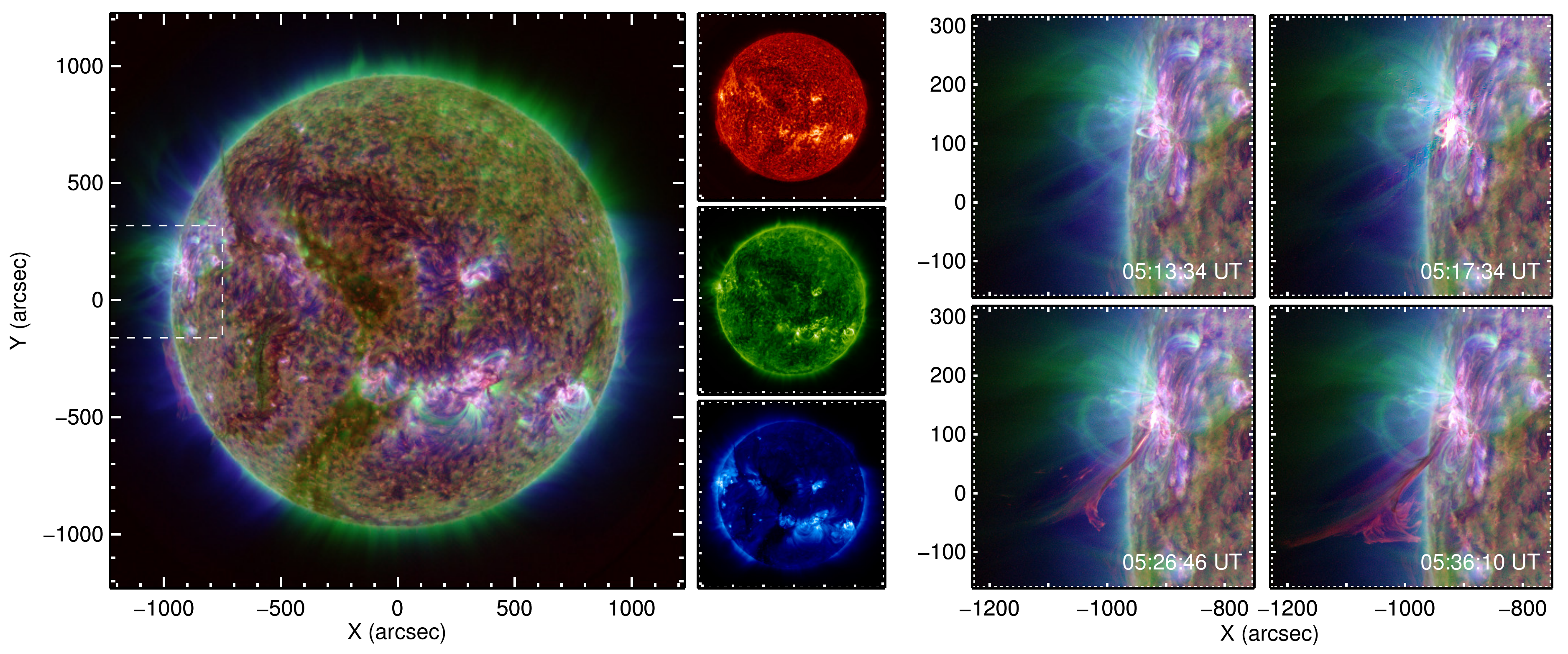}
 \caption{\footnotesize{}
An overview of the event seen by SDO/AIA using RGB composites of the 
304, 171, and 211 \AA{} channels. 
The top panels on the right half show nearly the same times as 
Figures~\ref{fig:preburst} (left) and~\ref{fig:burst} (right), with the rightmost panel corresponding to 
just before the SXR peak. 
The bottom-right panels show snapshots of the EUV jets that follow 
the radio bursts. 
 }
 \label{fig:aia}
 \end{figure*}


\subsection{\textit{Solar Dynamics Observatory} (SDO)}
\label{sdo}

The \textit{Solar Dynamics Observatory} (SDO, \citealt{Pesnell12}) is a satellite with three instrument suites, 
of which we use the \textit{Atmospheric Imaging Assembly} (AIA; \citealt{Lemen12}). 
We also indirectly use photospheric magnetic field observations from the \textit{Helioseismic and Magnetic Imager} (HMI; \citealt{Scherrer12}), 
which inform the synthetic images in Section~\ref{forward}. 
The AIA is a full-Sun imager consisting of four telescopes that observe 
in seven narrowband EUV channels with a 0.6''{} px$^{-1}$ spatial resolution and 12 s cadence, 
along with three UV bands with a lower cadence. 

Calibrated (``level 1") data are obtained from the Virtual Solar Observatory 
(VSO\footnote{VSO: \url{http://sdac.virtualsolar.org/}}, \citealt{Hill09}). 
The SSW routine \textsf{aia\_deconvolve\_richardsonlucy} is used to deconvolve the images 
with filter-specific point spread functions, and \textsf{aia\_prep} is used to co-align and uniformly 
scale data from the different telescopes. 
Figure~\ref{fig:aia} presents an overview of our event using RGB composites of the 304, 171, and 211 \AA{} channels. 
These bands probe the chromosphere, upper transition region / low corona, and corona, respectively, 
with characteristic temperatures of .05 (He II), 0.63 (Fe IX), and 2 MK (Fe XIV). 

The AIA observations show a fairly compact flare that produces several distinct EUV jets beginning 
just before the soft X-ray peak at 05:18 UT. This includes higher-temperature material visible in up to the 
hottest band (94 \AA{}, 6.3 MK), along with cooler ejecta at chromospheric temperatures that appears 
in emission at 304 \AA{} and in absorption at other wavelengths. 
These outflows reveal a complex magnetic field configuration south of the flare site, which we 
will explore in Section~\ref{persistence} and in Section~\ref{discussion1} with respect to the radio emission. 


\section{Analysis \& Results}
\label{analysis1}


\subsection{Quiescent Structure and Model Comparison}
\label{forward}

We examine model images of the coronal intensity at MWA frequencies to
qualitatively compare the expected and observed structures outside 
of burst periods. 
In the next subsection, we also use the predicted quiescent flux densities to obtain 
a rough flux calibration of our burst data. 
Synthetic Stokes I images are obtained using 
FORWARD\footnote{FORWARD: \url{https://www2.hao.ucar.edu/modeling/FORWARD-home}}, 
an SSW package 
that can generate a variety of coronal observables using different magnetic field and/or thermodynamic models. 
At radio wavelengths, FORWARD computes the expected contributions from thermal bremsstrahlung (free-free) 
and gyroresonance emission based on the modeled temperature, density, and magnetic field structure. 
Details on those calculations, along with the package's other capabilities, are 
given by \citet{Gibson16}.  


  \begin{figure*}\graphicspath{{chapter2/}}
\centering
\includegraphics[width = 1.0\textwidth]{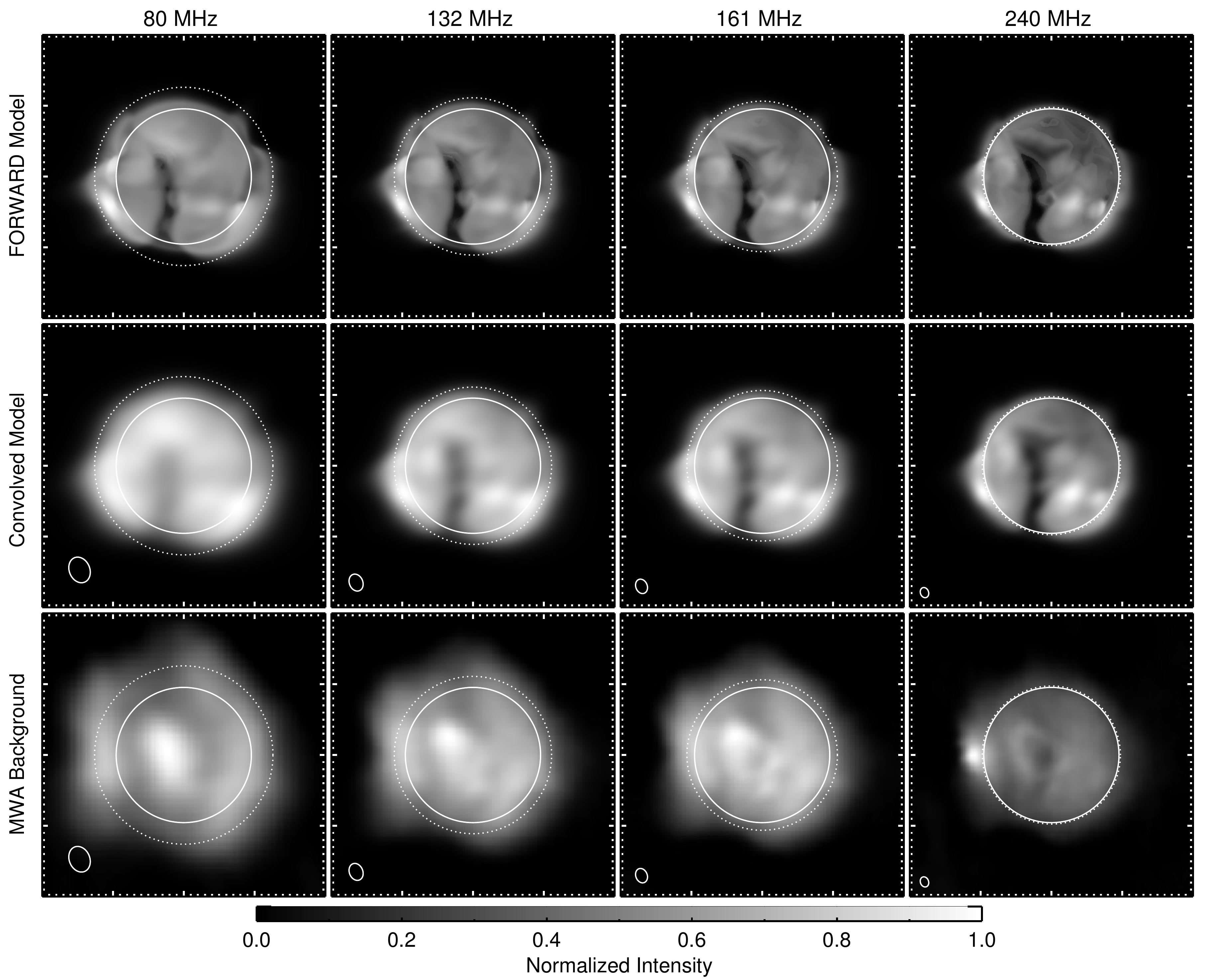}
 \caption{\footnotesize{}
\textit{Top}: Expected free-free and gyroresonance emission at four frequencies 
predicted by FORWARD based on the MAS thermodynamic MHD model.
\textit{Middle}: Model images convolved with the corresponding MWA beams. 
\textit{Bottom}: Median MWA emission outside burst periods over the first 4-min observation period, which is assumed to be 
the quiescent background for flux calibration. 
Plot axes and annotations are as in Figure~\ref{fig:preburst}. 
An animation with all 12 channels is available in the \href{https://iopscience.iop.org/article/10.3847/1538-4357/aa9cee}{online material}. 
 }
 \label{fig:forward1}
 \end{figure*}

Our implementation uses the Magnetohydrodynamic Algorithm outside a Sphere 
(MAS\footnote{MAS: \url{http://www.predsci.com/hmi/data\_access.php}}; \citealt{Lionello09}) 
medium resolution \\ (\textsf{hmi\_mast\_mas\_std\_0201}) model. 
The MAS model combines an MHD extrapolation of the coronal magnetic field (e.g. \citealt{Miki99}) based on 
photospheric magnetogram observations from the HMI with a heating model adapted from \citet{Schrijver04}. 
Comparisons between MAS-predicted images and data have been 
made a number of times for EUV and soft X-ray observations,  
with generally good agreement for large-scale structures (e.g. \citealt{Riley11,Reeves11,Downs12}). 
We make the first radio comparisons. 


  \begin{figure}\graphicspath{{chapter2/}}
\centering
\includegraphics[width = 0.7\textwidth]{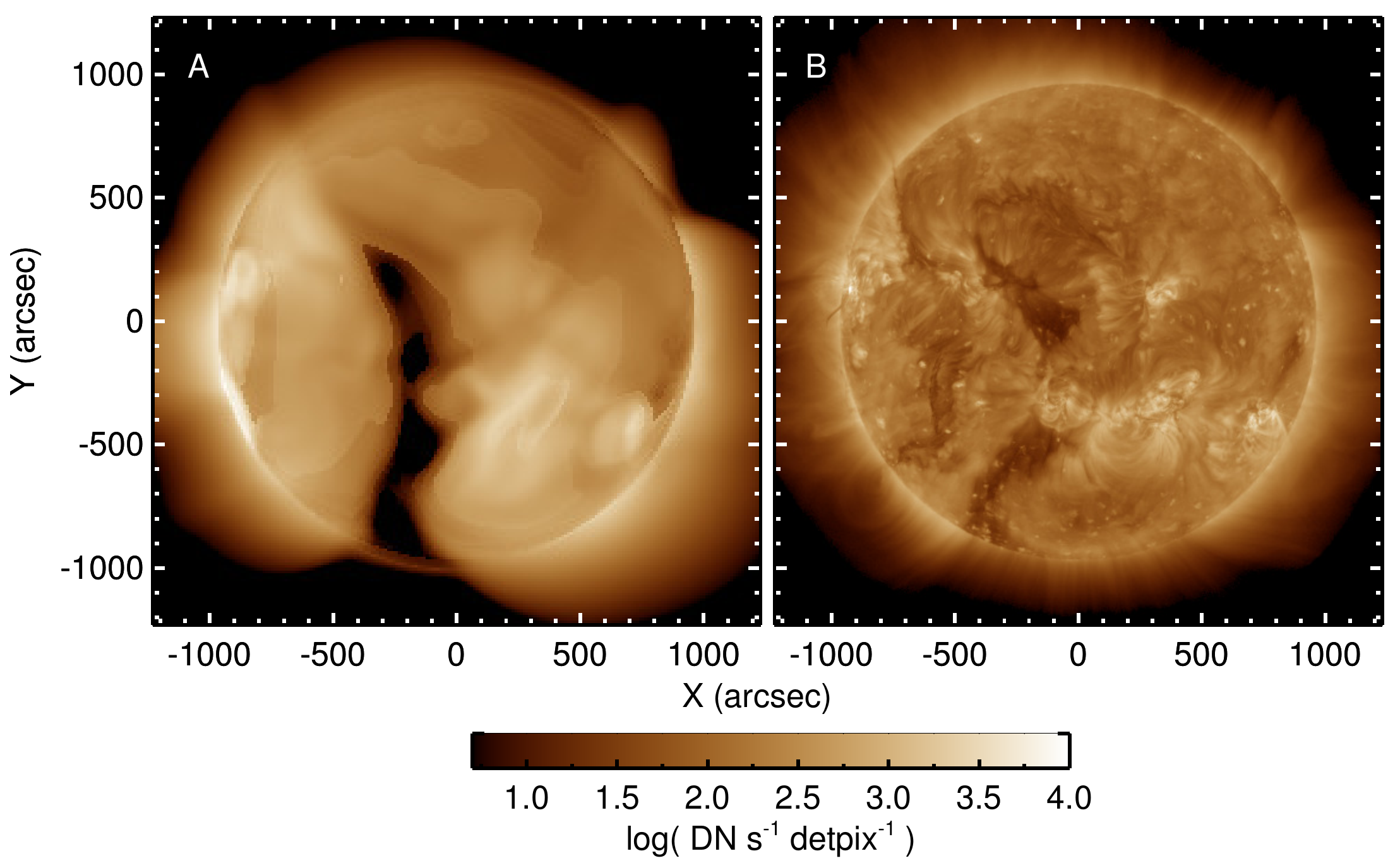}
 \caption{\footnotesize{}
193 \AA{} synthetic image (A) and SDO observation (B). The synthetic image applies the 
telescope response function so that both images are plotted on exactly the same scale 
in instrumental units (DN) per sec per detector pixel (detpix).  
 }
 \label{fig:forward_aia}
 \end{figure}

The top row of Figure~\ref{fig:forward1} shows synthetic images at four MWA frequencies.
Beam-convolved versions are shown in the middle row, but note that this does not 
account for errors introduced by the interferometric imaging process, such as effects 
related to deconvolving a mixture of compact and diffuse emission or to 
nonlinearities in the \textsf{clean} algorithm.  
MWA data are shown in the bottom row and 
reflect median pixel values over the first five-minute observation (05:13:33 to 05:18:20),   
excluding burst periods defined as when the total image intensities exceed 105\% of the first 0.5 s integration 
for each channel. 
An animation with all 12 channels is available in the 
\href{https://iopscience.iop.org/article/10.3847/1538-4357/aa9cee}{online material}. 
For context, we also show a comparison of a 193 \AA{} SDO observation and prediction  
using the same model in Figure~\ref{fig:forward_aia}. 
 
The agreement between the observed and modelled radio images is best at our highest 
frequencies ($\gtrsim$ 179 MHz), where the correspondence is similar to that of the EUV case. 
For both, the model reproduces structures associated with coronal holes near the central meridian and 
the large active region complexes in the southwest. 
The large-scale structure associated with the southern polar coronal hole 
is also well-modelled for the radio case.
A similar structure is predicted for the EUV but is disrupted by the 
observed polar plumes in the manner described by \citet{Riley11}. 
The modelled images also under-predict emission from EUV coronal holes, 
which may be due to contributions from low-temperature ($<$ 500,000 K) material 
ignored by the emissivity calculations.
Other contributing factors might be inaccuracies in the heating model, evolution 
of the magnetic boundary from that used for the simulation, or 193 \AA{} emission 
from non-dominant ions formed at low temperatures. 

A number of discrepancies between the model and MWA observations are also apparent, 
particularly with decreasing frequency. 
With the exception of the bright region on the east limb at 240 MHz, which we will revisit in 
Section~\ref{discussion1}, we suspect these differences underscore the importance of 
propagation effects to the appearance of the corona at low frequencies. 
In particular, refraction (ducting) of radio waves as they encounter low-density regions,  
as well as scattering by density inhomogeneities, can profoundly alter the observed source 
structure (see reviews by \citealt{Lantos99,Shibasaki11}). 
Both effects can increase a source's spatial extent, decrease its brightness, and   
alter its apparent location (e.g. \citealt{Aubier71,Alissandrakis94,Bastian94,Thejappa08,Ingale15}).
We likely see the effects of scattering and/or refraction in the increased radial extent of 
the observed emission at all frequencies compared to the beam-convolved model images, though 
an enhanced density profile may also contribute. Likewise, these propagation effects 
may be responsible for dispersing the signatures of the southwestern active regions, which 
are prominent in the synthetic images but only barely discernible in our  
observations. 

Most conspicuously, the disk-center coronal hole gradually transitions from a dark feature at high frequencies 
to a bright one at low frequencies in the observations but not in the synthetic data. 
This could be due to the diminished spatial resolution at low frequencies, 
meaning the coronal hole signature is  
swamped by emission from the bright region to the northeast. 
However, that effect should serve only to reduce the coronal hole contrast,  
as it does for the beam-convolved synthetic images. 
Indeed, another set of observations of a different disk-center coronal hole also show this 
dark-to-bright transition from high to low frequencies with even less ambiguity. 
In both cases, the transition is gradual and turns over around 120 MHz. 
Above the $\sim$120 MHz transition we observe, coronal holes are consistently 
reported as intensity depressions (e.g. \citealt{Mercier12}), which is expected given their low densities. 
At longer wavelengths, coronal holes have sometimes been seen in emission \citep{Dulk74,Lantos87}, as 
in our lower frequency channels. 
Again, scattering \citep{Riddle74,Hoang77} and/or refraction \citep{Alissandrakis94} may be able to explain 
low-frequency enhancements in low-density regions, but a satisfactory explanation 
has not been achieved, in part because of limited data.
The MWA appears to be uniquely poised to address this topic given that the transition of certain 
coronal holes between being dark or bright features occurs within the instrument's frequency range, 
but an analysis of this is beyond the scope of this paper. 


\subsection{Flux Calibration}
\label{flux}

Absolute flux calibration is challenging for radio data because of instrumental 
uncertainties and effects related to interferometric data processing. 
Astrophysical studies typically use catalogs of known sources to set the flux scale, 
and many MWA projects now use results from the GaLactic and Extragalactic All-sky MWA Survey (GLEAM; \citealt{Hurley17}).
We cannot take this approach because calibrator sources are not distinguishable in close proximity 
to the Sun given the dynamic range of our data. 
Even calibrators at sufficiently large angular separations from the Sun to be imaged are 
likely to be contaminated by solar emission due to the MWA's wide field of view (see Section~\ref{mwa1}). 

\begin{figure*}\graphicspath{{chapter2/}}
\centering
\includegraphics[width = 1.0\textwidth]{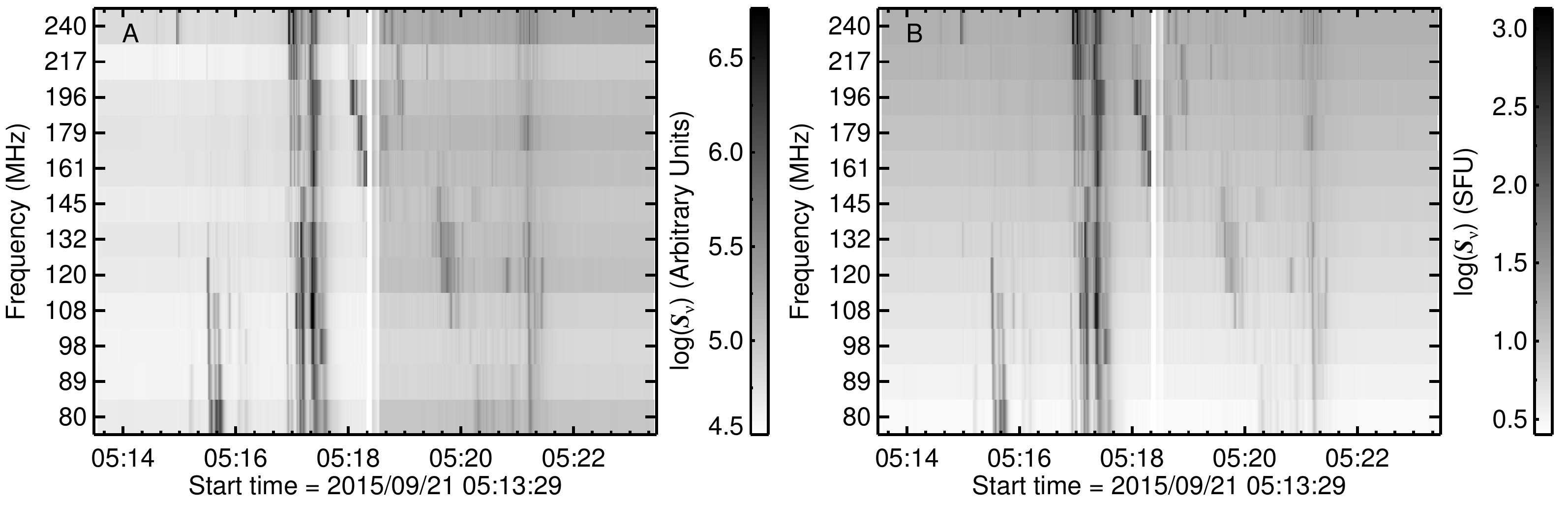}
 \caption{\footnotesize{}
Uncalibrated (A) and flux-calibrated (B) dynamic spectra generated from total image intensities. 
The Y axes intervals are not uniform; values refer to the 12 2.56-MHz-wide observing 
bandwidths separated by gaps of 9--23 MHz  (see Section~\ref{mwa1}). An interpolated dynamic spectrum with a uniform 
Y axis is shown in Fig~\ref{fig:spectra1}. 
 }
 \label{fig:fluxcal}
 \end{figure*}

To express our burst intensities in physical units, we 
take brightness temperature images from FORWARD and convert them to 
full-Sun integrated flux densities ($S_{\nu}$), which 
we then assume to be equal to the total flux density in the quiescent background images from Figure~\ref{fig:forward1}. 
From this comparison, we obtain a simple multiplicative scaling factor to convert between the uncalibrated image intensities  
and solar flux units (SFU; 1 SFU = 10\tsp{4} Jy = 10\tsp{-22} W m\tsp{-2} Hz\tsp{-1}). 
This procedure is performed separately for both observing periods, and 
Figure~\ref{fig:fluxcal} illustrates the result by plotting an uncalibrated dynamic spectrum next to the calibrated version. 


In the calibrated spectrum, we see that the quiescent intensities are coherently ordered in the pattern expected for 
thermal emission, with flux density increasing with frequency. 
Importantly, the adjacent MWA observing periods are also  
set onto very similar flux scales. 
We find an overall peak flux density of 1300 SFU at 240 MHz. 
Relative to the background, however, the burst series is most intense around 108 MHz, peaking at 680 SFU around 
140$\times$ the background level (see the log-scaled and then background-subtracted dynamic spectrum in Figure~\ref{fig:spectra1}). 
This makes our event of moderate intensity compared to those in the literature (e.g. \citealt{Saint-Hilaire13}). 

This technique provides a simple way to obtain reasonable flux densities for radio bursts in 
order to place them generally in context.
Given the differences between the observations and synthetic images, this method should 
not be applied if very accurate flux densities are important to the results, which is 
not the case here. 
It would also not be appropriate for analyzing quiet-Sun features, nor for cases where non-thermal 
emission from a particular active region dominates the Sun for the entire observation period.
However in this case, we see primarily thermal emission that we suspect is modulated 
by propagation effects not considered by FORWARD. 
These effects are not expected to dramatically affect the total intensity but may decrease it somewhat, which would 
cause our flux densities to be overestimated. 

A more sophisticated solar flux calibration method has recently been developed by 
\citet{Oberoi17}, who use a sky brightness model 
to subtract the flux densities of astronomical sources, leaving just that produced by the Sun. 
This method is applied to data from a single short baseline, yielding a total flux density 
that can be used to calibrate images with a scaling factor analogous to ours.  
This approach would be appropriate for quiet-Sun studies and preferable for burst studies that make 
significant use of the fluxes.
We note that our method yielded quiescent fluxes within a factor of 2 of those found by \citet{Oberoi17}
for a different day, after accounting for the different polarizations used. 
Future work will explicitly compare the two approaches. 


\subsection{Type III Source Structure and Motion}
\label{kinematics}

The Type III bursts begin around 05:15:30 UT during the early rise phase of the 
X-ray flare and continue at intervals through the decay phase. 
The two main bursts distinguishable in the \textit{Learmonth} and \textit{Culgoora} 
spectrographs are approximately 
coincident with the hard X-ray peak around 05:17 UT (Figure~\ref{fig:lightcurves}). 
The more sensitive and temporally-resolved MWA observations reveal these events to 
have a complicated dynamic spectrum structure that we interpret as the overlapping 
signatures of multiple electron injections in a brief period (Figure~\ref{fig:spectra1}).


  \begin{figure}\graphicspath{{chapter2/}}
\centering
\includegraphics[width = 0.7\textwidth]{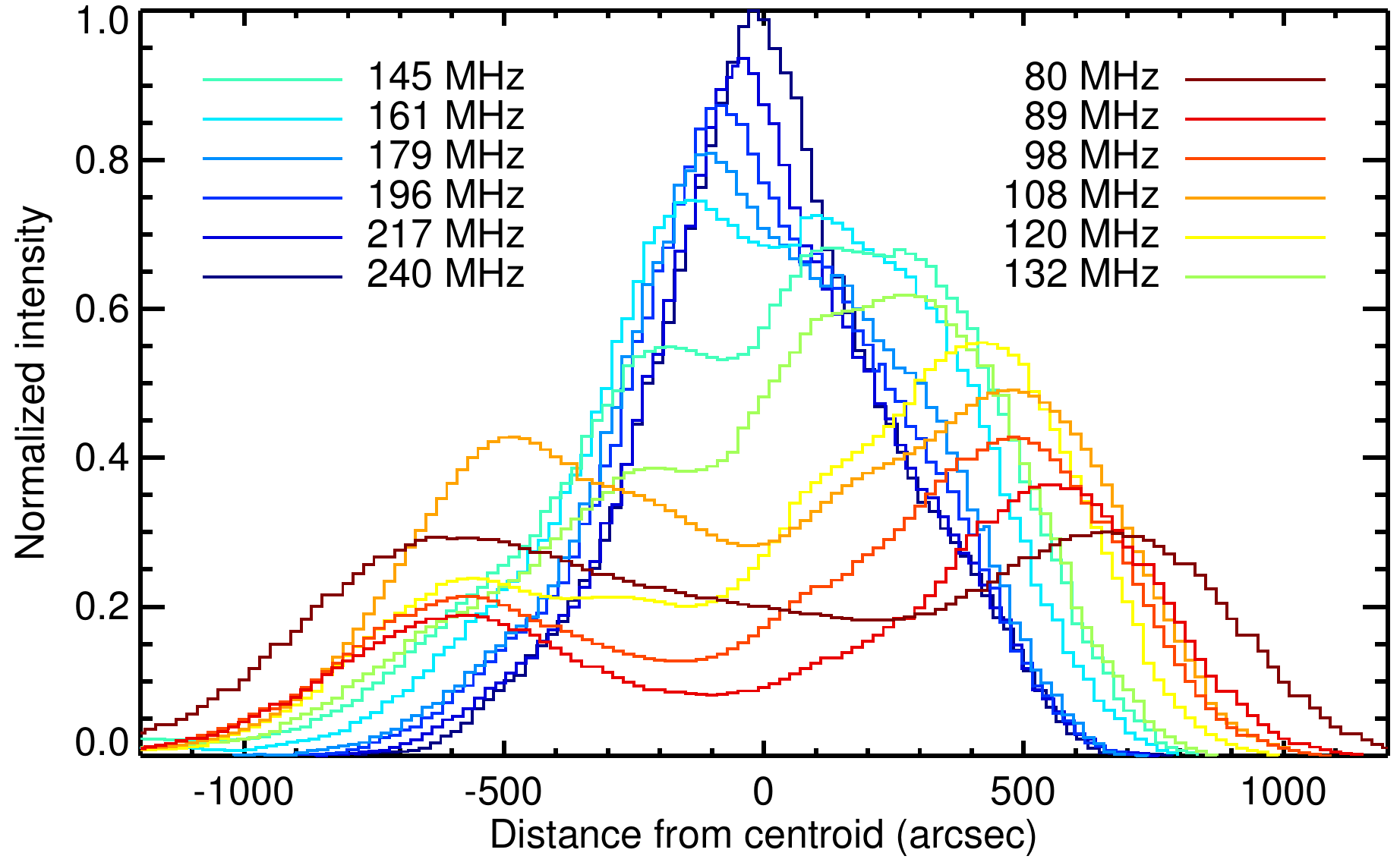}
 \caption{\footnotesize{}
Image slit intensities for each of the 12 MWA channels along the elongation axes of the individual burst source regions, 
illustrating the splitting of the source region from high to low frequencies.  
These data correspond to a period when the source regions are maximally extended at 05:17:26.6 UT. 
Each curve is normalized and multiplied by a scaling factor from 0.3--1.0 for clarity. 
 }
 \label{fig:elliptical_lightcurves}
 \end{figure}

Throughout all of the bursts, a consistent pattern emerges in both the spatial structure of the source 
regions as a function of frequency and in their motions at particular frequencies. 
At higher frequencies, the Type III source region is dominated by one spatial component 
with a much fainter component immediately to the north. 
Moving to lower frequencies and correspondingly larger heights, the two components 
separate along a direction tangent to the limb, reaching a peak-to-peak separation of 
1200'' (1.25 R$_{\odot}$) at 80 MHz. 
This structure is clear from the burst images in Figure~\ref{fig:burst} and is illustrated in 
further detail by Figure~\ref{fig:elliptical_lightcurves}. 

Figure~\ref{fig:elliptical_lightcurves} plots intensities extracted from image slits along the 
directions for which the emission is maximally extended. 
Slit orientations are determined by fitting ellipses to the overall  
source region in each channel after thresholding the images above 20\% of their peak intensities. 
Distances refer to that from the ellipse centers along their major axes, with values 
increasing from south to north. 
For clarity, the intensities are normalized and then multiplied by arbitrary scaling factors between 0.3 and 1.0 
from low to high frequencies.  
At least two Gaussian components are required to fit the curves at all frequencies, though 
the northern component is manifested only as a non-Gaussian shoulder on the dominant 
component at high frequencies. 
At some frequencies (e.g. 108 MHz), there are also additional weaker peaks between 
the two main components. 
Interpretation of the varying burst morphology as a function of frequency is given in Section~\ref{discussion1}. 


  \begin{figure*}\graphicspath{{chapter2/}}
\centering
\includegraphics[width = 1.0\textwidth]{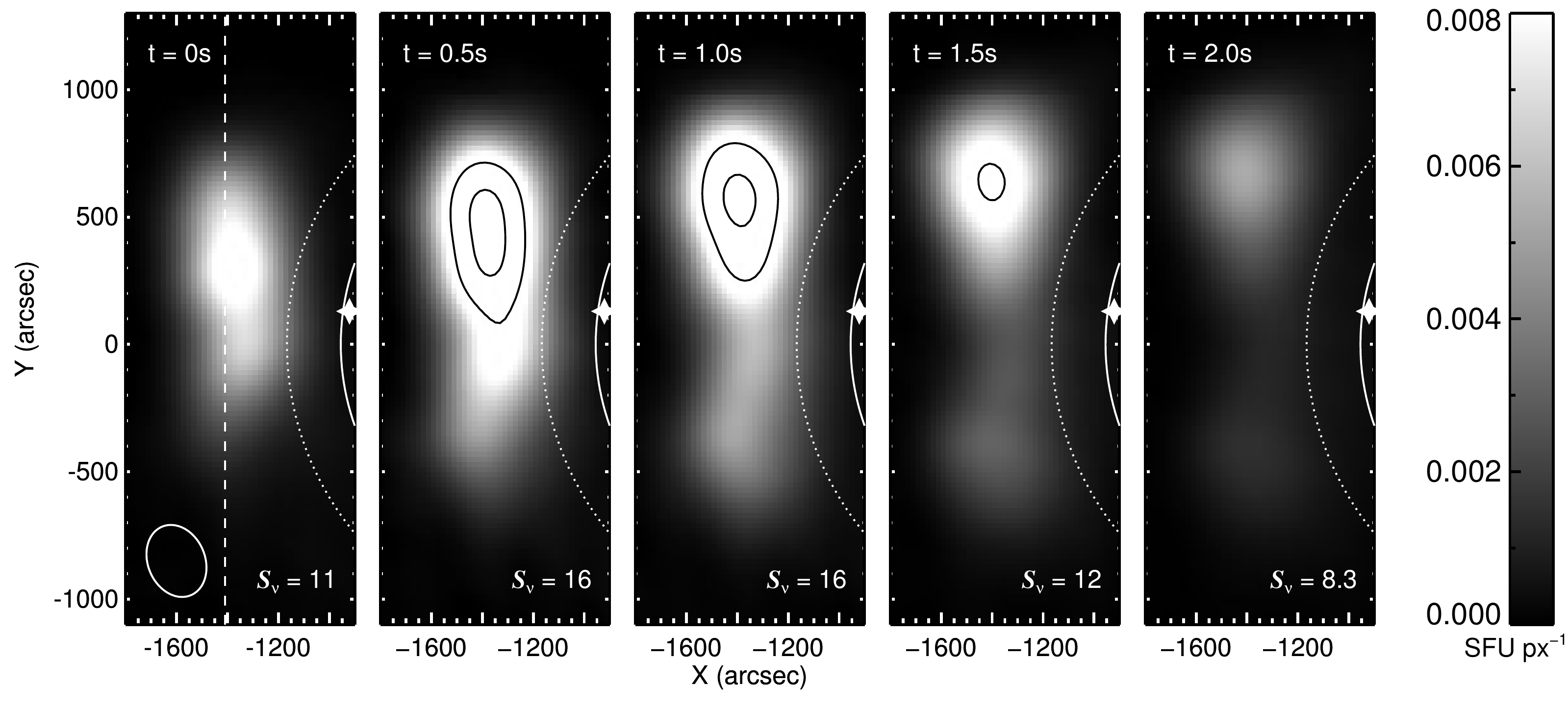}
 \caption{\footnotesize{}
Source splitting motion at 108 MHz, beginning at \edit{05:16:53.70} UT. The dashed line in the left 
panel denotes the slit used in Figure~\ref{fig:kinematics}. 
The two solid black contours in the source region are 
at 0.010 and 0.015 SFU px\tsp{-1}. 
Additional annotations are as in Figure~\ref{fig:burst}.
}
 \label{fig:splitting}
 \end{figure*}

The Type III source region components also 
spatially diverge as a function of time within single-channel observations below $\sim$132 MHz. 
At higher frequencies, for which there are one or two closely-spaced components, the source 
regions instead become increasingly elongated with time.
The direction of this motion is essentially the same as that of the frequency-dependent splitting, and 
the timescales for it are quite short, on the order of $\sim$2 s. 
This motion is repeated many times throughout the event, with each burst and corresponding 
``split" interpreted as a distinct particle acceleration episode. 
An example image set is shown in Figure~\ref{fig:splitting} for 108 MHz, 
the frequency that exhibits the highest intensities relative to the background. 

\begin{figure*}\graphicspath{{chapter2/}}
\centering
\includegraphics[width = 1.0\textwidth]{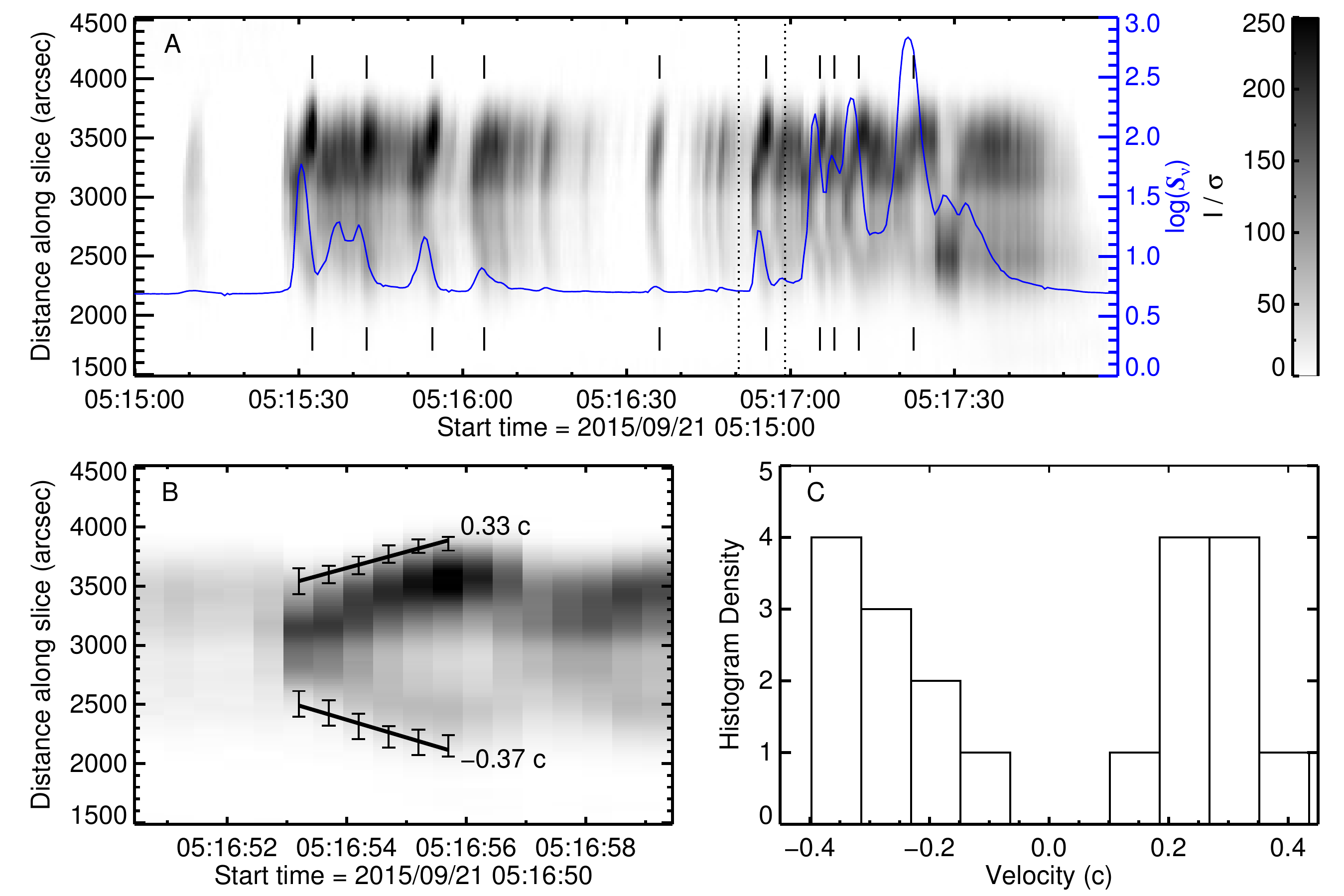}
 \caption{\footnotesize{}
An overview of the source splitting kinematics at 108 MHz. 
Panel A shows a distance-time plot using the slit shown in Figure~\ref{fig:splitting} 
along with a light curve of the total flux density in blue.
Dotted vertical lines demarcate the zoomed-in section in Panel B\edit{, 
which corresponds to the images shown in Figure~\ref{fig:splitting}.} 
Vertical ticks mark the 10 speed measurement periods whose results are collected 
in Panel C. Error bars in Panel B reflect the range of leading-edge estimates, 
obtained by thresholding the two components by 15--25\% of 
their maximum $I\cdot{}\sigma{}^{-1}$ values. 
 }
 \label{fig:kinematics}
 \end{figure*}

To quantify this behavior, we employ distance-time 
maps to track movement along a particular slice through the images. The emission along the slit shown 
in the left panel of Figure~\ref{fig:splitting} is extracted from each observation and stacked against 
those from adjacent images, such that each vertical column of Figures~\ref{fig:kinematics}a and 
~\ref{fig:kinematics}b represents the slit intensity at a given time. Slopes in the ``slit image" 
correspond to plane-of-sky velocity components in the slit direction. 
Figure~\ref{fig:kinematics}a shows the result of this analysis for the bursts during the 
first MWA observation period, lasting nearly 3 minutes 
after 05:15 UT. Intensities have been divided by the time-dependent noise level, defined as the standard 
deviation of values within a 5-pixel-wide border around the edge of each image 
(equivalent in area to a 75$\times$75 px, or 25$\times$25 arcmin, box).
Because the noise level is roughly 
proportional to the total intensity, which varies by 2--3 orders of magnitude, this operation flattens the 
dynamic range of the distance-time map and provides for the uniform thresholding scheme described 
next. 

\edit{Throughout the series, the bursts peak in intensity at around the midpoint in the splitting 
motion, which is illustrated by the blue light curve in Figure~\ref{fig:kinematics}a. 
When the motion ends, 
the source regions gradually fade into the background with constant morphologies, 
or they are supplanted by those of a subsequent burst.
This decay phase manifests as the flat region in the distance-time profile in Figure~\ref{fig:kinematics}b. 
Note that the time period for Figure~\ref{fig:elliptical_lightcurves} is chosen so that each of the frequencies are in the 
declining phase, which is possible in that case because a subsequent burst does not follow for several 
seconds.}


The leading edges of the two source regions (north and south) are defined and tracked independently by 
thresholding the slit image 
above a percentage of the peak signal-to-noise ratio (SNR) for each component.
Measurements are made for each burst using 11 integer thresholds between 
between 15 and 25\% of the peak SNR.
\edit{This corresponds to values of 40--67 $\sigma$ for the northern component and 19--32 $\sigma$ for the southern.}
Error bars in Figure~\ref{fig:kinematics}b represent the resulting range of leading edge locations, and 
corresponding speed uncertainties are on the order of 15\%. 
An SNR percentage is used instead of a single set of values for both sources 
because it expands the range of reasonable thresholds, better representing the measurement 
uncertainties compared to a more restrictive range that would be appropriate for both sources. 

\edit{We also explored quantifying the same motion by instead tracking the centroid positions of the 
two source components. 
This approach was ultimately discarded because of difficulties in reliably  
separating the two main components across the full time series, particularly when the region is most 
compact at the beginning of each burst.
Our results may be hindered somewhat by scattering of the 
type described in Section~\ref{forward}, which will be most pronounced near the source region perimeter. 
However, this would only affect the measured speeds if the scattering properties change significantly over 
the distance covered, and there appears to be little deviation of the leading edge slope from 
that of the overall source pattern in Figure~\ref{fig:kinematics}b.}

Vertical ticks in Figure~\ref{fig:kinematics}a mark the 10 bursts for which speed measurements were 
made at 108 MHz, and a histogram of the results is plotted in Figure~\ref{fig:kinematics}c. 
The time periods were chosen for particularly distinct source 
separation for which both components could be tracked. 
It is clear from Figure~\ref{fig:kinematics}a 
that the splitting motion occurs over a few additional periods for which measurements 
were precluded by confusion with adjacent events, faintness, or duration. 
We find speeds ranging between 0.11 and 0.40 c, averaging 0.26 c 
for the northern component and 0.28 c for the southern. The southern component is consistently 
faster for the 6 measurements before 05:16:55 UT and consistently slower after, but these differences 
are not statistically significant. 
These values cannot be straightforwardly interpreted as the exciter or 
electron beam speed \edit{(i.e. the average speed of accelerated electrons)} because that would require electrons traveling along flux tubes parallel  
to the limb in a manner inconsistent with the inferred magnetic field configuration (Section~\ref{persistence}).
In Section~\ref{discussion1}, we will argue that this motion is a projected time-of-flight effect such that the splitting speeds 
here exceed the beam speed by a factor of $\lesssim$ 1.2.

\begin{figure}\graphicspath{{chapter2/}}
\centering
\includegraphics[width = 0.6\textwidth]{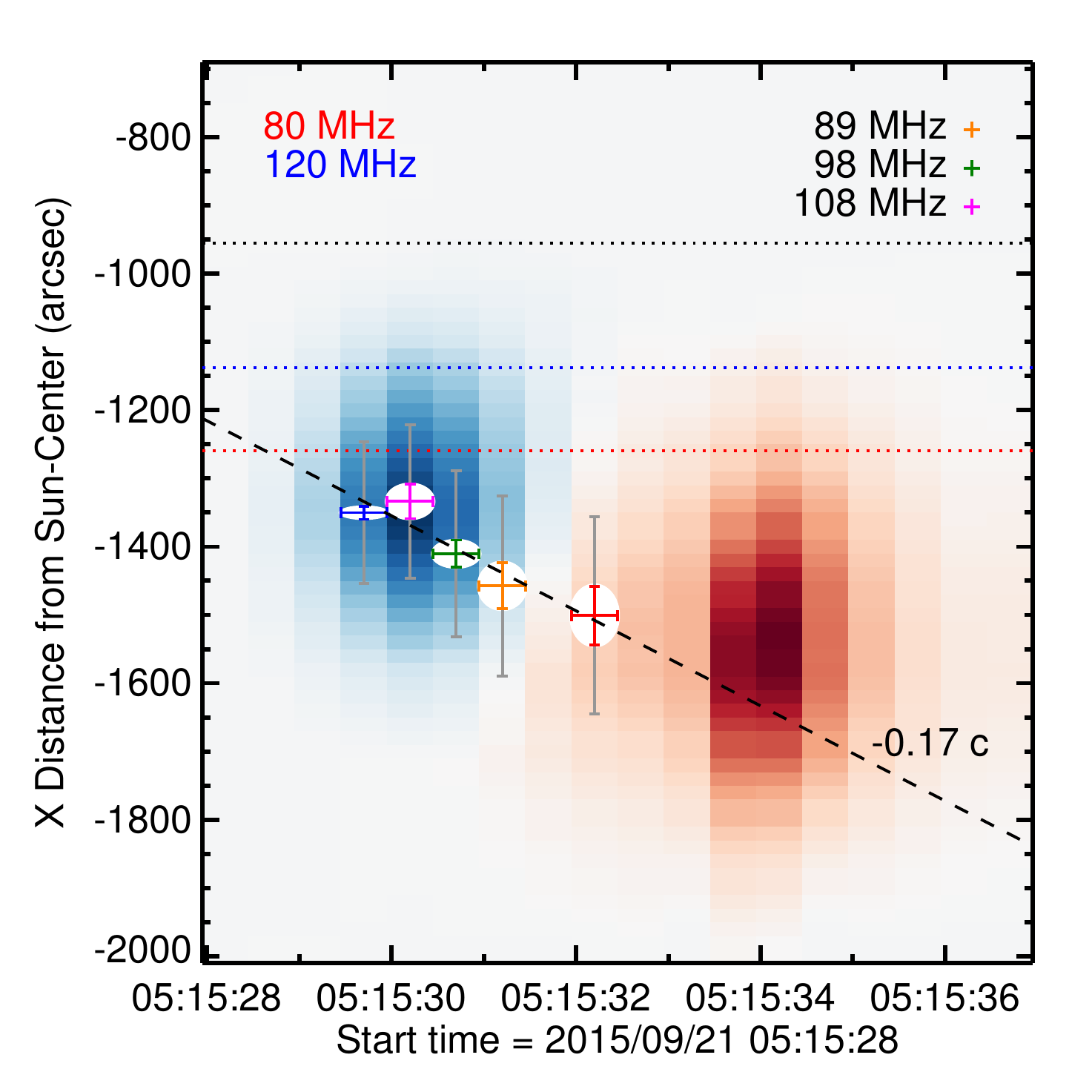}
 \caption{\footnotesize{}
Distance-time plot for burst emission from 05:15:28 to 05:15:37 UT. 
Red (80 MHz) and blue (120 MHz) images represent background-subtracted intensities averaged  
in the solar-Y direction, 
such that the slope reflects overall source motion in the solar-X direction. 
Crosshairs denote the burst onset times and centroid positions for each given frequency, where the 
onset is defined as exceeding 5$\times$ the background, 
Error bars correspond to the 0.5s time resolution (horizontal), the 3$\sigma{}$ variation in position 
over the burst period (vertical), and the minor synthesized beam axes (vertical, grey).  
Dotted horizontal lines represent the optical limb (black) and the Newkirk-model limbs at 
80 (red) and 120 (blue) MHz. 
}
 \label{fig:radial_extent}
 \end{figure}

\edit{The beam speed may be estimated more directly by examining} the burst location at different frequencies 
as a function of time. 
We do this in Figure~\ref{fig:radial_extent}, which shows a distance-time plot similar to Figure~\ref{fig:kinematics}. 
Instead of the emission along a particular slit, each column of Figure~\ref{fig:radial_extent} corresponds 
to the total image intensity binned down to a single row. 
Pixels with the same horizontal X 
coordinate are averaged, \edit{and} 
these \edit{Y-}averaged curves are stacked vertically against each other to show movement in the X direction. 
\edit{This is done so that the bidirectional vertical motion, 
which is primarily exhibited in single-channel observations (Figures~\ref{fig:splitting} \& \ref{fig:kinematics}), 
can be ignored to track the outward progression of the overall source region across frequency channels.}
Since our source regions are distributed on either side of the equator, this roughly corresponds to 
radial motion \edit{in the plane of the sky}. 


To \edit{quantify this motion}, we track the center position at the onset of the burst for 
each channel, which we define as 5$\times$ the background intensity.
\edit{We use the onset as opposed the times of peak intensity to avoid potential confusion 
between fundamental and second harmonic emission. Previous studies have shown from both 
observational \citep{Dulk84} and theoretical \citep{Robinson94} perspectives that emission at 
the fundamental plasma frequency arrives before associated harmonic emission, which may follow 
around the overall peak time after a frequency-dependent offset. 
Tracking the position at the onset of the burst thus ensures that we follow a coherent progression.
Note, however, that there is no standard in the literature. 
Estimates of Type III beam speeds 
using the frequency drift rate technique, which will be discussed in Section~\ref{discussion1}, 
have used both onset and peak times (see review by \citealt{Reid14}).
}

Center positions are determined by fitting a Gaussian to the relevant time column. 
\edit{We track center positions here because the same difficulties described 
for Figure~\ref{fig:kinematics} do not exist in this case and also because it mitigates the potential 
influence of frequency-dependent scattering.
Scattering may still impact our result if the source locations are modulated significantly 
as a function of frequency, but we cannot readily test that possibility.}
We choose to examine the earliest burst period, occurring from 05:15:29--05:15:35 UT 
at frequencies below $\sim$132 MHz, because 
that event can be easily followed from high to low frequencies, whereas the more intense bursts  
later appear to \edit{comprise} several overlapping events. 
Fitting a line to the resulting spatiotemporal positions in Figure~\ref{fig:radial_extent}, we find a 
speed of 0.17 c. This result reflects the average outward motion of the entire source, 
which can be taken as a lower limit to the exciter speed. 

In comparison, the 108 MHz splitting speed for the same period averages to 0.28 c for both components, which  
as we will discuss in Section~\ref{discussion1}, \edit{exceeds the beam speed by a small factor based on the field geometry.} 
Thus we have a range of 0.17--0.28 c for the burst from 05:15:29--05:15:35 UT. 
Note that although the speeds from Figures~\ref{fig:kinematics} and~\ref{fig:radial_extent} 
are measured in orthogonal directions, we cannot combine 
them in a quadrature sum as though they were components of one velocity vector. 
As we will explain next, this is because we interpret 
the source behavior in terms of several adjacent electron beams, each with a slightly 
different trajectory than the next, as opposed to one coherent system. 
\edit{Also note that in all cases, we are estimating two-dimensional (plane-of-sky) velocity components 
of three-dimensional motion, which has a somewhat greater magnitude depending on the projection geometry.
Given this event's position on the limb and the direction of the EUV jets considered in the next section, we assume that 
the line-of-sight component is much smaller than its plane-of-sky counterpart.}


\subsection{Magnetic Field Configuration}
\label{persistence}

Electron beams responsible for Type III bursts propagate along magnetic field lines from 
the reconnection site, and therefore understanding the magnetic field configuration is critical to understanding 
the radio source region behavior and vice versa. 
Active region 12420, where the flare occurs, had just rotated into visibility on the east limb at the 
time of this event. 
EUV jets that immediately follow the radio bursts after the flare peak reveal a complex 
magnetic field configuration that connects AR 12420 to a small, diffuse dipole to the 
south near the equator.
The southern region was just behind the limb during the flare, and based on its evolution in HMI 
magnetograms over the following days, appears to have been a decaying active region 
near the end of its evolution. 

Unfortunately, this system is a poor candidate for local magnetic field modeling because of its 
partial visibility and position on the limb, where magnetogram observations are hampered by projection effects. 
The east limb position prevents us from using data from a few days prior, which is a possibility for 
west-limb events, and the decay of the southern dipole, along with the emergence of a neighboring region, 
dissuades us from attempting any dedicated modeling using 
data from subsequent days. 
Fortunately, the EUV jets trace out the field structure to an extent that we believe is sufficient to 
understand our observations. 
Previous studies have also demonstrated that Type III electron beams are aligned with corresponding 
EUV and X-ray jets (e.g. \citealt{Chen13}), meaning that field lines traced out by the jets are 
preferentially those traversed by the accelerated electrons. 


  \begin{figure*}\graphicspath{{chapter2/}}
\centering
\includegraphics[width = 1.0\textwidth]{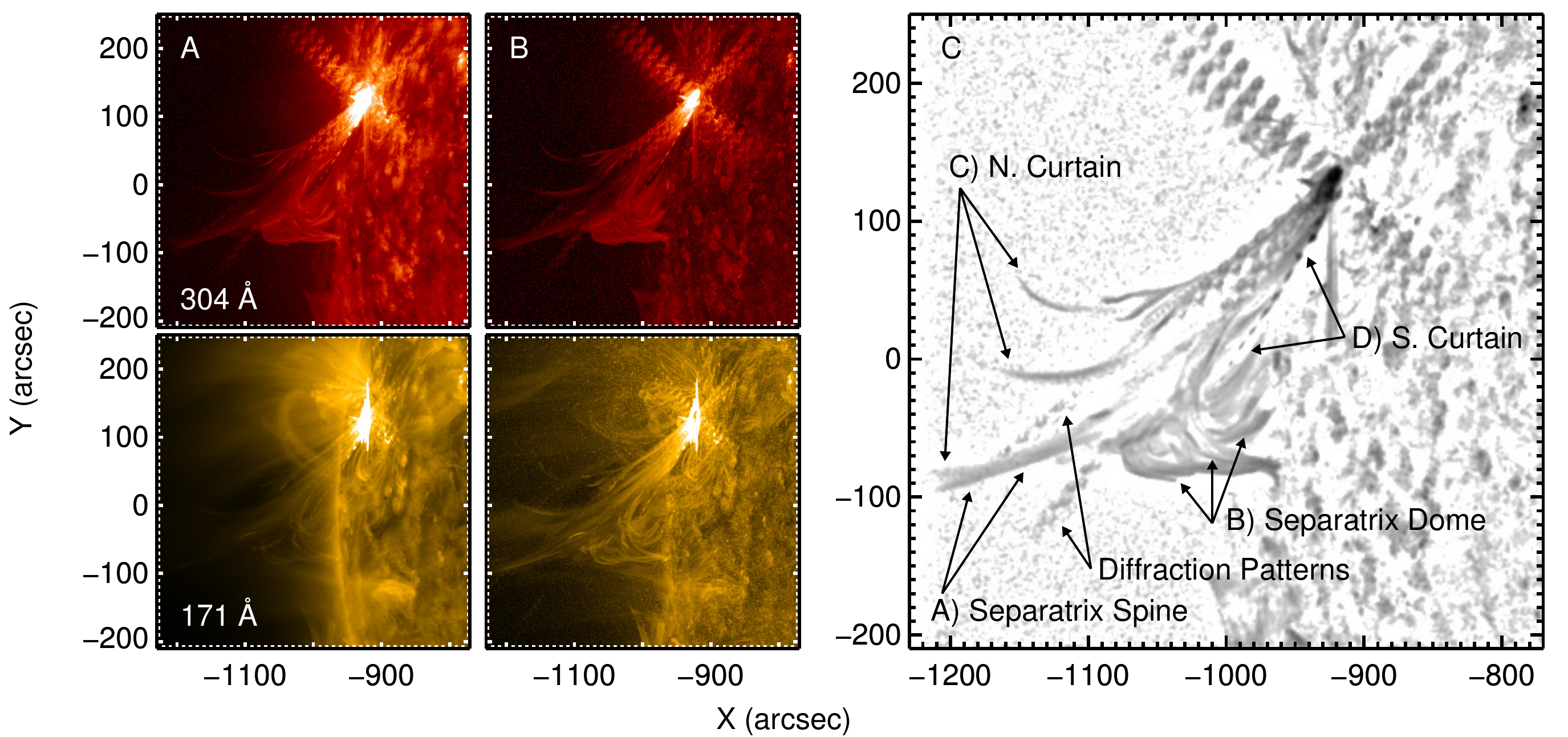}
 \caption{\footnotesize{}
A) Maximum value persistence maps for AIA 304 (top) and 171 \AA{} (bottom).
B) Column A subtracted by median backgrounds. 
C) Annotated 304-\AA{}, background-subtracted persistence map, further processed  
to accentuate features. See Section~\ref{persistence} for processing details and 
Figure~\ref{fig:cartoon} for a corresponding cartoon model. 
A corresponding movie is available in the \href{https://iopscience.iop.org/article/10.3847/1538-4357/aa9cee}{online material}.
 }
 \label{fig:pmap}
 \end{figure*}

We employ maximum-value persistence mapping to compile the separate EUV jet paths into one image. 
This style of persistence map refers simply to plotting the largest value a given pixel achieves over 
some period \citep{Thompson16}. 
Our maps cover from 05:18 to 05:39 UT, which corresponds to 
when the EUV jets begin around the peak flare time until they reach their full spatial extent visible to AIA around 
20 minutes later.
To further enhance the contrast, we subtract the persistence maps by a median-value background 
over the same period (i.e. $I_{\rm max} - I_{\rm med}$). 
Figures~\ref{fig:pmap}a and ~\ref{fig:pmap}b show maximum-value and background-subtracted 
persistence maps for both the 304 and 171 \AA{} channels, which are most sensitive to the jet material. 
Figure~\ref{fig:pmap}c shows a version of the 304 \AA{} map that has been Fourier filtered 
to suppress noise using a Hann window and then sharpened using an unsharp mask to accentuate 
the structure. 

The EUV jets trace out a toplogy, not apparent just 
prior to the flare, where the field connectivity changes rapidly. 
Such regions are generally known as \textit{quasi-separatrix layers} (QSLs; \citealt{Priest95,Demoulin96}), 
which are 3D generalizations of 2D separatrices that separate magnetic field connectivity domains. 
The key distinction is that the field linkage across a QSL is not discontinuous 
as in a true separatrix but instead changes drastically over a relatively small spatial scale, 
which can be quantified by the \textit{squashing factor} $Q$ \citep{Titov07}. 
QSLs are important generally because they are preferred sites for the development of current 
sheets and ultimately magnetic reconnection \citep{Aulanier05}. 
They are an essential part of 3D generalizations of the standard flare model \citep{Janvier13}, 
and modeling their evolution can reproduce a number of observed flare features (e.g. \citealt{Savcheva15, Savcheva16, Janvier16}). 
Here, we are less concerned with the dynamics of the flare site itself and focus instead on the 
neighboring region revealed by the EUV jets, which exhibits a topology associated 
with coronal null points. 
 
We first note that our observed structure is similar in several ways to that 
modeled by \citet{Masson12} and observed by \citet{Masson14}. 
The essential components are firstly the closed fan surface, or \textit{separatrix dome}, 
and its single \textit{spine} field line that is rooted in the photosphere and crosses the dome through 
the null point \citep{Lau90,Pontin13}. 
Open and closed flux domains are bounded above and below a separatrix dome, which can form 
when a dipole emerges into a preexisting open field region (e.g. \citealt{Torok09}). 
Above the dome and diverging around the null point is a vertical fan surface, or \textit{separatrix curtain}, 
comprised of field lines extending higher into the corona, with those closest to the separatrix spine likely being 
open to interplanetary space. 
Potential field source surface 
(PFSS\footnote{PFSS Software Package: \url{http://www.lmsal.com/~derosa/pfsspack/}}; \citealt{Schrijver03})
extrapolations (not shown) do predict open field in this region but do not reproduce other topological features, which 
is to be expected given the modeling challenges described above. 
Some openness to interplanetary space must also have been present to facilitate the corresponding 
interplanetary burst observed by \textit{Wind} and shown in Figure~\ref{fig:spectra1}. 

The separatrix dome, spine, and part of the curtain are clearly delineated by the EUV jets 
and are labeled in Figure~\ref{fig:pmap}c.
Note that some of the features, namely the closed field line associated with the southern 
portion of the separatrix curtain, are somewhat difficult to follow in Figure~\ref{fig:pmap}c but 
can be clearly distinguished in the corresponding movie available in the 
\href{https://iopscience.iop.org/article/10.3847/1538-4357/aa9cee}{online material}. 
In the following section, we will discuss how both types of source splitting described in Section~\ref{kinematics} are 
facilitated by this topology. 


  \begin{figure}\graphicspath{{chapter2/}}
\centering
\includegraphics[width = 0.75\textwidth]{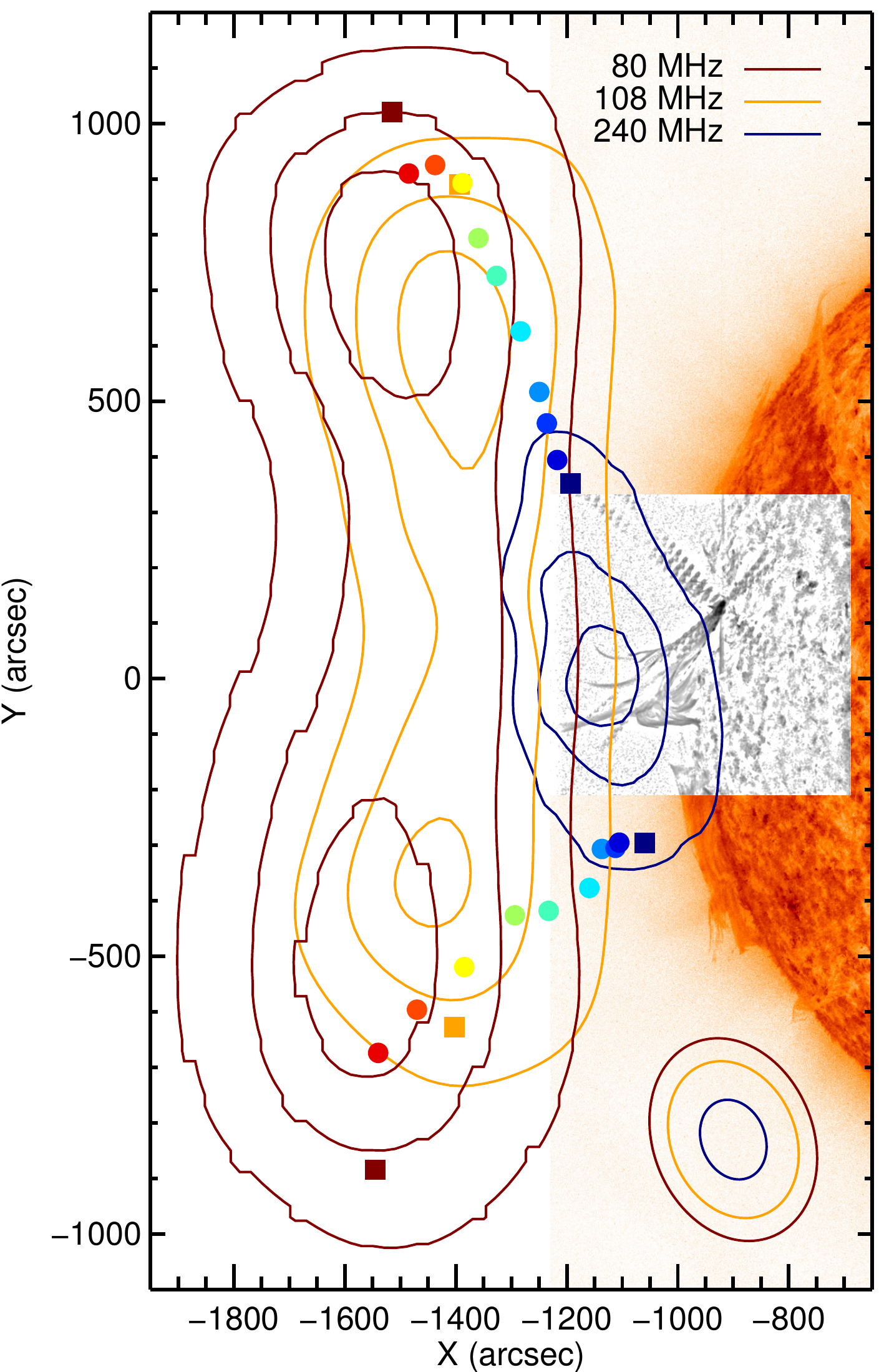}
 \caption{\footnotesize{}
MWA Type III burst contours overlaid on a 304 \AA{} SDO image.
The greyscale inset is the persistence map from Figure~\ref{fig:pmap}c. 
Pairs of colored dots represent the angular extent of the MWA source region in all 12 channels, 
with the squares from left to right corresponding to the \edit{reddish brown} (80 MHz), \edit{orange} (108 MHz), 
and \edit{dark} blue (240 MHz) contours, respectively. 
\edit{Contour levels are at 20, 50, and 80\% of the peak intensity.}
The MWA data are from a period when the source regions are maximally extended around 05:17:26.6 UT, and 
the SDO image combines data from the EUV jet period that follows (see Section~\ref{persistence}).  
 }
 \label{fig:pmap_overlay}
 \end{figure}


  \begin{figure}\graphicspath{{chapter2/}}
\centering
\includegraphics[width = 0.75\textwidth]{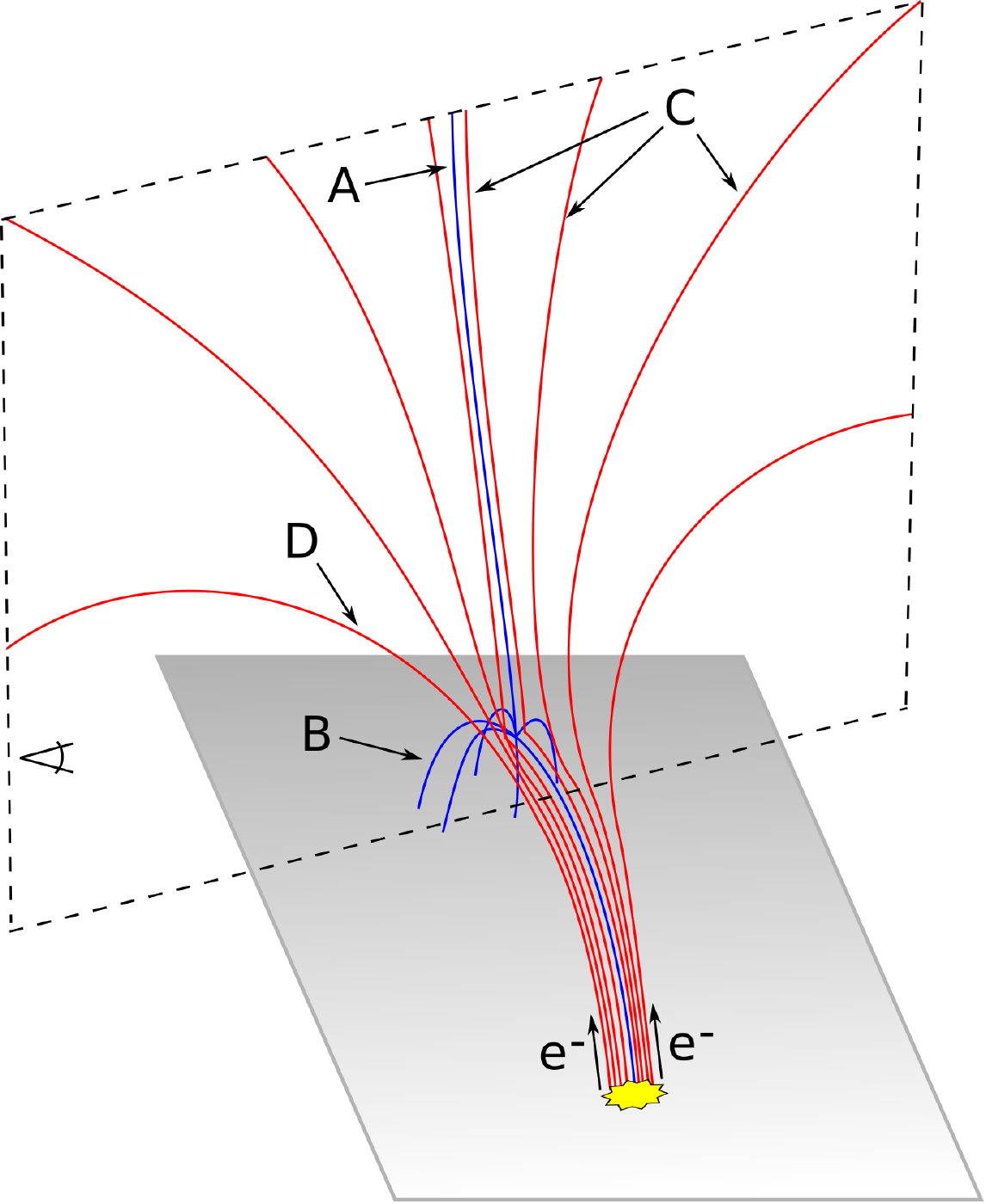}
 \caption{\footnotesize{}
Cartoon interpretation of the magnetic field configuration inferred from the EUV jet morphology and 
radio source regions (Figure~\ref{fig:pmap_overlay}). 
The yellow region denotes the flare site, which is connected to a neighboring region with open and closed 
QSLs. 
Red field lines form a separatrix curtain, with the field closest to the center being open to interplanetary space. 
The blue field lines represent the closed separatrix dome, with a single spine field line that crosses the dome 
through a magnetic null point. 
Electrons travel along the diverging field lines of the separatrix curtain to produce the radio source structure and motion. 
Capital letters correspond to features apparent in the EUV observations (Figure~\ref{fig:pmap}).  
 }
 \label{fig:cartoon}
 \end{figure}


\section{Discussion}
\label{discussion1}

When we overplot contours of the Type III burst emission on the persistence map of 
the EUV jets (Figure~\ref{fig:pmap_overlay}), 
we see that the 240 MHz emission 
is concentrated just above the separatrix dome. 
As we described in Section~\ref{kinematics}, the burst emission splits with decreasing frequency 
(increasing height) into two increasingly-separated components. 
Figure~\ref{fig:pmap_overlay} shows that the two components are distributed on 
either side of the separatrix spine. This implies a two-sided separatrix curtain with 
open field lines on either side of the spine, of which only the northern 
set are readily apparent in the EUV images. 
Given the position of the southern radio source and the closed field line that appears to form 
part of the southern curtain (D) in Figure~\ref{fig:pmap}, the southern half of the separatrix 
curtain seems to be oriented largely along the line of sight, which may explain why it is difficult 
to discern from the EUV jet structure. 
This two-sided separatrix curtain differs from the one-sided structure of \citet{Masson12,Masson14}, but a 
number of other studies consider somewhat similar topologies \citep{Maclean09,van12,Titov12,Craig14,Pontin15}. 

In Figure~\ref{fig:cartoon}, we sketch a 3D field configuration based on the aforementioned modeling 
studies that fits the EUV structure and extrapolates from there to satisfy the connectivity required by the 
radio source distribution.  
This cartoon can parsimoniously explain both the spatial splitting of the source from high to low frequencies 
and the source motion observed for individual frequency channels.
Type III bursts emit at the local plasma frequency or its second harmonic ($f \approx f_{\rm p}$ or $2f_{\rm p}$), 
which is proportional to the square of the ambient electron density. 
Thus, emission at a particular frequency can be associated with a particular height corresponding 
to the requisite background density. 
In our interpretation, electrons travel simultaneously along each of the red field lines 
in Figure~\ref{fig:cartoon}. 
The electron beams diverge on either side of the separatrix curtain, such that the beams are nearest 
to each other at lower heights (higher frequencies) and furthest apart at larger heights (lower frequencies). 
This produces the spatial source splitting and the dramatic increase of the overall angular extent toward 
lower frequencies, which is illustrated by the pairs of colored dots in Figure~\ref{fig:pmap_overlay}. 
\edit{The dots correspond to vertices of ellipses fit the overall source regions thresholded above 20\% of 
their peak intensities in the same manner and for the same time period used in Section~\ref{kinematics} for Figure~\ref{fig:elliptical_lightcurves}.} 

\begin{figure}\graphicspath{{chapter2/}}
\centering
\includegraphics[width = 0.5\textwidth]{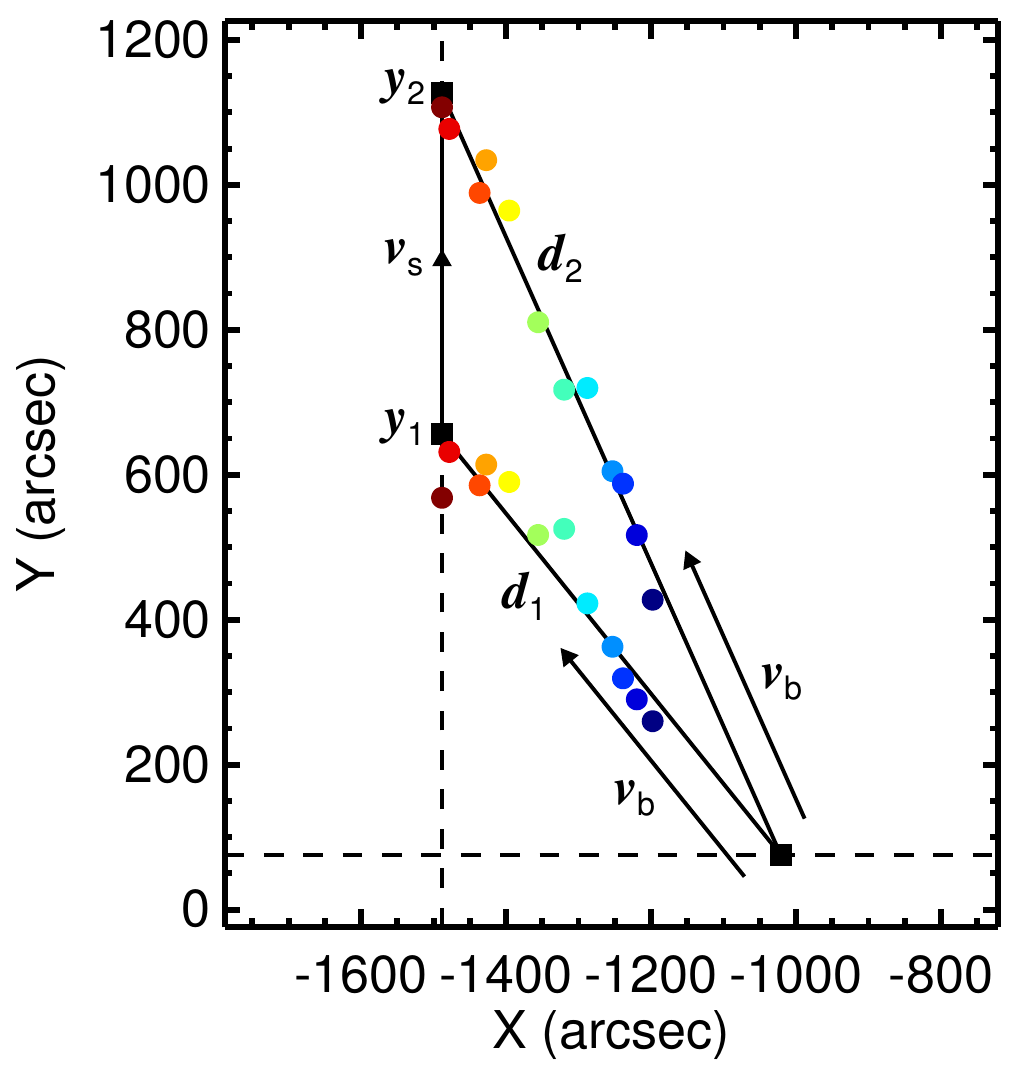}
 \caption{\footnotesize{}
Model schematic for the source splitting motion (Equation~\ref{eqn:split}).
Pairs of colored dots represent the average minimum and maximum vertical 
extents during each splitting episode; colors indicate frequency as in Figures~\ref{fig:elliptical_lightcurves} \& \ref{fig:pmap_overlay}.
The flux tubes along which the Type III beams travel are approximated by the solid fit lines,  
which intersect near the observed null point (Figure~\ref{fig:pmap}). 
Electrons take slightly longer to reach $y_2$ compared to $y_1$, which produces the apparent 
vertical motion with velocity $v_{\rm s}$. 
In reality, there would be a number of adjacent curved flux tubes between and below the two lines with 
nearby, but not identical, origins. 
 }
 \label{fig:split_speed}
 \end{figure}

The source motions illustrated by Figures~\ref{fig:splitting} and~\ref{fig:kinematics} can then 
be accounted for as a projected time-of-flight effect. 
Electrons moving along the increasingly curved outer field lines take slightly longer 
to reach the same height, producing emission at adjacent positions along the separatrix curtain at 
slightly later times for a given frequency. 
\edit{This assumes that adjacent field lines have roughly the same radial density gradient, which implies  
decreasing density gradients along the field lines themselves as path lengths to specific heights (densities) 
increase with distance from the separatrix spine.} 
Thus, the splitting speeds measured in Section~\ref{kinematics} are not the exciter or electron beam 
speeds. They are instead somewhat faster, depending on the difference 
in travel time to a given height along adjacent flux tubes. 
Adopting the geometry in Figure~\ref{fig:split_speed}, the 
expression for this is:

\begin{equation}
\label{eqn:split}
v_{\rm s} = \frac{y_2-y_1}{d_2-d_1}v_{\rm b}~, 
\end{equation}

\noindent{}where $v_{\rm s}$ is the apparent source splitting speed, $v_{\rm b}$ is the electron beam speed, 
$y_{1,2}$ are solar Y coordinates, and $d_{1,2}$ are the distances traveled along the 
field lines to reach $y_{1,2}$.


To estimate these parameters, we determine the average minimum and maximum vertical 
extents of the source regions for each frequency by fitting ellipses to every burst image, 
\edit{as was done for a single time step to illustrate the source region extents in Figures~\ref{fig:elliptical_lightcurves} and \ref{fig:pmap_overlay}.}
The X coordinates of the northern vertices are averaged, and the Y coordinates 
one standard deviation above and below the mean are averaged separately to 
obtain the pairs of colored dots in Figure~\ref{fig:split_speed}.
\edit{We take this approach rather than tracking the northern component's centroid 
because, along with the associated difficulties described in Section~\ref{kinematics}, 
it allows us to capture consistent information from the higher-frequency channels where there is only one component 
and also because it is 
similar to the leading edge method used to estimate $v_{\rm s}$ in Figure~\ref{fig:kinematics}.}
 
If we approximate the field lines as linear fits to these points, which intersect close 
to the observed null point (Figure~\ref{fig:pmap}), then the speed of the source 
motion is 1.16$\times$ the beam speed.
Taking each of the lower-frequency points individually, we find factors ranging from 1.14 at 120 MHz 
to 1.19 at 80 MHz. 
Slightly larger factors are found for lower frequencies because of the larger 
separations between $y_{1}$ and $y_{2}$ compared to the fit projection, 
which may be due to the field lines curving out with height.

\edit{As with the $v_{\rm s}$ estimates in Section~\ref{kinematics}, scattering may impact these results 
if the effect changes significantly between the colored dots in Figure~\ref{fig:split_speed}. 
Lower frequencies also tend to be more strongly scattered, which may enlarge the source regions 
as a function of decreasing frequency beyond the effect of the magnetic field divergence. 
Accounting for scattering would therefore preferentially decrease the Y-axis positions of the lower-frequency points 
in Figure~\ref{fig:split_speed}, which 
would flatten the slopes of both lines and slightly decrease the ratio $v_{\rm s}$/$v_{\rm b}$.
Including this effect would require an understanding of the local density structure and is beyond 
our scope. 
Also note that the model defined by Equation~\ref{eqn:split} and Figure~\ref{fig:split_speed} is 
specific to this magnetic field configuration and projection geometry. 
While the same basic effect may be observed for other events, different expressions may be needed to 
relate the observed motion to the beam speed.}

Using the 1.16 factor, \edit{the average speed ($v_{\rm s}$) from Figure~\ref{fig:kinematics}} 
corresponds to an average \edit{plane-of-sky beam speed ($v_{\rm b}$)} of 0.2 c. 
This value is consistent with and provides independent confirmation of beam speeds 
estimated from frequency drift rates, which is possible if one assumes a density model. 
Modest fractions of light speed are typical in the corona (e.g. \citealt{Alvarez73,Aschwanden95,Melendez99,Kishore17}), but some 
studies have found values in excess of 0.5 c \citep{Poquerusse94,Carley16} and even superluminal velocities 
given the right projection geometry \citep{Klassen03}. 
We also note that similar observations could be used to \edit{independently} probe the coronal 
density structure \edit{and beam speed} because our imaging capability allows us to estimate \edit{$v_{\rm b}$} without assuming a density model 
using time- and frequency-varying source positions in the manner illustrated by Figure~\ref{fig:radial_extent}. 
This particular event is not ideal for that analysis because of the complicated 
source structure, but a followup study is planned for a small ensemble of events that exhibit simple 
source structures without the type of motion described here. 
\edit{A similar study was also recently performed at lower frequencies (larger heights) by \citet{Morosan14} using Type III 
imaging from LOFAR. They found speeds ranging from 0.3--0.6 c and observed emission at significantly larger 
heights than would be expected from standard density models.}

A few other connections to the literature should be mentioned with respect to the observed 
radio structure and inferred field configuration. 
First, we see from Figure~\ref{fig:forward1} and in the movie associated with Figures~\ref{fig:preburst} 
and~\ref{fig:burst} that the source region of the bursts at 240 MHz is consistently 
enhanced and exhibits low-level burst activity outside of the intense burst periods. 
Figure~\ref{fig:pmap_overlay} demonstrates that this emission is concentrated just above the 
separatrix dome and associated null point.
These structures are interface regions between closed and open magnetic flux, where 
interchange reconnection may be ongoing (e.g. \citealt{Masson12,Masson14}). 
Such regions have previously been associated with radio enhancements and 
noise storms \citep{Wen07,Del11,Regnier13}. 

A few \textit{Nan\c{c}ay Radioheliograph} (NRH) observations exhibit characteristics reminiscent of  
those described here. 
For instance, \citet{Paesold01} conclude that the spatial separation of temporally adjacent 
Type III events predominantly resulted from different field line trajectories followed by the 
electron beams. 
\citet{Reid14b} show a number of elliptically extended Type III source regions that are represented 
as enveloping the diverging paths of electrons accelerated from the same site. 
Our observations that overlap in frequency with the NRH range ($\geq$150 MHz) are similarly extended to a larger degree before 
separating into two primary components at lower frequencies. 
\citet{Carley16} describe a ``radio arc" in their lowest-frequency images  
that is strikingly similar to our observations (e.g. Figure~\ref{fig:pmap_overlay}) but is suggested instead 
to trace the boundary of an erupting coronal mass ejection. 

\edit{We also} note that the complicated structure exhibited by the MWA dynamic spectrum 
(Figures~\ref{fig:spectra1} \&~\ref{fig:fluxcal}) may indicate the presence 
of other burst types. Classic Type III emission drifts from high to low frequencies as 
electron beams propagate outward into interplanetary space. 
If confined to closed field lines, the same beams may produce type U or J bursts 
for which the frequency drift rate switches signs as electrons crest the closed loops and 
propagate back toward the Sun \citep{Maxwell58,Aurass97,Reid17b}. 
We see hints of this in our dynamic spectrum at $\sim$196 MHz around 05:17:40 UT (Figure~\ref{fig:spectra1}), 
but it is difficult to interpret because of the MWA's sparse frequency coverage. 
Given that our interpretation of the magnetic field configuration (Figure~\ref{fig:cartoon}) includes 
closed field lines on either side of the separatrix curtain, such features in the dynamic spectrum 
would not be surprising. Our splitting motion could also be due partially to beams traveling largely 
tangent to the limb along such closed field lines, while adjacent beams make it to larger heights along 
field lines closer to the separatrix spine, but evidence for downward propagation is lacking in the images. 

\edit{Finally, the bursts in this series do not all exhibit 
the statistical tendency for increasing Type III flux densities with decreasing frequency 
(e.g. \citealt{Weber78, Dulk01, Saint-Hilaire13}), which is 
clear for the main event shown in Figure~\ref{fig:burst} and others visible in the flux-calibrated dynamic 
spectrum (Figure~\ref{fig:fluxcal}b).
Individual Type III bursts often deviate from this pattern, exhibiting enhancements at particular 
frequencies or breaks in the emission over a particular frequency range. 
This behavior may be attributed to, among other things, density turbulence along the beam path \citep{Li12, Loi14} and/or
variations in the ambient electron and ion temperatures \citep{Li11b,Li11}. 
Additionally, electrons streaming along closed field lines, as considered in the previous paragraph, 
may contribute to enhancements at particular frequencies.}


\section{Conclusion}
\label{conclusion1}

We have presented the first time series imaging study of MWA solar data.
Our observations reveal complex Type III burst source regions that exhibit previously 
unreported dynamics. 
We identify two types of source region splitting, one being a frequency-dependent 
structure and the other being source motion within individual frequency channels. 
For the former, the source regions splits from one dominant component at our highest frequency (240 MHz) 
into two increasingly separated sources with decreasing frequency down to 80 MHz. 
This corresponds to a straightforward 
splitting of the source region as a function of height, with larger separations at larger heights. 

With high time resolution imaging, we observe a splitting motion within the source regions at 
individual frequencies, 
particularly in the lower channels ($\lesssim$ 132 MHz), 
that is tangent to the limb in essentially the same direction as the source splitting 
from high to low frequencies. 
This motion is short-lived ($\sim$2 s), fast (0.1--0.4 c), and repetitive, occurring multiple times over a period 
of 7 min before, during, and after the X-ray flare peak. 
We interpret the repetitive nature as multiple electron beam injections
that produce distinct radio bursts with overlapping signatures in the dynamic spectrum, which 
is consistent with there being 
several distinct EUV jet episodes that immediately follow the radio bursts. 

The EUV jets, which are assumed to have very 
similar trajectories to the Type III electron beams, trace out a region where the 
magnetic field connectivity rapidly diverges over a small spatial scale. 
These types of configurations are broadly referred to as QSLs, 
and we argue that this field structure facilitates the radio source region splitting. 
Several common topological features associated with coronal null points are identifiable in 
persistence maps of the EUV outflows, including a separatrix dome, 
spine, and curtain. 
Electrons are accelerated simultaneously along adjacent field lines that connect 
the flare site to an open QSL, where their paths 
diverge to produce the source region splitting. 
At 240 MHz, the burst emission is concentrated just above 
the separatrix dome, a region that is consistently 
enhanced outside of burst periods. Moving to larger heights (lower frequencies), the source regions  
split on either side of the separatrix spine. 
\edit{The diverging field thereby enlarges the source regions at lower frequencies, an effect that 
may compound with angular broadening by refraction and scattering in this and other events.}
The northern radio component is consistent with field lines apparent from the EUV observations, 
but the southern component implies a two-sided separatrix curtain 
that is not obvious from the EUV observations. Thus, the radio imaging provides additional 
constraints on the magnetic field connectivity. 

The magnetic field configuration also offers a straightforward explanation for the radio source 
motion via a projected time-of-flight effect, whereby electrons moving along slightly longer outer field lines take slightly 
longer to excite emission at adjacent positions of roughly the same radial height. 
Given this interpretation, the speed of the source region is a factor of $\lesssim$ 1.2$\times$ greater than the 
electron beam speed. We estimate an average beam speed of 0.2 c, which is an independent confirmation of 
speeds estimated from frequency drift rates. 
We note that the same characteristics are observed in another Type III burst from the same  
region three hours earlier.
This implies that the field topology 
is stable at least on that timescale and strengthens our conclusion 
that the radio dynamics are caused by interaction with 
a preexisting magnetic field structure, as opposed to peculiarities of the flare process itself. 

Lastly, we motivate future studies of MWA solar observations. 
A survey of Type III bursts is underway. 
From preliminary results, we note that the dual-component splitting behavior described here is uncommon. 
However, analogous source region motion in one direction is common and could be explained in the 
same manner if coupled with a consistent picture of the particular field configurations. 
Similar events that occur near disk center or on the opposite (west) limb could be combined with magnetic field modeling 
to develop a more detailed topological understanding. 
The coronal density structure can also be probed by examining events with less complicated source structures. 
Finally, we showed a coronal hole that gradually transitions from dark to bright from high to low frequencies, turning over 
around 120 MHz. 
This adds a transition point to the small body of literature reporting coronal holes in emission at low frequencies, an 
effect that is not well-explained and could be addressed with additional MWA observations. \\


{\footnotesize{}\textbf{Acknowledgements}: PIM thanks Natasha Hurley-Walker for instruction on MWA data processing, 
Mike Wheatland and Yuhong Fan for discussions related to the magnetic field configuration, 
Emil Lenc for discussions related to polarization, 
and the Australian Government for supporting this work through an Endeavour Postgraduate Scholarship. 
\edit{We thank the anonymous referee for their constructive comments.}
JM, CL, and DO acknowledge support from the 
Air Force Office of Space Research (AFOSR) via grant FA9550-14-1-0192, 
and SEG acknowledges support from AFOSR grant FA9550-15-1-0030.
This scientific work makes use of the Murchison Radio-astronomy Observatory (MRO), operated by 
the Commonwealth Scientific and Industrial Research Organisation (CSIRO). 
We acknowledge the Wajarri Yamatji people as the traditional owners of the Observatory site. 
Support for the operation of the MWA is provided by the Australian Government's  
National Collaborative Research Infrastructure Strategy (NCRIS), 
under a contract to Curtin University administered by Astronomy Australia Limited. 
We acknowledge the Pawsey Supercomputing Centre, which is supported by the 
Western Australian and Australian Governments.
The SDO is a National Aeronautics and Space Administration (NASA) satellite, and 
we acknowledge the AIA and HMI science teams for providing open 
access to data and software. 
NCAR is supported by the National Science Foundation (NSF). 
This research has made use of NASA's Astrophysics Data System (ADS).}

%% file: chapter3/chapter3.tex
\pagestyle{fancy}

\providecommand{\edit}[1]{{\color{black}{#1}}}
\providecommand{\editb}[1]{{\color{black}{#1}}}



\chapter{Densities Probed by Coronal Type III Radio Burst Imaging}\label{ch3}

{\large{}Published as \citet{McCauley18}, \href{https://ui.adsabs.harvard.edu/abs/2018SoPh..293..132M/}{\textit{Solar Phys.}, 293:132}}

\fancyhead[OR]{~}\fancyhead[EL]{\bf Ch. \thechapter~Coronal Densities from Type III Burst Imaging} 

\section{Abstract}

We present coronal density profiles derived from low-frequency (80\,--\,240 MHz) imaging of 
three Type III solar radio bursts observed at the limb by the \textit{Murchison Widefield Array} (MWA). 
Each event is associated with a white-light streamer at larger heights and is plausibly associated 
with thin extreme-ultraviolet rays at lower heights. 
Assuming harmonic plasma emission, we find average electron densities 
of 1.8\e{8} cm\tsp{-3} down to 0.20\e{8} cm\tsp{-3} at heights of 1.3 to \rsolar{1.9}. 
These values represent approximately 2.4\,--\,5.4$\times$ enhancements over canonical background levels and 
are comparable to the highest streamer densities obtained from data at other wavelengths. 
Assuming fundamental emission instead would increase the densities by a factor of four. 
High densities inferred from Type III source heights can be explained by assuming that the exciting electron 
beams travel along overdense fibers or by radio propagation effects 
\edit{that may cause a source to appear at a larger height than the true emission site.}
We review the arguments for both scenarios in light of recent results.
We compare the extent of the quiescent corona to model predictions to estimate the 
impact of propagation effects, which we conclude can only partially explain the apparent density enhancements. 
Finally, we use the time- and frequency-varying source positions to estimate electron beam 
speeds of between 0.24 and 0.60 c.


\section{Introduction} %
\label{introduction2} %

Type III solar radio bursts are caused by semi-relativistic electrons 
streaming through and perturbing the ambient coronal or interplanetary plasma. 
A recent review is given by \citet{Reid14}.
The dominant theory, proposed by \citet{Ginzburg58}, invokes a two-step 
process beginning with the stimulation of Langmuir waves (plasma oscillations) 
in the background plasma by an electron beam. 
A small fraction of the Langmuir wave energy is then converted into electromagnetic 
radiation at either the local electron plasma frequency [$f_{\mathrm p}$] or its harmonic 
[2$f_{\mathrm p}$] (see reviews by \citealp{Robinson00,Melrose09}).
The emission frequency depends mainly on the ambient electron density [$n_{\mathrm e}$] because 
$f_{\mathrm p} \propto \sqrt{n_{\mathrm e}}$. 
This relationship produces the defining feature of Type III bursts, a rapid drift from high to low frequencies 
as the exciter beam travels away from the Sun through decreasing densities \citep{Wild50}. 

The rate at which the emission frequency drifts [${\mathrm d}f/{\mathrm d}t$] is therefore related to the 
electron beam speed, which can be obtained in the radial direction by assuming a density model $n_{\mathrm e}(r)$. 
Many authors have employed this technique for various events with various models, generally 
finding modest fractions of light speed (0.1\,--\,0.4 c; \textit{\textit{e.g.}} \citealp{Alvarez73,Aschwanden95,Mann99,Melendez99,Krupar15,Kishore17}).  
Alternatively, the coronal and/or interplanetary density gradient can be inferred by instead assuming a  
beam speed (\textit{e.g.} \citealp{Fainberg71,Leblanc98}) or by simply assuming that the beam speed is constant \citep{Cairns09}. 
While these methods can yield robust estimates for the density gradient, they cannot be converted into an explicit 
density structure [$n_{\mathrm e}(r)$] without normalizing the gradient to a specific value at a specific heliocentric distance. 
This normalization has typically been done using estimates from white-light polarized brightness data close to the Sun, 
\textit{in-situ} data in the interplanetary medium, or the observed height of Type III burst sources at various 
frequencies. 

Densities inferred from Type III source heights, particularly at lower frequencies, 
have frequently conflicted with those obtained from other methods. 
The earliest spatial measurements found larger source heights than would be expected from 
fundamental plasma emission, implying density enhancements of an order of magnitude or more \citep{Wild59}. 
This finding was confirmed by subsequent investigations (\textit{e.g.} \citealp{Morimoto64,Malitson66}), and along 
with other arguments, led many authors to 
two conclusions: 
First, that harmonic [2$f_{\mathrm p}$] emission likely dominates
(\textit{e.g.} \citealp{Fainberg71,Mercier74,Stewart76}). 
This brings the corresponding densities down by a factor of four, then implying only a moderate enhancement 
over densities inferred from white-light data. 
(Counterarguments for the prevalence of fundamental emission will be referenced in Section~\ref{density}.) 
Second, that the electron beams preferentially traverse overdense flux tubes
(\textit{e.g.} \citealp{Bougeret84}), a conclusion bolstered by 
spatial correlations between several Type III bursts and white-light streamers 
(\textit{e.g.} \citealp{Trottet82,Kundu84,Gopalswamy87,Mugundhan18}). 

The overdense hypothesis has been challenged by evidence that the large 
source heights can instead be explained by propagation effects.
If Type III emission is produced in thin, high-density structures, then it can escape relatively 
unperturbed through its comparatively rarefied surroundings.  
However, if the emission is produced in an environment near the associated plasma level 
(\textit{i.e.} with an average $n_{\mathrm e}$ corresponding to the radio waves' equivalent $f_{\mathrm p}$), then refraction and scattering 
by density inhomogeneities may substantially shift an observed source from its 
true origin (\textit{e.g.} \citealp{Leblanc73,Riddle74,Bougeret77}). 
\citet{Duncan79} introduced the term \textit{ducting} in this context, 
\edit{which refers to emission being guided to larger heights within a low-density structure though 
successive reflections against the high-density ``walls" of the duct.} 
This concept was generalized for a more realistic corona by \citet{Robinson83}, who showed that random 
scattering of radio waves by thin, overdense fibers has the \edit{same} net effect of elevating an observed source 
radially above its emission site.
\edit{Additional details on this topic, along with coronal refraction, will be given in Section~\ref{propagation}.}

Many authors came to favor propagation effects instead of the overdense structure interpretation 
for a few reasons. 
Despite the aforementioned case studies, Type IIIs did not appear to be statistically 
associated with regions of high average density in the corona \citep{Leblanc74,Leblanc77} or 
in the solar wind \citep{Steinberg84}. 
Interplanetary (kHz-range) Type III source regions are also so large as to demand angular broadening 
by propagation effects (\textit{e.g.} \citealp{Steinberg85,Lecacheux89}). 
Invoking propagation effects can also 
be used to explain apparent spatial differences between fundamental and harmonic sources (\textit{e.g.} \citealp{Stewart72,Kontar17})\edit{, 
along with large offsets between radio sources on the disk and their likely electron acceleration sites (\textit{e.g.} \citealp{Bisoi18}). 
These arguments are reviewed by \citet{Dulk00}, and further discussion with additional recent references will be presented 
in Sections~\ref{propagation} and \ref{discussion2}.} 

\edit{Both the interpretation of electron beams moving along overdense structures and of radio propagation effects 
elevating burst sources} rely on the presence of thin, high-density fibers. 
Either the electron beams are traveling within these structures or the Type III emission is being scattered 
by them. In this article, we will suggest that propagation effects are important but cannot entirely 
explain the density enhancements for our events.
Section~\ref{observations} describes our observations: Section \ref{mwa2} outlines our data reduction, 
Section \ref{events} details our event-selection criteria, and Section \ref{context} describes the multi-wavelength 
context for the selected Type III bursts. 
Section~\ref{analysis} describes our analysis and results: Section \ref{density} infers densities from Type III 
source heights, Section \ref{speed} estimates electron beam speeds from imaging data, and 
Section \ref{propagation} examines propagation effects by comparing the extent of the quiescent 
corona to model predictions. 
In Section~\ref{discussion2}, we discuss the implications of our results, along with other recent 
developments, on the debate between the overdense and propagation effects hypotheses. 
Finally, our conclusions are summarized in Section~\ref{conclusion2}. 

 
\section{Observations} %
\label{observations} %


\subsection{\textit{Murchison Widefield Array} (MWA)}
\label{mwa2}

The MWA is a low-frequency radio interferometer in Western Australia 
with an instantaneous bandwidth of 30.72 MHz that can be flexibly distributed 
from 80 to 300 MHz \citep{Tingay13}. 
Our data were recorded with a 0.5 second time cadence and a 40 kHz spectral 
resolution, which we average over 12 separate 2.56 MHz bandwidths 
centered at 
80, 89, 98, 108, 120, 132, 145, 161, 179, 196, 217, and 240 MHz.
We use the same data processing scheme as \citet{McCauley17}, and 
what follows is a brief summary thereof. 

Visibilities were generated with the standard MWA correlator \citep{Ord15} and the  
\textsf{cotter} software \citep{Offringa12,Offringa15}. 
Observations of bright and well-modelled calibrator sources were used 
to obtain solutions for the complex antenna gains \citep{Hurley14}, which 
were improved by imaging the calibrator and iteratively self-calibrating from there \citep{Hurley17}. 
\textsf{WSClean} \citep{Offringa14} was used to perform the imaging 
with a Briggs -2 weighting \citep{Briggs95} to maximize spatial resolution and minimize point spread function (PSF) sidelobes. 
The primary beam model of \citet{Sutinjo15} was used to produce Stokes \textit{I} 
images from the instrumental polarizations, 
and the SolarSoftWare (SSW\footnote{SSW: \url{www.lmsal.com/solarsoft/}}, \citealp{Freeland98}) 
routine \textsf{mwa\_prep} \citep{McCauley17} was used to translate the images onto solar coordinates. 
Flux calibration was achieved by comparison with thermal bremsstrahlung and gyroresonance emission predictions from 
FORWARD\footnote{FORWARD: \url{www2.hao.ucar.edu/modeling/FORWARD-home}} \citep{Gibson16} 
based on the Magnetohydrodynamic Algorithm outside a Sphere model
(MAS\footnote{MAS: \url{www.predsci.com/hmi/data\_access.php}}; \citealp{Lionello09}). 


\subsection{Event Selection}
\label{events}

These data are part of an imaging survey of many
Type III bursts observed by the MWA during 45 separate observing 
periods in 2014 and 2015. 
\citet{McCauley17} performed a case study of an event that exhibits 
unusual source motion, and future work will present statistical analyses. 
Burst periods during MWA observing runs were identified using the daily 
National Oceanic and Atmospheric Administration (NOAA) solar event 
reports\footnote{NOAA event reports: \url{www.swpc.noaa.gov/products/solar-and-geophysical-event-reports}}
based on observations from the \textit{Learmonth}  \citep{Guidice81,Kennewell03}
and \textit{Culgoora} \citep{Prestage94} solar radio spectrographs, which overlap with 
the MWA's frequency range at the low and high ends, respectively. 


 \begin{figure}\graphicspath{{chapter3/}}
 \centerline{\includegraphics[width=\textwidth,clip=]{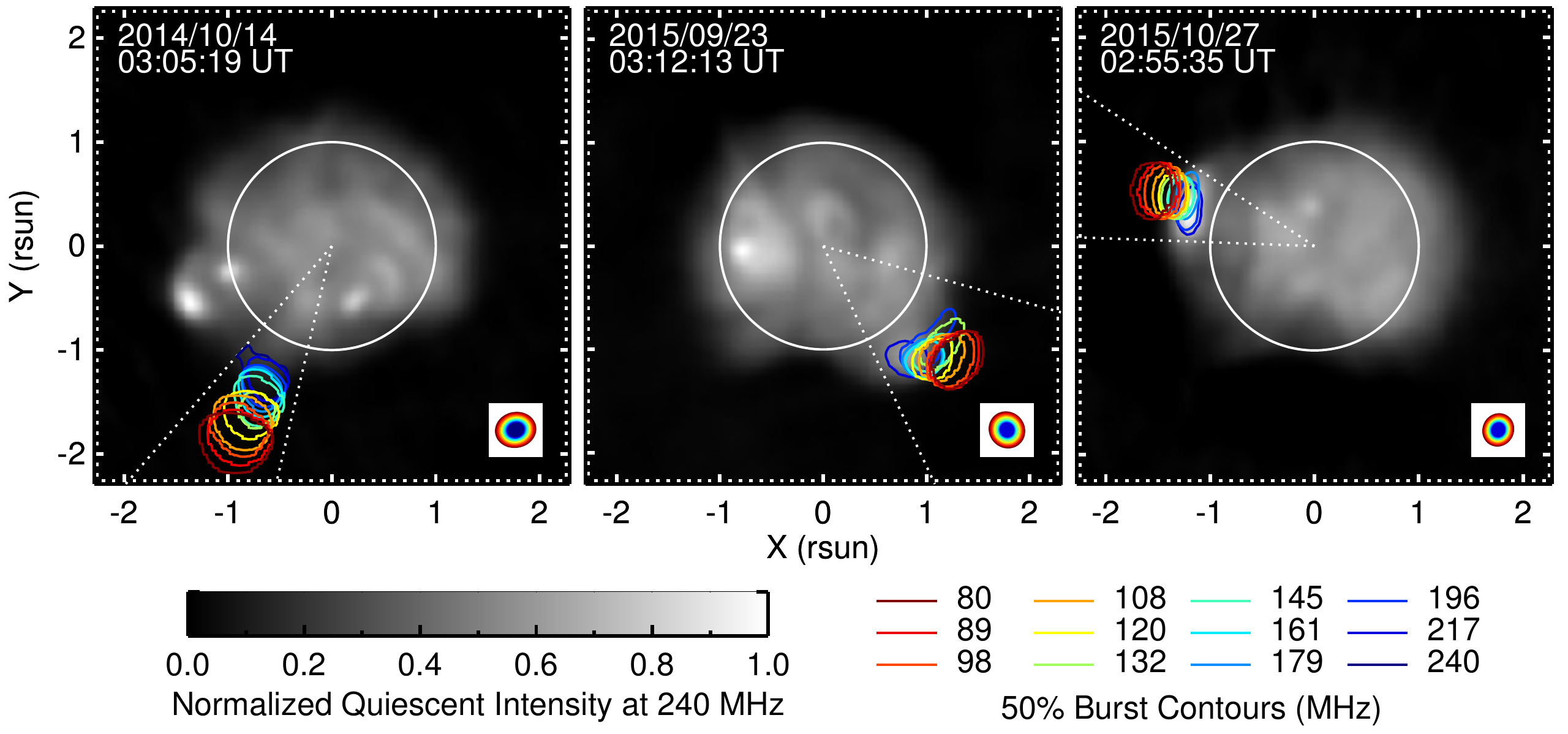}}
 \caption{\footnotesize{}MWA Type III burst contours at 50\% of the peak intensity 
 for each channel overlaid on 240 MHz images of the quiescent corona. 
 The solid circle represents the optical disk, and dotted lines bound the region included 
 in the dynamic spectra (Figures~\ref{fig:ds_comp} and \ref{fig:spectra}). 
 Colored ellipses in the lower-right corners show the synthesized beam sizes for each channel.}
 \label{fig:centroids}
 \end{figure}

Three events were selected from the full sample based on the following criteria. 
First, the burst sites needed to be located at the radio limb with roughly 
radial progressions across frequency channels. 
Limb events minimize projection effects, allowing us to reasonably approximate the 
projected distance from Sun-center as the actual radial height. 
Second, to eliminate potential confusion between multiple events and to maximize spectral coverage,
the bursts needed to be sufficiently isolated in time and frequency, with a 
coherent drift from high to low frequencies across the full MWA bandwidth. 
Third, the source regions needed to be relatively uncomplicated ellipses with 
little-to-no intrinsic motion of the sort described by \citet{McCauley17}. 
This again minimizes projection effects and ensures that we follow a single 
beam trajectory for each event. 


 \begin{figure}\graphicspath{{chapter3/}}
 \centerline{\includegraphics[width=\textwidth,clip=]{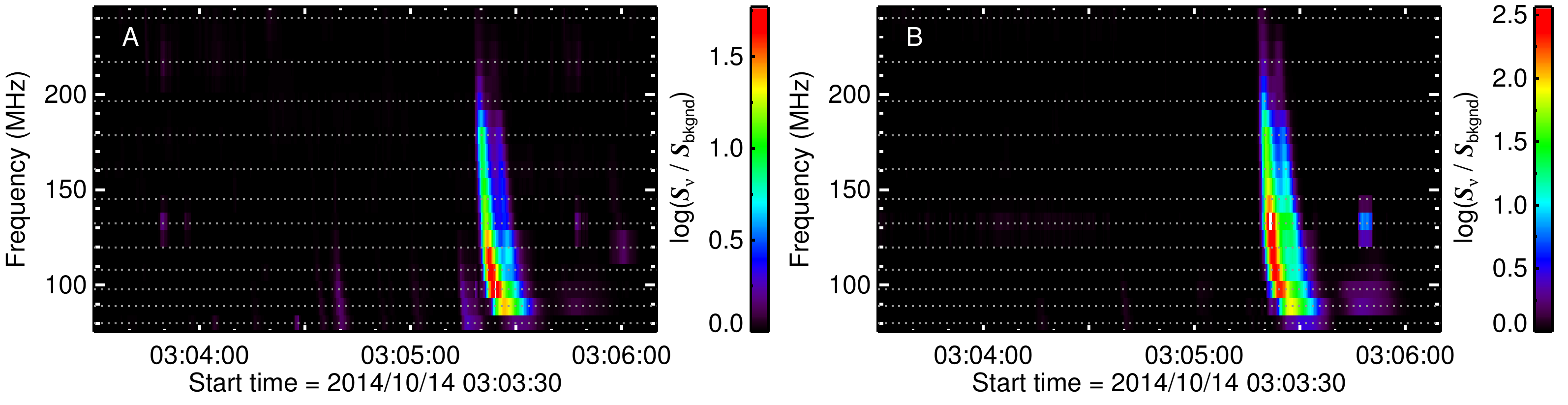}}
 \caption{\footnotesize{}Dynamic spectra constructed from image intensities 
 for the Type III burst near 03:05:20 UT on 14 October 2014. 
 Panel A includes the full FOV, while panel B includes only 
 the segment bounded by the dotted lines in Figure~\ref{fig:centroids}.
Dotted horizontal lines show the locations of the 12 channels, each having a spectral width of 2.56 MHz. 
Intensities have been divided by the background level and plotted on a logarithmic scale. 
 }
 \label{fig:ds_comp}
 \end{figure}


\begin{figure}\graphicspath{{chapter3/}}
\centerline{\includegraphics[width=\textwidth,clip=]{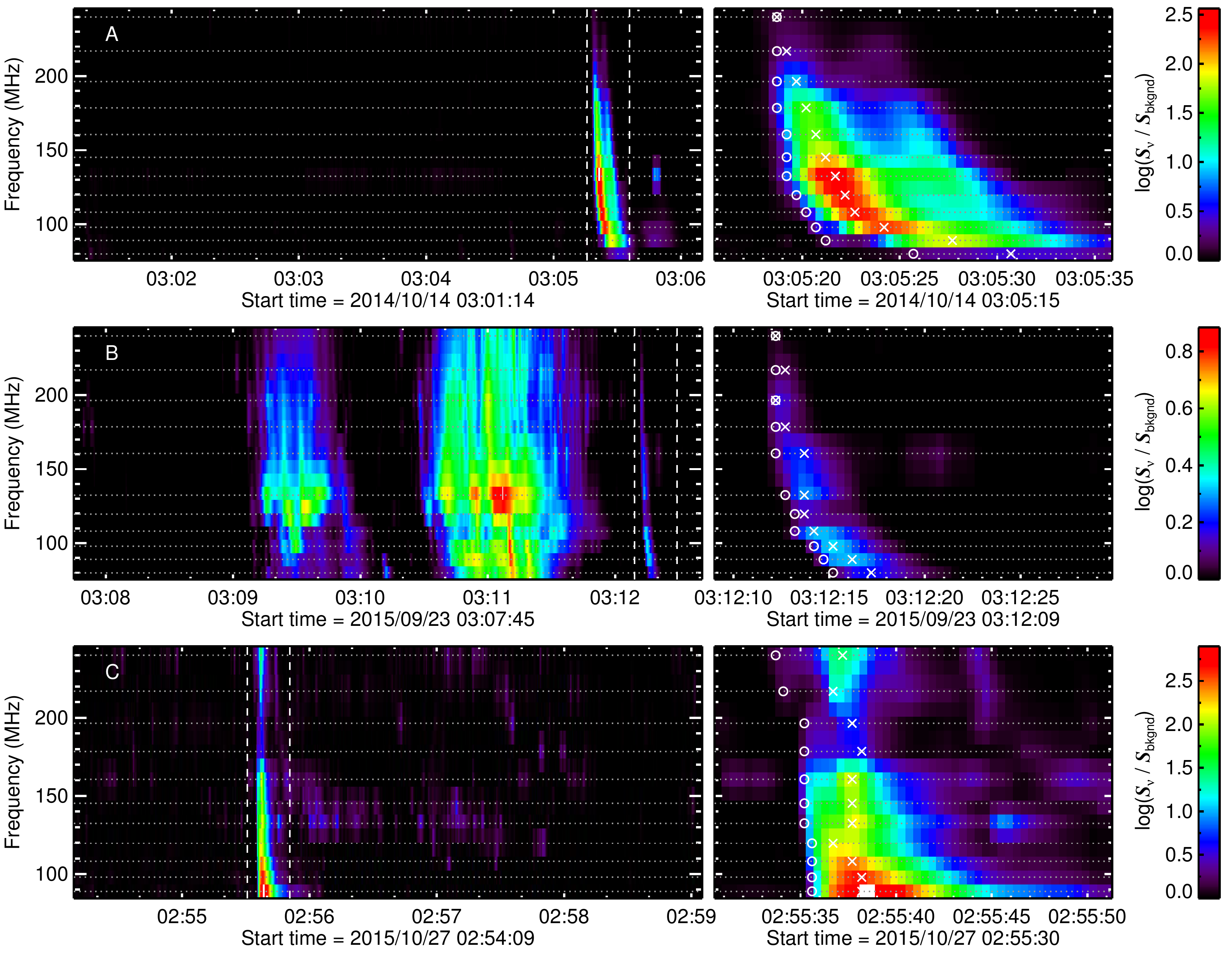}}
\caption{\footnotesize{}Dynamic spectra constructed from partial image intensities, including only the FOV segment bounded 
by the dotted lines in Figure~\ref{fig:centroids}. 
The left column shows the full five-minute observation intervals, while the right column shows 20-second  
periods surrounding the selected Type III bursts. 
Circles and crosses denote the onset and peak burst times for each channel.   
}
\label{fig:spectra}
\end{figure}

Figure~\ref{fig:centroids} shows the burst contours for each channel overlaid on 
quiescent background images at 240 MHz.
Each of the three events occurred on a different day, and we refer to them by the 
UTC date on which they occurred 
(SOL identifiers: SOL2014-10-14T03:05:19, SOL2015-09-23T03:12:12, and SOL2015-10-27T02:55:34). 
Figure~\ref{fig:ds_comp} shows dynamic spectra for the 14 October 2014 event, with the left panel  
covering the full Sun and the right panel including only the region demarcated by the dotted 
lines in Figure~\ref{fig:centroids}. 
The partial-Sun spectrum excludes a neighboring region that is active over the 
same period, allowing the Type III frequency structure to be more easily followed. 
This approach is similar to that of \citet{Mohan17}, who discuss the utility of spatially resolved dynamic spectra. 
Figure~\ref{fig:spectra} shows the masked spectra for all three events. 


\begin{figure}\graphicspath{{chapter3/}}
\centerline{\includegraphics[width=0.95\textwidth,clip=]{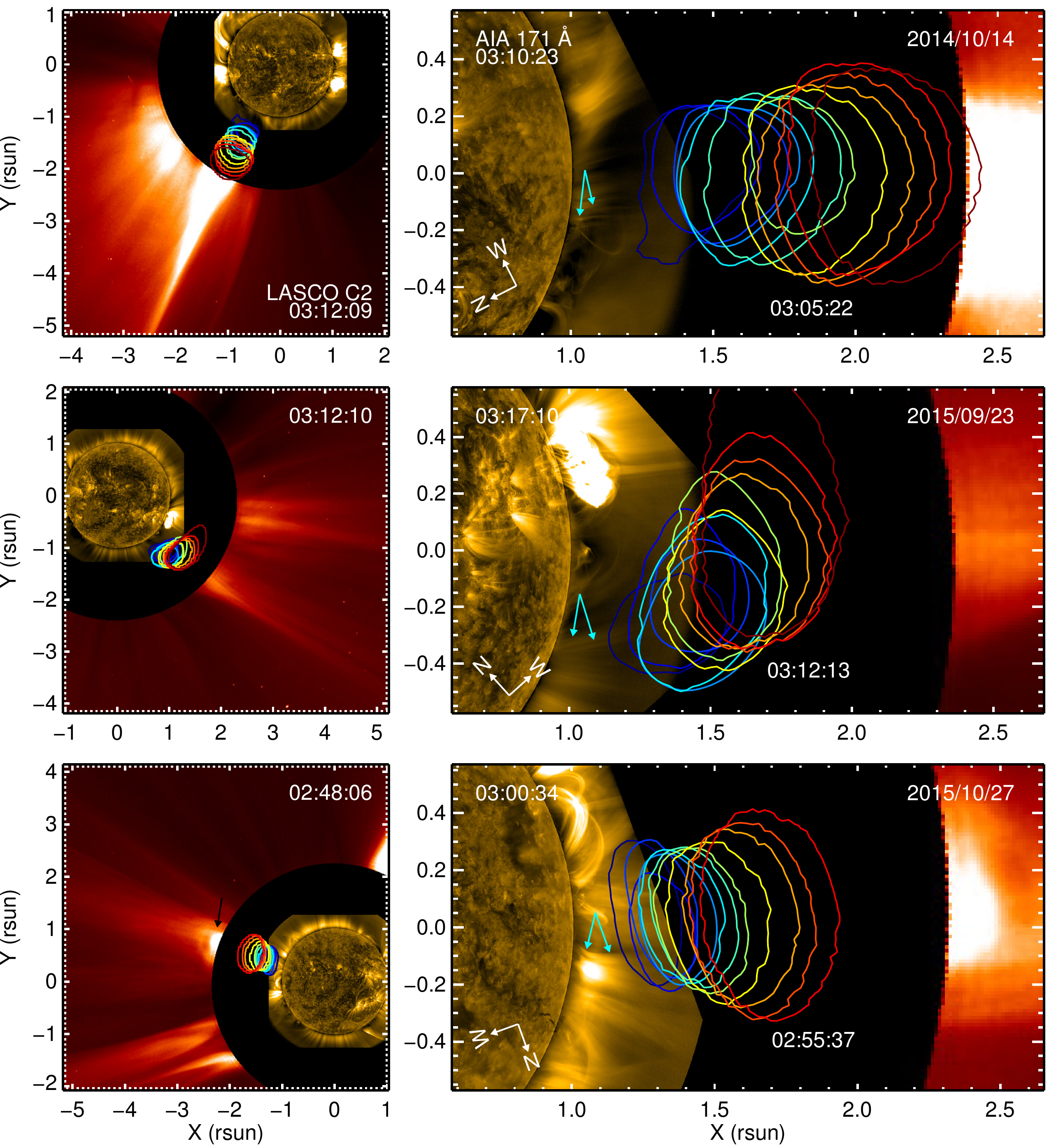}}
\clearpage
\caption{\footnotesize{}Overlays of the 50\% MWA burst contours onto 
AIA 171 \AA{} and LASCO-C2 images. 
Contour colors are for spectral channels from 80\,--\,240 MHz, as in Figure~\ref{fig:centroids}, where dark 
red represents the lowest frequency and dark blue represents the highest.
\edit{UTC observation times are shown for LASCO in the left panel, for AIA in the upper-left 
of the right panel, and for MWA in the middle of the right panel. 
The MWA times reflect the average peak time across frequency channels (see Figure~\ref{fig:spectra}). 
The AIA images are ten-minute (50-image) averages processed with a radial filter to accentuate off-limb features; 
times reflect the middle of these ten-minute windows, which begin at the burst onsets and cover the subsequent periods 
over which associated EUV signatures would be expected.}
Images are rotated in the right column such that the burst progression 
is roughly horizontal, which helps illustrate the extent to which each event 
progresses radially. 
Cyan arrows point to the EUV structures that exhibit activity  
during or just after the radio bursts. 
\edit{The black arrow in the lower-left panel points to a CME that originated behind the limb and 
passed the C2 occulting disk around 20 minutes prior to the Type III burst.} 
}
\label{fig:overlay}
\end{figure}

\subsection{Context}
\label{context}

In this section, we briefly describe the context for each of the radio bursts with 
respect to observations at other wavelengths and associated phenomena. 
Figure~\ref{fig:overlay} overlays the burst contours from Figure~\ref{fig:centroids} onto 
contemporaneous extreme-ultraviolet (EUV) and white-light data. 
The white-light images were produced by the 
\textit{Large Angle and Spectrometric Coronagraph C2} (LASCO-C2: \citealp{Brueckner95}) 
onboard the \textit{Solar and Heliospheric Observatory} (SOHO: \citealp{Domingo95}).  
\edit{C2 has an observing cadence of 20 min, and Figure~\ref{fig:overlay} 
includes the nearest images in time to our radio bursts.}

The EUV data come from the 
\textit{Atmospheric Imaging Assembly} (AIA: \citealp{Lemen12}) 
onboard the \textit{Solar Dynamics Observatory} (SDO: \citealp{Pesnell12}).
We use the 171 \AA{} AIA channel, which is dominated by Fe \textsc{ix} 
emission produced by plasma at around 0.63 MK, because it most 
clearly delineates the fine magnetic structures along which Type III beams 
are expected to travel. 
To further accentuate off-limb features, we apply a radial filter 
using the SSW routine \textsf{aia\_rfilter} \citep{Masson14}. 
Note that the apparent brightness of a given pixel in a radial filter image corresponds to  
its true intensity relative only to pixels of the same radial height (\textit{i.e.} equally bright structures 
at different heights do not have the same physical intensity).
\edit{AIA has an observing cadence of 12 seconds, and Figure~\ref{fig:overlay} uses ten-minute (50-image) 
averages that cover the periods during and immediately after the radio bursts. 
This temporal window is used because a potential EUV signature associated with a 
Type III burst will propagate at a much lower speed than the burst-driving electron beam and 
will likely be most apparent in the minutes following the burst (\textit{e.g.} \citealp{McCauley17,Cairns18}).}

In all cases, the radio bursts appear to be aligned with dense structures visible to AIA at lower 
heights and to LASCO-C2 at larger heights. 
The latter case is obvious, with each set of burst contours situated just below bright 
white-light streamers. 
Cyan arrows in the right panels of Figure~\ref{fig:overlay} identify the associated EUV structures, each 
of which exhibits a mild brightening and/or outflow during or immediately after the corresponding radio burst. 
This activity may be indicative of weak EUV jets, which are frequently associated 
with Type III bursts (\textit{e.g.} \citealp{Chen13,Innes16,McCauley17,Cairns18}), but robust outflows are not observed here. 
The alignment between the EUV and radio burst structure is particularly 
striking for the 23 September 2015 event in that both appear to follow roughly the same non-radial arc. 
A correspondence between EUV rays and Type III bursts was previously reported by \citet{Pick09}. 
 
Type III bursts are commonly, but not always, associated with X-ray flares (\textit{e.g.} \citealp{Benz05,Benz07,Cairns18}) and occasionally with 
coronal mass ejections (CMEs; \textit{e.g.} \citealp{Cane02,Cliver09}). 
Our 14 October 2014 event is not associated with either, but the other two are. 
On 23 September 2015, a weak B-class flare occurred just to the north of 
our radio sources from Active Region 12415.  
The flare peaked around 3:11 UT, which corresponds to a period of relatively 
intense coherent radio emission that precedes the weaker burst of interest here (see Figure~\ref{fig:spectra}). 
Given the radio source positions and associated EUV structure, we do not believe the flare site 
to be the source of accelerated electrons for our event, although the flare may have been responsible 
for stimulating further reconnection to the south. 

\edit{On 27 October 2015, a CME was ongoing at the time of the radio burst, and its leading edge, 
indicated by the black arrow in the lower left panel of Figure~\ref{fig:overlay}, 
can be seen just above the C2 occulting disk. 
Inspection of images from the Extreme Ultraviolet Imager (EUVI; \citealp{Howard08}) onboard the
\textit{Solar Terrestrial Relations Observatory A} (STEREO-A) 
spacecraft shows that the CME originated from a large active region close to the east limb 
but occulted by the disk from AIA's perspective. 
The CME was launched well before our Type III burst, but the region that produced it was very active 
over this period and is likely connected to the activity visible to AIA immediately after the radio burst 
along the structure indicated by the cyan arrows in the lower-right panel of Figure~\ref{fig:overlay}. 
So while we do not think the CME was directly involved in triggering the radio burst, it may have impacted 
the medium through which the Type III electron beam would later propagate, which is relevant 
to a hypothesis proposed by \citet{Morosan14} that will be discussed in Section~\ref{discussion2}.}


\section{Analysis and Results}
\label{analysis}

\subsection{Density Profiles}
\label{density}

Standard plasma emission theory expects Type III radiation  
at either the ambient electron plasma frequency [$f_{\rm p}$] or its harmonic 
[$2f_{\rm p}$]. 
The emission frequency [$f$] is related to electron density [$n_{\mathrm e}$] in the following way \edit{[cgs units]}:  

\edit{
\begin{equation} \label{eq:fp}
f = \textrm{N}f_{\rm p} = \textrm{N}\sqrt{\frac{e^2n_{\mathrm e}}{\pi{}m_{\mathrm e}}} 
~~~\Rightarrow{}~~~ 
n_{\mathrm e} = \pi{}m_{e}\left(\frac{f}{e\textrm{N}}\right)^2,
\end{equation} 
}

\noindent where  
$e$ is the electron charge, 
$m_{\mathrm e}$ is the electron mass, 
and \edit{N} is either 1 (fundamental) or 2 (harmonic). 
For frequencies in \edit{Hz} and densities in cm\tsp{-3}, $n_{\mathrm e} \approx 1.24\e{-8}f^2$ for fundamental 
and $3.10\e{-9}f^2$ for harmonic emission. 

Density can thus be easily extracted given the emission mode and location. 
Unfortunately, neither property is entirely straightforward. 
Harmonic emission is often favored in the corona because being produced above the ambient 
$f_{\mathrm p}$ makes it less likely to absorbed \citep{Bastian98} and because Type IIIs tend to be 
more weakly circularly polarized than expected for fundamental emission \citep{Dulk80}.
Harmonic emission also implies lower densities by a factor of four, which are easier to 
reconcile with the large heights often observed (see Section \ref{introduction2}). 
However, fundamental-harmonic pairs can be observed near our frequency range (\textit{e.g.} \citealp{Kontar17}),  
fundamental emission is expected to contribute significantly to interplanetary Type III burst spectra (\textit{e.g.} \citealp{Robinson98}), 
and fundamental emission is thought to be the more efficient process from a theoretical perspective (\textit{e.g.} \citealp{Li13b,Li14}). 
As described in Section~\ref{introduction2}, a source's apparent height may also be 
augmented by \edit{propagation effects}, which we will consider in Section~\ref{propagation}.


\begin{figure}\graphicspath{{chapter3/}}
\centerline{\includegraphics[width=\textwidth,clip=]{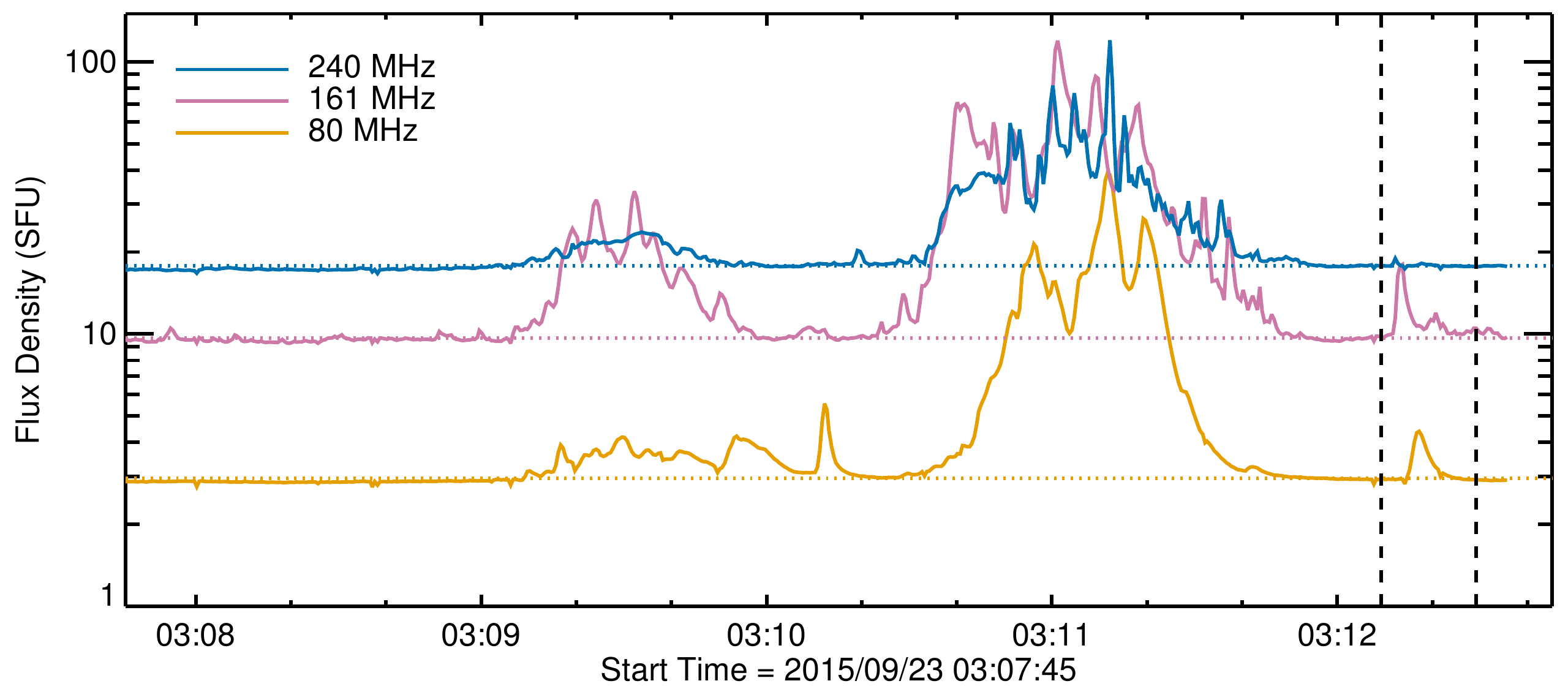}}
\caption{\footnotesize{}Light curves for the 23 September 2015 observation, shown to illustrate the background level determination. 
Backgrounds (dotted lines) are obtained by taking the median intensity, 
excluding points two standard deviations above that, 
and iterating until no more points are excluded. 
The dashed lines mark the burst period from the right column of Figure~\ref{fig:spectra}.}
\label{fig:baseline}
\end{figure}

We measure source heights at the onset of burst emission, which we define as 
when the total intensity reaches 1.3$\times$ the background level. 
Background levels are determined for each frequency by taking the median intensity, excluding 
points 2 standard deviations above that, and iterating until no more points are excluded. 
Figure~\ref{fig:baseline} shows the result of this baseline procedure for three frequencies from the 23 September 2015 event, 
which is shown because it exhibits the most complicated dynamic spectrum. 
Onset times are represented by circles 
in Figure~\ref{fig:spectra}, and centroids are obtained at these 
times from two-dimensional (2D) Gaussian fits.
As mentioned in Section \ref{events}, these events were chosen because they appear at the radio limb 
and thus the 2D plane-of-sky positions can reasonably approximate the 
physical altitude. 
Geometrically, these heights are lower limits to the true radial height, but  
propagation effects that increase apparent height 
are likely to be more important than the projection angle (see Section \ref{propagation}).   


\begin{figure}\graphicspath{{chapter3/}}
\centerline{\includegraphics[width=\textwidth,clip=]{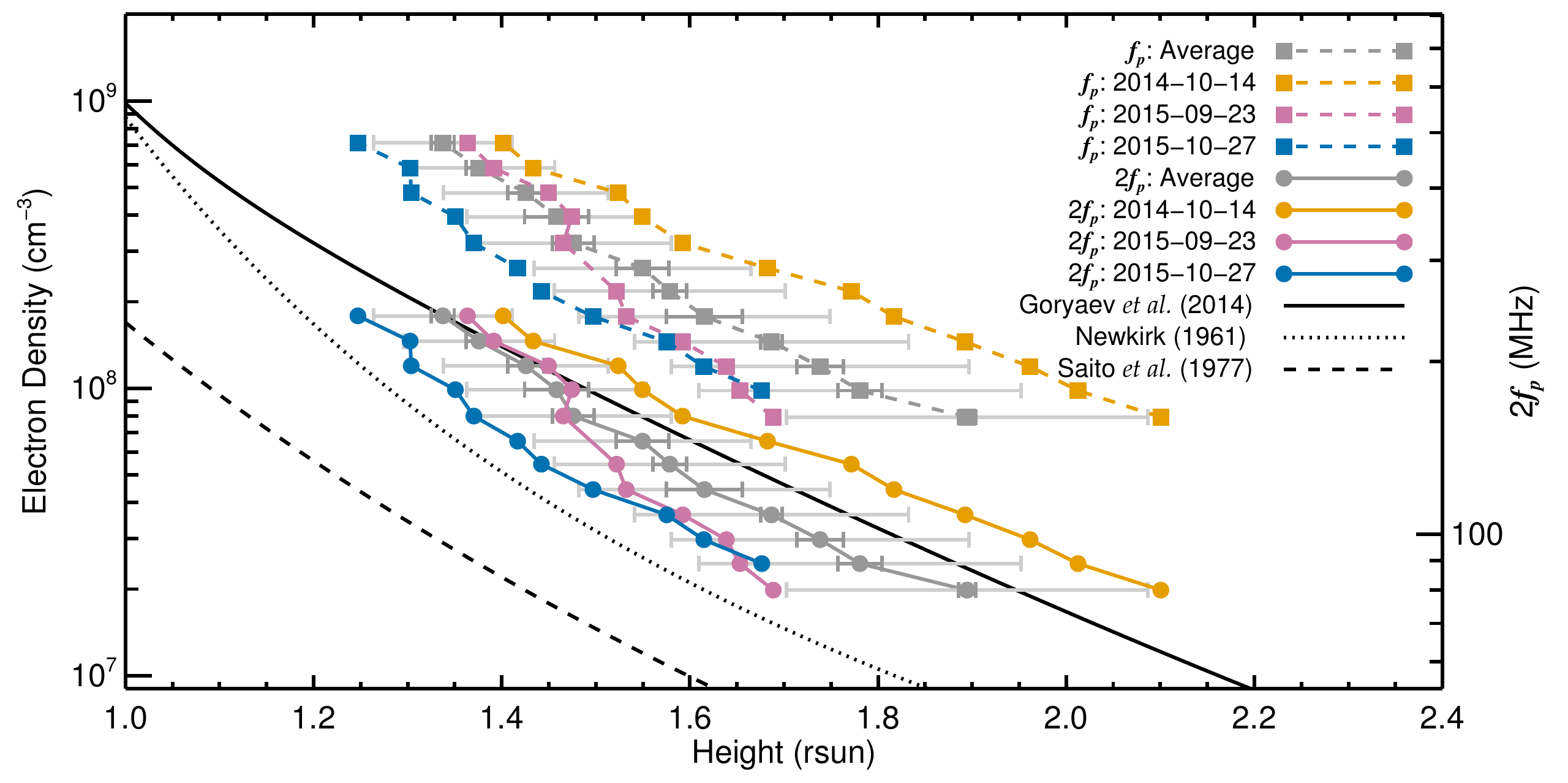}}
\caption{\footnotesize{}Densities inferred from the Type III source positions assuming fundamental [$f_{\mathrm p}$; dashed] 
or harmonic [$2f_{\mathrm p}$; solid] emission. 
Background coronal models based on white-light data near solar minimum \citep{Saito77} and 
maximum \citep{Newkirk61} are shown for comparison, along with a recent streamer model 
based on EUV data \citep{Goryaev14}.
Only the average uncertainties are shown for clarity; the dark-gray bars represent the one-$\sigma$ centroid  
variability over the full burst, and the light-gray bars represent the major axes of the synthesized beams. 
}
\label{fig:dens1}
\end{figure}  

Figure~\ref{fig:dens1} plots height \textit{versus} density for both the fundamental and harmonic assumptions. 
Two sets of height uncertainties are shown for the average density profiles. 
The smaller, dark-gray error bars reflect the one-$\sigma$ position variability over the full burst durations, and the larger, 
light-gray bars reflect the full width at half maximum (FWHM) of the synethesized beam major axes. 
Note that if the source is dominated by a single compact component, which would be a reasonable assumption here, 
then the FWHM resolution uncertainty can be reduced by a factor inversely proportional to the signal-to-noise ratio (SNR) \citep{Lonsdale18,Reid88}. 
\edit{Given our high SNRs, which average 217$\sigma$ at the burst onsets,} this ``spot mapping" approach \edit{typically} results in sub-arcsecond position uncertainties on the apparent source location.
However, spatial shifts may be introduced by changes in the ionosphere between the solar and calibration observation 
times, and more importantly, an apparent source may differ significantly from its actual emission site due to propagation effects (\textit{i.e.} refraction and scattering). 
For these reasons, we opt to show the more conservative uncertainties outlined above. 

For comparison, Figure~\ref{fig:dens1} includes radial density models from \citet{Saito77}, \citet{Newkirk61}, and \citet{Goryaev14}. 
The \citeauthor{Saito77} profile refers to the equatorial background near solar minimum based on white 
light polarized brightness data, while the \citeauthor{Newkirk61} curve is based on similar data near solar maximum 
and implies the largest densities among ``standard" background models. 
The \citeauthor{Goryaev14} model instead refers to a dense streamer and is based on a 
novel technique using widefield EUV imaging. 
This profile is somewhat elevated above streamer densities 
inferred from contemporary white-light (\textit{e.g.} \citealp{Gibson99}) 
and spectroscopic (\textit{e.g.} \citealp{Parenti00,Spadaro07}) measurements at similar heights, although 
some earlier white-light studies found comparably large streamer densities (\textit{e.g.} \citealp{Saito67}).
For additional coronal density profiles, see also \citet{Allen47,Koutchmy94,Guhathakurta96,Mann99,Mercier15,Wang17} 
and references therein. 

From Figure~\ref{fig:dens1}, we see that the Type III densities assuming fundamental emission are an average  
of 3\,--\,4$\times$ higher than the EUV streamer model. 
These values may be unreasonably large, meaning either that the fundamental emission hypothesis is 
not viable here or that fundamental emission \edit{originating} from a lower altitude 
\edit{was observed a larger height due to propagation effects} (see Section \ref{propagation}). 
Assuming harmonic emission, the 14 October 2014 burst implies electron densities of 1.8\e{8} cm\tsp{-3} (240 MHz) 
at \rsolar{1.40} down to 
0.20\e{8} cm\tsp{-3} (80 MHz) at \rsolar{2.10}. 
This represents a moderate ($\approx 1.4\times$) enhancement over the \citeauthor{Goryaev14} streamer model 
or a significant ($\approx 4.1\times$) enhancement over the \citeauthor{Newkirk61} background. 
The other two events fall between the EUV streamer and solar-maximum-background models, with the 
27 October 2015 source heights implying densities of 1.8\e{8} cm\tsp{-3} (240 MHz) at \rsolar{1.25} down to 
0.20\e{8} cm\tsp{-3} (80 MHz) at \rsolar{1.68}. 
Note that the 23 September 2015 burst implies an unusually steep density gradient that is not consistent with standard radial 
density models, perhaps because that event 
deviates significantly from the radial direction (see Figure~\ref{fig:overlay}).


\begin{figure}\graphicspath{{chapter3/}}
\centerline{\includegraphics[width=\textwidth,clip=]{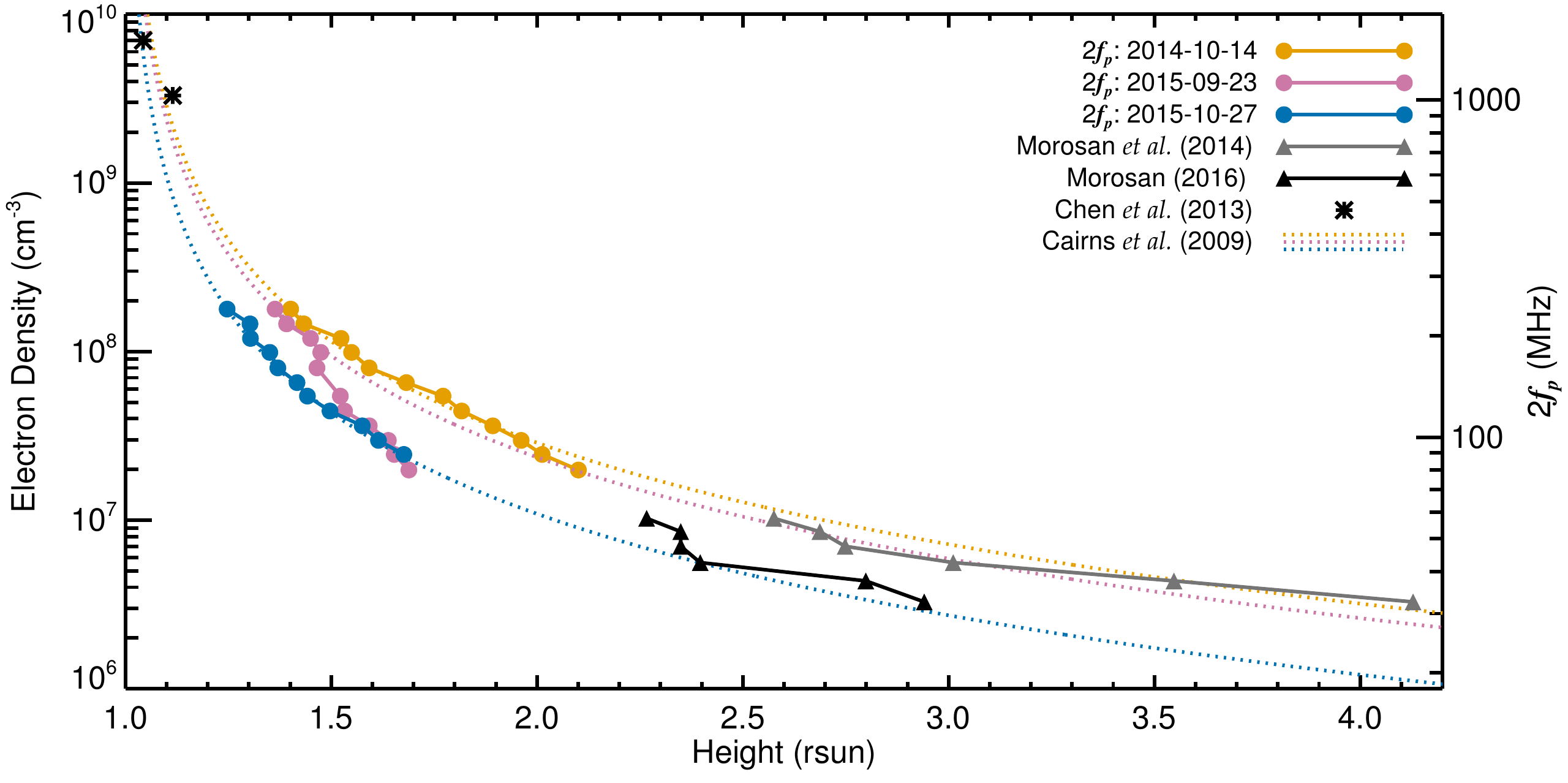}}
\caption{\footnotesize{}Densities assuming harmonic emission compared to recent Type III results at higher \citep{Chen13} 
and lower \citep{Morosan14,Morosan16} frequencies. 
\editb{The \citet{Morosan14} points (gray triangles) correspond to the same data as the \citet{Morosan16} values (black triangles), 
but the latter have been adjusted to account for ionospheric refraction.}
The dotted lines apply the $n_{\mathrm e}(r) = C(r - 1)^{-2}$ profile detailed by \citet{Cairns09}, where 
the constant $C$ has been normalized to the density implied by our 240 MHz source positions. 
}
\label{fig:dens2}
\end{figure} 

Figure~\ref{fig:dens2} shows how our results compare to densities inferred from recent Type III imaging at 
higher and lower frequencies, all assuming harmonic emission. 
The high-frequency (1.0\,--\,1.5 GHz) results come from \citet{Chen13}, who used the \textit{Very Large Array} (VLA) 
to find densities around an order of magnitude above the background. 
The low-frequency (30\,--\,60 MHz) points were obtained using the 
\textit{Low Frequency Array} (LOFAR) by \citet{Morosan14}, who also found large enhancements. 
We plot data from their ``Burst 2" (see Figures 3 and 4) because it began beyond 
our average radio limb height at 80 MHz. 
Their other two events exhibit 60 MHz emission near the optical limb, which may indicate 
that the 2D plane-of-sky positions significantly underestimate the true altitudes 
(\textit{i.e.} those electron beams may have been inclined toward the observer). 
\editb{The \citet{Morosan14} data are also reproduced in Chapter 4 of \citet{Morosan16}, 
who notes that ionospheric refraction likely contributed significantly to the observed source heights (D.E. Morosan, private communication, 2018), and 
adjusted values are also shown in Figure~\ref{fig:dens2}. 
As previously noted, ionospheric refraction may affect our positions as well if conditions changed significantly 
in the roughly 2 hours between the calibrator and solar observation times.
This effect would likewise shift the solar disk, and given that this is not noticeable (see Figure~\ref{fig:centroids}), we conclude that any 
positional shifts imparted by a changing ionosphere are within our conservative uncertainty estimates.}

Figure~\ref{fig:dens2} also includes density curves of the form $n_{\mathrm e}(r) = C(r - 1)^{-2}$, where 
$r$ is in solar radii and $C$ is normalized to match the densities implied by our 240 MHz source heights. 
This model was introduced by \citet{Cairns09} based on Type III frequency drift rates over 40\,--\,180 MHz 
and was subsequently validated over a larger frequency range by \citet{Lobzin10}. 
The \citeauthor{Cairns09} model is somewhat steeper than others over the MWA's height range 
($\approx$ 1.25\,--\,2.10\rsolar{}) but becomes more gently sloping at larger heights,  
effectively bridging the corona to solar-wind transition. 
From Figure~\ref{fig:dens2}, we see that this model is a good fit to the 14 October 2014 and 27 October 2015 data. 
The 23 September 2015 event is not well-fit by this or any other standard model, 
which may be attributed to its aforementioned non-radial structure.
Extending these gradients to larger heights matches the LOFAR data fairly well 
and likewise with the VLA data at lower heights, which come from higher frequencies than have been 
examined with this model previously. 

\subsection{Electron Beam Kinematics}
\label{speed}

Type III beams speeds are known primarily from frequency drift rates [${\mathrm d}f/{\mathrm d}t$] observed in dynamic spectra. 
Assuming either fundamental or harmonic emission, a given burst frequency can be straightforwardly 
converted into a radial height given a density model $n_{\mathrm e}(r)$, and ${\mathrm d}f/{\mathrm d}t$ then becomes ${\mathrm d}r/{\mathrm d}t$. 
The literature includes a wide range of values using this technique, reflecting the variability  
among models as well as any intrinsic variability in electron speed. 
Modest fractions of light speed are typically inferred from drift rates of coronal bursts ($\approx$ 0.1\,--\,0.4 c; 
\citealp{Alvarez73,Aschwanden95,Melendez99,Kishore17}), though speeds larger 
than 0.5 c have been reported by some studies \citep{Poquerusse94,Klassen03,Carley16}. 
Our imaging observations allow us to measure the exciter speed without assuming $n_{\mathrm e}(r)$ by following the 
apparent height progression of Type III sources at different frequencies. 

\begin{figure}\graphicspath{{chapter3/}}
\centerline{\includegraphics[width=\textwidth,clip=]{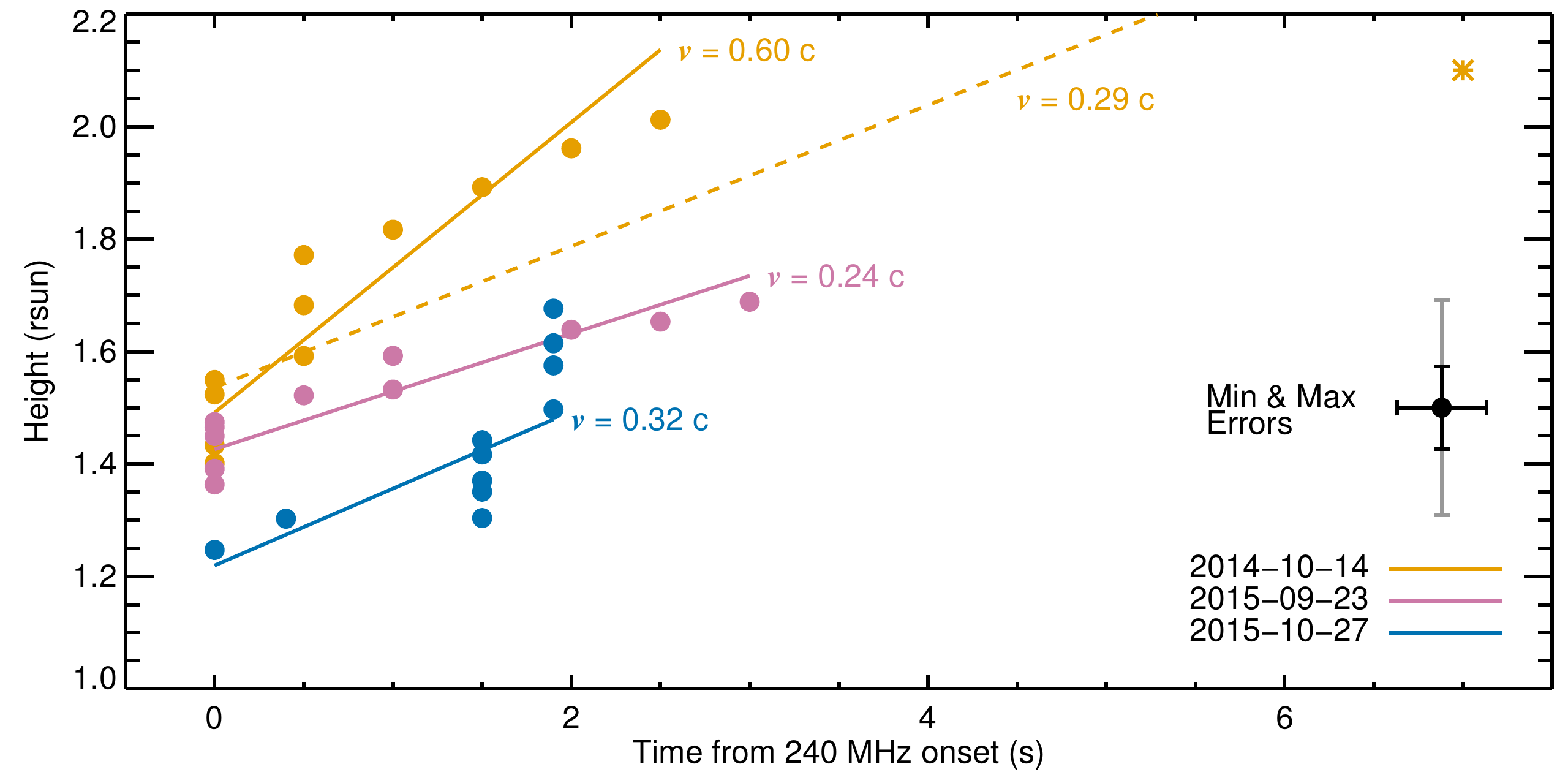}}
\caption{\footnotesize{}Exciter speed estimates from the time- and frequency-varying source positions.
The dashed orange line includes the high time outlier (orange asterisk). 
The uncertainties shown in the lower right are the same for a given frequency and reflect the time and spatial resolutions. 
The black bar represents the smallest synthesized beam size at 240 MHz (corresponding to the lower-left points), 
and the gray bar represents the largest beam size at 80 MHz (corresponding to the upper-right points). 
}
\label{fig:speeds}
\end{figure}

As in the previous section, we obtain radial heights from centroid positions at the 
onset of burst emission for each frequency. 
These data are plotted in Figure~\ref{fig:speeds} along with linear least-squares fits to the speed using   
the time and spatial resolutions as uncertainties. 
The 14 October 2014 event exhibits an anomalously \edit{late} onset time at 80 MHz 
\edit{(see the circles in Figure~\ref{fig:spectra}a and the orange asterisk in Figure~\ref{fig:speeds}). 
This is likely due to the diminished intensity at that frequency, which precludes an appropriate comparison 
to the onset times at higher frequencies where the burst is much more intense. 
Figure~\ref{fig:speeds} shows fits both including (0.29 c) and excluding (0.60 c) the 80 MHz point for 
the 14 October 2014 event, and the latter value is used in the discussion to follow because of the better overall fit. 
Note that while the onset of 80 MHz emission is at a later time than expected given the prior frequency progression, 
the source location is consistent with the other channels and thus 
its inclusion does not impact the inferred density profile from Figures~\ref{fig:dens1} and \ref{fig:dens2}.}


We find an average speed across events of 0.39 c, which is consistent with results from other imaging observations.
The same strategy was recently employed at lower frequencies by \citet{Morosan14}, who found an 
average of 0.45 c. 
\citet{McCauley17} indirectly inferred a beam speed of 0.2 c from MWA imaging. 
\citet{Chen13} also tracked centroid positions at higher frequencies, although 
in projection across the disk, finding 0.3 c. 
\edit{\citet{Mann18} recently examined the apparent speeds of three temporally adjacent Type III bursts imaged by 
LOFAR. They find that the sources do not propagate with uniform speed, with each burst exhibiting an  
acceleration in apparent height, and they conclude that the exciting electron beams must have broad velocity distributions. 
From Figure~\ref{fig:speeds}, we observe an apparent acceleration only for one event (27 October 2015), with the other two events 
exhibiting the opposite trend to some extent. 
However, our data are consistent with \citet{Mann18} in that a uniform speed is not a particularly good fit for any of our events, but 
the MWA's 0.5-second temporal resolution limits our ability to characterize the source speeds in great detail.}


\begin{table}
\footnotesize
\caption{\footnotesize{}Imaging Beam Speeds vs. ${\mathrm d}f/{\mathrm d}t$ Model Predictions}
\label{tab:speeds}
\begin{tabular}{l|c|ccc|c}
\hline
& \multicolumn{5}{c}{Beam Speed [c]} \\
&  & \multicolumn{3}{c|}{Assuming $f_{\mathrm p}$ \,--\, $2f_{\mathrm p}$ emission} &  \\
Event & Imaging & Goryaev \textit{et al.} & Newkirk & Saito \textit{et al.}\tabnote{$f_p$ case not viable because model does not include densities above $f_p \approx$ 116 MHz.} & Cairns \textit{et al.}\tabnote{Model normalized to match the densities implied by our 240 MHz heights.} \\
\hline
14 Oct 2014\tabnote{Excludes the 80 MHz outlier (orange asterisk in Figure 10).} & 0.60 $\pm$ 0.13 & 0.38 \,--\, 0.45 & 0.22 \,--\, 0.31 & *** \,--\, 0.30 & 0.58 \\
23 Sep 2015 & 0.24 $\pm$ 0.10 & 0.34 \,--\, 0.40 & 0.20 \,--\, 0.28 & *** \,--\, 0.27 & 0.50 \\
27 Oct 2015 & 0.32 $\pm$ 0.12 & 0.44 \,--\, 0.55 & 0.26 \,--\, 0.36 & *** \,--\, 0.48 & 0.40 \\
\hline
\end{tabular}
\tsp{a}$f_p$ case not viable because model does not include densities above $f_p \approx$ 116 MHz.\\
\tsp{b}Model normalized to match the densities implied by our 240 MHz heights.\\
\tsp{c}Excludes the 80 MHz outlier (orange asterisk in Figure~\ref{fig:speeds}).
\end{table}

Taken together, we see that speeds measured from imaging observations tend to produce values 
at the higher end of what is typical for ${\mathrm d}f/{\mathrm d}t$ inferences. 
We compare the two approaches for the same events in Table~\ref{tab:speeds} using the same 
models shown in Figure~\ref{fig:dens1}. 
We also include speeds derived using the \citet{Cairns09} model, normalized to the densities 
implied by our 240 MHz source heights.  
These values are separated from the others in Table~\ref{tab:speeds} because the normalization 
precludes direct comparisons to the other models. 
The ${\mathrm d}f/{\mathrm d}t$-inferred speeds are consistently smaller than the imaging results for the 14 October 2014 event, which 
was also true for the bursts studied by \citet{Morosan14}, but there is no major difference between the 
two approaches for our other events given the range of values. 
Note that this comparison is arguably a less direct version of the  
height \textit{versus} density comparison from the previous section in that 
the extent to which the imaging and model-dependent ${\mathrm d}f/{\mathrm d}t$ speeds agree unsurprisingly mirrors the extent 
to which the density profiles themselves agree.  
The 14 October 2014 speeds are closest to those derived using $n_{\mathrm e}(r)$ from \citeauthor{Goryaev14}, and 
the 27 October 2015 result is closest to the \citeauthor{Newkirk61}-derived speed, both assuming 
harmonic emission, because those density profiles are most closely matched in Figure~\ref{fig:dens1}.  
Likewise, the speeds from those events agree well with ${\mathrm d}f/{\mathrm d}t$ speeds obtained using the 
normalized \citeauthor{Cairns09} curves because a $C(r - 1)^{-2}$ gradient fits those data nicely. 
The 23 September 2015 speed is between the two values derived using the \citeauthor{Newkirk61} model assuming either 
fundamental and harmonic emission, but this may be coincidence given that the modeled and observed 
density profiles are widely discrepant. 
That event's non-radial 
profile may also prevent meaningful agreement with any simple $n_{\mathrm e}(r)$ model (see Figure~\ref{fig:overlay}). 


\subsection{Propagation Effects}
\label{propagation}

As described in Section \ref{introduction2}, a number of authors have argued that radio propagation 
effects, namely refraction and scattering, can explain the large source heights frequently exhibited by Type III bursts. 
\citet{Bougeret77} introduced the idea of scattering by overdense fibers in the 
context of radio burst morphologies, and \citet{Stewart74} suggested that Type III 
emission may be produced in underdense flux tubes as a way of explaining observed 
harmonic--fundamental ratios. 
These two concepts were combined by \citet{Duncan79}, who introduced the 
term \textit{ducting} to refer to radiation that is produced in an underdense environment and subsequently 
guided to a larger height by reflections against a surrounding ``wall" of much 
higher-density material, which eventually becomes transparent with sufficient altitude.   
\edit{While plausible, this concept generalizes poorly in that electron beams are not 
expected to be found preferentially within coherent sets of low-density structures that would be conducive to ducting.}

\edit{\citet{Robinson83} addressed this by showing}
that random reflections against overdense fibers can 
have the same effect of elevating an observed burst site above 
its true origin, but without requiring any peculiarities of the emission site (\textit{i.e.} low density). 
\edit{Because the high-density fibers known to permeate the corona are not randomly 
arranged and are generally radial, random scattering against them does not randomly modulate 
the aggregate ray path, but it instead tends to guide the emission outward to larger heights in a manner 
that is analogous to the classic ducting scenario.
For this reason, other authors (\textit{e.g.} \citealp{Poquerusse88}) have chosen to retain \textit{ducting} 
to refer to the similar but more general impact of scattering, without implying that the emission is 
guided within a particular density structure as originally proposed by \citet{Duncan79}. 
Here, we will simply refer to \textit{scattering} to avoid potential confusion between the two concepts.} 

\edit{After being scattered for the last time upon reaching a height with sufficiently low densities, a radio wave will 
then be refracted through the corona before reaching an observer, further shifting the source location. 
As the coronal density gradient generally decreases radially, radio waves will tend to refract toward 
to the radial direction such that a source originating at the limb will appear at a somewhat lower height than its origin, which  
could be either the actual emission site (\textit{e.g.} \citealp{Stewart76}) or, more likely, the point of last scatter (\textit{e.g.} \citealp{Mann18}). 
Accounting for the refractive shift, which becomes larger with decreasing frequency, therefore requires that the emission be generated at 
or scattered to an even larger height than is implied by the observed source location. 
Recent results on this topic from \citet{Mann18} will be discussed in the next section.} 
 
Propagation effects are also thought to be important to the observed structure of the quiescent corona, 
where the dominant emission mechanism is thermal 
bremsstrahlung (free-free) radiation at MWA frequencies.   
Outside of coronal holes, this emission is expected to be in or close to the optically thick regime (\textit{e.g.} \citealp{Kundu82,Gibson16}), which means that the observed brightness temperature 
should be the same as the coronal temperature.
However, well-calibrated 2D measurements have generally found lower brightness 
temperatures than expected from temperatures derived at other wavelengths (see review by \citealp{Lantos99}). 
Additionally, the size of the corona appears to be larger than expected at low frequencies 
(\textit{e.g.} \citealp{Aubier71,Thejappa92,Sastry94,Subramanian04,Ramesh06}). 
The prevailing explanation for these effects is also scattering by density inhomogeneities 
(\textit{e.g.} \citealp{Melrose88,Alissandrakis94,Thejappa08})\edit{, although the refractive effect described in the 
previous paragraph is also important \citep{Thejappa08}.}

Thus, the \edit{scattering} process that may act to elevate Type III sources also 
affects quiescent emission, increasing the apparent size of the corona.
We will take advantage of this by using the difference in extent between 
observed and modeled quiescent emission 
as a proxy for the net effect of propagation effects on our Type III source heights. 
\edit{This approach is limited in that, although both are related to scattering, the extent 
to which the magnitudes of these two phenomena are related is unclear. 
In particular, previous studies on the broadening of the radio Sun by scattering 
have considered random density inhomogeneities as opposed to the more realistic 
case of high density fibers capable of producing the ducting-like effect for Type III sources.}
 

\begin{figure}\graphicspath{{chapter3/}}
\centerline{\includegraphics[width=\textwidth,clip=]{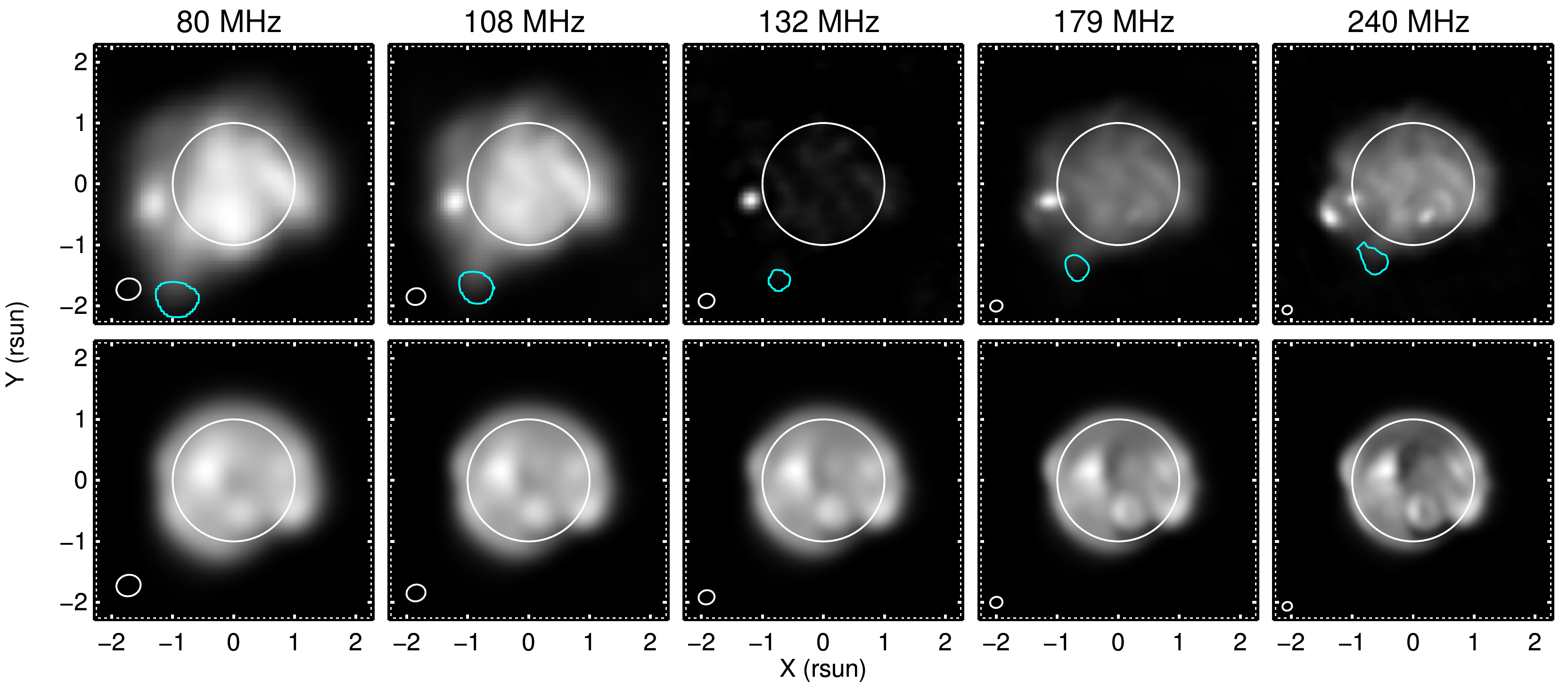}}
\caption{\footnotesize{}MWA background images for the 14 October 2014 event (top) and corresponding MAS-FORWARD synthetic 
images convolved with the MWA beam (bottom). 
Beam ellipses are shown in the lower-left corners, 
and the cyan curves are the 50\% burst contours from Figures~\ref{fig:centroids} \& \ref{fig:overlay}. 
This day is shown because thermal emission is only barely distinguishable at 132 MHz, precluding 
the Figure~\ref{fig:offset} analysis at that frequency, which was not the case for any other event-channel combination.
}
\label{fig:forward}
\end{figure}  

Figure~\ref{fig:forward} shows the observed background emission \textit{versus} synthetic images  
based on a global MHD model. 
The MWA images are obtained by averaging every frame with a total intensity less than 
two $\sigma$ above the background level, which is determined via the procedure shown in Figure~\ref{fig:baseline}.
Synthetic images are obtained using
FORWARD \citep{Gibson16}, a 
software suite that calculates the expected bremsstrahlung and gyroresonance emission 
given a model atmosphere. 
We use the Magnetohydrodynamic Algorithm outside a Sphere 
(MAS; \citealp{Lionello09}) medium resolution (\textsf{hmi\_mast\_mas\_std\_0201}) model, which extrapolates 
the coronal magnetic field from photospheric magnetograms (\textit{e.g.} \citealp{Miki99}) 
and applies a heating model adapted from \citet{Schrijver04} to compute density and temperature. 


\begin{figure}\graphicspath{{chapter3/}}
\centerline{\includegraphics[width=\textwidth,clip=]{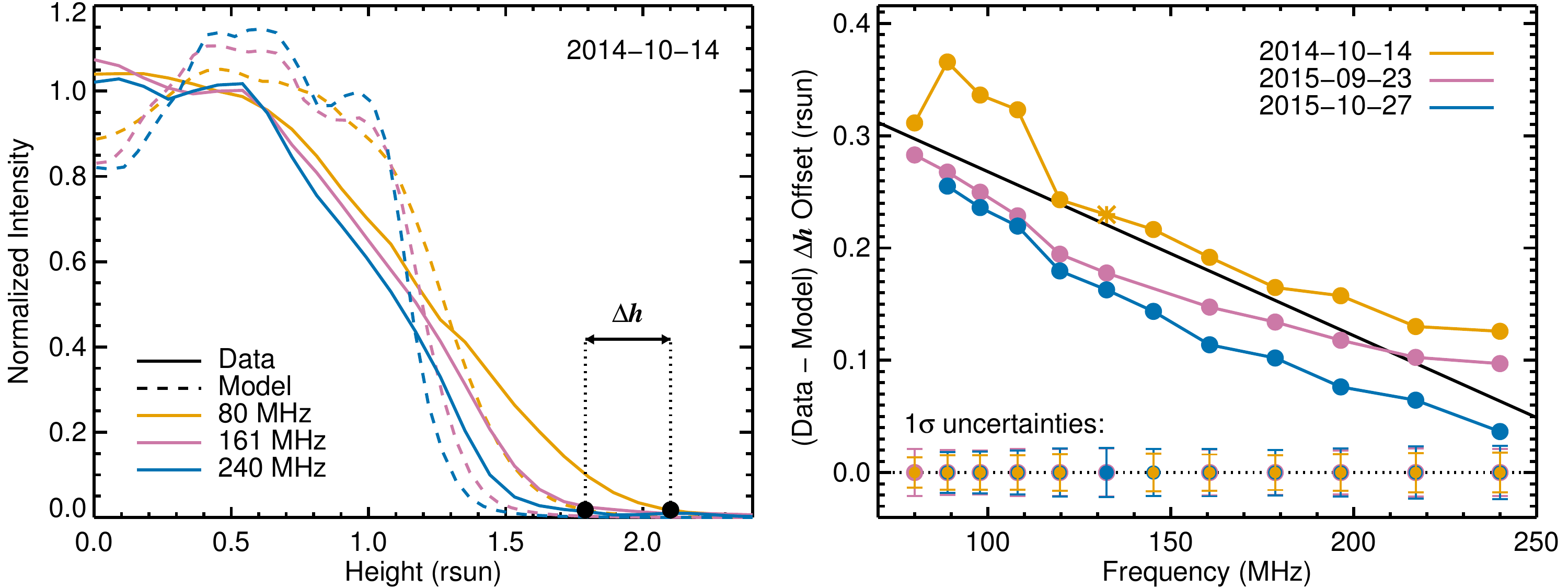}}
\caption{\footnotesize{}\textit{Left}: Average intensity \textit{versus} radial distance obtained from the Figure~\ref{fig:forward} images 
and normalized by the median value below \rsolar{1}.  $\Delta{h}$ refers to the height offset between the observed and 
modeled intensity profiles at the apparent Type III burst height at 80 MHz.  
\textit{Right}: $\Delta{h}$ for each frequency and event. 
An orange asterisk marks the one instance where data was available but a measurement could not be made 
because the thermal background emission was not well-detected (see Figure~\ref{fig:forward}), so an average of the 
adjacent points is used. 
The uncertainties reflect the sensitivity of $\Delta{h}$ to the normalization choice in the left panel (see Section~\ref{propagation}).  
}
\label{fig:offset}
\end{figure}   

\citet{McCauley17} established the use of these model images for flux calibration and included a qualitative 
comparison to MWA observations. 
As in the aforementioned literature, 
the radial extent of the corona is somewhat larger in the observations than 
in the beam-convolved model images. 
To quantify this difference, we divide both image sets into concentric rings about Sun-center. 
The average intensity within each ring is plotted against its radial distance in 
the left panel of Figure~\ref{fig:offset}, where the intensities have been normalized by the 
median value below one solar radius. 
We then measure the offset $\Delta{h}$ between the observed and modeled profiles at the 
heights obtained from the Type III positions. 
$\Delta{h}$ is sensitive to how the intensity curves are normalized, and we quantify 
this uncertainty by repeating the procedure for ten different normalization factors that 
reflect the median intensities within radial bins of width \rsolar{0.1} from 0 to \rsolar{1}.
The right panel of Figure~\ref{fig:offset} plots the $\Delta{h}$ results for each event,   
which have one-$\sigma$ uncertainties of less than $\pm$\rsolar{0.025}.
The offset appears to depend roughly  
linearly on frequency, with larger offsets at lower frequencies. 
Fitting a line through all of the points, we find that: 

\begin{equation}
\label{eq:dh}
\Delta{h} \approx{} -1.5\e{-3}f + 0.41; ~80\leq{} f \leq{240}~\rm{MHz}
\end{equation}

\noindent where $\Delta{h}$ is in solar radii and $f$ is in MHz. 
This yields \rsolar{0.30} at 80 MHz and \rsolar{0.06} at 240 MHz. 
We do not expect this expression to be relevant much outside of the prescribed frequency range, 
but extrapolating slightly, we obtain \rsolar{0.32} at 60 MHz. This value is a bit more than 
half of the $<$ \rsolar{0.56} limit found by \citet{Poquerusse88}. 

\citeauthor{Poquerusse88}, and others who have quantified the scattering effect (\textit{e.g.} \citealp{Robinson83}), 
obtained their results by computing ray trajectories through a model corona. 
That approach allows a fuller understanding of the propagation physics, 
but the result is dependent on the 
assumed concentration and distribution of high-density fibers, which are not well constrained. 
Our critical assumption is that emission produced at significantly lower heights would be absorbed, as 
would be the case in our optically thick model corona. 
However, low coronal brightness temperatures could also be 
explained by lower opacities (\textit{e.g.} \citealp{Mercier09}) or a low filling factor, which would allow burst emission 
to escape from lower heights and lead us to underestimate the potential impact of \edit{propagation effects}. 


\begin{figure}\graphicspath{{chapter3/}}
\centerline{\includegraphics[width=\textwidth,clip=]{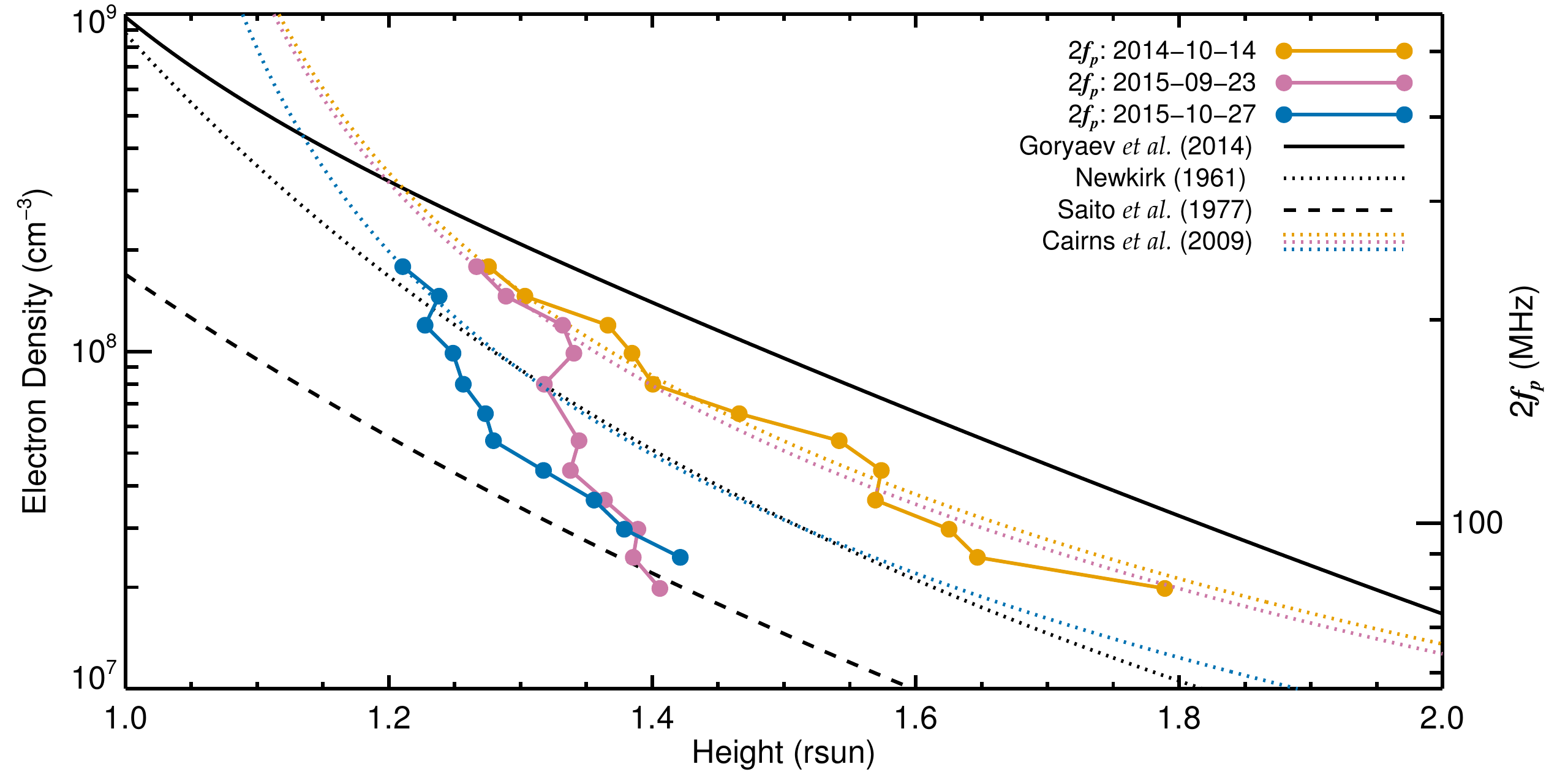}}
\caption{\footnotesize{}Imaging density profiles after applying the $\Delta{h}$ offsets from Figure~\ref{fig:offset} and 
assuming harmonic emission. 
The solid \citep{Goryaev14}, dotted \citep{Newkirk61}, and dashed \citep{Saito77} black curves are density 
models corresponding to a dense streamer, the solar-maximum background, and the solar-minimum 
background, respectively (see Section~\ref{density}). 
The dotted color lines apply the $n_{\mathrm e}(r) = C(r - 1)^{-2}$ profile detailed by \citet{Cairns09}, where 
the constant $C$ has been normalized to the density implied by the corresponding 240 MHz source position. 
}
\label{fig:dens3}
\end{figure}  

Figure~\ref{fig:dens3} shows how the Figure~\ref{fig:dens1} density results change after subtracting the 
height offsets from Figure~\ref{fig:offset}. 
The 14 October 2014 harmonic [$2f_{\mathrm p}$] profile remains reasonable with the offsets, lying just below the 
\citet{Goryaev14} model instead of just above it, while the 
fundamental emission densities for that event would still be quite large. 
Given that the \citeauthor{Goryaev14} model is among the highest-density streamer models 
in the literature, we conclude that harmonic emission from a beam traveling 
along an overdense structure is consistent with 14 October 2014 data. 
Our assessment for this event is that  
propagation effects may contribute to some but not all the apparent density enhancement. 

The other two events exhibit unusually steep density profiles once the offsets are subtracted.  
That was true also for the original 23 September 2015 results, which we attributed to its non-radial 
trajectory in Section~\ref{density}. 
However, the original 27 October 2015 densities were well-matched to the 
\citet{Cairns09} $C(r - 1)^{-2}$ gradient but become too steep to match any standard density gradient
with the inferred offsets. 
This may simply reflect the intrinsic density gradient of the particular structure.
Alternatively, it is possible that we have over- or underestimated the impact of \edit{propagation effects} at the low or high end of our 
frequency range, respectively.  
However, the frequency dependence of scattering \edit{and refraction} means that any treatment will  
steepen the density gradient. 
Aside from their slopes, the offsets bring the densities implied by both bursts to generally within the 
normal background range assuming harmonic emission or to a moderately enhanced level 
assuming fundamental emission.  

 
\section{Discussion} %
\label{discussion2} %

The previous section suggests that propagation effects can partially explain 
the apparent density enhancements implied by our Type III source heights. 
Assuming harmonic emission, our estimates for the potential magnitude of \edit{propagation effects} bring 
the densities to within normal background levels for two events, while one 
event remains enhanced at a level consistent with a dense streamer. 
In Section~\ref{density}, we showed that our original density profiles are 
consistent with those found from recent Type III burst studies at lower  
and higher frequencies, which together are well-fit by the 
\citet{Cairns09} $C(r - 1)^{-2}$ gradient. 
Both the low- and high-frequency studies conclude that their large densities 
imply electron beams traveling along overdense structures, 
but neither consider the impact of propagation effects.  

\citet{Morosan14} propose a variation of the overdense hypothesis based 
on their 30\,--\,60 MHz LOFAR observations. They suggest that 
the passage of a CME just prior to an electron beam's arrival may compress streamer plasma 
enough to facilitate Type III emission at unusually large heights. 
This was consistent with their events being associated with a CME and could be relevant to 
our 23 September 2015 event, \edit{which was also preceded by a CME} (see~Section \ref{context}). 
While this interpretation is plausible, 
we do not think such special conditions need to be invoked given that the densities inferred from the 
\citeauthor{Morosan14} results are consistent with ours (Figure~\ref{fig:dens2}) and are broadly consistent with the large  
Type III source heights found using the previous generations of low-frequency instruments (\textit{e.g.} \citealp{Wild59,Morimoto64,Malitson66,Stewart76,Kundu84}). 
\edit{Propagation effects} seem particularly likely to have contributed (at least partially) to their inferred density enhancement, as the 
effects become stronger with decreasing frequency. 

\edit{Recently, \citet{Mann18} also examined the heights of Type III sources observed at the limb by LOFAR. 
After accounting for the refractive effect described in Section~\ref{propagation}, and relying on scattering to direct 
emission toward the observer at large heights prior to being refracted, 
their results imply a density enhancement of around 3.3$\times$ over the \citet{Newkirk61} density 
model, assuming fundamental emission. 
Incorporating our offsets from Section~\ref{propagation} gives us an average enhancement of 4.6$\times$ 
over the same model across our three events, also assuming fundamental emission.
Our results are therefore consistent with those of \citet{Mann18}, though our attempts to quantify the impact
of propagation effects are quite different.}

\citet{Chen13} also found large densities using VLA data at higher frequencies (1.0\,--\,1.5 GHz).
Scattering is also thought to be important at high frequencies given the apparent lack 
of small-scale structure \citep{Bastian94}, but the 
extent to which \edit{scattering may also elevate radio sources} in that regime has 
not been addressed to our knowledge. 
\citeauthor{Chen13} observed an on-disk event, from which they obtain source heights by comparing 
their projected positions to stereoscopic observations of an associated EUV jet, which is assumed 
to have the same inclination as the Type III electron beam. 
This method would also be impacted by any source shifting caused by \edit{scattering}. 
Although these shifts would be much smaller than at low frequencies, the background 
gradient is much steeper, so a reasonably small shift may still strongly influence the 
inferred density relative to the background. 

If we accept the densities obtained at higher frequencies, albeit from just one example, 
then their consistency with low-frequency observations in general is striking. 
As we describe in Section~\ref{introduction2}, the community largely came to favor propagation effects 
over the overdense hypothesis in the 1980s, and 
the topic has not had much consideration since. 
If new observations at low heights (high frequencies) also suggest beams moving 
preferentially along dense structures, then it elicits the question of whether or not that interpretation is again  
viable at larger heights (lower frequencies). 
In that case, this would need to be reconciled with the 
fact that electron beams have not been found to be preferentially associated 
with particularly high-density regions in \textit{in-situ} solar wind measurements \citep{Steinberg84}, along with 
the evidence for other impacts of scattering such as angular broadening 
(\textit{e.g.} \citealp{Steinberg85,Bastian94,Ingale15}).

Selection effects may be relevant, as radiation produced well above the ambient plasma frequency is 
less likely to be absorbed before reaching the observer. 
Thus, coronal Type III bursts may imply high densities because beams traveling along dense structures are 
more likely to be observed. 
Type III bursts also have a range of starting frequencies, 
which has been interpreted in terms of a range in acceleration (\textit{i.e.} reconnection) heights that 
are often larger than those inferred from X-ray observations \citep{Reid14b}.
Alternatively, a beam may be accelerated at a smaller height than is implied by 
the resultant Type III starting frequency due to 
unfavorable radiation-escape conditions (absorption) below the apparent starting height. 
Simulations also suggest that electron beams may travel a significant distance before producing 
observable emission \citep{Li13b,Li14} and that they may be radio loud at 
some frequencies but not at others due to variations in the ambient 
density \citep{Li12,Loi14} and/or temperature \citep{Li11b,Li11}.

The magnetic structures along which electron beams travel also evolve with distance from 
the Sun. 
A popular open flux tube model is an expanding funnel that is thin at the base of the corona 
and increasingly less so into the solar wind (\textit{e.g.} \citealp{Byhring08,He08,Pucci10}). 
Such structures may allow a dense flux tube to become less dense relative 
to the background as it expands with height. 
Moreover, a beam following a particular magnetic field line from the corona into the high-$\beta$ solar wind may not 
necessarily encounter a coherent density structure throughout the heliosphere. 
Turbulent mixing, corotation interaction regions, CMEs, and other effects influence 
solar wind density such that it is not obvious that an electron beam traversing an overdense structure near the Sun 
should also be moving in an overdense structure at large heliocentric distances. 

We also note that one of the main conclusions from many of the Type III studies referenced here is unchanged 
in either the overdense or \edit{propagation effects} scenarios. 
Both cases require a very fibrous corona that can supply dense structures 
along which beams may travel and/or dense structures capable of \edit{scattering} radio emission \edit{to larger heights}.  
This is consistent with the fine structure known from eclipse observations (\textit{e.g.} \citealp{Woo07}) that has more recently 
been evidenced by EUV observations. 
For instance, analyses of a sungrazing comet \citep{Raymond14} and of EUV spectra \citep{Hahn16} 
independently suggest large density contrasts ($\gtrsim$ 3\,--\,10) between neighboring flux tubes in 
regions where the structures themselves are undetected. 
As our understanding of such fine structure improves, better constraints can be placed on them 
for the purpose of modeling the impact of \edit{propagation effects} on radio sources. 

 
\section{Conclusion} %
\label{conclusion2} %

We presented imaging of three isolated Type III bursts observed at the 
limb on different days using the MWA.
Each event is associated with a white-light streamer 
and plausibly associated with EUV rays that exhibit activity around the 
time of the radio bursts. 
Assuming harmonic plasma emission, density profiles derived from 
the source heights imply enhancements of 
approximately 2.4\,--\,5.4$\times$ over background levels. 
This corresponds to electron densities 
of 1.8\e{8} cm\tsp{-3} (240 MHz) 
down to 0.20\e{8} cm\tsp{-3} (80 MHz) at average heights of 1.3 to \rsolar{1.9}.
These values are consistent with the highest streamer densities inferred from other 
wavelengths and with the large radio source heights found using older instruments. 
The densities are also consistent with recent Type III results at higher and lower frequencies, 
which combined are well-fit by a $C(r - 1)^{-2}$ gradient. 
By comparing the extent of the radio limb to model predictions, we estimated that 
\edit{radio propagation effects, principally the ducting-like effect of random scattering by high-density fibers,} 
may be responsible for 0.06\,--\,0.30\rsolar{} 
of our apparent source heights. 
This shift brings the results from two of our three events to within a standard range of background densities. 
We therefore conclude that propagation effects can partially explain the apparent density enhancements but 
that beams moving along overdense structures cannot be ruled out. 
We also used the imaging data to estimate electron beam speeds of 0.24\,--\,0.60 c.\\ 

{\footnotesize{}\textbf{Acknowledgements}: Support for this work was provided by the Australian Government through an Endeavour Postgraduate Scholarship. 
We thank Stephen White and Don Melrose for helpful discussions 
\edit{and the anonymous referee for their constructive comments}.
This scientific work makes use of the Murchison Radio-astronomy Observatory (MRO), operated by 
the Commonwealth Scientific and Industrial Research Organisation (CSIRO). 
We acknowledge the Wajarri Yamatji people as the traditional owners of the Observatory site. 
Support for the operation of the MWA is provided by the Australian Government's  
National Collaborative Research Infrastructure Strategy (NCRIS), 
under a contract to Curtin University administered by Astronomy Australia Limited. 
We acknowledge the Pawsey Supercomputing Centre, which is supported by the 
Western Australian and Australian Governments.
SDO is a National Aeronautics and Space Administration (NASA) spacecraft, and 
we acknowledge the AIA science team for providing open 
access to data and software. 
The SOHO/LASCO data used here are produced by a consortium of the Naval Research Laboratory (USA), 
Max-Planck-Institut f\"ur Aeronomie (Germany), Laboratoire d'Astronomie (France), 
and the University of Birmingham (UK). 
SOHO is a project of international cooperation between ESA and NASA.
This research has also made use of NASA's Astrophysics Data System (ADS), 
along with JHelioviewer \citep{Muller17} and the Virtual Solar Observatory (VSO: \citealp{Hill09}).}

%% file: chapter4/chapter4.tex
\pagestyle{fancy}

\providecommand{\edit}[1]{{\color{black}{#1}}}
\providecommand{\redit}[1]{{\color{black}{#1}}}




\chapter[The Low-Frequency Solar Corona in Circular Polarization]{The Low-Frequency Solar Corona in \\ Circular Polarization}\label{ch4}

{\large{}Published as \citet{McCauley19}, \href{https://ui.adsabs.harvard.edu/abs/2019SoPh..294..106M}{\textit{Solar Phys.}, 294:106}}

\fancyhead[OR]{~}\fancyhead[EL]{\bf Ch. \thechapter~Spectropolarimetric Imaging of the Corona} 

\section{Abstract}

We present spectropolarimetric imaging observations of the solar corona 
at low frequencies (80\,--\,240 MHz) using the \textit{Murchison Widefield Array} (MWA). 
These images are the first of their kind, and we introduce an algorithm to mitigate 
an instrumental artefact \edit{by} which the total intensity 
signal contaminates the polarimetric images due to calibration errors. 
We then survey the range of circular polarization 
(Stokes $V$) features detected in over 100 observing runs near solar maximum \redit{during quiescent periods}. 
\edit{First,} we detect around \redit{700} compact polarized sources across our dataset with polarization 
fractions ranging from less than 0.5\% to nearly 100\%. 
These sources exhibit a positive correlation between polarization fraction and 
total intensity, and we interpret them as a continuum of plasma emission noise storm 
(Type I burst) \redit{continua} sources associated with active regions. 
\edit{Second,} we report a characteristic ``bullseye" structure observed for many low-latitude coronal holes 
in which a central polarized component is surrounded by a ring of the opposite sense. 
The central component does not match the sign expected from thermal bremsstrahlung emission, and 
we speculate that propagation effects or an alternative emission mechanism may be responsible. 
\edit{Third,} we show that the large-scale polarimetric structure at our lowest frequencies is reasonably well-correlated 
with the line-of-sight (LOS) magnetic field component inferred from a global potential field source surface (PFSS) model. 
The boundaries between opposite circular polarization signs are generally aligned with polarity 
inversion lines in the model at a height roughly corresponding to that of the radio limb. 
\edit{This is not true at our highest frequencies, however, where the LOS magnetic field direction and polarization sign 
are often not straightforwardly correlated.}


\section{Introduction} %
\label{introduction} %

Radio emission in a magnetized plasma is produced in one or both of two modes, the 
ordinary [$o$] and extraordinary [$x$], which are each 100\% circularly polarized with opposite senses 
in the quasi-circular approximation generally used for the solar corona \citep{Zheleznyakov77,Melrose80}.  
The $x$-mode refers to when the electric field vector of the electromagnetic wave rotates in the same direction 
as the gyromotion of electrons around the magnetic field where the emission was generated. 
A net circular polarization arises when the two modes are \edit{received} unequally, 
which is characterized by the degree [$r_{\rm c}$] of circular polarization [Stokes $V$] relative to the total intensity [Stokes $I$].
In detail,  

\begin{equation}
	r_{\rm c}  = \frac{T_{{\rm b},x}-T_{{\rm b},o}}{T_{{\rm b},x}+T_{{\rm b},o}}~, 
\end{equation}

\noindent where $T_{{\rm b},x}$ and $T_{{\rm b},o}$ refer to the brightness temperatures of the $o$ and 
$x$ modes, respectively \citep{Dulk85}. 
The quantity $r_{\rm c}$, also labeled dcp or $V/I$, depends on the emission 
mechanism and plasma parameters, along with a number of effects such as mode coupling and refraction 
that may modulate the polarization state or separate the two modes during propagation. 
Low-frequency (meter-wave) emission from the solar corona is dominated by two mechanisms,
thermal bremsstrahlung and plasma emission (\textit{e.g.} \citealp{Dulk85,White99,Aschwanden05}). 
Other mechanisms are also important in specific contexts, such as (gyro)synchrotron emission 
in coronal mass ejections, but \edit{these} will not be discussed in detail here. 

Bremsstrahlung emission is produced by the conversion of kinetic energy into radiant energy that 
occurs when a charged particle accelerates, and  
thermal bremsstrahlung refers to 
a plasma in thermal equilibrium for which free electrons are deflected by the Coulomb fields of ions and atomic nuclei.
This is often referred to as free-free radiation for a fully-ionized plasma like the corona because 
the particles are not in bound states throughout the entire process. 
Emission \edit{at} a particular frequency is generated only by plasma with electron densities [$n_{\rm e}$] equal to 
or below that corresponding to the \edit{local} fundamental electron plasma frequency 
[$f_{\rm p} \approx 9\e{-3}\sqrt{n_{\rm e}}$~MHz, for $n_{\rm e}$ in cm\tsp{-3}]. 
Lower-frequency emission therefore corresponds to lower-density material at 
generally larger heights above the surface, meaning that the corona appears 
larger with decreasing frequency. 
\edit{Canonical coronal background density models (\textit{e.g.} \citealp{Newkirk61,Saito77}) correspond to frequencies of 
below $\approx$ 300 MHz, but dense coronal structures may produce free-free emission well into the GHz range.}

Thermal bremsstrahlung slightly favors the $x$-mode to a degree that depends primarily on the 
line-of-sight (LOS) magnetic field strength. 
The opacity, $\kappa$, can be written as

\begin{equation} 
	\label{eq:brem}
	\kappa = 0.2\frac{n_{\rm e}^2}{T_{\rm e}^{1.5}(f \pm f_B|\cos{\theta}|)^2}~{\rm cm}^{-1}, 
\end{equation}

\noindent where $T_{\rm e}$ is the electron temperature, $f$ is the 
emission frequency, $f_B$ is the electron gyrofrequency [$f_B = 2.8\e{6}B_{\rm gauss}$ Hz], and 
$\theta$ is the angle between the line of sight and the magnetic field direction \citep{Dulk85,Gelfreikh04,Gibson16}. 
The plus sign refers to the $o$-mode, the minus sign refers to the $x$-mode, and the difference 
between the two modes produces the net circular polarization. 
\redit{Equation~\ref{eq:brem} is a quasi-linear (QL) approximation that is valid for most angles $\theta$. 
Values of $\theta$ close to 90$\degree$, for which the propagation direction is nearly perpendicular to the magnetic field 
orientation, are referred to as quasi-transverse (QT) propagation and produce linear polarizations \citep{Zheleznyakov70,Ryabov04}. 
Circularly-polarized emission that passes through a QT region may also experience polarization state changes, which will be 
discussed in Section~\ref{discussion}}.
Equation~\ref{eq:brem} also assumes that $f \gg f_B$. 
This condition means that the difference between the two modes, and therefore the polarization 
fraction, will always be fairly small, generally a few percent or less \edit{at the low frequencies considered in this article} \citep{Sastry09}. 
For a homogenous, optically-thin plasma, $r_c \approx 2 \cos{\theta}(f_B/f)$, \edit{while}  
for the optically-thick case, a temperature gradient is required for the two modes to be produced unequally \edit{\citep{Dulk85,Gibson16}}.  

Thermal bremsstrahlung radiation generates a continuous background  
that slowly varies as the corona evolves. 
This may be slightly or dramatically augmented by transient emission associated with 
nonthermal electrons that are accelerated through a variety of mechanisms underpinned 
either by magnetic reconnection or shock waves. 
These electron streams produce oscillations in the background plasma known as 
Langmuir waves, which then deposit energy into radio emission through scattering 
by ion sound waves or by other nonlinear Langmuir wave processes \citep{Ginzburg58,Robinson00,Melrose09}. 
These are typically coherent mechanisms, often grouped together under the term ``plasma emission," for which the 
intensity is related nonlinearly to the energy of the nonthermal electrons. 
Plasma emission is responsible for most types of solar radio bursts \citep{Dulk85}, which may 
exceed the thermal background by several orders of magnitude, but it is also likely 
the source of very weak nonthermal emissions that enhance the background only slightly \citep{Suresh17,Sharma18}. 
Like thermal bremsstrahlung, plasma emission is tied to the ambient 
density though the electron plasma frequency. 
However, in this case, the emission frequency is highly localized to just above the plasma frequency or 
its harmonic.  

The polarization of plasma emission depends firstly on the harmonic number. 
For fundamental [$f_{\rm p}$] emission, the circular polarization fraction should be 100\% in the sense of 
the $o$-mode because\edit{, for frequencies expressed in Hz,} $f_{\rm p}$ is above the cutoff for $x$-mode production, meaning that $x$-mode radiation  
begins only at frequencies slightly lower than the plasma frequency \citep{Melrose09}. 
Polarization fractions approaching 100\% are indeed sometimes observed for Type I bursts (\textit{e.g.} \redit{\citealp{Kai62,Tsuchiya63,Dulk84,Aschwanden86,Mugundhan18}}). 
\edit{However,} this is almost never true for other radio burst types that are also attributed to fundamental plasma emission (\textit{e.g.} \citealp{Wentzel84,Reid14,Kaneda15}). 
\edit{The} reason for this remains an open question, but a common explanation is that scattering of the 
radio emission by other wave modes or by sharp density gradients tends to have a depolarizing effect (\textit{e.g.} \citealp{Wentzel86,Melrose89,Melrose06,Kaneda17}). 
The polarization fraction of harmonic [$2f_{\rm p}$] emission is more complicated because it depends on the 
angular distribution of the Langmuir waves. 
Polarization in the sense of the $o$-mode is still generally expected, assuming that the Langmuir waves are confined to relatively small 
angles with respect to the magnetic field, which is generally assumed to be true because of the associated magnetic field strengths \citep{Melrose78}. 
However, it is possible for the $x$-mode to dominate in specific\edit{, and likely less common,} contexts \citep{Willes97}. 
Thus, for the same LOS magnetic field direction, the two dominant low-frequency emission mechanisms generally produce opposite 
circular polarization signatures. 
 
Radio polarimetry has long been a powerful tool for diagnosing solar magnetic fields, particularly using 
high-frequency observations of gyroresonance emission \citep{Akhmedov82,White97} and, more recently, 
bremsstrahlung emission \citep{Grebinskij00}.
Low-frequency polarimetry has generally been restricted to radio bursts  
because their high intensities and large polarization fractions are easiest to detect. 
An early review on the polarization of metric bursts and their utility as magnetic field probes is given by \citet{Dulk78}.
Very few instruments have been capable of making two-dimensional polarimetric measurements of the low-frequency Sun, 
and until now, none have been sensitive enough to detect the weak polarization signatures during quiescent periods. 
In recent decades, this type of analysis could be done with two instruments, the Nan{\c c}ay Radioheliograph (NRH; \citealp{Kerdraon97}), 
which operates between 150 and 450 MHz, 
and the Gauribidanur Radioheliograph (GRH; \citealp{Ramesh98}), which \edit{usually} operates at 80 MHz. 

A few studies have ulilized the polarimetric imaging capabilities of the NRH to examine spatial variation in radio bursts. 
For example, \citet{Mercier90} showed that Type III bursts have different spatial characteristics 
in circular polarization compared to the total intensity, and 
\citet{Bouratzis16} investigated similar differences in spike bursts as a function of time. 
Several others have examined source positions and structures in total intensity NRH observations, while using the polarization 
information to help discriminate between emission mechanisms (\textit{e.g.} \citealp{Gopalswamy94,Tun13,Kong16,Liu18}).
The radioheliograph at Gauribidanur does not have a polarimetric capability itself, but several one-dimensional 
polarimeters have been installed alongside it \citep{Ramesh08,Sasikumar13,Kishore15}. 

GRH imaging and simultaneous polarimeter observations have been used for studies of 
Type I noise storms \citep{Ramesh11,Ramesh13,Mugundhan18}, Type II bursts \citep{Hariharan14,Hariharan15,Kumari17}, 
Type III bursts \citep{Ramesh10b,Sasikumar13b,Kishore17}, Type IV bursts \citep{Hariharan16}, and 
gyrosynchrotron emission from CMEs \citep{Sasikumar14}. 
Most of these results include estimates of the associated magnetic field strength assuming a particular emission mechanism. 
Additionally, \citet{Ramesh10} report polarized emission from streamers that is attributed to thermal bremsstrahlung, though 
the polarization fraction ($\approx$ 15\%) is unusually large for bremsstrahlung emission. 
Moreover, the polarized source cannot be localized beyond assuming that it comes from the dominant total intensity source, and 
as we will show, polarized emission from the low-frequency corona is often not straightforwardly correlated with total intensity, particularly 
during quiescent periods. 

This article presents the first spectropolarimetric imaging observations \edit{of the Sun} from the \textit{Murchison Widefield Array} (MWA; \citealp{Tingay13}). 
These are the first circular polarization images of the low-frequency corona that are sensitive enough to detect the \edit{polarimetric} signatures 
associated with thermal bremsstrahlung emission and very weak plasma emission outside of major burst periods. 
We will survey the range of features detected in over 100 observing runs near solar maximum and motivate future studies with these novel data. 
Section~\ref{mwa} describes the MWA instrument, and Section~\ref{leakage} introduces an algorithm used to mitigate 
an important calibration artefact. 
Section~\ref{ar} discusses active region noise storm sources, Section~\ref{ch} characterizes the polarimetric signature of 
coronal holes, and Section~\ref{qs} details the large-scale quiescent structure. 
We discuss the implications of our results and motivate future studies in Section~\ref{discussion}. 
Our conclusions are summarized in Section~\ref{conclusion}. 


\section{\textit{Murchison Widefield Array} (MWA)}
\label{mwa}

The MWA is a low-frequency radio interferometer \edit{located} in Western Australia \citep{Lonsdale09,Tingay13}, and  
heliophysics is among the instrument's principal science themes alongside 
astrophysical topics \citep{Bowman13}. 
Direct solar observations have characterized the weakest nonthermal 
emissions reported to-date \citep{Suresh17,Sharma18}, provided definitive evidence 
for the standard theory of Type III bursts \citep{Cairns18}, detailed new radio 
burst dynamics \citep{McCauley17,Mohan18}, used radio bursts to probe the 
coronal density structure \citep{McCauley18}, characterized the low-frequency 
signature of coronal holes \citep{Rahman19}\edit{, and provided evidence for coronal heating 
via weak particle acceleration episodes \citep{Mohan19}.  
Solar imaging with the MWA has also motivated advances in data processing techniques 
related to flux calibration \citep{Oberoi17}, spatially resolved dynamic spectra \citep{Mohan17}, and high dynamic range imaging \citep{Mondal19}.}
Additionally, widefield interplanetary scintillation observations may be used for studies 
of the solar wind and of coronal mass ejections (CMEs) propagating through the heliosphere \citep{Kaplan15,Morgan18}. 

The MWA is comprised of \edit{4096} dipole antennas 
arranged in 128 aperture arrays called ``tiles". 
This refers to the Phase I array used here, which began observing in \edit{2013}.
An expanded Phase II array began full operations in 2018 with twice as many tiles\edit{, of 
which 128 can be used simultaneously in different configurations \citep{Wayth18}.}
The MWA has an instantaneous bandwidth of 30.72 MHz that can be distributed 
between 80 and 300 MHz in various configurations.
Our data utilize a ``picket fence" mode with 12 contiguous 2.56 MHz bandwidths 
centered at 80, 89, 98, 108, 120, 132, 145, 161, 179, 196, 217, and 240 MHz.
The data were recorded with a 0.5 sec time \edit{resolution} and a 40 kHz spectral resolution, 
but the observations presented here are averaged over each 2.56 MHz bandwidth before 
imaging and then time-averaged to different degrees after imaging. 
\redit{The spatial resolution is defined by the synthesized beam sizes, which have 
major axes of around 6.4 arcmin (\rsolar{0.40}) at 80 MHz and 2.5 arcmin (\rsolar{0.16}) at 240 MHz. 
The beam sizes and orientations, shown in the lower-left corners of each image, vary somewhat 
between observations due to pointing differences and occasional antenna failures.}

We use the same data processing scheme as \citet{McCauley17} and \citet{McCauley18}, and 
what follows is a brief summary thereof. 
Visibilities were generated with the standard MWA correlator \citep{Ord15} and the  
\textsf{cotter} software \citep{Offringa12,Offringa15}. 
Observations of bright and well-modelled calibrator sources were used 
to obtain solutions for the complex antenna gains \citep{Hurley14}, which 
were improved by imaging the calibrator and iteratively self-calibrating from there \citep{Hurley17}. 
All of our observations were calibrated using either Centaurus A or Hercules A. 
\textsf{WSClean} \citep{Offringa14} was used to perform the imaging 
with a Briggs -2 weighting \citep{Briggs95} to emphasize spatial resolution and minimize point spread function (PSF) sidelobes. 
The primary beam model of \citet{Sutinjo15} was used to produce Stokes $I$ and $V$
images from the instrumental polarizations, 
and the SolarSoftWare (SSW; \citealp{Freeland98}) 
routine \textsf{mwa\_prep} \citep{McCauley17} was used to translate the images onto solar coordinates. 
\redit{The data presented here are not flux calibrated on an absolute scale. 
Intensities are expressed either relative to the Stokes $I$ background level or in units of signal-to-noise.}

The next section will describe further steps required to calibrate the polarization images. 
\edit{We use the International Astronomical Union (IAU) and Institute of Electrical and Electronics Engineers (IEEE) convention on circular polarization, which defines 
positive as being right-handed (clockwise) from the source's perspective \citep{IEEE69,IAU73}, 
where right-handed refers to the rotation of the electric field vector of the electromagnetic wave about the 
orthogonal direction of motion. 
This convention is convenient here because it means that a net polarization in the sense of the $x$-mode 
will match the sign of the line-of-sight magnetic field component [\redit{$B_{\rm LOS}$}], 
where positive is outward.}

Each observation period lasted around 5 minutes, and a total of 111 such periods in 2014 and 2015 were reduced. 
52 of these were imaged at the full 0.5-sec time resolution and 59 \edit{were sampled} at a 4-sec cadence. 
Our objective is to survey the longer-lived features that 
are present in the corona on timescales of at least minutes, outside of transient radio burst periods.  
All of the images presented in this article are median averages of the individual 0.5-sec integrations 
\edit{with total intensities that are within two standard deviations of the background level during} each 5-min observing window.
\edit{Depending on the sources present, these averaged background images} may still contain significant nonthermal emission. 
Identifying which images to include in the average is done automatically using the 
baseline procedure illustrated by Figure 5 of \citet{McCauley18}. 
This involves finding the total intensity in each image, excluding times for which the intensity 
is greater than two standard deviations above the median, and iterating until no more images 
are excluded. 
Each pixel in the output image then contains the median of the corresponding pixels 
in those low-intensity images. 
The consideration of time dependent behavior on scales of less than 5 minutes will be a topic 
of future work.


\section{An Algorithm to Mitigate the Leakage of Stokes \textit{I} into \textit{V}}
\label{leakage}

To obtain useful polarimetric images, it is necessary to account for possible ``leakage" of the Stokes $I$ signal into 
the other Stokes \edit{parameters}. 
\redit{The MWA uses dual-polarization dipole antennas arranged in 4$\times$4 grids, or ``tiles", where the 
signals for each tile component are combined in an analog beamformer that produces two outputs representing orthogonal 
$X$ and $Y$ linear polarizations \citep{Tingay13}. 
The beamformer outputs are correlated into products that fully describe 
the polarization state in ``instrumental" polarizations ($XX$, $YY$, $XY$, and $YX$) 
that may be converted into the standard Stokes parameters ($I$, $Q$, $U$, and $V$) using a model of the MWA beam pattern. 
However, there are significant differences between the analytic beam pattern and that measured empirically 
by imaging known sources \citep{Sutinjo15}. 
These differences between the actual instrumental response and the complex primary beam model lead to 
``leakage" errors in the Stokes images where some fraction of $I$ contaminates the other parameters.}

A \redit{more} detailed description of this problem and of MWA polarimetry in general is given by \citet{Lenc17}. 
\redit{Sources of discrepancy between the beam model and the true response include} imperfections in the model itself along with instrumental effects, such as 
individual dipole failures during a particular observing run, that may cause the true response to 
vary from an otherwise perfect beam model. 
\edit{Importantly, the polarimetric response is also affected by a source's position within the beam 
and zenith angle, which means that the response changes somewhat between the observation used to calibrate 
the array and the solar observation. 
Changes in the ionosphere over the 
$\lesssim$ \edit{5} hours between the calibrator and solar observations may also degrade the calibration solution.}
Our data were reduced using the \citet{Sutinjo15} beam model, which dramatically reduced leakage 
from Stokes $I$ into Stokes $Q$ but somewhat increased the leakage from $I$ into $V$ compared to previous beam models. 

The leakage fraction also varies with a source's position on the sky and its position within the field-of-view \edit{for a given calibration solution} \citep{Sutinjo15,Lenc17,Lenc18}. 
Sources observed at lower elevations and/or near the edge of the field tend to exhibit higher leakage fractions. 
It is possible to reduce the leakage by means of iterative self-calibration on the source of interest, but 
this may affect the polarimetric calibration in ways that are difficult to understand, and self-calibration can also be 
difficult to effectively apply to diffuse sources like the Sun. 
Instead, the leakage effect can be mitigated with an empirical correction if there are sources within the field for which 
the polarization fractions are known. 
For the very large fields typical of many astrophysical MWA observations, a two-dimensional fit to the leakage 
fraction may be obtained from the known sources scattered throughout the field. 
The leakage fraction may vary by as much as 8\% across a 25 deg$^2$ patch \citep{Lenc17}, 
but we do not expect significant variations across the spatial extent of the radio Sun ($\lesssim{}$1.2$^{\degree}$ at 80 MHz) 
or over the duration of a typical observation ($\approx$ 5 min). 

For the solar observations presented here, it is not possible to simultaneously observe background sources alongside the Sun due 
to limited dynamic range. 
Recent advances in calibration techniques may enable this capability \citep{Mondal19}, but those methods
\edit{cannot yet be used for polarimetry and are not used here.}
In other words, the astronomical sources that may be present are too faint to be observed in close proximity to the Sun 
and cannot be used to characterize the Stokes $I$ into $V$ leakage.
We also do not know what the polarization fraction of any particular region on the Sun should be at any given time, as 
the polarization fraction may vary considerably depending on the dominant emission mechanism and local plasma parameters. 
However, outside of radio bursts, solar emission at low frequencies is dominated by the thermal bremsstrahlung (free-free) 
process. The importance of this is that under normal quiet-Sun conditions at MWA wavelengths, 
we can expect to see bremsstrahlung radiation in most locations that is only 
slightly polarized, and we can use this statistical information to estimate the leakage fraction with an algorithm that minimizes 
the number of pixels with polarization fractions greater than some threshold (\textit{\textit{i.e.}} $|V/I| > r_{\rm c, thresh}$). 

To determine this threshold, we generated synthetic Stokes $V/I$ images for each of our observing periods and frequencies using 
the forward modeling code \textsf{FORWARD} \citep{Gibson16} in SolarSoft IDL. 
\edit{FORWARD calculates the Stokes $I$ and $V$ intensities expected from thermal bremsstrahlung emission using Equation~\ref{eq:brem}, 
with the temperature, density, and magnetic field parameters taken in this case from the month-averaged 
Magnetohydrodynamic Algorithm outside a Sphere (MAS; \citealp{Lionello09}) global coronal model.}
On average 50\% of pixels in these images with Stokes $I$ brightness temperatures greater than 100,000 K have fractional polarizations of 
less than 0.3\%.
We choose a slightly larger threshold of 0.5\% because we wanted to implement our procedure uniformly, and 
the noise level in some of our observations makes a lower threshold impractical.
This value is also consistent with the predictions of \citet{Sastry09}, and the
effect of varying the threshold is folded into $V/I$ uncertainty estimates presented in Section~\ref{ch}. 

Our algorithm therefore assumes that most of the pixels in our images of the quiescent corona should exhibit polarization 
fractions of less \edit{than} 0.5\% and determines the leakage fraction that minimizes the number of pixels with $V/I$ values 
greater than 0.005. The algorithm can be expressed formally as:

\begin{equation}
\label{eq:alg1}
f(L) = \sum_{k=1}^{n} \left[ \left| \frac{V_k - L\cdot{}I_k}{I_k} \right| > r_{\rm c, thresh} \right]
\end{equation}

\begin{equation}
\label{eq:alg}
L_{\rm min} = \argminA_{L\in(-1,1)} f(L),
\end{equation}

\noindent \edit{where $f(L)$ is the number of pixels with polarization fractions greater than $r_{\rm c, thresh}$ as a function 
of $L$, the constant fraction of Stokes $I$ that is assumed to have leaked into Stokes $V$.}
The aim is to find the value \edit{$L_{\rm min}$} that minimizes \edit{$f(L)$}, where $k$ is a given pixel in an 
image and $n$ is the number of pixels to be considered. 
We consider only pixels for which a Stokes $I$ signal is detected above 5 $\sigma$. 
The square brackets \edit{in Equation~\ref{eq:alg1}} refer to the Iverson bracket notation, meaning that their contents evaluate to 1 if 
the condition is satisfied and 0 otherwise. 
In this case, that simply means that a pixel is counted if its polarization fraction is greater than 0.005 (0.5\%). 
Equation~\ref{eq:alg} is evaluated using an adaptive grid search with increments in $L$ of 0.1, 0.01, and 0.001.  
Note that this strategy is not the same as minimizing the total polarized intensity, which is not advisable because the two senses 
may not be equally represented and specific regions may have large polarized intensities that would bias the result if 
one were to simply find $L$ that minimizes the total polarization fraction in the image. 


\begin{figure*}\graphicspath{{chapter4/}}
    \centering
    \includegraphics[width=1.0\textwidth]{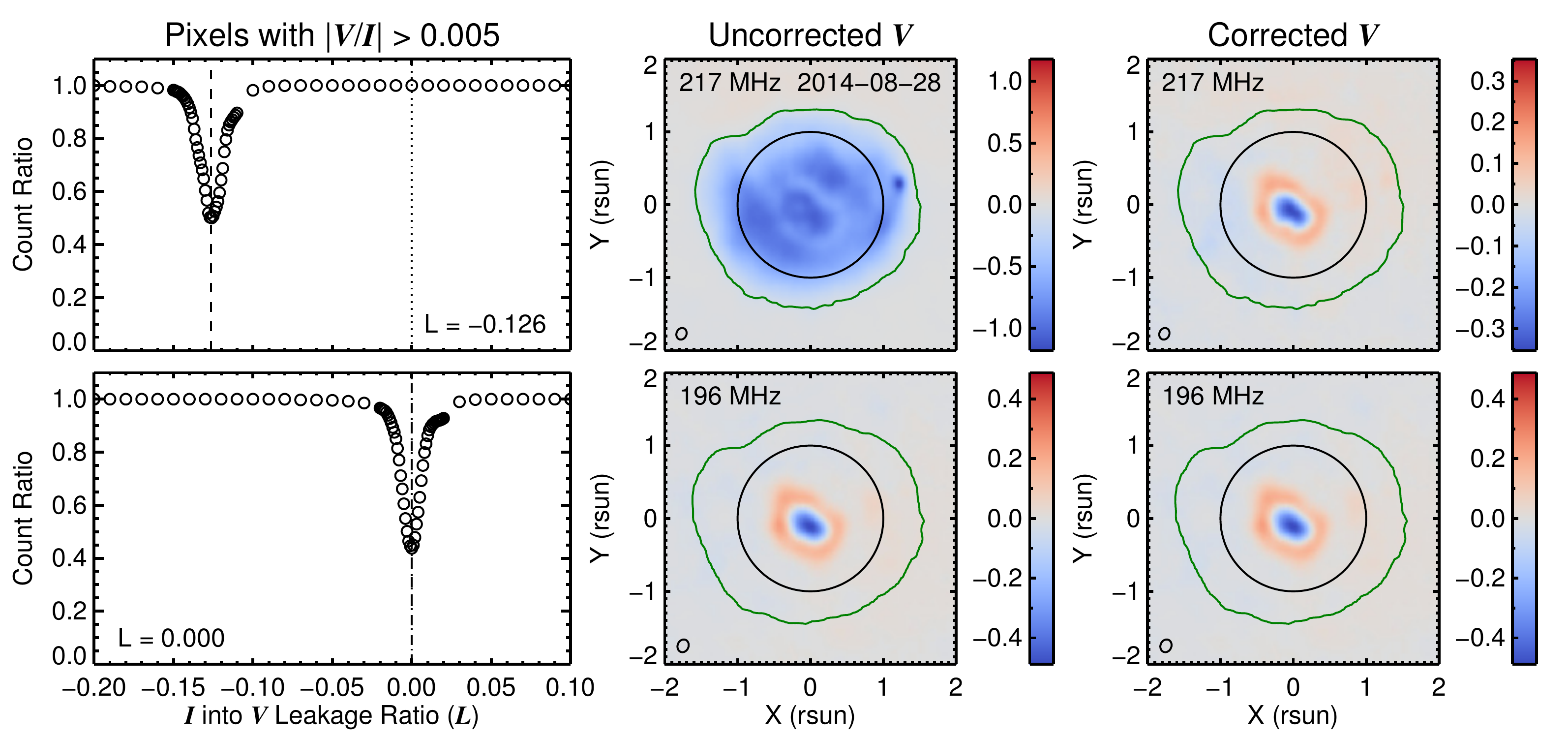}
    \caption{\footnotesize{
	An illustration of the leakage subtraction algorithm. The left panels show the implementation of 
	Equation~\ref{eq:alg} to find the $L$ that minimizes the number of pixels with $|V/I| > 0.005$. 
	At 217 MHz (top), we estimate that -12.5\% of the Stokes $I$ signal leaked into Stokes $V$, and 
	at 196 MHz, the same procedure estimates there to be no leakage. The middle panels show 
	the uncorrected Stokes $V$ images, and the right panels show the corrected images 
	(\textit{\textit{i.e.}} $V - L\cdot{}I$). \redit{The corrected image at 217 MHz is shown as a function of $L$  
	in an animated version of this figure available in the \href{https://link.springer.com/article/10.1007\%2Fs11207-019-1502-y}{online material}.} 
       }}
    \label{fig:lsub1}
\end{figure*}


\begin{figure*}\graphicspath{{chapter4/}}
    \centering
    \includegraphics[width=1.0\textwidth]{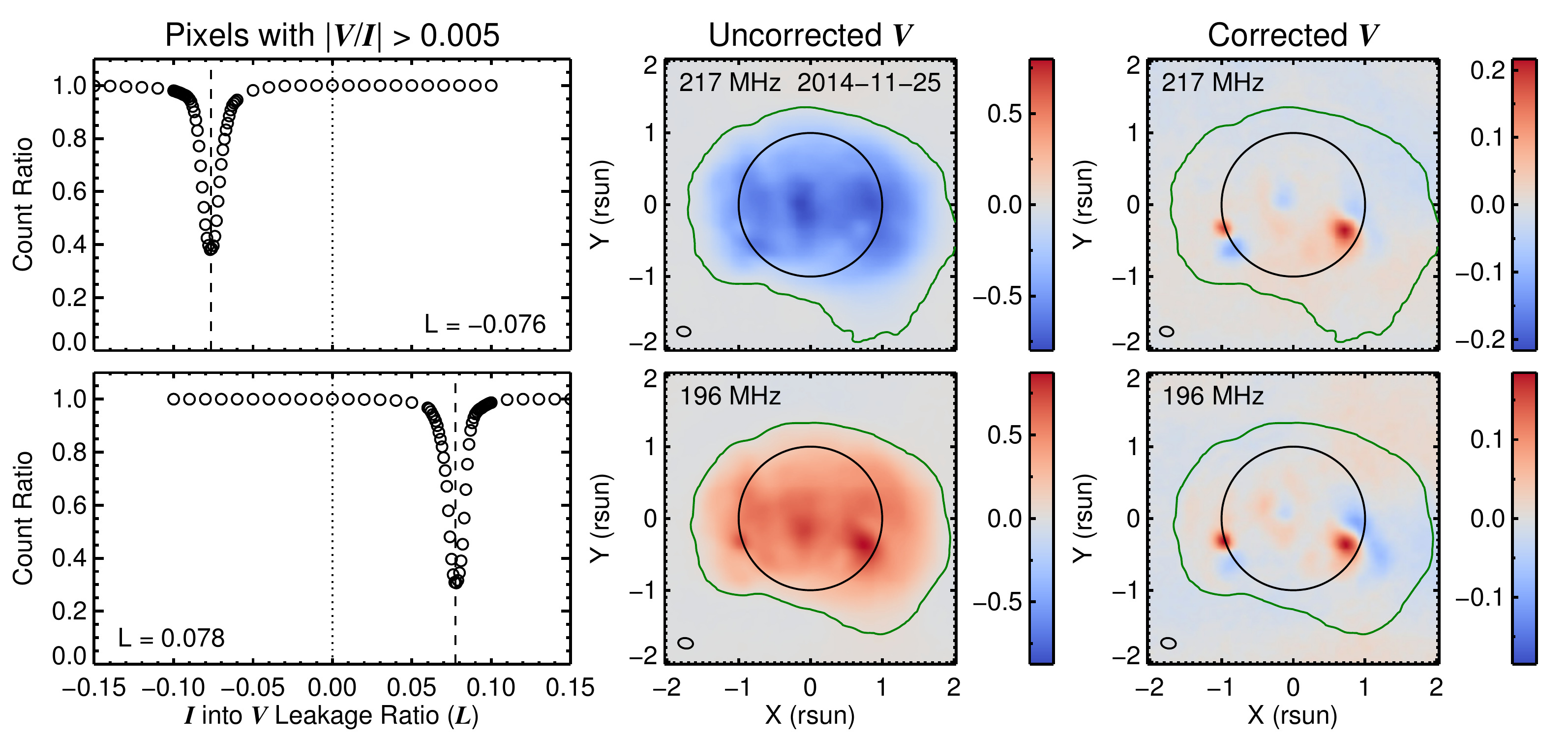}
    \caption{\footnotesize{
	Same as Figure~\ref{fig:lsub1} but for a different observation with different structures and a different leakage 
	behavior. 
       }}
    \label{fig:lsub2}
\end{figure*}

Figures~\ref{fig:lsub1} and \ref{fig:lsub2} show the result of applying this algorithm to images from two frequency channels 
on two different days. In Figure~\ref{fig:lsub1}, we see that the uncorrected Stokes $V$ image at 217 MHz \edit{would imply} that the 
entire corona is highly circularly polarized with a single sense, while the uncorrected 196 MHz image suggests a very 
different structure with a mixture of opposite signs. 
We know that entire corona at 217 MHz should not be polarized with a single sense to the extent implied by the uncorrected 
217 MHz image. 
Our algorithm suggests that $L$ = -0.125 in this case, and applying that correction recovers the same structure that is apparent in the 
196 MHz image. 
Importantly, the same procedure can also be applied to observations for which there is little or no leakage, as is illustrated by the 
196 MHz example in Figure~\ref{fig:lsub1}. 
Figure~\ref{fig:lsub2} shows another example for which the two frequency channels shown are impacted by significant leakage of opposite signs, but 
once corrected, they exhibit very similar structures.  

\begin{figure*}\graphicspath{{chapter4/}}
    \centering
    \includegraphics[width=1.0\textwidth]{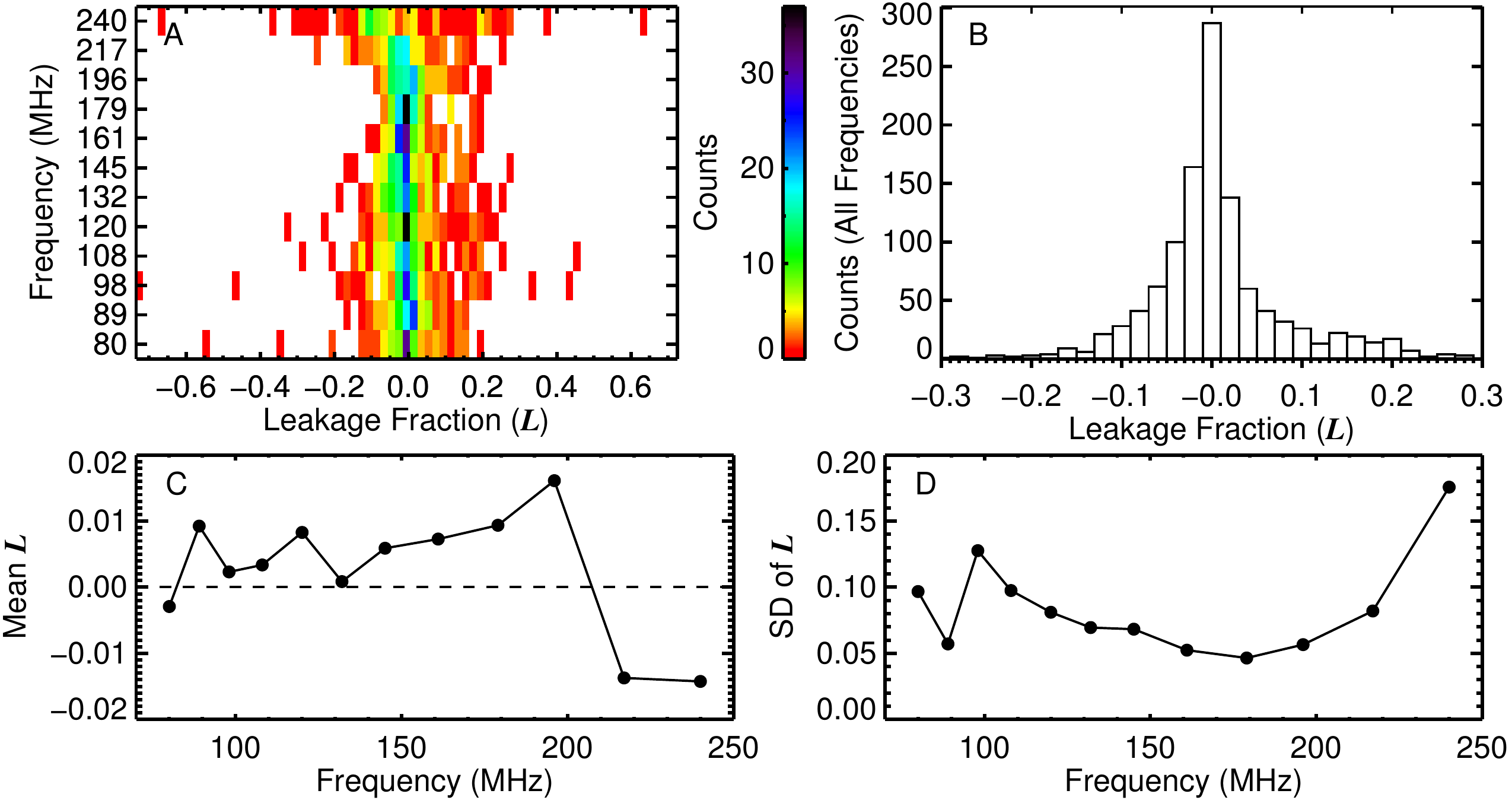}
    \caption{\footnotesize{
    	Summary of results from applying the Equation~\ref{eq:alg} algorithm to 1144 different observations. 
	\textbf{(A):} Two-dimensional histogram showing the leakage fraction [$L$] as a function of frequency with a 
	bin size of 0.02 in $L$. 
	\textbf{(B):} Histogram of $L$ across all frequencies.  
	\textbf{(C):} The average $L$ for each frequency channel. 
	\textbf{(D):} The standard deviation of $L$ for each frequency channel. 
	Around 26\% of observations have $|L| < 0.01$, and around 84\% have $|L| < 0.1$. 
	Panels A and D show that the leakage is most severe and variable at the extremes of 
	the bandwidth.
       }}
    \label{fig:lhist}
\end{figure*}

The examples in Figures~\ref{fig:lsub1} and \ref{fig:lsub2} are cases for which the leakage is fairly severe. 
Figure~\ref{fig:lhist} summarizes the leakage behavior across our dataset.
\redit{83\% of the images exhibit leakage fractions less than or equal to 10\%. 
$L \leq$ 0.05 and 0.01 for 66\% and 25\% of the images, respectively.} 
Figure~\ref{fig:lhist}d shows that the standard deviation of $L$ is lowest at 179 MHz, which is consistent with the 
astrophysical MWA literature that indicates the leakage tends to be worst near the ends of the bandwidth. 
This summary includes 106 different sets of spectroscopic imaging observations for a total of 1144 images to which the algorithm could be applied.  
In Section~\ref{mwa}, we stated that 111 observing periods were analyzed, which would imply 1332 images given our 12 frequency 
channels.
Some images are rejected for polarimetry because they do not contain enough pixels detected above 5 $\sigma$ in Stokes $I$, 
generally because a nonthermal active region source is so intense as to elevate the noise floor above the level of the thermal 
disk. 
\redit{In other words, there is insufficient dynamic range to simultaneously detect both the thermal and 
nonthermal components present at those times.}

\redit{These observations must be excluded because the algorithm relies on the statistical expectation that most pixels are 
dominated by thermal bremsstrahlung emission, and in} these cases, there are not enough ``thermal" pixels with 
sufficient signal-to-noise ratios for the algorithm to function.
\redit{Images were rejected for polarimetry if they contained fewer pixels above 5 $\sigma$ in 
Stokes $I$ than that enclosed by a circle of radius equal to the height of the plasma frequency layer at a given frequency given a 
3$\times$ \citet{Newkirk61} density model, which roughly approximates the height of the radio limb in our observations.
Around 14\% of our data failed this test and are excluded from further analysis.}
\redit{It is important to note that this} introduces a bias in the next section on active region sources because the most intense sources tend to be the most 
highly polarized, but the leakage artefact cannot be constrained using our method for the brightest among them. 
The intense and highly-polarized population is therefore very likely to be underrepresented. 


\redit{Leakage may also occur from Stokes $U$ into $V$ \citep{Lenc17}, but this is not a concern here because linear polarizations from the 
corona are negligible at our frequencies and observing bandwidths (\textit{e.g.} \citealp{Gibson16}).
Of potential concern, however, is possible leakage from $V$ into $I$. 
This could add to or subtract from the Stokes $I$ levels, decreasing the reliability of measured polarization fractions. 
Unlike leakage from Stokes $I$ into the other parameters, 
the reverse case has not been investigated for the MWA because the polarization fractions of astrophysical sources 
are generally so low as to make this effect very difficult to characterize and unlikely to significantly impact the results. 
However, solar radio bursts may have large circular polarization fractions, meaning that Stokes $V$ into $I$ leakage 
could be a significant contaminant in some cases. 
We currently have no way to assess or mitigate this contamination, but 
we anticipate that the effect should occur at a similar or lower level than leakage from Stokes $I$ into $V$, 
as the mechanism would be similar but with generally lower magnitudes.}

\redit{Of concern are sources with high polarization fractions and large leakage fractions. 
This is relevant mainly for the next section, which focuses on nonthermal active region sources. 
Assuming that $V$ into $I$ leakage may occur at up to the same level estimated for $I$ into $V$, 
this introduces an uncertainty in $V/I$ of less than 1\% for 79\% of the sources 
and an uncertainty of less than 5\% for 95\% of the sample described by Figure~\ref{fig:ars_hist}. 
The remainder have uncertainties of 10\% on average, and up to 19\% for one event, due to this effect. 
We have chosen to represent this latter population with a different symbol in Figure~\ref{fig:ars_hist} 
to indicate that their polarization fractions should be treated with additional skepticism.
For Figure~\ref{fig:ch_slices} in Section~\ref{ch}, we estimate uncertainties on $V$/$I$ of $\lesssim$ 3\% 
by combining several effects for two cases where different observations on the same day with different values of $L$ could be compared. 
These observations correspond to thermal or very weak nonthermal emission for which the polarization 
fractions are lower than 5\%.
Potential $V$ into $I$ leakage is therefore not a significant concern in that case 
and constitutes an average of 14\% of the total 
error bars in Figure~\ref{fig:ch_slices}.}


\section{Active Region Noise Storm Sources}
\label{ar}

The most common features in these images are compact polarized sources, 
the most intense of which are identified here as \redit{noise storm continua associated 
with Type I bursts.}
This is apparent from the variability in their associated dynamic spectra along with their high polarization fractions. 
However, as we will see, there are also very weak and weakly-polarized sources for 
which the source type and emission mechanism is less obvious. 

Noise storms are periods of extended burstiness that are associated with active regions and may 
persist for several days as an active region transits the disk. 
They are characterized by many distinct, narrowband \redit{Type I} bursts, often with enhanced continuum 
emission around the same frequency range \citep{Elgar77,Klein98}. 
As our data reflect the background levels during each observation period, our detections correspond to 
the continuum enhancement, along with any burst periods that could not be filtered out by our baseline 
procedure because they occurred on timescales less than the 0.5-sec time resolution. 
Despite decades of study, there are a number of unanswered questions about the  
nature of Type I bursts. 
Not all active regions that are productive at other wavelengths produce noise storms, and 
the non-radio signatures are often scant \citep{Willson05,Iwai12,Li17}, unlike Type II and III bursts, which have obvious 
associations with CMEs and flares \citep{Cairns03,Reid14}. 
There is general agreement that \redit{both the burst and continuum components of} noise storms are produced by plasma emission, largely due 
to their often high circular polarizations \citep{Aschwanden86,Mugundhan18}, but what accelerates the electrons is still debated.
Small-scale reconnection events \citep{Benz81} or weak shocks associated with upward-propagating waves \citep{Spicer82} are the 
two leading ideas, and recent work has favored persistent interchange reconnection between open and closed 
fields at the boundaries of active regions \citep{Del11,Mandrini15} or reconnection driven by moving magnetic features \citep{Bentley00,Li17}.

To automatically detect these features in the Stokes $V$ images, we developed a simple 
algorithm that begins \edit{with suppressing any diffuse polarized emission that may be present 
by applying a Butterworth bandpass filter to the fast Fourier transform (FFT) of each image. 
 \redit{The filter aims to flatten the frequency response over a particular passband, in this case the FFT frequencies corresponding to larger spatial scales, without 
 producing sharp discontinuities between the filtered and unfiltered frequencies \citep{Butterworth30}.} 
The filtered FFT is transformed back, and the resulting} image is thresholded into two binary masks, one for each polarization 
sense, that include pixels with values above the larger of 10 $\sigma$ or 20\% of the maximum value. 
Ellipses are fit to all of the contiguous regions in the masks, and several criteria are imposed to 
obtain the final detections. 
These criteria include ensuring that 
1) the signal-to-noise ratio of pixels pulled from the 
filtered image are above 10~$\sigma$ in the original image and are of the same polarization sense,
2) the areas of the fitted ellipses are within 0.75\,--\,1.5$\times$ that of the corresponding  
synthesized beam for a given frequency, 
3) the fitted ellipses have aspect ratios no more than 1.1$\times$ that of the \edit{synthesized} beams, 
and 4) the masked regions are sufficiently elliptical, which we defined as filling at least 95\% of the fitted ellipse. 


\begin{figure*}\graphicspath{{chapter4/}}
    \centering
    \includegraphics[width=1.0\textwidth]{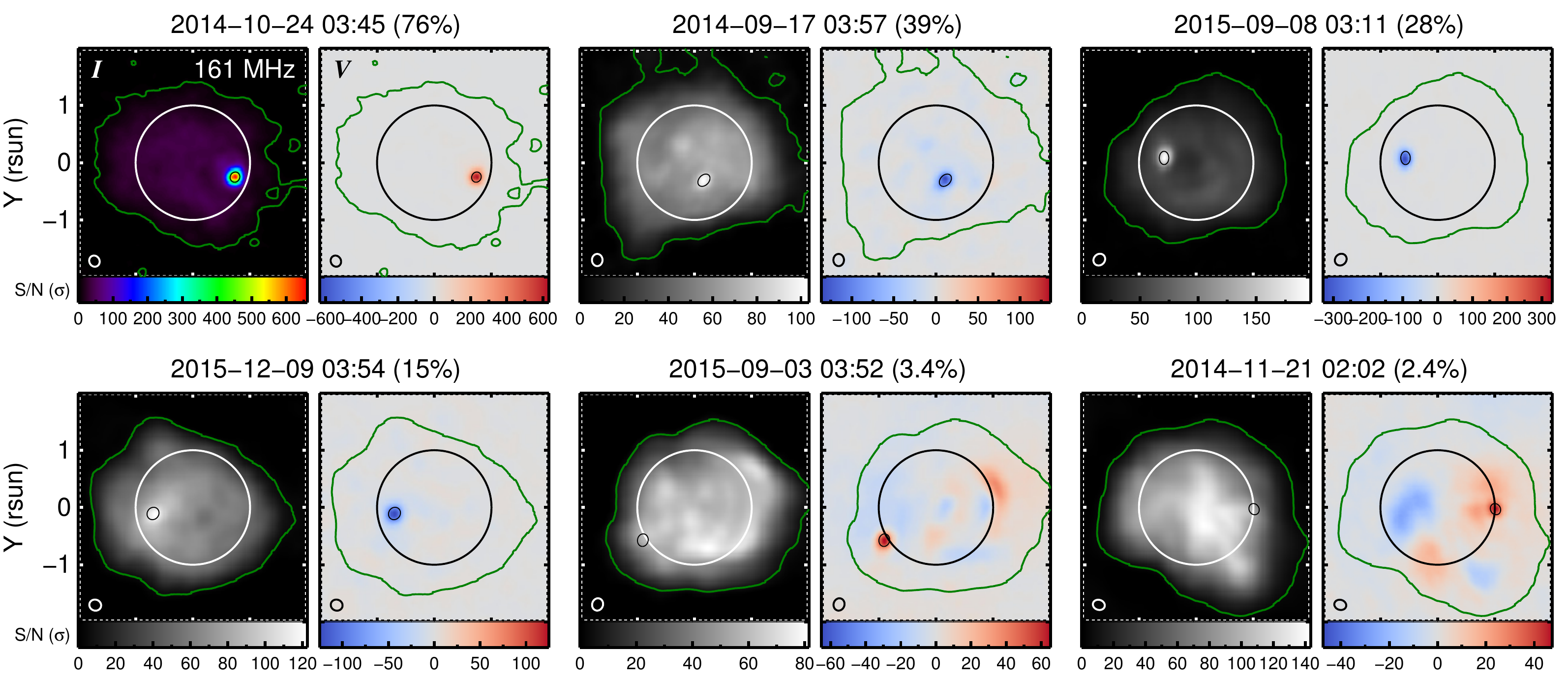}
    \caption{\footnotesize{
	Six randomly chosen examples of compact polarized sources detected at 161 MHz, sorted by 
	polarization fraction.
	The titles correspond to UTC times followed by the peak polarization fraction in parentheses.
	The color scales are linear, and the first example is plotted with a 
	different color scheme to better reflect the dynamic range of that observation. 
	Color bar intensities are expressed in units of signal-to-noise [$\sigma$], and the green 
	contour reflects the 5-$\sigma$ level in Stokes $I$. 
	Black ellipses around the sources show the region identified by the source finder algorithm. 
	The large solid circles represent the optical disk, and the ellipses in the lower-left corners 
	represent the synthesized beam sizes. 
       }}
    \label{fig:ars}
\end{figure*}

These criteria all serve to eliminate false positives that arise from the bandpass filtering, which may amplify 
noise, introduce artefacts near very bright sources, and/or not entirely suppress the large-scale diffuse emission. 
Adjusting the tolerance parameters of this algorithm can satisfactorily extract sources from any given image, 
but finding a set of defaults that could serve the entire dataset was somewhat difficult. 
We opted to aggressively tune the parameters to eliminate false positives at the cost 
of excluding false negatives. 
This procedure is run independently for all of the frequency channels in a given observation. 
The detections are then grouped across frequencies by checking for overlap among the fitted 
ellipses.
Only sources that are detected in at least three frequency channels are kept and incorporated into the 
following plots.
\edit{We find 693 sources with this method from 112 separate regions, and at least one source 
is found on 64 out of 82 days (78\%). 
Solar Cycle 24 peaked in April 2014, and our data correspond to between August 2014 and December 2015, 
meaning that we are examining the early part of the declining phase in the solar cycle. 
As these features are associated with active regions, we would likely have found a higher fraction of days with at 
least one noise storm if our observations were shifted one year earlier and a lower fraction if the observations were taken in subsequent years.} 

Figure~\ref{fig:ars} shows six randomly-selected examples of these sources at the center 
of our bandwidth, 161 MHz. 
They exhibit polarization fractions ranging from 2.4 to 76\% and well represent the range of sources found.
Most sources in the full sample are unipolar and are fairly isolated in the polarization images, and those with very low polarization fractions 
are sometimes embedded in diffuse emission of the same sign. 
A small number of bipolar sources were also found. 
This is somewhat inconsistent with \citet{White92}, who found that bipolar sources were nearly as common as unipolar 
sources in 327 MHz Very Large Array (VLA) observations. 
Bipolar sources are presumably less common in our observations because we are looking at lower frequencies for 
which the emission is generated at a larger height and the spatial resolution is lower. 
Preliminary analysis has revealed interesting potential anti-correlations in the intensities of the two components of 
one bipolar source, and this sort of time variability may be explored in future work. 


\begin{figure*}\graphicspath{{chapter4/}}
    \centering
    \includegraphics[width=1.0\textwidth]{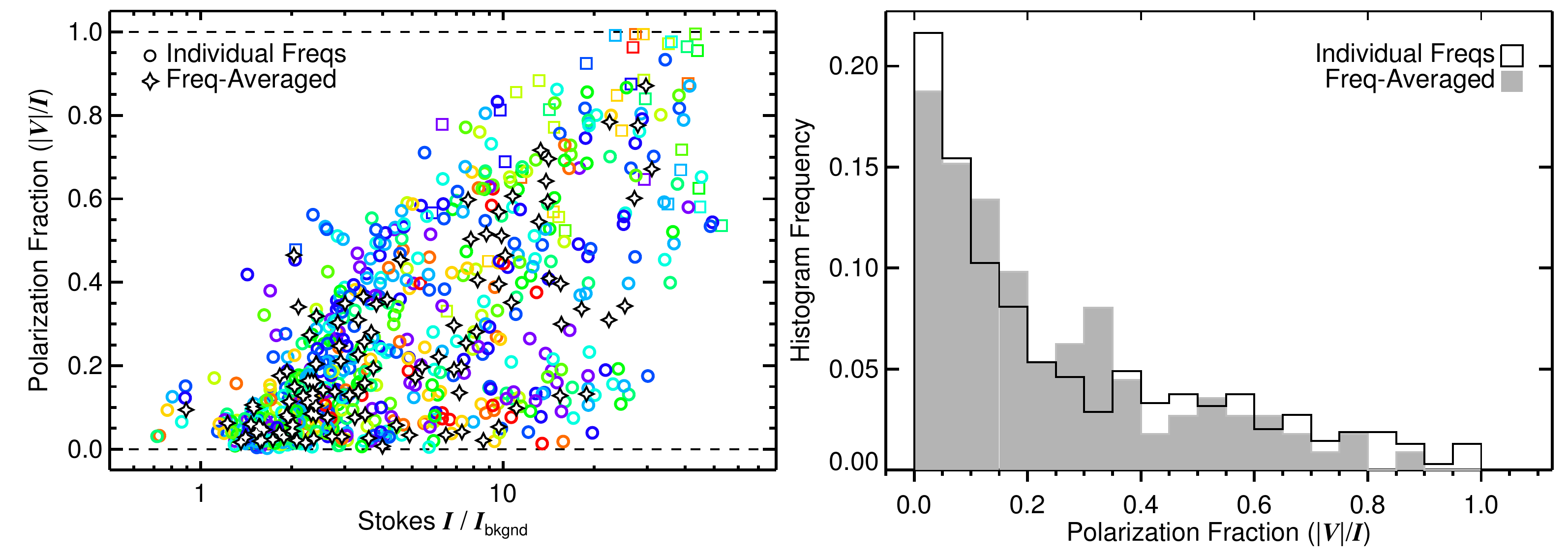}
    \caption{\footnotesize{
	\textit{Left}: Scatterplot of \redit{polarization fraction [$|V|$/$I$] versus} Stokes $I$ intensity for \edit{693}  
	compact polarized sources. 
	Values are measured at the location of peak Stokes $V$ intensity, and the Stokes $I$ intensities 
	are normalized by the median intensity of pixels detected above 5-$\sigma$ [$I_{\rm bkgnd}$]. 
	Colored circles represent measurements from individual frequency channels; purple refers to the 
	highest frequency (240 MHz) and red refers to the lowest (80 MHz). 
	All sources are detected in at least three channels, \edit{representing 112 separate regions}, 
	and the stars represent averages across frequency \edit{for a given region}.
	\redit{Squares indicate the 5\% of sources for which the $V/I$ uncertainty is large ($\approx$ 10\%) 
	due to possible calibration errors that could not be accounted for (see Section~\ref{leakage}).} 
	\textit{Right}: Histogram of $|V|$/$I$, where the white region corresponds to the colored circles 
	and the shaded region corresponds to the stars from the left panel. 
       }}
    \label{fig:ars_hist}
\end{figure*}

Figure~\ref{fig:ars_hist} shows a scatterplot of polarization fraction [$|V|$/$I$] versus the total intensity divided by 
the background level [$I$/$I_{\rm bkgnd}$] in the left panel, along with a simple histogram of $|V|$/$I$ in the right panel. 
These are plotted both for each frequency channel independently and for averages of the same source detected 
in multiple channels. 
The background is defined as the median intensity in pixels detected above 5 $\sigma$, and the noise level 
[$\sigma$] is defined as the standard deviation within a 1-pixel border (1156 pixels) that run along the edge of the
289$\times$289-pixel  ($\pm$ \rsolar{3}) field-of-view. 
We find a very broad range of source intensities, ranging from slightly below the background level 
to 50 times greater, with polarization fractions ranging from a few tenths of a percent to nearly 100\%. 
\redit{The average source has a Stokes $I$ intensity of 7.6$\times$ the background level and a 
polarization fraction of 27\%.}
The most striking aspect of Figure~\ref{fig:ars_hist} is the relationship between total intensity \edit{over the background} and polarization 
fraction\edit{, which are positively correlated with a Pearson correlation coefficient [$r$] of 0.64.}


\begin{figure*}\graphicspath{{chapter4/}}
    \centering
    \includegraphics[width=1.0\textwidth]{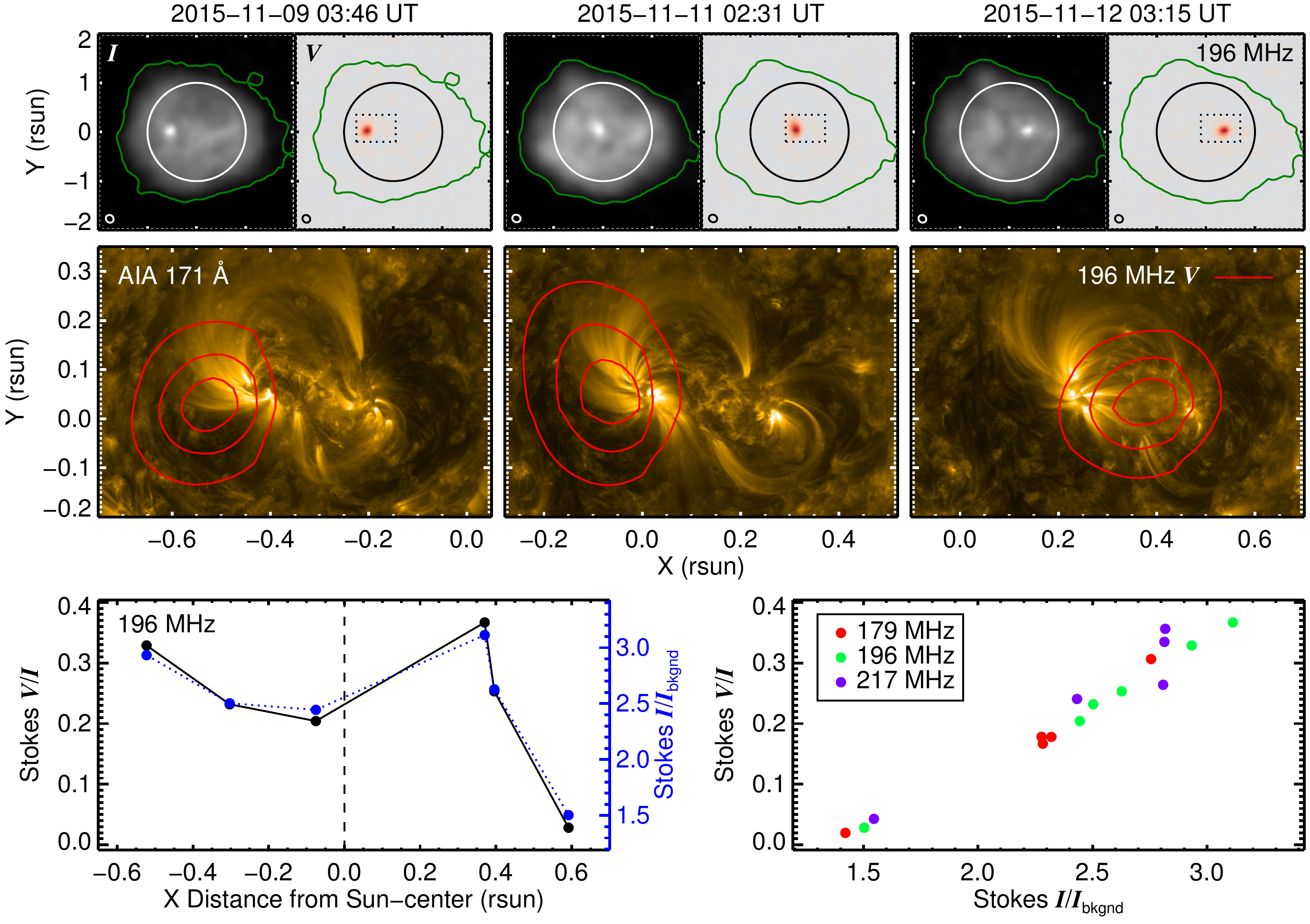}
    \caption{\footnotesize{
	\textit{Top}: The same compact source observed on consecutive days at 196 MHz. 
	\textit{Middle}: Overlays of the 196 MHz circular polarization signal onto 171 \AA{} images from AIA. 
	The field-of-view is marked by the dotted region in the top row, and the contours are at 20, 50, and 80\% 
	of peak intensity. 
	\textit{Bottom}: The left panel plots polarization fraction [$V$/$I$] and Stokes $I$/$I_{\rm bkgnd}$ 
	as a function of horizontal distance from Sun-center at 196 MHz \redit{for the same source over five days from 9 to 13 November 2015}.
	The right panel plots the same two parameters against each other for three different frequency 
	channels. 
       }}
    \label{fig:ars_lon}
\end{figure*}

Figure~\ref{fig:ars_lon} displays results from one of the \redit{few} sources in our sample for which we have observations 
on several consecutive days and for which \redit{a compact radio source appears in association with the same active region on each day.} 
The source is polarized between 3\% and 37\% in the same sense in six observations between \edit{9 and \redit{13} November 2015}. 
Our goal here is to investigate a potential relationship between distance from Sun-center and polarization fraction, 
as previous studies have found noise storms to exhibit higher polarization fractions near disk center. 
We do not find such a relationship \edit{in Figure~\ref{fig:ars_lon}}, and instead this exercise further reinforces the \edit{positive correlation} between 
total intensity and polarization fraction, which apparently becomes stronger if one considers several observations 
of the same source. 
Recall from Section~\ref{mwa} that the observations presented in this paper all represent the baseline intensity over 
5-min observing periods, constructed from the averages of images with the lowest total intensities.
These sources do fluctuate in intensity, so it may be possible to control for the intensity relative to the background 
and then recheck if the polarization fraction has a longitudinal dependence in a future study, ideally with 
more observations of individual sources detected on consecutive days. 

Figure~\ref{fig:ars_lon} also overlays circular polarization contours onto 171 \AA{} images from the 
\textit{Atmospheric Imaging Assembly} (AIA; \citealp{Lemen12}) onboard the \textit{Solar Dynamics Observatory} (SDO; \citealp{Pesnell12}). 
The radio source is associated with a large active region (AR 12448) and is located over fan-loop
structures that represent the bases of loops extending to larger heights. 
This make sense, as the accelerated electrons that are presumably responsible for this radiation must be able to escape to 
sufficiently large heights, corresponding to densities that are sufficiently low for low-frequency emission. 
It is also interesting to note that when the source is east of disk-center, the radio source is associated with the trailing 
sunspot. 
\redit{Noise storms have long been associated with active regions (\textit{e.g.} \citealp{LeSqueren63,Gergely75,Alissandrakis85}), and} 
previous observations have found noise storm sources to be more often associated with the leading spot \citep{White92}. 
A natural followup would be to investigate that aspect systematically for the sources detected here. 
The site of radio emission within the active region also shifts somewhat in time, with the apparent Carrington longitude 
jumping by 12 degrees between \edit{11 and 12 November 2015}. 
This is likely due to evolution in the active region shifting the region where the energetic electrons 
either originate or are able to \edit{reach} in height. 

The main question posed by these results, particularly the scatterplot in Figure~\ref{fig:ars_hist}, is 
whether or not these sources all represent the same basic phenomenon. 
\redit{Our average polarization fraction (27\%) is lower than previous measurements of noise storm continua. 
Most studies report similar polarization levels for the Type I burst and continuum components of noise storms, which generally 
exceed 80\% \citep{Elgar77}, but sources with lower polarization fractions have also been reported. 
\citet{Dulk84} observed noise storm continua with polarization fractions of $\approx$ 40\%, around 15\,--\,20\% lower 
than the associated bursts. 
As discussed in Section~\ref{leakage}, the most intense and likely highly-polarized sources in our dataset could not be 
included because the leakage mitigation algorithm could not be applied. 
This diminishes the overall average, but more importantly, our sample includes a large number of weak and weakly-polarized sources 
that could not have been characterized by previous instruments.
For example, the 3 September 2015 source ($V/I$ = 3.4\%) shown in Figure~\ref{fig:ars} is prominent in the polarization map but 
is visually indistinguishable from the quiet Sun in total intensity.}
Given that there does not appear to be any separation into distinct populations \redit{in Figure~\ref{fig:ars_hist}}, we suggest that 
\redit{the data represent} a continuum of plasma emission noise storm \redit{continua} sources with intensities and polarization fractions down to 
levels that were not previously detectable.

\edit{For sources with relatively low total intensities (\textit{e.g.} $I/I_{\rm bkgnd} \lesssim 5$; 61\% of the population),}
\redit{where the nonthermal component is not entirely dominant,}
very low polarization fractions can be explained by there being a mixture of thermal and nonthermal emission 
within the same resolution element. 
Recall from Section~\ref{introduction} that the thermal bremsstrahlung and nonthermal plasma emission mechanisms 
generally produce opposite polarization signs for the same magnetic field orientation, but the plasma emission 
component is much more highly polarized. 
Therefore, a pixel may be dominated in total intensity by bremsstrahlung emission while the polarized intensity 
is dominated by plasma emission.
The polarization fraction then rises with intensity relative to the background because the relative contribution from 
plasma emission increases. 

Filling factors and beam dilution are also likely to be important, as the thermal component is likely to fill the resolution 
element while the nonthermal component \edit{may come} from a sub-resolution structure. 
This \edit{would mean} that the more highly-polarized nonthermal signal is diluted, which would 
further bring down the polarization fraction. 
\edit{Nonthermal emission sources may not necessarily be intrinsically smaller than the beam size, however.
For instance, \citet{Mohan18} found the scattering-deconvolved sizes of type III burst sources to be significantly larger than the PSF.
As the total intensity becomes much larger than the background and the nonthermal component becomes entirely 
dominant, physical effects related to the emission mechanism and radio wave propagation become increasingly 
important to the interpretation of relatively low polarization fractions in plasma emission sources.
As described in Section~\ref{introduction}, scattering by density inhomogeneities may reduce 
the polarization fraction, as can other propagation effects such as mode coupling.} 
These ideas are discussed further in Section~\ref{discussion} in the context of our other results.


\section{Coronal Holes}
\label{ch}

\begin{figure*}\graphicspath{{chapter4/}}
    \centering
    \includegraphics[width=1.0\textwidth]{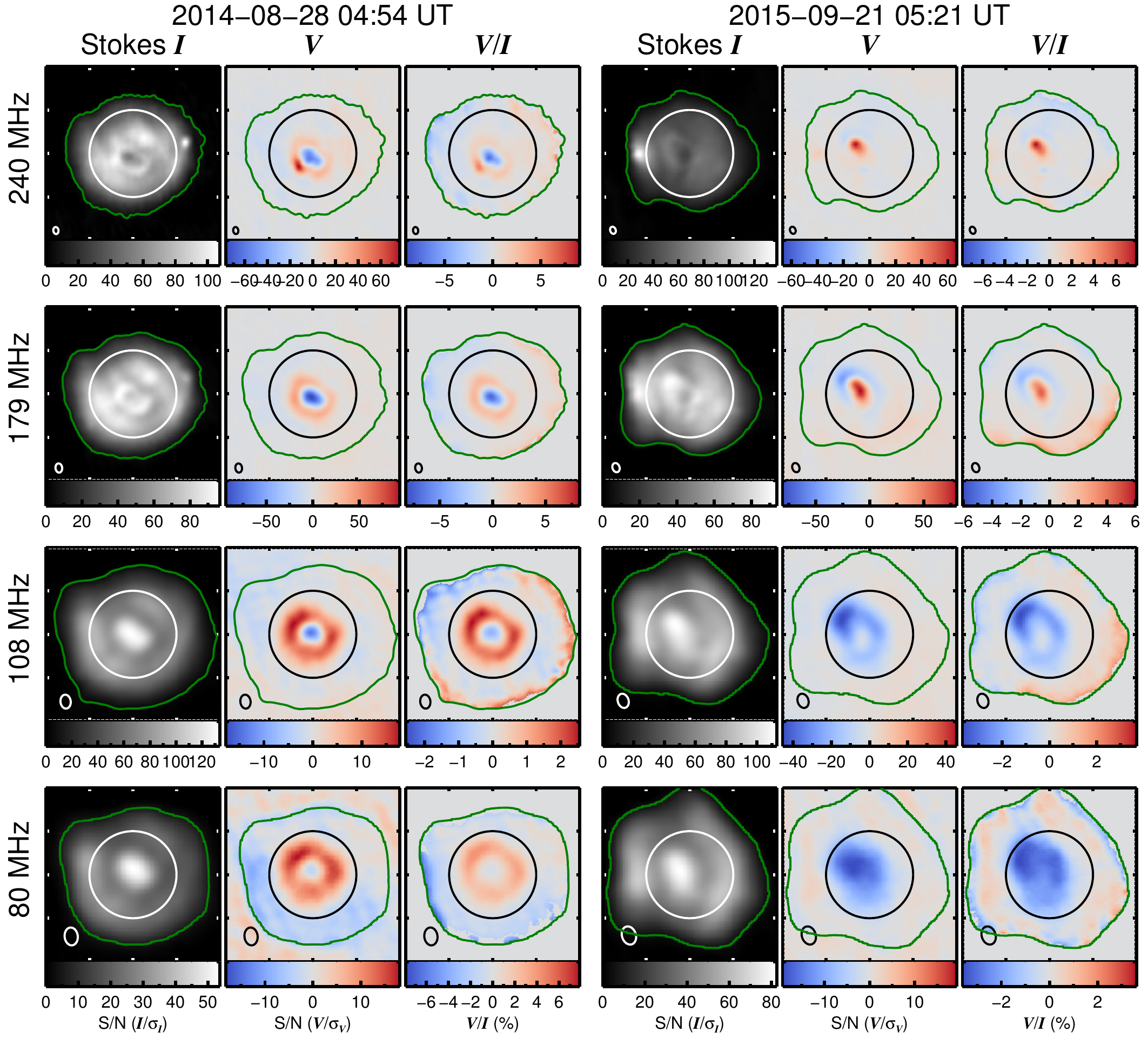}
    \caption{\footnotesize{
	Stokes $I$, $V$, and $V$/$I$ at four frequencies across our bandwidth for two different coronal holes with 
	opposite polarization signatures.
	Color bar units are in signal-to-noise [S/N] for $I$ and $V$ and percent for polarization fraction [$V$/$I$]. 
	The green contours represent the 5 $\sigma$ level in Stokes $I$, the solid circles represent the optical disk, and 
	the ellipses in the lower-left corners represent the synthesized beam sizes. 
       }}
    \label{fig:ch}
\end{figure*}

Perhaps the most surprising finding to immediately emerge from these data is a 
\edit{characteristic ``bullseye"} structure that is frequently exhibited by low-latitude coronal holes and, more 
generally, that coronal holes are the Sun's most prominent features in circular 
polarization at low frequencies in the absence of intense noise storm emission. 
Coronal holes are regions where the magnetic field is open, allowing material 
to freely flow into interplanetary space to form the fast solar wind \citep{Cranmer09}. 
Because the plasma is not confined by closed fields, the densities inside coronal 
holes are considerably lower than in the surrounding corona, and they 
are correspondingly fainter in the soft X-ray and extreme ultraviolet (EUV) observations that are typically 
used to characterize them. 
This is also true at our highest frequencies, which can be seen for two different 
coronal holes in the Stokes $I$ images shown in the upper row of Figure~\ref{fig:ch}. 


As frequency decreases across our bandwidth, many coronal holes transition from 
being relatively dark to relatively bright with respect to their surroundings. 
This effect had been known from a few previous observations \redit{\citep{Dulk74,Lantos87,Lantos99,McCauley17}} and 
was recently characterized in more detail using MWA observations \citep{Rahman19}. 
The mechanism that produces this increase in brightness is unclear, but 
\redit{different authors have suggested} that refraction near the coronal hole boundary   
\redit{may systematically redirect} emission generated outside of the coronal hole to the interior from an 
observer's perspective \redit{\citep{Lantos87,Alissandrakis94,Rahman19}. 
As discussed by \citet{Rahman19},} 
this leads to a ring of enhanced emission around the coronal hole edge, which is apparent 
in our \edit{higher-frequency images} but cannot be distinguished at the lower frequencies, likely due to the 
lower spatial resolution. 
A corollary of this effect is a ring of diminished intensity in the regions from which 
the refracted emission originated, which we see prominently in the low-frequency images 
at the bottom of Figure~\ref{fig:ch}. 


\begin{figure*}\graphicspath{{chapter4/}}
    \centering
    \includegraphics[width=0.7\textwidth]{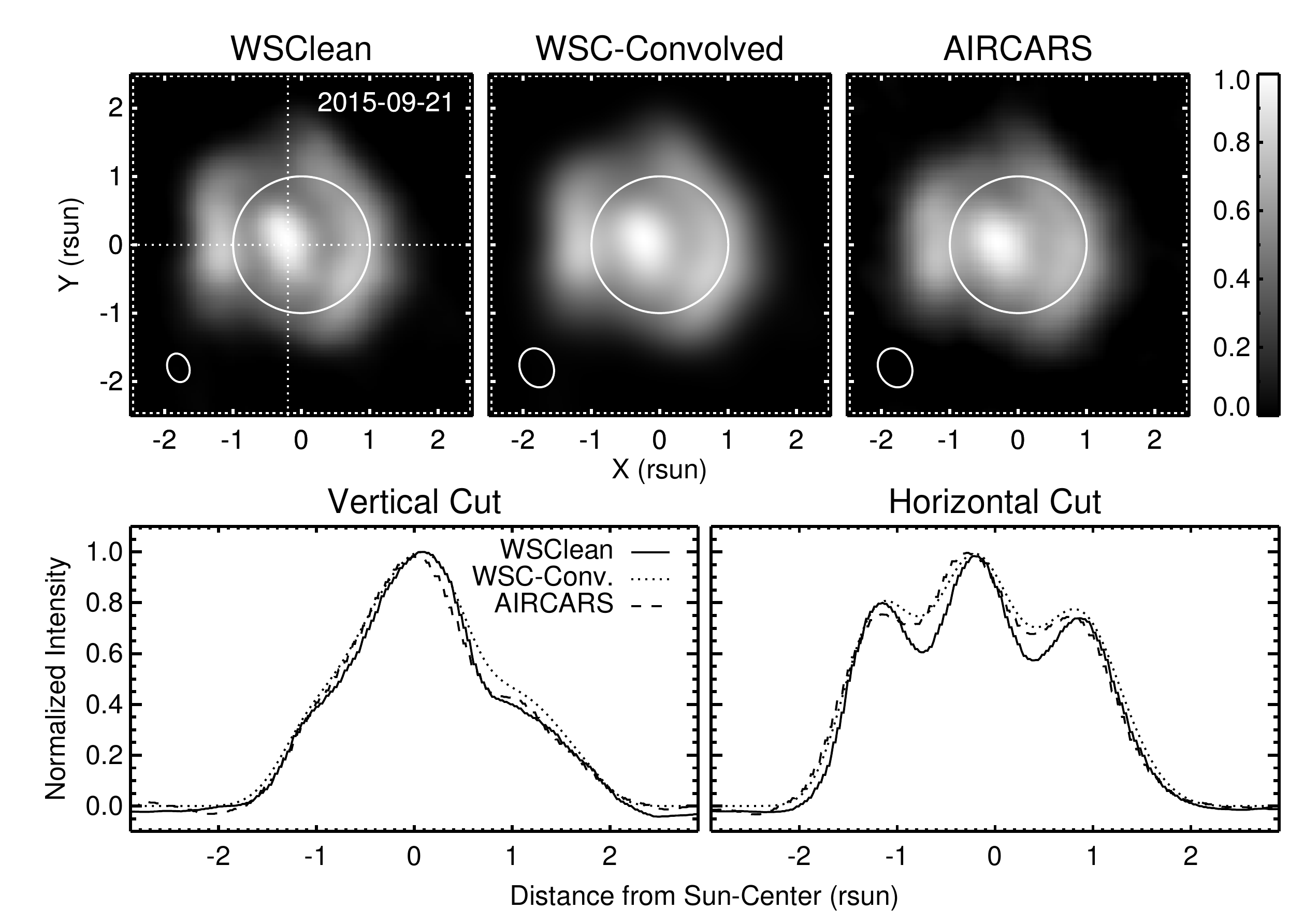}
    \caption{\footnotesize{
	A comparison of the same dataset independently reduced with different calibration techniques 
	and different implementations of the \textsf{CLEAN} algorithm. 
	The Stokes $I$ image at 80 MHz is shown for the \edit{21 September 2015} coronal hole from Figure~\ref{fig:ch}.
	Our reduction on the left, the  \textsf{AIRCARS} reduction on the right, and the  \textsf{WSClean} image convolved 
	with the \textsf{AIRCARS} beam is in the middle. 
	Dotted lines in the left panel indicate the cuts shown in lower two panels. 
	The convolved \textsf{WSClean} and \textsf{AIRCARS} images appear nearly identical, 
	and the cuts demonstrate that this \edit{is} also true quantitatively. 
       }}
    \label{fig:aircars}
\end{figure*}

The second and third columns of Figure~\ref{fig:ch} show the corresponding 
Stokes $V$ and $V$/$I$ images, respectively. 
Both coronal holes in Figure~\ref{fig:ch} exhibit a central polarized component 
of one sense surrounded by a ring of the opposite sense. 
The outer ring grows in area as the frequency decreases, while the central component 
shrinks until it may or may not be completely gone by 80 MHz. 
This bullseye structure is peculiar \edit{in that coronal holes have unipolar line-of-sight (LOS) 
magnetic field configurations that we expected to result in unipolar Stokes $V$ maps across our 
entire observing band.
We therefore first consider if the feature might be an instrumental or calibration artefact.
While we cannot validate this signature with a completely independent 
observation and data reduction procedure, we have strong evidence to believe that 
this structure is real for the reasons outlined below.}

First, we can validate the structure seen in Stokes $I$ with an independent reduction. 
Figure~\ref{fig:aircars} compares our 80 MHz image of the \edit{21 September 2015} coronal hole, which 
was produced using \textsf{WSClean}, to an image produced using the  \textsf{AIRCARS} pipeline \citep{Mondal19}. 
 \textsf{AIRCARS} uses an entirely different calibration scheme through iterative self-calibration on the Sun itself 
without the need to observe a separate calibrator source. 
This approach is advantageous in that the calibration is tuned to the specific observation of interest, 
which may greatly improve the dynamic range, but it cannot yet be used for polarimetry. 
Our method determines the calibration solutions solely from a known calibrator source, generally 
observed before or after an observing campaign of several hours. 
Figure~\ref{fig:aircars} shows that we obtain nearly identical results from the two pipelines. 
Both methods do use the \textsf{CLEAN} algorithm for deconvolution, as is the standard, 
although implemented through different software packages. 
However, we can be confident that we are not seeing an artefact of the \textsf{CLEAN} algorithm because the 
features of interest are also present in the \edit{undeconvolved} ``dirty images". 


\begin{figure*}\graphicspath{{chapter4/}}
    \centering
    \hspace{-0.7cm}
    \includegraphics[width=1.0\textwidth]{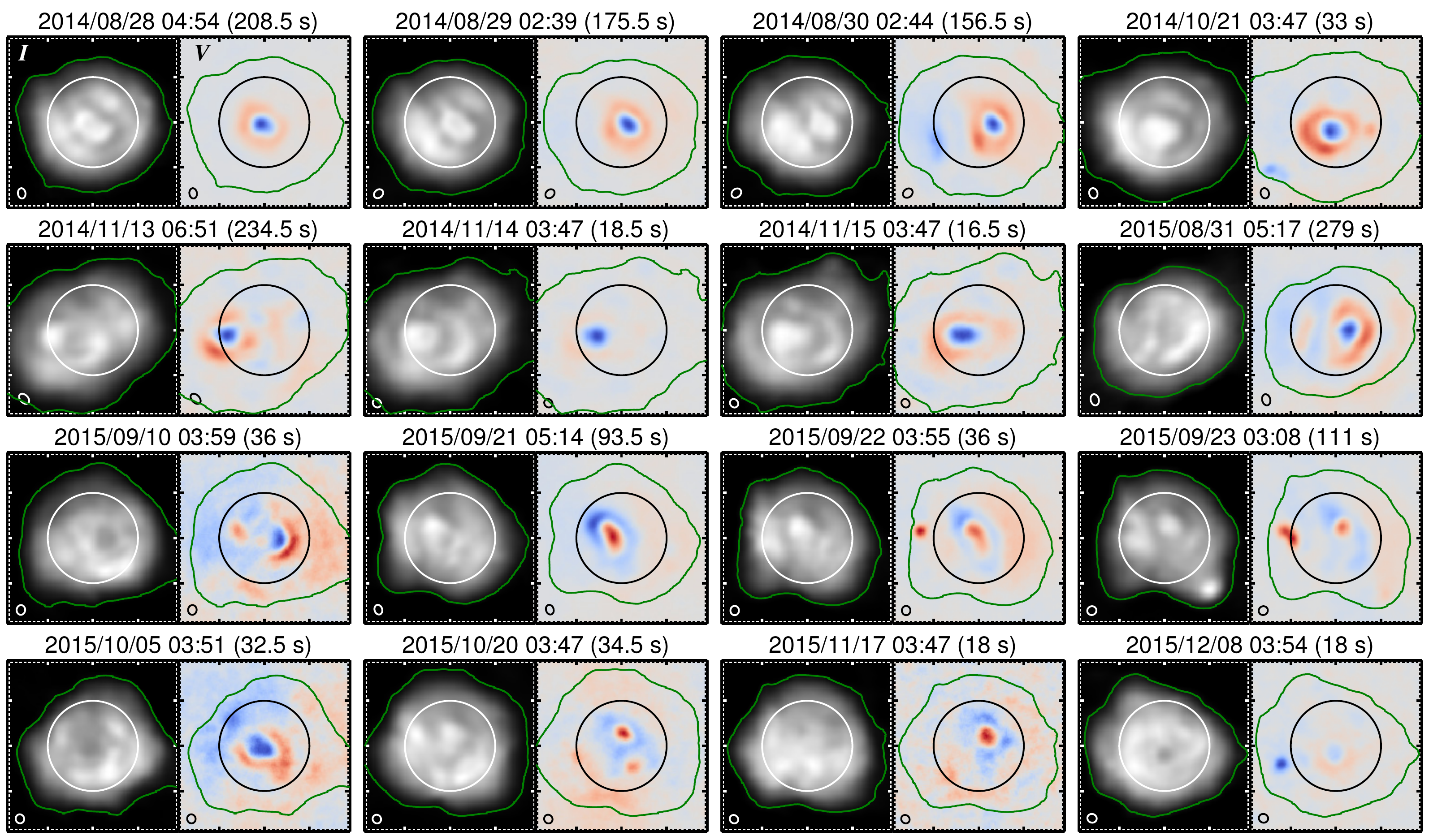}
    \caption{\footnotesize{
	Mosaic of coronal hole observations at 161 MHz, with Stokes $I$ on the left and Stokes $V$ on the right. 
	The green contours show the 5-$\sigma$ levels for Stokes $I$, the solid circles denote the optical 
	limb, and ellipses in the lower-left corners reflect the synthesized beam sizes. 
	The titles correspond to UTC times followed by the effective integration times in parentheses.
	Linear color scales are as in Figure~\ref{fig:ch}, with red, blue, and gray 
	corresponding to positive, negative, and zero Stokes $V$ intensity, respectively.  
	\redit{Images for all 12 frequency channels can been seen in an animated version of 
	this figure available in the \href{https://link.springer.com/article/10.1007\%2Fs11207-019-1502-y}{online material}. 
	Note that some dates do not have observations at all frequency channels because of calibration failures, and 
	two observation periods exhibit bright nonthermal sources at higher frequencies that prevent the coronal 
	hole structure from being visible due to limited dynamic range.  }
       }}
    \label{fig:ch_mosaic}
\end{figure*}

\redit{We are also confident that the Stokes $I$ into $V$ leakage subtraction method has not introduced this feature.  
As described in Section~\ref{leakage}, the leakage fraction ($L$) is assumed to be constant across the relatively small 
spatial scale of the Sun based on results from widefield astronomical studies \citep{Lenc17}.
Because $L$ is constant, varying it changes the fractional polarization level without changing the qualitative structure.
Figure~\ref{fig:lsub1} illustrates how the subtraction algorithm works for one of the same coronal hole observations shown in 
Figure~\ref{fig:ch}, and the animated version of Figure~\ref{fig:lsub1} (available \href{https://link.springer.com/article/10.1007\%2Fs11207-019-1502-y}{online}) shows how the ``corrected" images look as 
a function of $L$.
The animation shows that the polarization reversal bullseye pattern remains for all values of $L$ until reaching the extremes, 
where 80\,--\,100\% of pixels across the Sun are too highly polarized of the same sense to be believable. 
Further, varying $L$ by just 1\,--\,2\% on either side of the value obtained from the correction algorithm quickly pushes 
into this extreme case. 
And even at the extremes, the qualitative ring pattern remains as a sharp change in polarization fraction instead of a reversal.}


\begin{figure*}\graphicspath{{chapter4/}}
    \centering
    \includegraphics[width=1.0\textwidth]{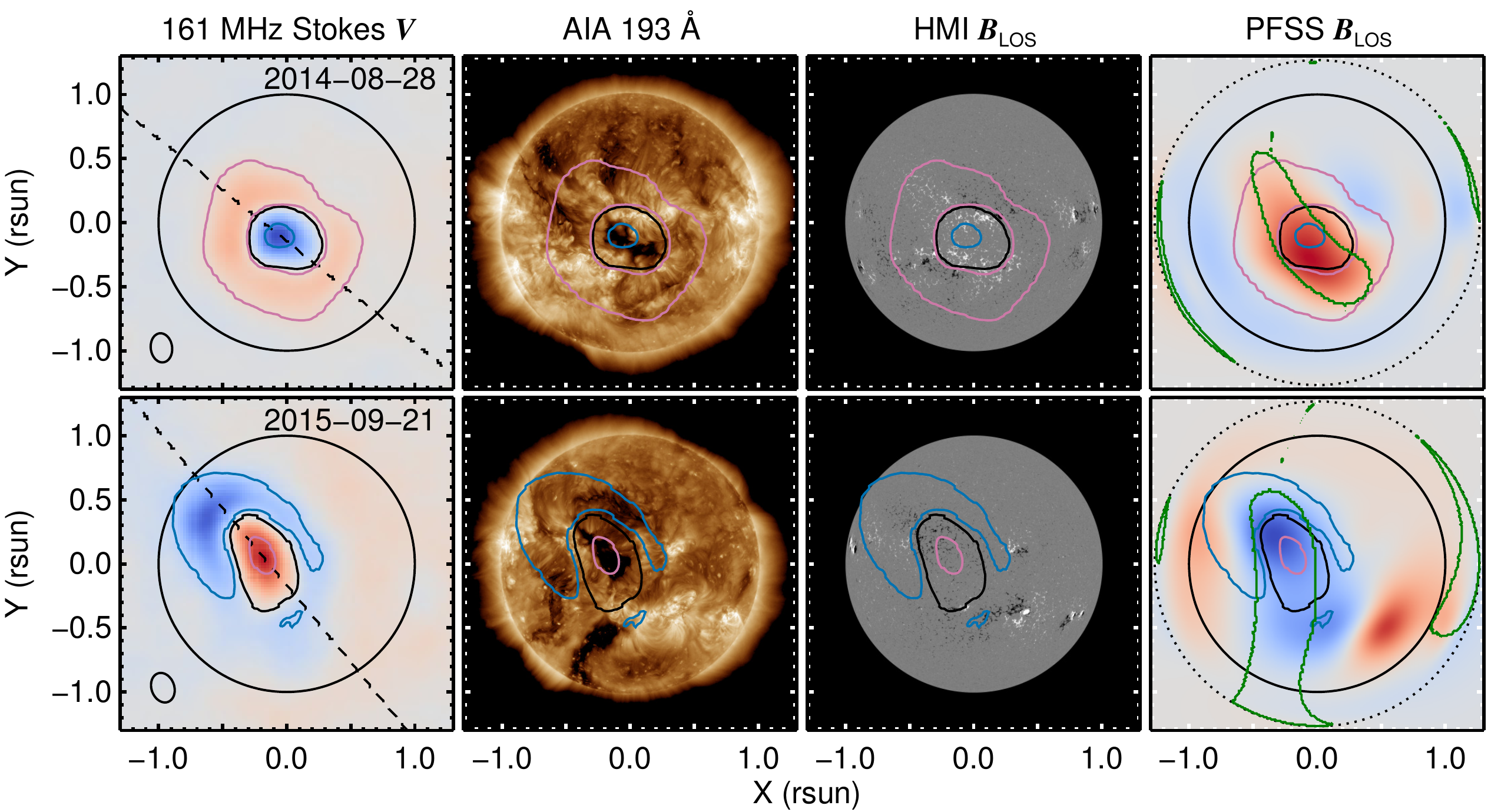}
    \caption{\footnotesize{
	The first column shows Stokes $V$ at 161 MHz for the two coronal holes from Figure~\ref{fig:ch}. 
	The solid black circle represents the optical disk, the black contour is where $V$ = 0 in 
	that region, the blue contour is at 75\% of the central components' maxima, and the pink 
	contour is at 25\% of the ring components' maxima. 
	The dashed lines denote the slits used in Figure~\ref{fig:ch_slices}.  
	The second and third columns plot AIA 193 \AA{} and HMI line-of-sight [\redit{$B_{\rm LOS}$}] magnetograms 
	with the $V$ contours from the first column. 
	The last column shows the \redit{$B_{\rm LOS}$} component of the PFSS model 
	at a height roughly corresponding to that of the radio limb (\rsolar{1.27}). 
	Red, blue, and gray colors represent positive, negative, and zero \redit{$B_{\rm LOS}$}, respectively, 
	and green contours represent open field regions in the model at that height. 
       }}
    \label{fig:ch_overlay}
\end{figure*}

\redit{Next}, this polarization ring structure is not rare and seems to be characteristic 
of low-latitude coronal holes across our dataset. 
We have 28 separate observations of 13 different 
coronal holes in 2014 and 2015 that exhibit this effect. 
A mosaic of examples is shown in Figure~\ref{fig:ch_mosaic} at 161 MHz, the center 
of our \edit{observing band}\redit{, and the other channels can be seen in the corresponding animation.}
Several coronal holes in the mosaic are shown on consecutive days, and they move 
with the solar rotation as expected. 
This structure is not observed in association with other solar features, despite 
noise storms often having similar appearances in total intensity at the 
lowest frequencies. 
\edit{Therefore, from both a data reduction perspective and with respect to its association with solar features, 
the bullseye feature does not appear to be consistent with an instrumental effect.}
Moreover, while this feature is a surprise \edit{to us}, we do find various points of 
consistency between the observations and our expectations that will be discussed later. 

Figure~\ref{fig:ch_overlay} overlays contours of the 161 MHz polarimetric signal from the two 
coronal holes shown in Figure~\ref{fig:ch} onto 193 \AA{} images from the 
AIA and LOS magnetograms from the \textit{Helioseismic and Magnetic Imager} (HMI; \citealp{Scherrer12}) 
onboard the SDO.  
These two coronal holes were chosen for this exercise because they have opposite polarization signatures, 
which is consistent with them having opposite magnetic field 
configurations at the photosphere, and because we have two observations of each on the same day, 
which we will use to estimate the uncertainty in $V/I$. 
The last column of Figure~\ref{fig:ch_overlay} overlays the polarization contours onto 
the LOS component [\redit{$B_{\rm LOS}$}] of a potential field source surface (PFSS; \citealp{Schrijver03}) model at a 
height roughly corresponding to that of the radio limb. 
The models were obtained from the PFSS module in SolarSoft IDL and manipulated using the \textsf{FORWARD} codes. 
A height of \rsolar{1.27} is used, which corresponds to the height of the plasma frequency layer at 161 MHz 
in a 3$\times$ \citet{Newkirk61} density model. 


\begin{figure*}\graphicspath{{chapter4/}}
    \centering
    \includegraphics[width=1.0\textwidth]{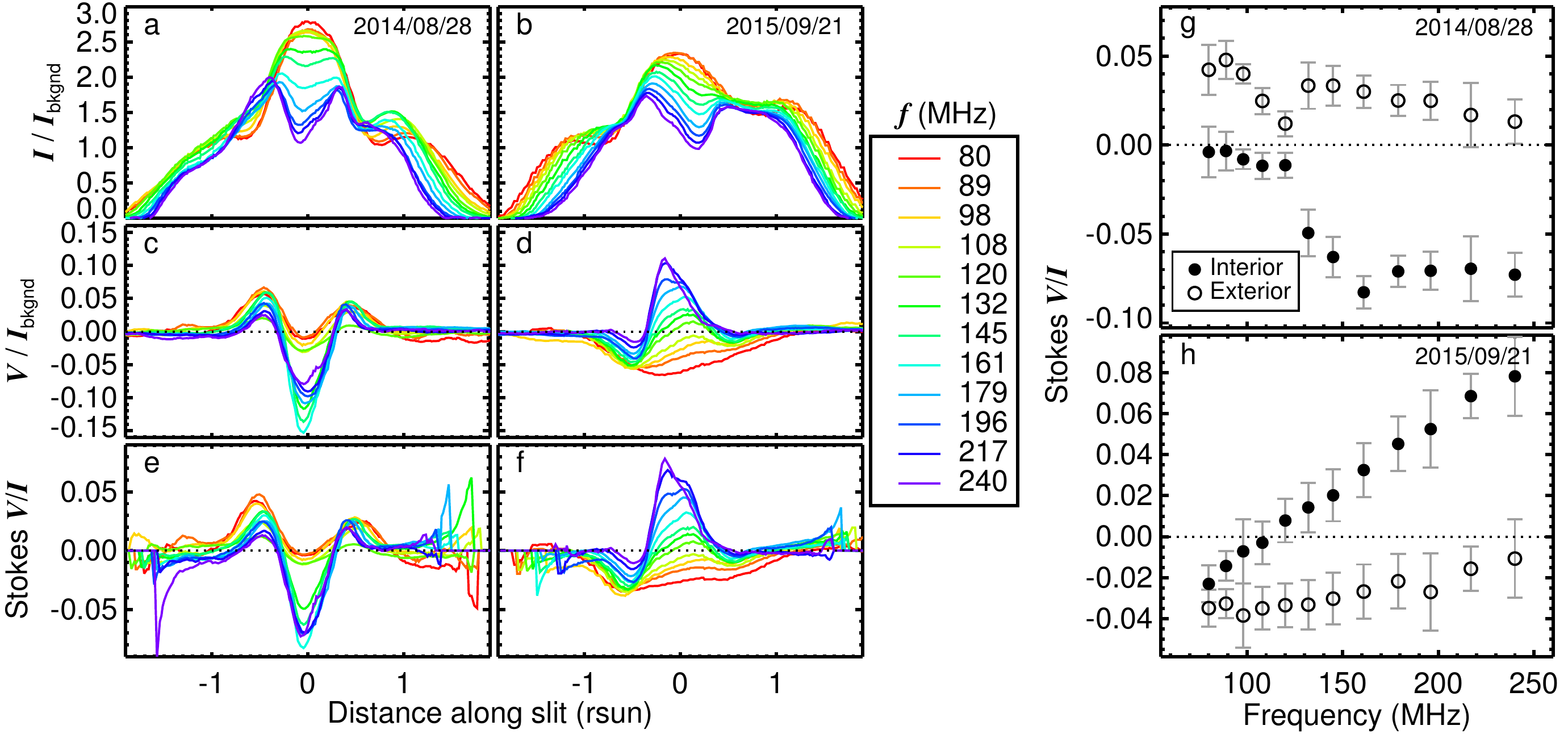}
    \caption{\footnotesize{
	\textbf{(a\,--\,b):} Stokes $I$ intensity for each frequency channel along the slits shown in Figure~\ref{fig:ch_overlay} 
	through two coronal holes on \edit{28 August 2014} (left) and \edit{21 September 2015} (right). 
	\textbf{(c\,--\,d):} Stokes $V$ intensities along the same slits. 
	Panels a\,--\,d are shown in normalized units, where the normalization factor corresponds to the 
	median Stokes $I$ intensity of pixels detected above 5 $\sigma$ (\textit{\textit{i.e.}} the background level). 
	\textbf{(e\,--\,f):} The corresponding Stokes $V/I$ fractional polarization levels.  
	\textbf{(g\,--\,h):} Peak Stokes $V/I$ for the coronal holes' interior (\edit{filled circles}) and exterior ring 
	(open circles) components as a function of frequency. 
       }}
    \label{fig:ch_slices}
\end{figure*}

Crucially, the central polarized components do not match the orientation of \redit{$B_{\rm LOS}$}, as 
would be expected from the thermal bremsstrahlung process that is assumed 
to be the dominant, if not sole, emission mechanism in coronal holes. 
This effect is further characterized in Figure~\ref{fig:ch_slices}, which plots cuts through 
the same two coronal holes in Stokes $I$, $V$, and $V$/$I$. 
The $I$ and $V$ cuts are normalized by the background intensity in Stokes $I$, which we 
define as the median pixel intensity for pixels detected above 5-$\sigma$. 
The gradual transition with decreasing frequency of both coronal holes from being dark to 
bright relative to the background is nicely illustrated by the Stokes $I$ curves for both examples, 
as is the oppositely-oriented ring structure in Stokes $V$. 
Note that the large spikes in $V$/$I$ near the ends of the slits are in locations where both the 
total intensity and polarized signals approach the noise level, making the fractional polarizations unreliable. 

While the overall pattern is similar for both coronal holes, the behavior of the central component is 
somewhat different in each case. 
The rightmost panels (g and h) of Figure~\ref{fig:ch_slices} show the peak $V$/$I$ for the central 
and ring components. 
The central component is most highly polarized ($\approx$ 5\,--\,8\%) at our highest frequency (240 MHz) 
and gradually decreases in polarization fraction with decreasing frequency. 
For the \edit{28 August 2014} example, the central component falls to nearly 0\% polarization at 80 MHz but remains 
of the same sign at all frequencies, whereas the \edit{21 September 2015} examples crosses 0\% around 108 MHz and gradually 
approaches the same polarization level as the ring component. 
The latter scenario is somewhat more common in our experience. 
That is, by 80 MHz, the entire source is typically polarized in the same sense expected by bremsstrahlung emission and at a similar level, often with 
a small dip in polarization fraction at the center where the source is oppositely-polarized at higher frequencies.

The uncertainties in panels g and h of Figure~\ref{fig:ch_slices} are the combination of measurement noise 
and \redit{three} effects related to the leakage subtraction algorithm described in Section~\ref{leakage}. 
The first is the range of values found by varying the 
minimization parameter $r_{\rm c,thresh}$ in Equation~\ref{eq:alg1} between 0.3 and 0.8\%, along with varying the pixels 
included in the operation between those detected above 5 $\sigma$ and those detected above 15 $\sigma$. 
The second \redit{is the difference in} polarization fraction \redit{at} the same \redit{locations} in two observations separated by 2\,--\,3 
hours\redit{, and the third is the potential for unnacounted for leakage of Stokes $V$ into $I$ at up to the same level as that 
measured for $I$ into $V$.}
\redit{These combined uncertainties in $V$/$I$ average $\pm$ 1.2\% and are as large as $\pm$ 2.9\%}. 
In both cases, the sign of the leakage fraction flips for each frequency channel between the two observations as the \edit{Sun} moves 
to different locations in the primary beam with respect to the phase center. 
(The MWA does not continuously track an object and instead has a set of discrete pointings that may be changed 
after every $\approx$ 5-min observing period.) 

Despite the sign change in the leakage artefact between observations separated by 2\,--\,3 hours, after implementing the subtraction algorithm, 
the polarization fraction remains consistent to within 1\% for a given location and frequency channel. 
However, a sharp discontinuity remains between the 120 and 132 MHz channels in the \edit{28 August 2014} observation. 
The leakage is more severe in this observation as compared to the \edit{21 September 2015} data, 
and the sign of the leakage also changes between those two channels. 
The discontinuity \redit{in the polarization fraction trend shown in Figure~\ref{fig:ch_slices}g} 
is therefore likely to be a calibration artefact that cannot be removed by uniformly implementing our correction algorithm. 
\redit{This suggests an additional source of uncertainty in the polarization fraction that is not accounted for by the methods described 
in the previous paragraph. However,}
note that \redit{leakage} affects the polarization level uniformly across the image and cannot warp the qualitative structure observed because the 
leakage does not vary on the small \edit{angular} scale of the Sun, given what we know from widefield astrophysical observations. 

The puzzle with respect to this feature is again the fact that the polarization of the central component does not match the sign expected 
from thermal bremsstrahlung emission. We will discuss possible interpretations for this in Section~\ref{discussion}. 


\section{The Large-Scale Quiescent Structure}
\label{qs}


\begin{figure*}\graphicspath{{chapter4/}}
    \centering
    \includegraphics[width=1.0\textwidth]{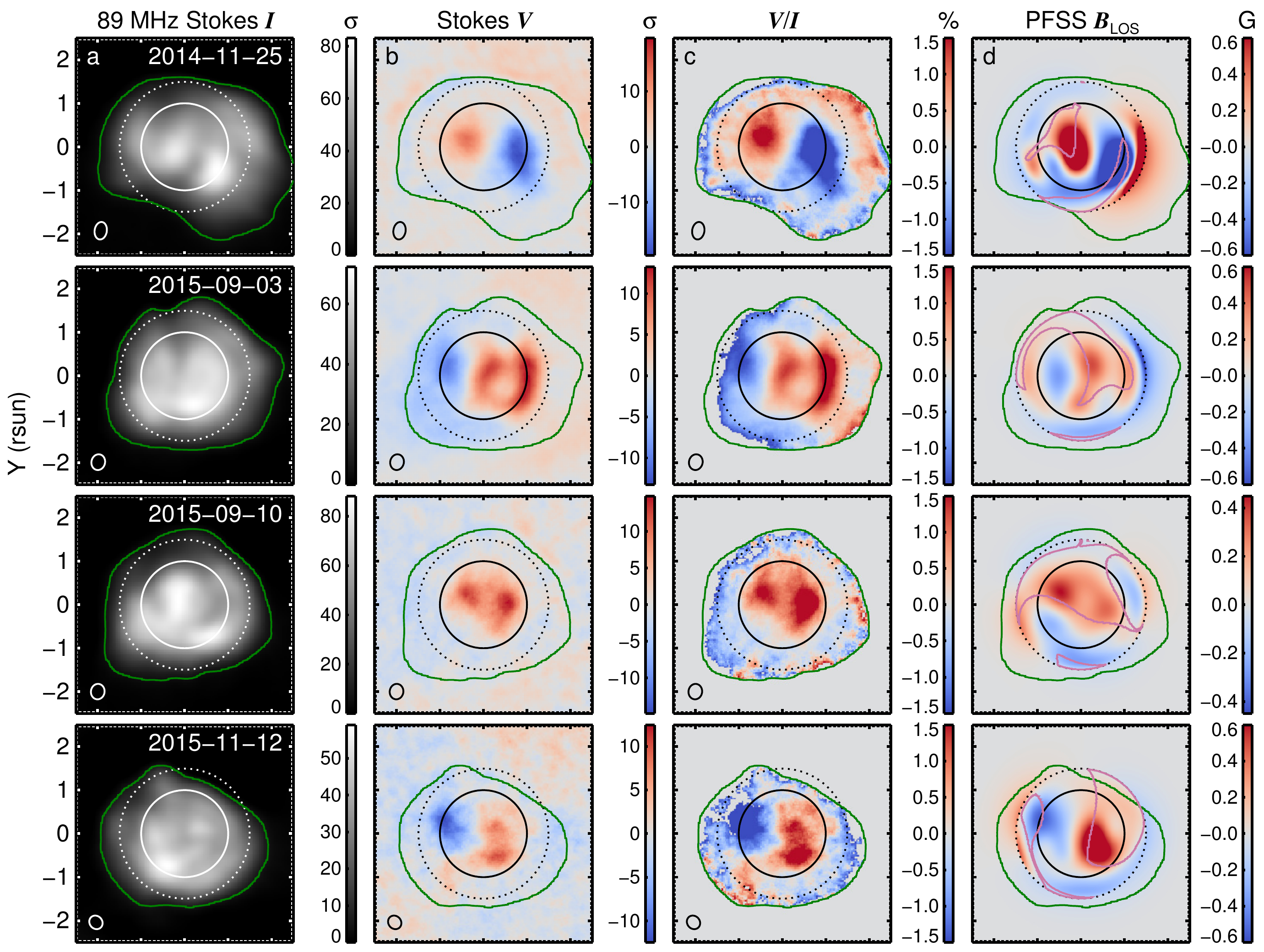}
    \caption{\footnotesize{
	Columns \textbf{a\,--\,c} show Stokes $I$, $V$, and $V/I$ images at 89 MHz on four different days for which 
	a single compact region does not dominate the total intensity. 
	The solid circles denote the optical limb, and the green contours show the 5 $\sigma$ level for Stokes $I$. 
	Ellipses in the lower-left corners reflect the synthesized beams. 
	Column \textbf{d} shows the line-of-sight magnetic field strength [\redit{$B_{\rm LOS}$}] in the PFSS model at a height of \rsolar{1.49}, which 
	roughly corresponds to the height of the radio limb at 89 MHz and is indicated by the dotted circles in each panel. 
	The pink contours indicate open field regions in the model.  
	Color bar units are in signal-to-noise [$\sigma$] for Stokes $I$ and $V$, percent for $V/I$, and Gauss [G] for the field model. 
	\redit{An animated version of this figure that shows all 12 frequency channels is available in the \href{https://link.springer.com/article/10.1007\%2Fs11207-019-1502-y}{online material}.}
       }}
    \label{fig:large_low}
\end{figure*}

As mentioned in Section~\ref{introduction}, the thermal bremsstrahlung process that dominates quiescent coronal emission 
at low frequencies produces a slight circular polarization signature in a magnetized plasma that depends primarily on the 
line-of-sight magnetic field strength. 
Absent of other emission mechanisms, a positive LOS field should produce a positive Stokes $V$ signature of up to a few 
percent that depends on the field strength. 

Figure~\ref{fig:large_low} shows Stokes $I$, $V$, and $V/I$ images on four different days for which the polarimetric signature 
is not dominated by a bright noise storm source or disk-center coronal hole. 
These days were also selected to have a mixture of positive and negative Stokes $V$ 
regions so that we can compare the structure to that of the LOS field. 
The fourth column of Figure~\ref{fig:large_low} shows the LOS magnetic field direction 
and strength in the corresponding PFSS model. 
The dotted circle indicates the height at which the model LOS field is shown, which is chosen to be roughly that 
of the radio limb, and the plane-of-sky field is shown beyond the dotted circle. 
This height is \rsolar{1.49} at 89 MHz and corresponds to the height of the plasma 
frequency layer in a 3-fold \citet{Newkirk61} density model. 
Pink contours indicate the open-field regions in the model at the same height, 
which were determined using the ``topology" keyword in the \textsf{FORWARD} code. 
\edit{It is immediately apparent that the Stokes $I$ and $V$ maps 
show very different morphologies in general. 
Regions with the highest polarized intensities are often not straightforwardly correlated with 
those of highest total intensity.}
It is \edit{also} interesting to note that larger polarized intensities are often associated with open 
field regions, which is also consistent with the coronal hole observations from the previous section.  
While we have not investigated this effect systematically, it may be due to there being lower 
densities and lower density contrasts between adjacent regions in open field regions, which 
then reduces the depolarizing effect of scattering by density inhomogeneities. 


\begin{figure*}\graphicspath{{chapter4/}}
    \centering
    \includegraphics[width=1.0\textwidth]{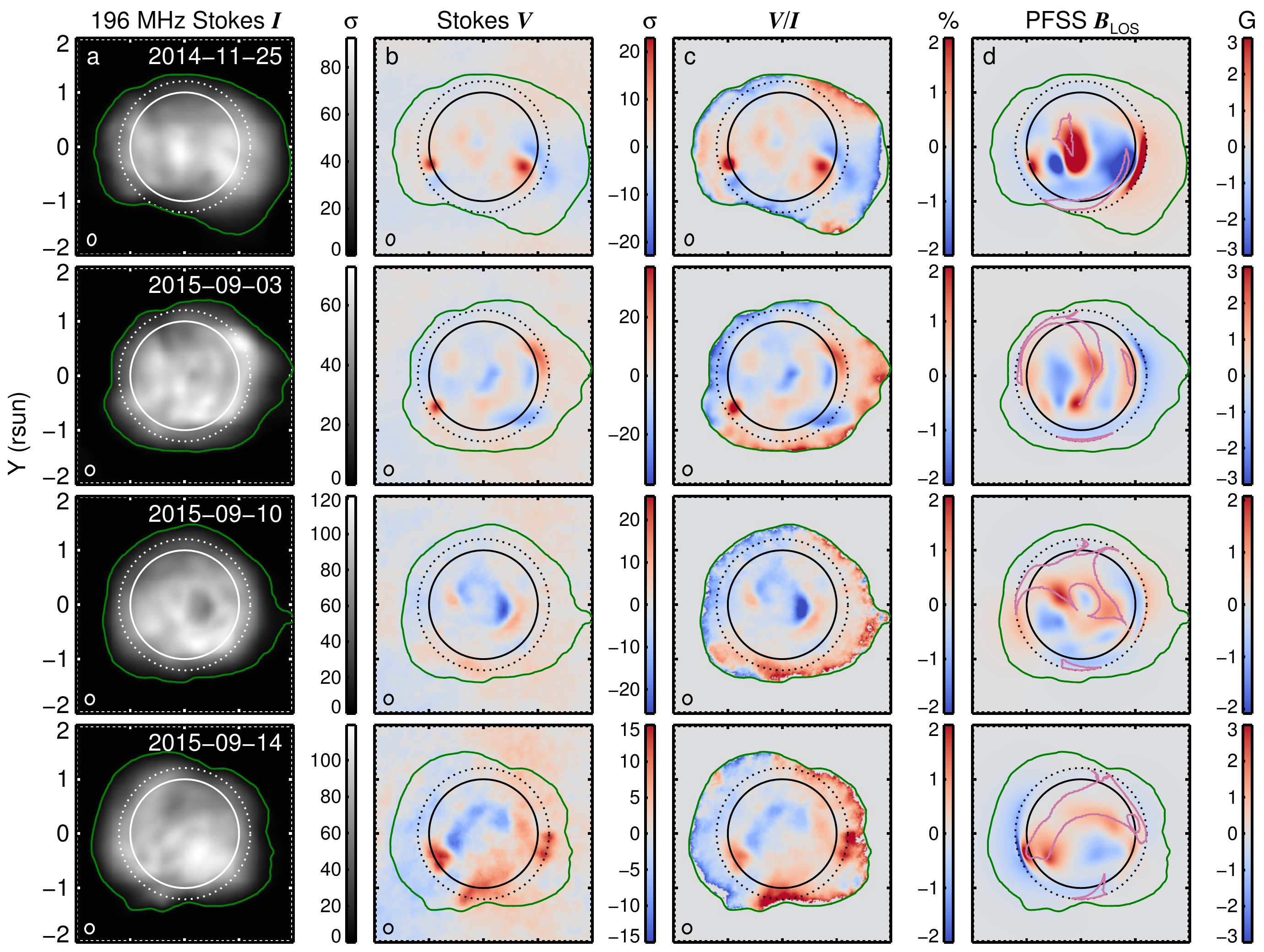}
    \caption{\footnotesize{
	Columns \textbf{a\,--\,c} show Stokes $I$, $V$, and $V/I$ images at 196 MHz on four different days for which 
	a single compact region does not dominate the total intensity. 
	The days are the same as Figure~\ref{fig:large_low} except for the last row. 
	The solid circles denote the optical limb, and the green contours show the 5 $\sigma$ level for Stokes $I$. 
	Ellipses in the lower-left corners reflect the synthesized beams. 
	Column \textbf{d} shows the line-of-sight magnetic field strength [\redit{$B_{\rm LOS}$}] in the PFSS model at a height of \rsolar{1.21}, which 
	roughly corresponds to the height of the radio limb at \edit{196} MHz and is indicated by the dotted circles in each panel. 
	The pink contours indicate open field regions in the model.  
	Color bar units are in signal-to-noise [$\sigma$] for Stokes $I$ and $V$, percent for $V/I$, and Gauss [G] for the field model. 
       }}
    \label{fig:large_high}
\end{figure*}

Figure~\ref{fig:large_low} demonstrates that the Stokes $V$ structure at our lowest frequencies 
is generally well-matched to the LOS field, at least near disk center\edit{, and that the 
sign of Stokes $V$ is broadly consistent with that expected from thermal bremsstrahlung emission given the LOS field orientation}. 
The boundaries between opposite polarization signs are roughly aligned with polarity inversion 
lines in the model. 
The agreement tends to diminish with distance from disk center, which is likely due to two effects. 
First, low-frequency radio emission is heavily influenced by propagation effects, namely refraction, scattering, and mode coupling, 
that can influence the polarization sign and fraction, and these effects 
become more pronounced near the limb \citep{Shibasaki11}. 
Second, although the polarization fraction is expected to be highest off the limb \citep{Sastry09}, the 
intensity is much lower there. 
The third column of Figure~\ref{fig:large_low} shows that we often do find relatively high polarization 
fractions toward the radio limb, but these pixels are very close to the noise level in Stokes $V$ and 
we do not regard them as reliable. 
Missing pixels in the Stokes $V/I$ images inside the green 5 $\sigma$ Stokes $I$ contour 
are censored because they have polarization fractions greater than 1.5\% but Stokes $V$ 
signals below 5 $\sigma$. 

Figure~\ref{fig:large_high} shows the same parameters as Figure~\ref{fig:large_low} but at 196 MHz. 
The same observation periods are used in the first three rows, but the fourth row is different because 
the \edit{12 November 2015} observation becomes dominated by a noise storm source at higher frequencies. 
\redit{First, we note that the polarized quiet Sun emission is more localized at 196 MHz compared to 89 MHz, 
which is likely the combination of at least three effects. 
First, the spatial resolution is simply lower at lower frequencies. 
Second, lower frequency emission is produced at larger heights where the corona is somewhat less 
finely structured, with smaller contrasts in magnetic field and density between adjacent regions. 
Third, lower-frequency emission is more strongly scattered, 
which leads to increased angular broadening with decreasing frequency.}

\redit{Figure~\ref{fig:large_high} also exhibits} much less straightforward agreement between the LOS field orientation and 
the polarization structure. 
This is a gradual transition with increasing frequency that we cannot attribute to instrumental effects 
because we never observe such an inversion for the noise storm sources that are detected across our 
entire \edit{observing band} from 80 to 240 MHz. 
Some of the differences between the Stokes $V$ sign and that expected from the LOS field orientation 
can likely be attributed to the same unknown effect present in the coronal hole observations from 
Section~\ref{ch}.
For instance in the \edit{10 September 2015} and \edit{14 September 2015} observations, there are coronal holes near the 
west limb and north pole, respectively, that exhibit this effect. 
However, those regions aside, there is still not the same alignment between opposite polarization 
signs and LOS field polarity inversion lines that we see at low frequencies. 

One possible explanation for the discrepancies at higher frequencies is simply the accuracy and resolution 
of the potential field model. 
Higher frequencies correspond to lower heights, and the true coronal magnetic 
field becomes increasingly non-potential closer to the surface, with larger contrasts between adjacent regions. 
This is also true for the density and temperature, which also affect the polarization signal to some extent 
and will be important for future forward modeling efforts. 
However, our impression is that physical effects are also likely to be important, 
and we discuss some possibilities in the next section. 

 
\section{Discussion} %
\label{discussion} %

These data offer an opportunity to probe the coronal magnetic field at heights and 
scales that are not easily accessed with other instruments. 
The intent of this article is to introduce the data and survey the range of features observed 
in this new regime. 
In a forthcoming study, we will directly compare the magnetic field strength and structure 
implied by our observations to model predictions of the thermal bremsstrahlung emission 
implied by different global models.  
Preliminary results suggest that we can successfully generate synthetic Stokes $V/I$ images 
that reproduce the low-frequency polarization structure near disk center, but we have 
yet to explicitly compare the field strengths implied by the observed polarization fractions.

While these preliminary results are encouraging, there are a number of questions 
that will require deeper investigation. 
The most perplexing of these is perhaps the bullseye structure in Stokes $V$ described in 
Section~\ref{ch}, which is found for many low-latitude coronal holes. 
At our higher frequencies ($\leq$ 240 MHz), low-latitude coronal holes often exhibit a central 
circularly-polarized component surrounded by a full or partial ring of the opposite sense, 
and the central component is of the opposite sign than would be expected from thermal 
bremsstrahlung, the presumed emission mechanism. 
With decreasing frequency, the central component diminishes and \edit{the} ring expands. 
By 80 MHz, there may be some trace of the central component remaining or the region may 
be entirely of the sign expected by bremsstrahlung emission. 

We have two suggestions for what may be responsible for this effect. 
The first has to do with differential refraction of the $x$- and $o$-modes. 
The two modes have slightly different group velocities within the plasma and therefore 
slightly different refractive indices, as is often observed in ionospheric propagation experiments at Earth \citep{Melrose86}.
It may be possible to separate the two modes sufficiently via refraction so as to produce a 
polarization sign that is the opposite of that expected by bremsstrahlung emission. 
\edit{For example, the refractive index of the $x$-mode is further below unity than the $o$-mode. 
Emission generated by denser plasma at the edge of a coronal hole may refract at the coronal hole boundary, with 
the $x$-mode refracting more strongly in the radial direction, producing an enhancement of $x$-mode around 
the perimeter and $o$-mode near the center.}
Propagation effects do seem to be particularly important in coronal holes, as they are the 
preferred explanation for why some coronal holes are significantly brighter than their 
surroundings at low frequencies \citep{Alissandrakis94,Rahman19}. 
However, this later question has not been resolved, and it is not obvious to us how 
this polarization signal would be produced. 
Ray tracing simulations, such as those done by \citet{Benkevitch12} or \citet{Vocks18}, 
are likely needed to investigate this further. 

A second possibility that we regard as less likely is that the central polarized component 
is produced by weak plasma emission, which generally produces the opposite sign in circular 
polarization compared bremsstrahlung for the same line-of-sight field. 
Because plasma emission may be up to 100\% circularly polarized, it is possible 
for the total intensity to be dominated by thermal bremsstrahlung emission while the polarized component 
is dominated by plasma emission. 
This is what we expect to be happening in the low-intensity and weakly-polarized active region sources described 
in Section~\ref{ar}. 
For coronal holes, the driver of this plasma emission might be the transition region network jets 
discovered by the Interface Region Imaging Spectrograph \citep{Tian14}. 
These jets are continuously generated throughout the transition region, but they are 
particularly common and intense inside coronal holes \citep{Narang16}. 
They strongly resemble the coronal X-ray jets associated with Type III bursts, 
and some attribute the network jets to small-scale reconnection events \citep{Tian14,Kayshap18}. 
However, the network jets also resemble chromospheric Type II spicules 
and may instead be driven by shocks \citep{Cranmer15} or heating fronts \citep{Pontieu17}. 
If these jets are associated with reconnection or shocks, then nonthermal electrons capable of producing 
plasma emission may be expected at least to some extent by analogy with solar radio bursts. 
We may see this emission only in coronal holes because the densities there are low 
enough that the plasma levels associated with our highest frequencies are in, 
or very close to, the transition region. 
This idea could be explored observationally through examining variability in the 
polarization signal and comparing that to the typical jet timescales. 

A mixture of plasma emission and thermal emission may also help to explain 
why our higher-frequency Stokes $V$ maps are not well-correlated with the 
LOS field structure. 
One of the MWA's main contributions thus far has been to demonstrate 
the prevalence of very weak nonthermal emissions \citep{Suresh17}. 
\citet{Sharma18} report that up to 45\% of the total intensity outside of nominal burst periods 
may be nonthermal during moderately active periods, and preliminary imaging analyses  
suggest that nonthermal components are present to varying degrees in every environment. 
These nonthermal emissions are attributed to plasma emission, which again is generally 
much more highly polarized and of the opposite sign compared to bremsstrahlung. 
A relatively minor total intensity contribution from plasma emission would therefore have a 
much greater impact on the circularly-polarized intensity and may be capable of reversing 
the observed sense from that expected from thermal emission. 
The polarization fraction could also potentially be used to disentangle the contributions of 
both mechanisms. 

Mode coupling effects associated with quasi-transverse (QT) regions are also likely to be 
important at least in some regions and may contribute to differences between the \edit{observed} polarization 
sense and that straightforwardly expected from a particular emission mechanism\edit{, along 
with reductions in the polarization fraction expected from plasma emission sources.} 
QT regions refer to when the magnetic field orientation is nearly perpendicular to that of 
the emission region, such that there is no magnetic field component along the ray path. 
Passing through such a region may cause the circular polarization 
sign to reverse if the emission frequency is below a certain threshold that depends on the 
plasma properties \citep{Cohen60,Zheleznyakov64,Melrose94}.
This concept is very important to the interpretation of polarization reversals and associated 
magnetic field diagnostics in high-frequency microwave observations (\textit{e.g.}  \citealp{Ryabov99,Altyntsev17,Shain17,Sharykin18}).
At lower frequencies, QT regions are also invoked to explain the polarization properties of noise storms \citep{Suzuki80,White92}, 
Types U and N bursts \citep{Suzuki80,Kong16}, and zebra patterns in Type IV bursts \citep{Kaneda15}.
A natural place to start investigating the importance of QT regions in our observations would be 
by comparing the polarization sense of the active region noise storm sources from Section~\ref{ar} to 
that expected \edit{from} the magnetic field orientation assuming $o$-mode polarization from plasma emission. 

Lastly, it would be useful to explore improving our calibration approach by imposing 
constraints specific to solar observing. 
We have introduced a strategy, adapted from the astrophysical literature, 
to mitigate an artefact referred to as ``leakage," whereby the polarimetric signal is 
contaminated by some fraction of the total intensity signal. 
This is described in Section~\ref{leakage}. 
While we have demonstrated that our approach is reasonably effective, 
it is clearly not perfect given the discrepancies occasionally observed between adjacent 
frequency channels, as illustrated by Figure~\ref{fig:ch_slices}g. 
A better solution may be available by imposing constraints based on the expectation that  
we can generally assume that the linear polarizations (Stokes $Q$ and $U$) are zero 
because Faraday rotation destroys the linear polarization signal over most observing bandwidths. 
Assuming $Q$ and $U$ are zero implies that the XX and YY instrumental polarizations are 
equal, and this constraint may be applied for each antenna in visibility space, 
allowing for a direction-dependent polarization calibration.
We do not yet know the feasibility of this approach, and new software tools would 
need to be developed to implement it. 

 
\section{Conclusion} %
\label{conclusion} %

We have presented the first spectropolarimetric imaging of the Sun using the MWA. 
These are the first imaging observations of the low-frequency corona that are capable of 
measuring the weak polarization signals outside of intense burst periods. 
We reviewed the two dominant emission mechanisms, thermal bremsstrahlung and plasma 
emission, and how their expected polarization signatures relate to our observations. 
Our data were taken from over 100 observing runs near solar maximum, and we surveyed the 
range of features detected \redit{in quiescent periods}. 
These observations can be used to diagnose the coronal magnetic field at heights and scales 
for which the available data constraints are limited, and this will be the focus of future work. 
Our contributions are as follows: 

\begin{itemize}

\item{We introduced an algorithm to mitigate an instrumental artefact known as ``leakage," whereby some fraction of 
the total intensity [Stokes $I$] signal contaminates the circular polarization [Stokes $V$] images (Section~\ref{leakage}). 
Leakage occurs due to \edit{differences between the actual instrumental response and} 
the primary beam model used to convert images of the instrumental polarizations into 
Stokes \edit{images}. These errors may be due to imperfections in the beam model itself or to other effects that change the instrument's  
effective response, such as individual antenna failures or, importantly, the practice of applying calibration solutions from a calibrator 
source at one pointing to the target source at another pointing. 
We adapted an approach used for astrophysical MWA studies, which show that the leakage varies negligibly over the 
spatial scale of the Sun. 
Given that most of the pixels in our images should be very weakly polarized based on our 
expectations \edit{for} thermal bremsstrahlung emission, we determined the leakage fractions with an algorithm that minimizes the number 
of pixels with polarization fractions [$|V/I|$] greater than 0.5\%.}

\item{We developed and employed a source finding algorithm that detected around 700 compact sources in the Stokes $V$ images \edit{(Section~\ref{ar})}. 
Only sources detected in a least three frequency channels were analyzed further, corresponding to 112 distinct sources found at 
multiple frequencies. The intensities of these sites ranged from slightly below the background level to 60$\times$ greater than 
the background. Their polarization fractions ranged from less than 0.5\% to nearly 100\%. 
\edit{At least one of these sources was present on 78\% of our observing days, and we found a positive 
correlation between the total intensity over the background and the polarization fraction ($r$ = 0.64).} 
The high-intensity sources with large polarization fractions are \redit{noise storm continua sources} produced by 
plasma emission and associated with active regions. 
As there is no obvious separation of these sources into distinct populations, we suggest that they represent a continuum 
of plasma emission sources down to intensities and polarization fractions that were not previously observable in imaging observations. 
Although the plasma emission theory predicts 100\% circular polarization for fundamental emission, very low polarizations can be explained 
in this context through three effects. First, \edit{the weaker sources} may still be dominated by thermal bremsstrahlung emission with a 
minor contribution from plasma emission that then dominates the polarized component. 
Second, the plasma emission sites may often be considerably smaller than the beam size, leading to beam dilution that smears the 
polarized signal across a larger area.
Third, scattering by density irregularities may also reduce the polarization fraction\edit{, even for very intense sources.}}

\item{We reported the discovery of a ``bullseye" polarization structure often associated with low-latitude coronal holes in which one polarization 
sense is surrounded by a full or partial ring of the opposite sense \edit{(Section~\ref{ch})}. The polarization of the central component is of the opposite sign 
from that expected from thermal bremsstrahlung, the presumed emission mechanism. Moving from our highest frequency (240 MHz) 
to our lowest (80 MHz), the central component diminishes and the ring expands. Some coronal holes continue to exhibit the ring 
structure at 80 MHz, while others are unipolar in Stokes $V$ with a sense that matches that expected \edit{for} bremsstrahlung emission.
This effect was observed in 28 separate observations of 13 different coronal holes. 
We validated the Stokes $I$ structure with an independent data reduction, and we noted that similar total intensity structures associated 
with noise storms never exhibit this effect. 
We speculated that the structure may be the result of propagation effects, namely refraction, that separate the $x$- and $o$-modes, 
but ray tracing simulations are needed to test this. 
Alternatively, we suggested that the polarization signature may be produced by weak plasma emission produced by the recently-discovered 
transition region network jets that are particularly prevalent inside coronal holes. 
}

\item{We showed that at our lowest frequencies, the large-scale Stokes $V$ structure is reasonably well-correlated with the 
line-of-sight magnetic field structure \edit{obtained from} a global potential field source surface model \edit{(Section~\ref{qs})}. 
The boundaries between opposite polarization signs are generally aligned with polarity inversion lines in the model\edit{, with 
the polarization sign matching the expectation from thermal bremsstrahlung emission given the LOS field orientation.}
The correspondence is best near disk center and diminishes toward the limb, where propagation effects become 
increasingly important and the signal-to-noise decreases. 
At our highest frequencies, there is little straightforward agreement between the LOS field orientation 
and the polarization sign. 
This may be due to the limited accuracy of the potential model, as the coronal field becomes increasingly non-potential 
at lower heights where higher-frequency emission generated. 
However, we suspect that physical effects are also important. 
These may include a mixture of thermal and non-thermal emission in the same region, along with propagation effects such 
as refraction and polarization reversals due to quasi-transverse (QT) regions.
}

\end{itemize}

{\footnotesize{}\textbf{Acknowledgements}: \edit{This work was primarily supported} by the Australian Government through an Endeavour Postgraduate Scholarship.  
\edit{P. McCauley acknowledges the Asian Office of Aerospace Research and Development (AOARD) of the 
United States Air Force Office of Scientific Research (AFOSR) for travel support through the Windows on Science (WOS)
program.}
\redit{I. Cairns and J. Morgan acknowledge support from AFOSR grants FA9550-18-1-0671 and FA9550-18-1-0473, respectively.}
We thank Don Melrose and Sarah Gibson for helpful discussions. 
\redit{We also thank the anonymous referee for a careful reading and constructive comments.}
This scientific work makes use of the Murchison Radio-astronomy Observatory (MRO), operated by 
the Commonwealth Scientific and Industrial Research Organisation (CSIRO). 
We acknowledge the Wajarri Yamatji people as the traditional owners of the Observatory site. 
Support for the operation of the MWA is provided by the Australian Government's  
National Collaborative Research Infrastructure Strategy (NCRIS), 
under a contract to Curtin University administered by Astronomy Australia Limited. 
We acknowledge the Pawsey Supercomputing Centre, which is supported by the 
Western Australian and Australian Governments.
The SDO is a National Aeronautics and Space Administration (NASA) satellite, and 
we acknowledge the AIA and HMI science teams for providing open 
access to data and software. 
This research has also made use of NASA's Astrophysics Data System (ADS) 
and the Virtual Solar Observatory (VSO, \citealt{Hill09}).}

%% file: chapter5/chapter5.tex
\pagestyle{fancy}



\chapter{Conclusions and Future Work} \label{ch5}

\fancyhead[OR]{~}\fancyhead[EL]{\bf Ch. \thechapter~Conclusions and Future Work} 

This thesis has presented novel low-frequency observations of the solar corona, 
and the major results of each chapter are briefly summarized in the next section. 
The subsequent sections elaborate on certain conclusions and outline potential 
next steps for future work on several open questions. 
As noted in Chapter~\ref{ch1}, millions of images were reduced in the course of this work, 
which was the first attempt to process large amounts of MWA solar data with 
supercomputing facilities.    
The science presented here is only a faction of what can be extracted 
from this dataset, which itself is a small fraction of the total MWA solar archive, and 
most of the following ideas for future work can be addressed with the existing images.   
 
\section{Results Summary}

In Chapter~\ref{ch2} \citep{McCauley17}, 
new dynamics of Type III solar radio bursts were identified, characterized, and interpreted as 
resulting from a divergent magnetic field configuration that was evidenced by 
contemporaneous EUV observations. 
The radio burst images were used to develop a more complete picture of the magnetic field topology, 
showing that these data are valuable probes of the field connectivity at heights that are not  
easily accessed by other data. 
A rough flux calibration method was also developed, and the structure of the quiescent corona was 
compared to model predictions. 
Notably, this comparison showed a coronal hole transitioning from being a relatively dark 
to a relatively bright structure 
moving from high to low frequencies in the data, which was not found in the model. 

In Chapter~\ref{ch3} \citep{McCauley18}, Type III bursts were used to estimate the coronal density profile. 
At its heart, this was a repetition of a classic experiment with updated instrumentation 
that confirmed earlier results for the existence of significant apparent density 
enhancements over standard models. 
The cause for these enhancements has been debated for many years, and the arguments for the two 
primary interpretations were reviewed in light of recent results. 
Some authors suggest that observable Type III bursts are produced by electron beams moving along 
overdense structures relative to their surroundings (i.e., the density enhancements are real), 
while others suggest that propagation effects, namely scattering, cause the bursts to be observed 
at larger heights than they were produced at (i.e., the enhancements are not real). 
Here, a novel comparison between the observed extent of the quiescent corona and model predictions was 
used to conclude that the apparent density enhancements can largely but not entirely be explained 
by propagation effects for the events studied in Chapter~\ref{ch3}.

Chapter~\ref{ch4} \citep{McCauley19} 
presented the first spectropolarimetric imaging of the quiescent corona at low frequencies.
This required first developing an algorithm to mitigate calibration errors that cause 
a fraction of the total intensity signal to contaminate the polarization images. 
Over 100 observing runs near solar maximum were used to survey the principal features of 
the low-frequency corona in circular polarization (Stokes $V$). 
Around 700 compact polarized sources were detected with a range of polarization 
fractions from less than 0.5\% to nearly 100\%. 
These were interpreted as a set of plasma emission noise storm continua sources 
that ranged from intense and highly-polarized down to  
having intensities and polarization levels that could not be 
observed by previous instruments. 
Coronal holes were also found to dominate the polarization maps in the absence of 
intense noise storms, often exhibiting a characteristic ``bullseye" structure with one 
polarization sense surrounded by a ring of the opposite sense. 
Finally, the large-scale polarized component of thermal bremsstrahlung emission was 
mapped for the first time, and 
good agreement was found between the 
polarization structure and that of the line-of-sight magnetic field in a global potential field model 
at the lowest MWA frequencies. 

Some elaborations and suggestions for future work related to these conclusions are 
described in the following sections. 

\section{Type III Burst Source Motions}

Chapter~\ref{ch2} \citep{McCauley17} presented one example of motion exhibited by Type III bursts. 
The burst sources were observed to repetitively and rapidly split into two components 
at relatively low frequencies or simply elongate in two directions at higher frequencies. 
A model for this motion was developed based on a divergent quasi-separatrix layer 
magnetic field structure that was traced out by contemporaneous EUV jet observations: 
Simultaneously-accelerated electrons travel along neighboring field lines that are 
immediately adjacent to each other at the flare site but diverge with height, causing 
electrons to reach the requisite height 
to produce emission of a particular frequency at slightly different times.  
This produces an apparent motion that is nearly perpendicular to that of the actual electron beams. 

While this model is well-supported by the available data for that event, it is 
not clear how broadly-applicable it is to other events. 
Based on a cursory examination of around 50 bursts, splitting motions are relatively 
uncommon but significant motion in at least one direction is observed for around half of the bursts. 
Can these motions be entirely explained by divergent magnetic field structures in essentially the same way?
Such structures are known to be common, particularly in and around the active regions that 
generally produce Type III bursts. 
However, in at least one event (2015-08-25 03:15 UT), the direction of motion does not seem consistent with 
the magnetic field structure inferred from EUV observations and a potential field extrapolation.  
If a deeper investigation concludes that the \citet{McCauley17} model is not viable 
for that event, what else could produce such motion, given that it is very 
unlikely to correspond to the actual magnetic-field-aligned motion of the 
electron beam(s) moving through an iso-density layer? 

One possibility relates to scattering. 
As detailed Chapter~\ref{ch3} \citep{McCauley18}, scattering in the corona 
may not randomly modulate the ray paths because the density inhomogeneities      
responsible for scattering are not necessarily randomly oriented. 
Instead, they are often field-aligned high-density fibers that are generally close to radial. 
Scattering by these structures tends to guide emission outward to larger 
heights before it can escape unimpeded to the observer. 
This is analogous to classic ducting, where radiation is actually guided within a 
coherent low-density structure. 
Scattering may behave like a ``leaky duct," where burst radiation scatters 
outward but escapes at slightly different heights at slightly different times. 
Depending on the geometries, perhaps this could also generate an apparent motion. 

Another investigation using these data, currently being conducted by 
a collaborator, is to examine how the polarimetric structure of bursts 
differs between events that exhibit motion and 
events that do not (Rahman et al., in preparation, 2019). 
Preliminary results suggest that events with motion exhibit asymmetric 
Stokes $V$ profiles and that the leading edge in the direction of motion 
tends to be somewhat more highly polarized. 
Scattering seems likely to play some role in this behavior given that 
it is strongly suspected to depolarize burst emission. 
However, how this relates (if at all) to the mechanism(s) responsible for  
source motions is unclear. 

\section{Coronal Hole Peculiarities} 

There are multiple open questions surrounding the appearance of 
coronal holes in low-frequency radio observations. 
As is observed in soft X-ray and EUV data, coronal holes are naively 
expected to be darker than their surroundings in radio observations
because their lower densities imply lower emissivities. 
However, below around 120 MHz, many low-latitude coronal 
holes transition from being dark structures at higher frequencies 
to progressively brighter structures at lower frequencies. 
In fact, in the absence of intense noise storm emission, coronal holes may far 
exceed the brightness of nearby active regions.

This effect was first noted in a few isolated single-channel 
observations by \citet{Dulk74} and \citet{Lantos87}. 
The first observations to simultaneously observe a coronal hole 
on either side of this dark-to-bright transition were presented 
in Chapter~\ref{ch2} \citep{McCauley17}. 
This motivated a followup MWA study that characterized  
coronal hole intensities in more detail, including  
presenting multiple examples along with counterexamples  
that do not exhibit significant low-frequency 
enhancements \citep{Rahman19}. 
Following suggestions from \citet{Lantos87} and \citet{Alissandrakis94}, 
\citet{Rahman19} proposed a qualitative model in which emission generated 
outside a coronal hole is systematically redirected into it from 
an observer's perspective via refraction at the coronal hole boundary. 
However, no detailed quantitative modeling of this scenario has 
yet been done.

In Chapter~\ref{ch4} \citep{McCauley19}, low-latitude coronal holes were found to  
exhibit another peculiar effect. 
In circular polarization images, many coronal holes exhibit a ``bullseye" 
pattern with one polarization sense surrounded by a ring of the opposite sense. 
The central component does not match the sign expected from 
thermal bremsstrahlung, the presumed emission mechanism, given 
the orientation of the line-of-sight magnetic field. 
Two speculative ideas were proposed to explain this signature. 
The first is that the different refractive indices 
of the $o$- and $x$-modes causes the two modes to 
separate somewhat during propagation to the observer. 
If the refractive effect is large enough, this may lead to 
an unexpectedly large polarization fraction and/or a reversal 
in the polarization sense. 

The second idea is that the emission mechanism is not 
entirely thermal, instead including a small amount of 
plasma emission driven by the recently-discovered network 
jets that pervade coronal holes. 
This would explain the polarization sign of the central component, 
as plasma emission generally produces the opposite sign compared 
to bremsstrahlung for the same magnetic field orientation. 
However, given that refraction is already thought to be important in 
the total intensities of coronal holes and that the potential for 
plasma emission from network jets is uncertain, the latter 
idea was regarded as less plausible.  

To test the refraction idea, ray tracing simulations are needed. 
These would examine the propagation of the $o$- and $x$-modes 
through models of the density structure in and around a coronal hole to determine 
if refraction can produce the bullseye pattern in circular polarization 
and/or the low-frequency enhancements in total intensity.

\section{Probing the Large-Scale Magnetic Field} 
\label{field}

In Chapter~\ref{ch4} \citep{McCauley19}, the large-scale circular polarization structure 
at the lowest MWA frequencies was shown to be reasonably well-matched 
to the line-of-sight magnetic field structure inferred from a global 
potential field model at a height roughly corresponding to that of the radio limb. 
That is, the boundaries between opposite polarization signs were 
generally aligned with polarity inversion lines in the model, and 
the polarization sign was generally consistent with expectations 
assuming thermal bremsstrahlung emission. 
The same was not true at higher frequencies, which may be due to a 
number of effects that are outlined in Chapter~\ref{ch4}. 
Perhaps the most significant import of these data 
for the broader solar physics community is the potential capability to 
probe the coronal magnetic field at heights and scales 
that cannot be accessed by any other data type.  
This can be done by comparing the observations 
to forward models that predict the polarized intensities 
given models of the global plasma parameters. 

\begin{figure*}\graphicspath{{chapter5/}}
    \centering
    \includegraphics[width=0.9\textwidth]{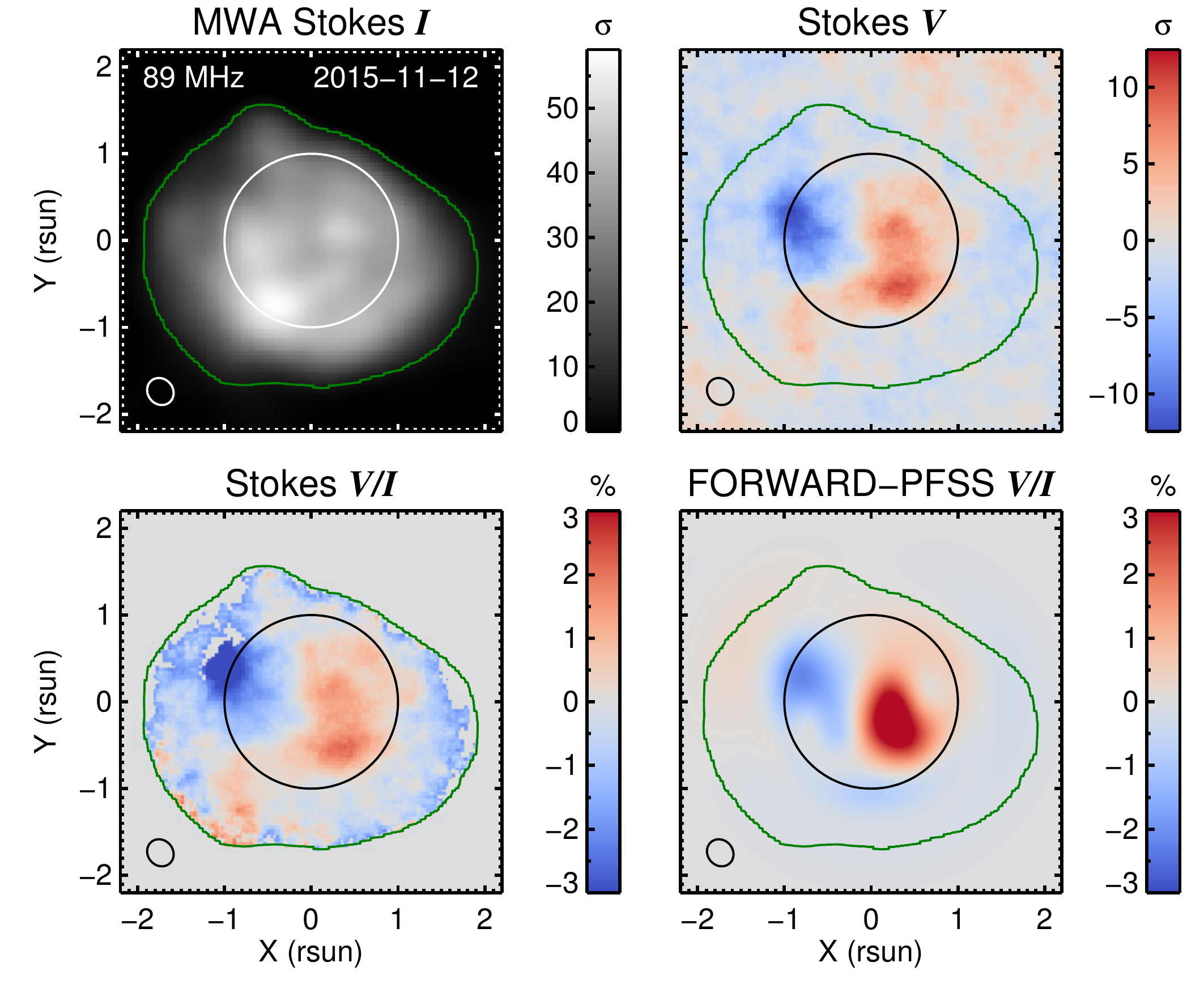}
    \caption{\footnotesize{}\textit{Top}: MWA Stokes $I$ and $V$ images with color bars 
    in units of signal-to-noise [$\sigma$]. 
    \textit{Bottom}: Observed and predicted polarization fractions [$V/I$].
    Solid circles represent the optical disk, the ellipses in the lower-left 
    corners represent the synthesized beam, and the green contours reflect 
    the 5-$\sigma$ level in Stokes $I$. 
    The synthetic image predicts $V/I$ from thermal bremsstrahlung emission 
    assuming a potential magnetic field model and a spherically-symmetric 
    hydrostatic atmosphere model.}
    \label{fig:forward_pol}
\end{figure*}

We\footnote{The preliminary results presented in Section~\ref{field} 
were produced in collaboration with Stephen White, Sarah Gibson, 
and Iver Cairns.}
have recently upgraded the \textsf{FORWARD} software 
suite \citep{Gibson16} to improve the radiative transfer computation for 
bremsstrahlung emission. 
Figure~\ref{fig:forward_pol} compares a synthetic image of 
the polarization fraction at 89 MHz to an MWA observation 
that was presented in Chapter~\ref{ch4}. 
There is a clear correspondence in the figure between the modeled 
and observed emission, both qualitatively and quantitatively.
The agreement is best near disk center, where propagation effects 
not considered by the model are likely to be least important.  
The synthetic image is calculated using the potential field 
model from Chapter~\ref{ch4} and an isothermal, 
spherically-symmetric, hydrostatic density model. 
The next steps are, first, to explicitly estimate the magnetic field 
strength implied by the observed polarization fractions 
at specific locations given assumed plasma parameters 
 and, second, to examine how uniformly varying 
the model field strength by a constant factor affects the 
quantitative agreement between the observed and synthetic 
images over some larger region. 
The same comparison can then be made for a more sophisticated coronal  
model, such as the global MHD models used in Chapters~\ref{ch2} and \ref{ch3}. 

A longterm goal for these data is to make 
regular estimates of the global magnetic field strength  
that can routinely be used to improve coronal magnetic field models. 
However, significant advances 
are needed to achieve this.  
First, the polarization maps are intrinsically affected 
by the coronal density and temperature, meaning 
that the reliability of magnetic field strength estimates depends on 
the reliability of the density and temperature models used.
Second, the polarization maps are influenced, likely significantly, 
by propagation effects (i.e. refraction, scattering, and mode coupling) 
that have never been adequately modeled in this context. 
Third, weak nonthermal emission may have a significant 
influence on the polarization fraction, particularly at 
higher frequencies, and assessing the 
contribution of possible nonthermal components may be 
challenging. 
Finally, the data quality must still be improved, specifically 
the polarimetric calibration and dynamic range. 
As described in Chapter~\ref{ch4}, the measured polarization fractions have 
sources of uncertainty that cannot yet be characterized or eliminated. 
Further, the dynamic ranges of the polarization images are often insufficient 
to detect the polarized component of thermal emission if a bright 
nonthermal source is present, which is often the case, although 
this may be overcome by adapting recently-developed 
high-dynamic-range calibration techniques for polarimetry \citep{Mondal19}. 

These are longterm challenges that will require the attention of many 
researchers before low-frequency polarimetry can be used to advance 
operational coronal magnetic field models. 
Chapter~\ref{ch4} and the future steps outlined in this section represent 
some of the early steps in this process. 

\begin{figure*}\graphicspath{{chapter5/}}
    \centering
    \includegraphics[width=1.0\textwidth]{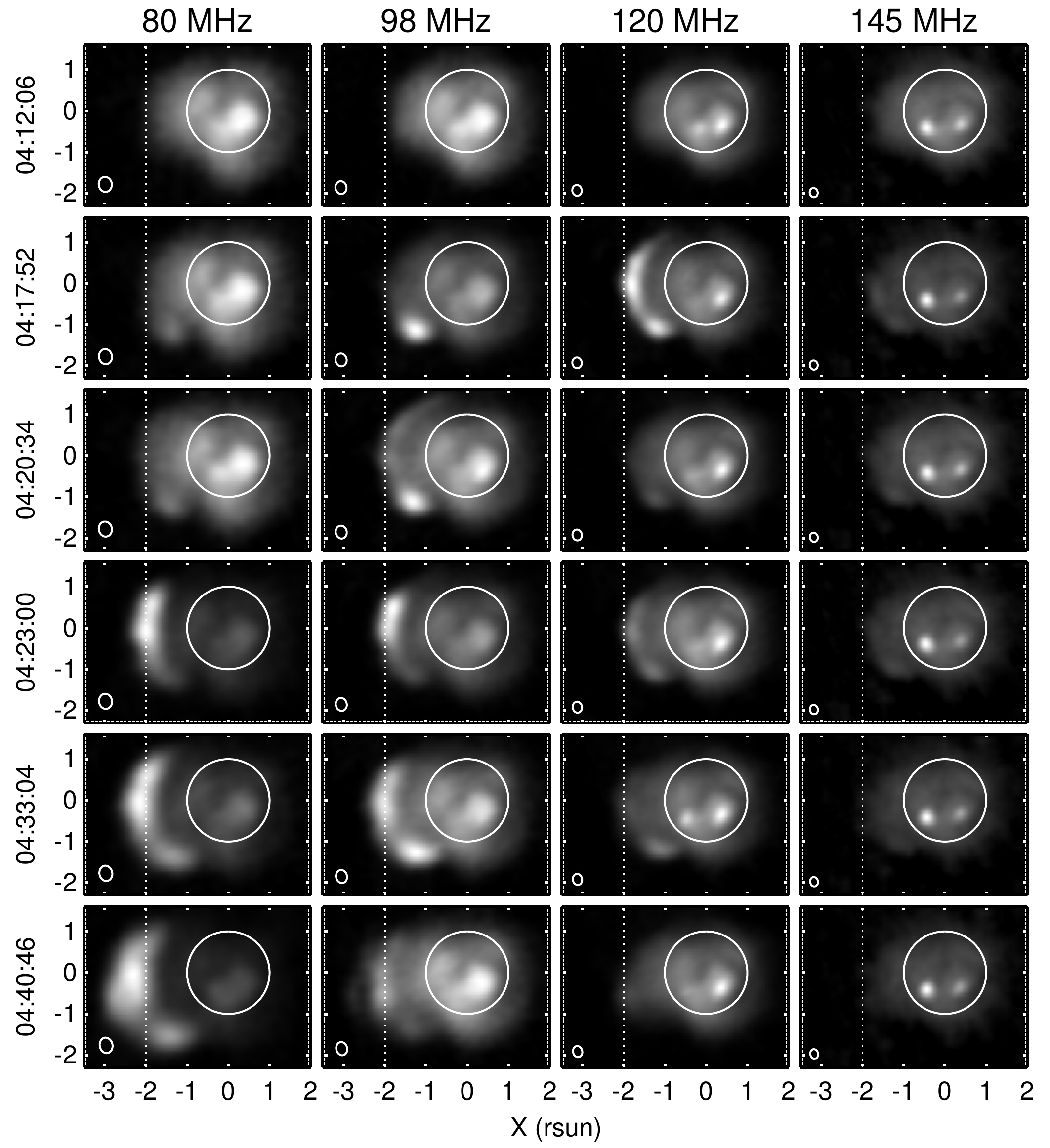}
    \caption{\footnotesize{}2014-09-26 radio CME evolution at four MWA frequencies (columns) 
    and six time steps (rows). Solid circles represent the optical disk and 
    ellipses in the lower-left corners represent the synthesized beam sizes. 
    A dotted line at $x =$ \rsolar{-2} is included to help illustrate the source motion.}
    \label{fig:cme1}
\end{figure*}

\section{Coronal Mass Ejections}
\label{cme}

While none of the research chapters in this thesis are directly-related to 
coronal mass ejections (CMEs), the image archive developed for this 
thesis produced at least one serendipitous CME observation.
This event was recently imaged over its full duration, and additional 
CME data are currently being reduced. 

\begin{figure*}\graphicspath{{chapter5/}}
    \centering
    \includegraphics[width=1.0\textwidth]{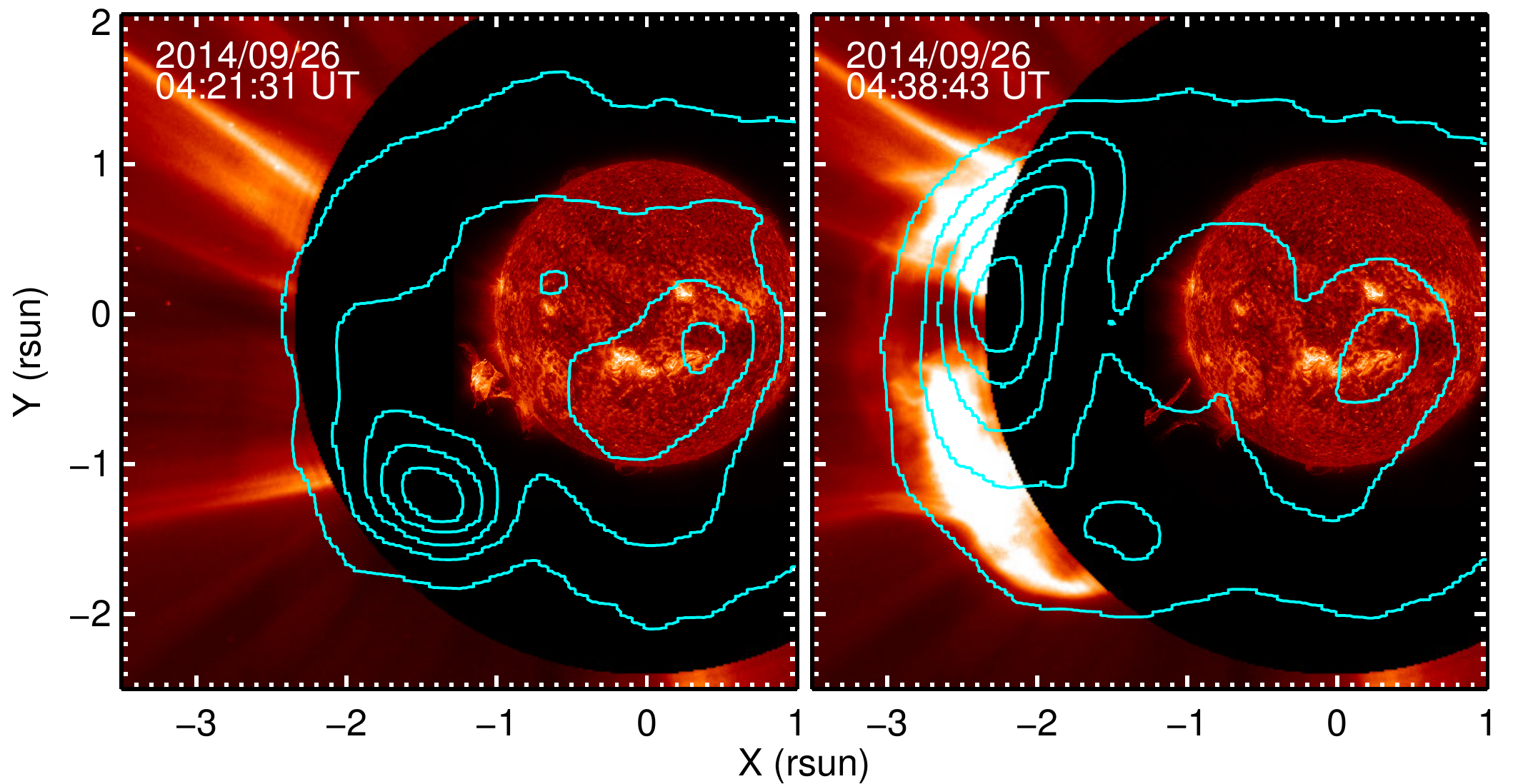}
    \caption{\footnotesize{}White light (LASCO C2; outer red), EUV (AIA 304 \AA{}; inner red) 
    and radio (MWA 80 MHz; cyan contours) CME observations. 
    The left panel shows a compact radio source ahead of the erupting filament in the southeast, 
    and the right panel shows an arc of radio emission across the white light CME front. }
    \label{fig:cme2}
\end{figure*}

Figure~\ref{fig:cme1} shows the time evolution of a CME structure 
seen at four MWA frequencies, and Figure~\ref{fig:cme2} overlays 
contours of the 80 MHz emission on EUV and white light observations. 
The event occurred on 26 September 2014 with a far-side prominence 
eruption that becomes visible in the SDO/AIA 304 \AA{} data around 4:05 UT. 
The left panel of Figure~\ref{fig:cme2} shows the development of a compact and highly-variable 
radio source just ahead of the erupting filament, and 
the right panel shows the white-light and radio structures shortly 
after the CME emerged beyond the occulting disk of the coronagraph. 

An arc of 80 MHz emission appears to trace out the CME front structure 
seen in white light. 
In Figure~\ref{fig:cme1}, this arc can be seen first developing 
at higher frequencies (lower heights) and then progressively forming 
at lower frequencies (larger heights) as the CME propagates. 
Specifically, at 04:17:52 UT, the arc is dominant at 120 MHz but later 
becomes brightest at 98 and 80 MHz near 04:20:34 and 04:23:00, respectively. 
Moreover, the arc is not stationary at each individual frequency, instead expanding 
outward at a speed of around 480 km s\tsp{-1} in the 80 MHz images. 
Importantly, the bandwidth of emission at a given time and location 
appears to be relatively broad; 
for instance, near ($x = -2$, $y = 0$) \rsolar{}, emission is enhanced between 80 and 120 MHz 
from roughly 04:23 to 04:40 UT. 
This will not be characterized here, but the relative intensity across the radio arc also 
fluctuates somewhat in time, with occasional compact bursts at its southern extent, 
and substructures within the radio and white-light sources appear to be correlated. 

To the best of our knowledge, intense radio emission that is aligned and morphologically 
similar to the CME front seen in white light, as in Figure~\ref{fig:cme2}, has not been observed before. 
Preliminary results from other MWA observations suggest that while 
compact emission associated with the CME core can routinely be observed given sufficient dynamic range, 
the bright arc structure from Figures~\ref{fig:cme1} and \ref{fig:cme2} is uncommon. 
Preliminary results from other events also suggest that the 
compact core emission is likely to be gyrosynchrotron emission from the central 
CME flux rope (Mondal et al., in preparation, 2019). 
We suspect that gyromagnetic emission is also the most likely mechanism 
for the arc structure, but no analysis has yet been done to determine this. 

As described in Section~\ref{ch1}, the primary low-frequency signatures 
of CMEs are Types II and IV radio bursts. 
There are very few reports of radio counterparts to the white light CME emission. 
The first example was presented by \citet{Bastian01} using the 
\textit{Nan\c{c}ay Radioheliograph} at frequencies above 164 MHz. 
These observations show emission primarily associated with the lower portions of 
the large CME loops, with a very slight signature from loop tops at the 
CME front.
\citet{Bastian01} attributes the radiation to synchrotron emission from electrons injected 
into the expanding magnetic fields of the CME. 

Our observations differ from those of \citet{Bastian01} in that the arc-shaped 
loop structure is most intense near the apex at the leading edge of the CME, with 
comparatively less emission toward the footpoints. 
If the emission mechanism is the same in our case, the different morphology 
may be related to the lower observation frequencies 
or differences in the location and nature of the electron acceleration process. 
The frequency of gyromagnetic emission depends on the magnetic field strength, 
with lower field strengths producing lower-frequency emission. 
As the magnetic field strength is weaker toward the apexes of the expanding 
CME loops, lower-frequency emission might be expected to be more intense there, though 
the lower portions of the loops may also simply be occulted given that 
the event originated on the far-side.
However, a detailed analysis of the MWA emission spectrum is first 
needed to determine the emission mechanism.

%% file: bibliography/bibliography.tex
\newcommand{\adv}{    {\it Adv. Space Res.}} 
\newcommand{\annG}{   {\it Ann. Geophys.}} 
\newcommand{\aap}{    {\it Astron. Astrophys.}}
\newcommand{\aaps}{   {\it Astron. Astrophys. Suppl.}}
\newcommand{\aapr}{   {\it Astron. Astrophys. Rev.}}
\newcommand{\ag}{     {\it Ann. Geophys.}}
\newcommand{\aj}{     {\it Astron. J.}} 
\newcommand{\apj}{    {\it Astrophys. J.}}
\newcommand{\apjl}{   {\it Astrophys. J. Lett.}}
\newcommand{\apjlett}{   {\it Astrophys. J. Lett.}}
\newcommand{\apss}{   {\it Astrophys. Space Sci.}} 
\newcommand{\cjaa}{   {\it Chin. J. Astron. Astrophys.}} 
\newcommand{\gafd}{   {\it Geophys. Astrophys. Fluid Dyn.}}
\newcommand{\grl}{    {\it Geophys. Res. Lett.}}
\newcommand{\ijga}{   {\it Int. J. Geomagn. Aeron.}}
\newcommand{\jastp}{  {\it J. Atmos. Solar-Terr. Phys.}} 
\newcommand{\jgr}{    {\it J. Geophys. Res.}}
\newcommand{\jgrsp}{    {\it J. Geophys. Res. (Space Phys.)}}
\newcommand{\mnras}{  {\it Mon. Not. Roy. Astron. Soc.}}
\newcommand{\nat}{    {\it Nature}}
\newcommand{\pasp}{   {\it Pub. Astron. Soc. Pac.}}
\newcommand{\pasj}{   {\it Pub. Astron. Soc. Japan}}
\newcommand{\pre}{    {\it Phys. Rev. E}}
\newcommand{\solphys}{{\it Solar Phys.}}
\newcommand{\sovast}{ {\it Soviet  Astron.}} 
\newcommand{\ssr}{    {\it Space Sci. Rev.}} 
\newcommand{\araa}{    {\it Ann. Rev. Astron. Astrophys.}} 
\newcommand{\pasa}{    {\it Pub. Astron. Soc. Austral.}} 
\newcommand{\apjs}{   {\it Astrophys. J. Supp.}}
\newcommand{\bain}{   {\it Bull. Astron. Inst. Netherlands}}
\newcommand{\aplett}{   {\it Astrophys. Lett.}}
\newcommand{\prl}{   {\it Phys. Rev. Lett.}}
\newcommand{\baas}{   {\it Bull. Amer. Astron. Soc.}}
\newcommand{\lrsp}{   {\it Living Rev. Sol. Phys.}}
\newcommand{\jpp}{   {\it J. Plasma Phys.}}
\newcommand{\pop}{   {\it Phys. Plasmas}}
\newcommand{\srep}{   {\it Sci. Rep.}}
\newcommand{\ptrsi}{   {\it Philos. Trans. Royal Soc.}}
\newcommand{\ptrs}{   {\it Philos. Trans. Royal Soc. A}}
\newcommand{\prsl}{   {\it Proc. Royal Soc. Lond.}}
\newcommand{\pieee}{   {\it Proc. IEEE}}
\newcommand{\ajp}{   {\it Aust. J. Phys.}}
\newcommand{\jaa}{   {\it J. Astrophys. Astron.}}
\newcommand{\raa}{   {\it Res. Astron. Astrophys.}}

\renewcommand\bibname{References}

\clearpage

\fancyhead[OR]{~}\fancyhead[EL]{\bf References}

\bibliographystyle{bibliography/mcagu08}
\refstepcounter{chapter}
\addcontentsline{toc}{chapter}{\bibname}
{\footnotesize{}\bibliography{bibliography/bibliography}}